\documentclass[manuscript=article, journal=jpcbfk, layout=twocolumn]{achemso}

\usepackage{achemso}
\usepackage[english]{babel}
\usepackage[utf8]{inputenc}
\usepackage[colorinlistoftodos, color=green!40, prependcaption]{todonotes}
\usepackage{listings}
\usepackage{amsthm}
\usepackage{bm}
\usepackage{mathtools}
\usepackage{physics}
\usepackage{xcolor}
\usepackage{graphicx}
\usepackage{adjustbox}
\usepackage{placeins}
\usepackage[T1]{fontenc}
\usepackage{lipsum}
\usepackage{csquotes}
\usepackage{amsfonts}

\usepackage[pdftex, pdftitle={Article}, pdfauthor={Author}]{hyperref} 

\usepackage[colorinlistoftodos]{todonotes}

\SectionNumbersOn

\newcommand{\bc}{\begin{center}}
\newcommand{\ec}{\end{center}}
\newcommand{\be}{\begin{equation}}
\newcommand{\ee}{\end{equation}}
\newcommand{\bea}{\begin{eqnarray}}
\newcommand{\eea}{\end{eqnarray}}

\newcommand{\vR}{{\bf R}}

\newcommand{\vecr}{\ensuremath{\mathbf{r}}}

\title[The CP2K Program Package]{The CP2K Program Package Made Simple}

\author{Marcella Iannuzzi}
\affiliation{Department of Chemistry, University of Zurich, Switzerland}
\author{Jan Wilhelm}
\affiliation{Regensburg Center for Ultrafast Nanoscopy (RUN) and Institute of Theoretical Physics, University of Regensburg, Germany}
\author{Frederick Stein}
\affiliation{Center for Advanced Systems Understanding (CASUS), Helmholtz Zentrum Dresden-Rossendorf, Germany}
\author{Augustin Bussy}
\affiliation{Swiss National Supercomputing Centre (CSCS), ETH Zurich, Lugano, Switzerland}
\author{Hossam Elgabarty}
\affiliation{Department of Chemistry, Paderborn University, Germany}
\author{Dorothea Golze}
\affiliation{Department of Chemistry and Food Chemistry, Technische Universit\"at Dresden, Germany}
\author{Anna Hehn}
\affiliation{Department of Chemistry, Christian-Albrechts-University Kiel, Germany}
\author{Maximilian Graml}
\affiliation{Regensburg Center for Ultrafast Nanoscopy (RUN) and Institute of Theoretical Physics, University of Regensburg, Germany}
\author{Stepan Marek}
\affiliation{Regensburg Center for Ultrafast Nanoscopy (RUN) and Institute of Theoretical Physics, University of Regensburg, Germany}
\author{Beliz Sertcan G\"okmen}
\affiliation{Department of Chemistry, University of Zurich, Switzerland}
\author{Christoph Schran}
\affiliation{Cavendish Laboratory, Cambridge University, United Kingdom}
\author{Harald Forbert}
\affiliation{Lehrstuhl für Theoretische Chemie, Ruhr-Universität Bochum, Germany}
\author{Rustam Z. Khaliullin}
\affiliation{Department of Chemistry, McGill University, Canada}
\author{Anton Kozhevnikov}
\affiliation{Swiss National Supercomputing Centre (CSCS), ETH Zurich, Lugano, Switzerland}
\author{Mathieu Taillefumier}
\affiliation{Swiss National Supercomputing Centre (CSCS), ETH Zurich, Lugano, Switzerland}
\author{Rocco Meli}
\affiliation{Swiss National Supercomputing Centre (CSCS), ETH Zurich, Lugano, Switzerland}
\author{Vladimir Rybkin} 
\affiliation{HQS Quantum Simulations GmbH, Karlsruhe, Germany}
\author{Martin Brehm} 
\affiliation{Department of Chemistry, Paderborn University, Germany}
\author{Robert Schade}
\affiliation{Paderborn Center for Parallel Computing (PC2), Paderborn University, Germany}
\author{Ole Schütt} 
\affiliation{CP2K Foundation, Zurich, Switzerland}
\author{Johann V. Pototschnig}
\affiliation{Center for Advanced Systems Understanding (CASUS), Helmholtz Zentrum Dresden-Rossendorf, Germany}
\author{Hossein Mirhosseini}
\affiliation{Center for Advanced Systems Understanding (CASUS), Helmholtz Zentrum Dresden-Rossendorf, Germany}
\author{Andreas Kn\"upfer}
\affiliation{Center for Advanced Systems Understanding (CASUS), Helmholtz Zentrum Dresden-Rossendorf, Germany}
\author{Dominik Marx}
\affiliation{Lehrstuhl für Theoretische Chemie, Ruhr-Universität Bochum, Germany}
\author{Matthias Krack}
\affiliation{PSI Center for Scientific Computing, Theory and Data, Villigen, Switzerland}
\author{J\"urg Hutter}
\affiliation{Department of Chemistry, University of Zurich, Switzerland}
\author{Thomas D. K\"uhne}
\affiliation{Center for Advanced Systems Understanding (CASUS), Helmholtz Zentrum Dresden-Rossendorf, Germany}
\alsoaffiliation{Institute of Artificial Intelligence, Technische Universit\"at Dresden, Germany}
\email{tkuehne@cp2k.org}

\date{\today} 
\keywords{simulation package, electronic structure calculations, molecular dynamics}

\begin{document}

\begin{abstract}
CP2K is a versatile open-source software package for simulations across a wide range of atomistic systems, from isolated molecules in the gas phase to low-dimensional functional materials and interfaces, as well as highly symmetric crystalline solids, disordered amorphous glasses, and weakly interacting soft-matter systems in the liquid state and in solution. This review highlights CP2K's capabilities for computing both static and dynamical properties using quantum-mechanical and classical simulation methods. In contrast to the accompanying theory and code paper [J. Chem. Phys. \textbf{152}, 194103 (2020)], the focus here is on the practical usage and applications of CP2K, with underlying theoretical concepts introduced only as needed.
\end{abstract}

\maketitle

\section{Introduction} \label{sec:Introduction}

The aim of CP2K is to predict average properties, such as those that arise in statistical mechanics and thermodynamics, of any substance made of interacting electrons and nuclei from first-principles. Although with the advent of quantum mechanics this became possible by solving the non-relativistic many-body Schr\"odinger equation, early on Dirac recognized \textit{that the exact application of these laws leads to equations much too complicated to be soluble}~\cite{Dirac1929}. 
Yet, assuming the so-called Born-Oppenheimer approximation, 
which is underlying most of the methods within CP2K, the full many-body Schr\"odinger equation can be separated into its electronic and nuclear counterparts. 

Consequently, the general structure of CP2K entails a wide variety of different classical and quantum mechanical energy and force methods that can be arbitrarily combined with geometry optimization and transition state search techniques~\cite{Billeter2003, Schlegel2011}, as well as (rare event) sampling approaches, such as Monte Carlo (MC), Meta- and molecular dynamics (MD)~\cite{Metropolis1953, laio2002escaping, PhysRev.136.A405}, to name just a few. The latter approaches dealing with the nuclear motion are specified in the \texttt{\&MOTION} section, whereas the former force methods, which are due to the electronic structure of matter, are detailed in the \texttt{\&FORCE\_EVAL} section.  However, since we have elected to limit ourselves to computational schemes that make CP2K unique, the techniques of the \texttt{\&MOTION} section, which are well described in various textbooks~\cite{Tuckerman2010, Frenkel2023}, will be mostly neglected. The same also applies to \textit{ab-initio} MD (AIMD)~\cite{Marx2009, Hutter2011}, in particular the second-generation Car-Parrinello method~\cite{Kuehne2007, Prodan2018}, which has been extensively covered elsewhere~\cite{Kuehne2014, Kuehne2020, hutter2024ab}.

Due to the user-focused nature of the present review, the paper is organized by the property to be calculated instead of the computational methods used, which will be introduced in the relevant sections where necessary. Hence, first how to compute the total energy and nuclear forces of a system using density functional theory (DFT)~\cite{RevModPhys.71.1253} and quantum chemical models~\cite{RevModPhys.71.1267} is presented in section~\ref{sec:TotalEnergy}. Thereafter, the calculation of the electronic band structure using the pseudopotential plane wave (PP-PW)~\cite{Ihm1979}, projector augmented-wave (PAW)~\cite{Bloechl1994}, full-potential linearized augmented plane wave (FP-LAPW)~\cite{PhysRevB.12.3060}, as well as single-particle Green's function and the screened Coulomb interaction (\textit{GW}) methods~\cite{Gunnarsson1998}, is detailed in section~\ref{sec:BandStructure}. Embedding schemes such as implicit solvation and mixed classical-quantum mechanical approaches, as well as purely quantum mechanical embedding theories, are the content of section~\ref{sec:Embedding}. Nuclear magnetic resonance (NMR) and electron paramagnetic resonance (EPR) spectroscopies are described in section~\ref{sec:NMR+EPR}, whereas optical spectroscopy methods such as linear-response time-dependent DFT (LR-TDDFT) and the Bethe-Salpeter equation (BSE) are the subject matter of section~\ref{sec:OpticalSpectroscopy}. Section~\ref{sec:ExcitedStateDynamics} is dedicated to excited state dynamics in terms of real-time TDDFT (RT-TDDFT), Ehrenfest dynamics and real-time BSE (RT-BSE), whereas X-ray absorption (XAS) and X-ray emission (XES) spectroscopies are described in section~\ref{sec:XraySpectroscopy}. Energy decomposition analysis (EDA) based on absolutely localized molecular orbitals (ALMO) for condensed phase systems is the focus of section~\ref{sec:ALMO-EDA}. Finite temperature effects using machine learning potentials (MLPs) including nuclear quantum effects (NQEs) by means of path-integral MD (PIMD) simulations are presented in section~\ref{sec:FiniteTemperature}, before concluding the paper on the extended topic of vibrational spectroscopy.  
\section{Total Energy and Force Methods} \label{sec:TotalEnergy}

Even though total energies and atomic forces are rarely relevant observables for themselves, they are the basis for the theoretical computation of all properties considered here. In the spirit of CP2K, to simulate all possible kinds of matter from isolated molecules to the extended condensed phase, all energy and force methods can not only be employed \textit{in vacuo} (0D), but also in the presence of arbitrary periodic boundary conditions (PBC), i.e. 1D (chains and nanotubes), 2D (surfaces and interfaces) and 3D (solids and liquids). Moreover, to explicitly include finite temperature and pressure effects by means of dynamical sampling methods, analytical gradients are generally available for all classical and QM electronic structure methods, even in combination with PBCs. 

This is made possible by the \textsc{Quickstep} module~\cite{Krack2004, VandeVondele2005, Kuehne2020}, which is based on the Gaussian and plane wave (GPW) method and its all-electron counterpart denoted as the Gaussian and augmented plane wave (GAPW) approach~\cite{Lippert1997, Lippert1999}, and contains all electronic structure methods described in Section~\ref{sec:TotalEnergy}, i.e. in particular DFT and Hartree-Fock (HF)~\cite{slater1951simplification}, but also hybrid-DFT and post-HF schemes such as second-order M{\o}ller–Plesset perturbation theory (MP2)~\cite{moller1934note} and the random phase approximation (RPA)~\cite{bohm1953collective, gell1957correlation}.

\subsection{Density Functional Theory}

The unique aspect of the GPW and GAPW methods is their usage of a mixed Gaussian and an auxiliary plane wave (PW) basis set, thereby unifying the computational efficiency of a compact localized basis set with the simplicity of PWs for periodic electronic structure calculations. 
Within CP2K/\textsc{Quickstep}, the Kohn-Sham (KS) orbitals $\phi_i(\mathbf{r})$ are represented by atom-centered Gaussian functions, 
whereas the electron density is expanded in PWs on a regular grid. This is to say that the molecular orbitals (MO) are represented by the common linear combination of atomic orbitals (LCAO)
\begin{equation}
  \phi_i(\mathbf{r}) = \sum_j c_{ji} \, \varphi_j(\mathbf{r}), 
\end{equation}
where $c_{ji}$ are the so-called orbital coefficients, and the atomic orbitals (AO) 
\begin{equation}
  \varphi_j(\mathbf{r}) = \sum_k d_{kj} \, g_k(\mathbf{r})
\end{equation}
are themselves expanded in terms of primitive Gaussian functions $g_k(\mathbf{r})$ and their corresponding contraction coefficients $d_{kj}$. 

The electron density $\rho(\mathbf{r})$ can either be written as 
\begin{equation}
  \rho(\mathbf{r}) = \sum_{\mu \nu} P^{\mu \nu} \varphi_{\mu}(\mathbf{r}) \, \varphi_{\nu}(\mathbf{r}), 
\end{equation}
or alternatively in an auxiliary PW basis as 
\begin{equation}
  \tilde{\rho}(\mathbf{r}) = \frac{1}{\Omega} \frac{}{} \sum_{\mathbf{G}} \rho(\mathbf{G}) \, e^{i \mathbf{G} \cdot \mathbf{r}}, 
\end{equation}
where $\Omega$ is the volume of the unit cell and the expansion coefficients $\rho(\mathbf{G})$ are such that $\rho(\mathbf{r}) = \tilde{\rho}(\mathbf{r})$. 

Exploiting the efficiency of the fast Fourier transform (FFT), the KS matrix construction including the periodic long-range electrostatics can be performed in $\mathcal{O}(M \log M)$ with $M$ being the number of basis functions, via 
\begin{eqnarray}
  \mathbf{P} \, \rightarrow \, \rho(\mathbf{r}) \:\, &\!\!\!\! \xrightarrow{\mbox{FFT}}
  \rho(\mathbf{G}) \rightarrow &4\pi\frac{\rho(\mathbf{G})}{G^2} \nonumber \\
  &&\qquad \parallel \\ 
  \quad\;\;\;\mathbf{V} \leftarrow V_{\rm H}(\mathbf{r}) &\xleftarrow{\mbox{FFT$^{-1}$}}
  &V_{\rm H}(\mathbf{G}) \nonumber
  \label{HartreeFFT}
\end{eqnarray}
$V_{\rm H}(\mathbf{G})$ is the Hartree potential that is the solution of the Poisson equation $\nabla^2 V_{\rm H}(\mathbf{G}) = -4\pi \rho(\mathbf{r})$. By substituting $\rho(\mathbf{r})$ with the total charge density, which consists of the electron density plus Gaussian-smeared nuclear charges, all long-range 
electrostatic energy terms can be coalesced into a single Hartree-like expression that has to be augmented by compensating overlap and self-energy terms generated by these Gaussian distributions. Similarly, the exchange and correlation (XC) functional is evaluated on the very same uniform density grid as the Hartree energy, so that starting from $\rho(\mathbf{r})$, the combined potential for the long-range electrostatic and XC energies can be computed using FFT techniques~\cite{Lippert1997, Lippert1999, Kuehne2020}. 

\subsubsection{Basis Set Convergence using the \textsc{Quickstep} Method} \label{BS_section}

This is why in a typical \textsc{Quickstep} calculation, which is activated via \texttt{\&FORCE\_EVAL\%METHOD QUICKSTEP}, the basis has to be defined in two sections. On the one hand, the localized Gaussian basis set is defined in the \texttt{\&FORCE\_EVAL\%SUBSYS\%KIND} section and, on the other hand, the PW density cutoff in the \texttt{\&FORCE\_EVAL\%DFT\%MGRID} section. It is important to recognize, however, that the latter is a density cutoff and therefore 4 times higher than a comparable energy cutoff typically specified in PW codes, which is a manifestation of Nyquist's sampling theorem together with the fact that the electron density is the squared modulus of the wavefunction (WF). Also, contrary to conventional PW codes, the complete basis set limit cannot be reached just by increasing the density cutoff of the auxiliary PW basis only. Instead, by increasing \texttt{CUTOFF} the limit of the finite Gaussian set is approached, which is why the convergence of the auxiliary PW basis depends on the underlying Gaussian basis and both have to be balanced and increased concurrently to yield the complete basis set limit. 

An important aspect of the 
\textsc{Quickstep} method is the usage of real-space integration grids to represent the electron density and product Gaussian functions in conjunction with multigrid techniques, so that wide and smooth Gaussian functions are mapped onto a coarser grid than narrow and sharp Gaussians. However, the electron density is always mapped onto the finest grid. Hence, choosing a fine enough integration grid is crucial for any calculation to obtain highly accurate results as efficiently as possible. All settings related to multigrids are controlled within the \texttt{\&MGRID} subsection. The number of multigrid levels is defined by \texttt{NGRIDS}, with 5 typically being a suitable value for most applications. As alluded to above, the keyword \texttt{CUTOFF} defines the PW density cutoff (in Rydbergs, Ry), the corresponding cutoffs for subsequent grid levels (from finer to coarser) are defined via 
\begin{equation}
  E^i_{\rm cut} = \frac{E^1_{\rm cut}}{\alpha^{i-1}},
\end{equation}
where $i$ is the corresponding multigrid level and $\alpha$ the progression factor controlled by \texttt{PROGRESSION\_FACTOR}. Hence, the higher the value of \texttt{CUTOFF}, the finer the grids of all multigrid levels. To determine which product Gaussians are mapped onto which multigrid level, \texttt{REL\_CUTOFF} defines the PW cutoff of a reference grid covered by a Gaussian with unit standard deviation, i.e. $\exp(|\mathbf{r}|^2)$. Therewith, a Gaussian is mapped onto the coarsest level of the multigrid, on which the function will require a number of grid points greater than or equal to the number of grid points $\exp(|\mathbf{r}|^2)$ will cover on a reference grid defined by \texttt{REL\_CUTOFF}. In this way, CP2K tries to map each Gaussian onto a grid such that the number of grid points covered by the Gaussian, no matter how wide or narrow, is roughly the same. Hence, the two most important keywords affecting real-space integration grids and, as such, the convergence of the auxiliary PW basis are \texttt{CUTOFF} and \texttt{REL\_CUTOFF}, respectively. A prototypical CP2K input section, using the GPW method, reads as follows:
\begin{verbatim}
&FORCE_EVAL
  METHOD QUICKSTEP
  &DFT
    &QS
      METHOD GPW
    &END QS
    &MGRID
      CUTOFF 400
      REL_CUTOFF 50
      NGRIDS 5
      PROGRESSION_FACTOR 3.0
    &END MGRID
  &END DFT
&END FORCE_EVAL
\end{verbatim}

\subsubsection{Pseudopotentials and Basis Sets}
\label{PP+BS}
Besides the auxiliary PW basis, the primary Gaussian basis set and optionally also the employed pseudopotentials (PP) are essential for the accuracy of any \textsc{Quickstep} calculation. 
For that purpose, in recent years the so-called ``UZH protocol'' has been developed that contains both molecularly optimized Gaussian basis sets for all-electron and PP calculations~\cite{VandeVondele2007b}, as well as corresponding separable dual-space Goedecker-Teter-Hutter (GTH) PPs that are highly transferable and norm-conserving~\cite{Goedecker1996, Hartwigsen1998, Krack2005}. 

In addition to accelerating the calculation of the electronic structure by reducing the number of electrons and the basis set size, the usage of PPs also allows for the inclusion of relativistic effects. The separable dual-space GTH PPs of the ``UZH protocol'' are fully non-local with Gaussian-type projectors because of their simplicity and efficiency due to the usage of analytic integrals and FFTs. The local part, however, is represented by a Gaussian form in real-space, which can be well combined with Gaussian basis sets. Hence, the fully analytical form of the GTH PPs requires only a small set of parameters for each element that is optimized
with respect to an atomic all-electron WF of scalar relativistic DFT reference calculations~\cite{reiher2012relativistic}, as obtained by the CP2K \texttt{ATOM} code. 
The ``UZH protocol'' contains GTH PPs for the periodic table up to Rn for generalized gradient approximation (GGA) (PBE~\cite{Perdew1996}),
meta-GGA (SCAN~\cite{sun2015strongly}), and hybrid functionals (PBE0~\cite{adamo1999toward}).
Special purpose GTH PPs for lanthanides and actinides~\cite{lu2019norm, lu2021norm, lu2022norm, lu2024norm}, as well as including nonlinear core corrections and spin-orbit coupling parameters, are also available at \url{https://github.com/cp2k/cp2k/tree/master/data}~\cite{willand2013norm}. In addition, via the libgrpp library to evaluate molecular integrals over Gaussian functions~\cite{oleynichenko2023libgrpp}, most effective core potentials (ECP) provided by the EMSL Basis Set Exchange can be employed~\cite{Pritchard2019}.

The basic idea of the molecular-optimized MOLOPT basis set is to use generally contracted Gaussian basis sets, including diffuse primitives, fully optimized on molecular calculations~\cite{VandeVondele2007b}. The absence of lone diffuse functions ensures a low condition number of the overlap matrix, leading to better self-consistent field (SCF) convergence and a sparser density matrix, whereas the inclusion of diffuse primitive functions entails a particularly small basis set superposition error. By molecularly optimizing the basis functions, small but accurate basis sets are obtained, suitable for large-scale isolated gas and condensed phase calculations. As with PPs, in addition to the standard general purpose ``UZH protocol'', more specialized MOLOPT basis sets for lanthanides and actinides~\cite{lu2019norm, lu2021norm, lu2022norm, lu2024norm}, solids~\cite{zijlstra2008optimized, peintinger2013consistent, vilela2019bsse, laun2021bsse, laun2022bsse}, and for the later described auxiliary density matrix method (ADMM)~\cite{Guidon2010}, \textit{GW}~\cite{HedinGW,Aryasetiawan1998}, and resolution of the identity approaches (RI)~\cite{whitten1973coulombic, dunlap1979some}, such as RI-MP2 and RI-RPA, are also available in the same repository. In addition, the ``UZH protocol'' also contains all-electron MOLOPT basis sets, although all other basis sets that can be downloaded at \url{http://basissetexchange.org} in CP2K format are also supported~\cite{Pritchard2019}. 

A typical section for gold as an example looks as follows:
\begin{verbatim}
&FORCE_EVAL
  &DFT
    BASIS_SET_FILE_NAME BASIS_MOLOPT_UZH
    POTENTIAL_FILE_NAME POTENTIAL_UZH
  &END DFT
  &SUBSYS
    &KIND Au
      BASIS_SET TZV2P-MOLOPT-GGA-GTH-q19
      POTENTIAL GTH-GGA-q19
    &END KIND
  &SUBSYS
&END FORCE_EVAL
\end{verbatim}

The accuracy of the ``UZH protocol'' can best be assessed by means of all-electron FP-LAPW calculations at the complete basis set limit using the CP2K-integrated PW code \textsc{SIRIUS}, which is described in Section~\ref{sec:SIRIUS}. 
\begin{figure}
\includegraphics[width=.98\linewidth]{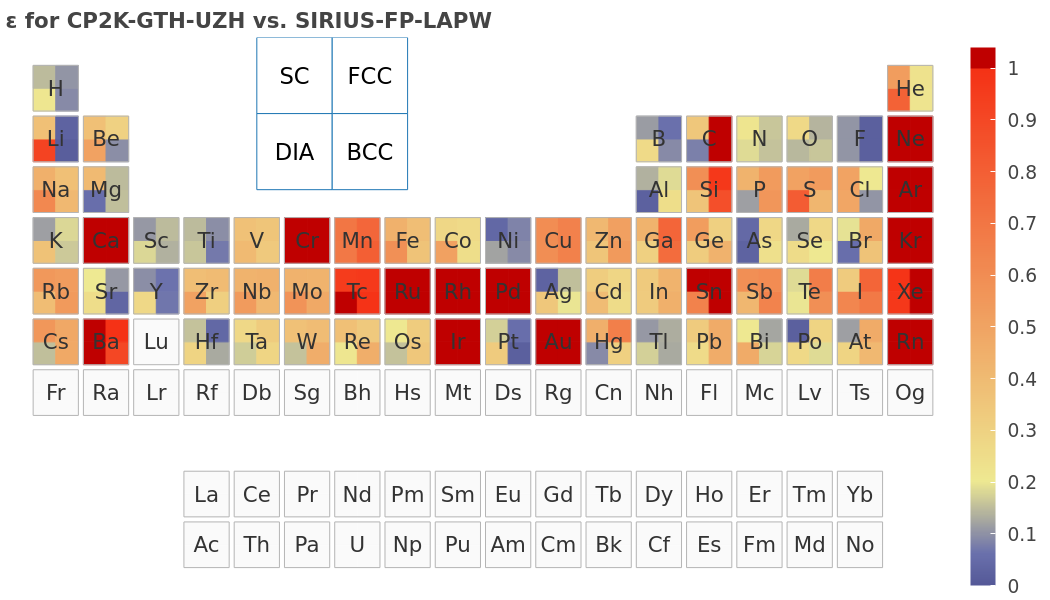}
\caption{\label{fig:comparison_1} 
Values of the comparison metric $\varepsilon$ for CP2K/\textsc{Quickstep} using the ``UZH protocol'' with respect to
all-electron FP-LAPW calculations using the CP2K/\textsc{SIRIUS} code. For each element up to Rn four mono-elemental cubic crystals were considered~\cite{bosoni2024verify}. 
}
\end{figure}
The deviation between \textsc{SIRIUS} and \textsc{Quickstep} employing a TZV2P MOLOPT basis set and corresponding GTH PPs is shown in Fig.~\ref{fig:comparison_1} for a large number of different unary crystal structures and quantified in terms of the previously suggested metric 
\begin{equation}
\varepsilon(a,b) = \sqrt{
    \frac{  \sum_i [E_{a}(V_i) - E_{b}(V_i)]^2}
    {\sqrt{ \sum_i [E_{a}(V_i) - \langle E_{a} \rangle]^2 ~ \sum_i [ E_{b}(V_i) - \langle E_{b} \rangle]^2}}},
    \label{eq:epsilin_discrete}
\end{equation}
where the index $i$ runs over multiple calculations of $E(V)$ within a given space group~\cite{bosoni2024verify}. Since this quantity is insensitive to the magnitude of the bulk modulus, it provides a uniform metric across a large variety of structural and chemical environments.

\subsubsection{All-Electron Calculations using the GAPW Method}
All-electron calculations with CP2K/\textsc{Quickstep} necessitate the use of the GAPW method~\cite{Lippert1999, Krack2000}, because a description of the core electrons with PWs becomes rapidly computationally prohibitive. Likewise, small-core and sometimes even medium-core PPs can only be used with GAPW efficiently. There are various possibilities to devise a PP for heavy elements such as uranium with 92 electrons:
\begin{itemize}
    \item Large-core PP ([86 core] 6 valence electrons)\\
          $\lbrack\text{Rn}\rbrack\,7s^2\,5f^3\,6d^1$   
    \item Medium-core PP ([78 core] 14 valence electrons)\\
          $\lbrack\text{Xe}\;4f^{14}\,5d^{10}\rbrack\,6s^2\,6p^6\,7s^2\,5f^3\,6d^1$
    \item Small-core PP ([60 core] 32 valence electrons)\\
          $\lbrack\text{Kr}\;4d^{10}\,4f^{14}\rbrack\,5s^2\,5p^6\,5d^{10}\,6s^2\,6p^6\,7s^2\,5f^3\,6d^1$
\end{itemize}
The large-core PP for uranium is known to be inaccurate due to the non-negligible overlap of the outer valence orbitals with the semi-core $6s$ and $6p$ orbitals, which does not allow keeping these semi-core orbitals as frozen. Medium-core PPs for uranium are mostly used by PW codes~\cite{Dorado2009} and with GPW~\cite{Krack2015}, because these PPs are sufficiently accurate and the semi-core $6s$ and $6p$ orbitals are not too hard for an efficient expansion by PWs.
However, this is not possible for small-core uranium PPs that require GAPW.

Furthermore, the calculation of several properties, especially spectroscopic properties involving core electrons, requires the explicit inclusion of the core electrons~\cite{Krack2002, muller2019impact}.
The GAPW method must be activated explicitly in the \texttt{\&DFT\%QS} section of the input, i.e.
\begin{verbatim}
&QS
  METHOD GAPW
  EPSFIT 1.0E-6
  EPSRHO0 1.0E-8
  EPSSVD 1.0E-12
&END QS
\end{verbatim}
The GAPW specific parameters \texttt{EPSFIT}, \texttt{EPSRHO0} and \texttt{EPSSVD} are optional and their default values are $10^{-4}$, $10^{-6}$ and $10^{-8}$, respectively. These parameters are used to control the accuracy of a GAPW calculation as indicated by the electron count printed after SCF convergence has been achieved. \texttt{EPSFIT} controls the split of Gaussian basis sets into a hard and a soft basis set for each atomic kind in the system. Smaller values for \texttt{EPSFIT} promote the inclusion of larger Gaussian exponents in the soft basis. Only the part of the electron density defined by the soft basis set is expanded in PWs, whereas the hard basis set describes the hard (frozen) part of the electron density close to the nuclei. A GAPW setup seamlessly turns into a GPW one when all exponents are included in the soft basis sets and all hard basis sets are ``empty''. In this case, the default \texttt{METHOD GPW} can be used, since GAPW is not required. Gaussian exponents up to 5 should always be included in the soft basis set. Check the values in the first column of the Gaussian basis set definition or the basis set printout in the CP2K output for \texttt{PRINT\_LEVEL MEDIUM} or higher. The electron count should show at least an accuracy of \texttt{1.0E-6}. For improved accuracy, Gaussian exponents in the range 5 to 10 can be added to the soft basis by lowering the value of \texttt{EPSFIT}. The inclusion of larger exponents should always be accompanied by an adequate increase of the density \texttt{CUTOFF} in \texttt{\&MGRID}. \texttt{EPSRHO0} and \texttt{EPSSVD} improve the numerical precision of $V(\rho_0-\rho_0^\text{soft})$ and the singular value decomposition (SVD) of the PAW projector matrix, respectively. In this way, an electron count of \texttt{1.0E-8} or even less can be achieved. Gaussian exponents up to about 20 can still be expanded in PWs using large \texttt{CUTOFF} values, but the expansion of larger exponents quickly becomes computationally prohibitive or even impossible.\\
A few atomic kind-specific settings complete a GAPW all-electron setup:
\begin{verbatim}
&KIND O
  BASIS_SET ORB TZVPP-MOLOPT-GGA-ae
  POTENTIAL ALL
  LEBEDEV_GRID 400
  RADIAL_GRID 100
&END KIND
\end{verbatim}
The atomic \texttt{POTENTIAL} must be set to \texttt{ALL} and an all-electron \texttt{ORB}ital basis set has to be selected. 
Note that most of these all-electron basis sets are made for isolated (0D) systems such as molecules or clusters in the gas phase, but there are also a few all-electron basis sets suited for condensed phase calculations with CP2K/\textsc{Quickstep}, as described in Section~\ref{PP+BS}. For heavier elements, ECPs are employed that require GAPW like small core GTH PPs, since an all-electron description for condensed phase systems with heavy elements quickly becomes computationally intensive. There are usually Gaussian basis sets specifically optimized for such ECPs.

Optionally, the size of the atom-centered spherical Lebedev grids can be increased to improve the accuracy of the GAPW electron count for atomic densities and, therefore, also for total energies and nuclear forces. The number of \texttt{RADIAL\_GRIDS} can be increased from the default value 50 to 100 or more, whereas only spherical Lebedev grids with certain sizes exist. The atomic grids are automatically generated in an adaptive manner~\cite{Krack1998}. The default value for the maximum \texttt{LEBEDEV\_GRID} is 50 and thus is rather small. The following Lebedev grids are implemented in CP2K: 6, 14, 26, 38, 50, 86, 110, 146, 194, 302, 434, 590, 770, 974. The grids close to and far from the nuclei are small because the electron density is rather spherical at these distances, while the selected maximum grid is employed in the valence regions. \texttt{LEBEDEV\_GRID} 400 limits the maximum employed Lebedev grid for an atomic kind to 302.

\subsubsection{Wavefunction Optimization} \label{WFopt}

Since the Hamiltonian of effective single-particle theories, such as DFT, HF, semi-empirical quantum chemistry (SQC) and tight-binding (TB), depends on the electron density, i.e. on their own solution, the corresponding equations have to be solved iteratively using the SCF approach. The ability to converge this SCF cycle and to achieve self-consistency depends on multiple factors such as the band gap and dimensionality of the system, as well as the locality of the employed basis functions. 

CP2K/\textsc{Quickstep} provides two general techniques to yield self-consistency: traditional diagonalization and the orbital transformation (OT) method~\cite{VandeVondele2003}. The former approach entails the direct diagonalization of the Hamiltonian followed by a mixing strategy, such as the Kerker, Pulay, multisecant, and Broyden techniques~\cite{kerker1981efficient, Pulay1980, marks2008robust, Broyden1965}. In these damped preconditioned fixed-point iteration schemes, the electron density of the next step is constructed in terms of a linear combination of previous densities, in order to achieve a robust SCF convergence with the fewest number of steps. For hardly converging systems, such as metallic solids for which finite electron temperature smearing and ${\mathbf k}$-point sampling may also be required, Broyden mixing is typically the best choice, and the associated settings in the following input snippet are a good starting point: 
\begin{verbatim}
&QS
  EPS_DEFAULT 1.0E-12
&END QS
&SCF
  MAX_SCF 100
  EPS_SCF 1.0E-6
  SCF_GUESS RESTART
  ADDED_MOS 10
  &MIXING
    METHOD BROYDEN_MIXING
    ALPHA 0.2
    BETA 0.8
    NBROYDEN 10
  &END MIXING
  &SMEAR
    METHOD FERMI_DIRAC
    ELECTRONIC_TEMPERATURE [K] 300
  &END SMEAR
&END SCF
\end{verbatim}
Therein, the keyword \texttt{EPS\_SCF} denotes the SCF convergence threshold $\| \mathbf{P}_i-\mathbf{P}_{i-1} \|_{\max}$, where $\mathbf{P}_i$ is the density matrix of the i-th iteration. Most other threshold values of the \texttt{\&DFT\%QS} section are connected to \texttt{EPS\_DEFAULT}, whose value should be approximately that of \texttt{EPS\_SCF} squared to ensure a sufficiently high numerical accuracy in order to reach the desired SCF convergence threshold. For metallic systems, finite electron temperature or smearing the density of states, which is specified in the \texttt{SMEAR} section, is often used to improve the convergence with respect to Brillouin zone sampling that is described in Section~\ref{kp_dft}. Besides the temperature, however, the occupation function must be given, whose natural choice is the physical Fermi-Dirac distribution that would lead to the grand-canonical extension to DFT of Mermin~\cite{Mermin1965}.

However, for gapped systems, a computationally much more efficient way to locate the electronic ground state is by variationally minimizing the total energy with respect to the orbital coefficient matrix $\mathbf{C}$ subject to the orthonormality constraint $\mathbf{C}^T \mathbf{SC}=\mathbf{I}$ since electrons are fermions. Inspired by the exponential transformation~\cite{hutter1994exponential}, the occupied orbitals are parameterized in terms of an auxiliary variable $\mathbf{X}$, i.e. 
\begin{equation}
  \mathbf{C}(\mathbf{X}) = \mathbf{C} \cos(\sqrt{\mathbf{X}^T\mathbf{SX}}) + \frac{\mathbf{X}}{\sqrt{\mathbf{X}^T\mathbf{SX}}}\sin(\sqrt{\mathbf{X}^T\mathbf{SX}}), 
\end{equation}
allowing for an unconstrained optimization using the OT method, given that $\mathbf{X}$ obeys the linear constraint $\mathbf{X}^T\mathbf{SC}=\mathbf{0}$. 

In addition to the minimization scheme employed, the choice of a preconditioner is essential. In this regard, the ability to converge the KS equations has to be carefully balanced against the number of necessary minimization steps, which should not be confused with SCF iterations, and the computational effort to construct the preconditioner. 
Hence, for well converging large-scale systems, preconditioners based on the Cholesky inversion of $\mathbf{H}-\mathbf{\epsilon_0 S}$ (\texttt{FULL\_SINGLE\_INVERSE})~\cite{VandeVondele2003}, or $\mathbf{T}+\mathbf{\epsilon_0 S}$ (\texttt{FULL\_KINETIC})~\cite{gan2001preconditioned}, where $\epsilon_0$ is an estimate of the highest eigenvalue of $\mathbf{C}^T \mathbf{HC}$ and $\mathbf{T}$ the kinetic energy matrix, are typically offering the computationally best price/performance ratio. However, contrary to the aforementioned SCF approach, the convergence criterion employing OT is tested in terms of the preconditioned mean gradient deviation, which is why it has to be chosen approximately one order of magnitude tighter to achieve a comparable convergence than using direct diagonalization. In case convergence is not foreseeable after approximately 20 OT minimization steps, an additional loop can be introduced via an \texttt{\&OUTER\_SCF} section, whose purpose is not to conduct an outer-loop optimization, but solely to update the preconditioner. For even more difficult cases, the diagonalization-based \texttt{FULL\_ALL} preconditioner should be used, which relies strongly on an estimate of the HOMO-LUMO gap. Since this must not overestimate the true band gap, \texttt{ENERGY\_GAP} can hardly be set too low.

Among the various implemented optimization techniques, the direct inversion of the iterative
subspace (DIIS) scheme is the method of choice as long as the fixed-point iteration is convergent~\cite{Pulay1980}. The \texttt{SAFE\_DIIS} option automatically switches to the steepest descent (SD) approach whenever a DIIS step points away from the minimum. But, since this leads to an unreasonably slow convergence behavior, this should be prevented from the outset by setting \texttt{STEPSIZE 0.1} or even smaller, i.e.
\begin{verbatim}
&OT
  MINIMIZER DIIS
  SAFE_DIIS TRUE
  PRECONDITIONER FULL_ALL
  ENERGY_GAP 0.0001
  STEPSIZE 0.1
&END OT
&OUTER_SCF
  EPS_SCF 1.0E-7
  MAX_SCF 10
&END OUTER_SCF
\end{verbatim}
If this is not successful, it is advisable to shift to the conjugate gradient (CG) method in conjunction with \texttt{LINESEARCH ADAPT}, and otherwise to the traditional diagonalization alluded to above.

\subsubsection{Brillouin-Zone Sampling using \texorpdfstring{$\mathbf{k}$}{k}-Points}\label{kp_dft}

It is possible to run periodic calculations with ${\mathbf k}$-point sampling in CP2K, with both the GPW and GAPW methods. For each ${\mathbf k}$-point in the first Brillouin zone, a generalized complex Hermitian eigenvalue problem must be solved, i.e. 
\begin{equation}\label{fock_kp}
\mathbf{K^k C^k} = \mathbf{S^k C^k}\boldsymbol{\varepsilon}^\mathbf{k},
\end{equation}
where $\mathbf{K^k}$, $\mathbf{S^k}$, $\mathbf{C^k}$ and $\boldsymbol{\varepsilon}^\mathbf{k}$ are the ${\mathbf k}$-point dependent KS matrix, overlap matrix, MO coefficients and eigenvalues, respectively. In CP2K/\textsc{Quickstep}, the KS and overlap matrices are first calculated in real-space, before being Fourier transformed to reciprocal-space. Hence, the overlap matrix elements read as 
\begin{equation}
    S_{\mu\nu}^{\mathbf{k}} = \sum_\mathbf{R} e^{i\mathbf{k}\cdot\mathbf{R}}\ S^\mathbf{R}_{\mu\nu},
\end{equation}
with
\begin{equation}
    S_{\mu\nu}^{\mathbf{R}} = \int  \varphi_\mu(\mathbf{r})\varphi_\nu(\mathbf{r}-\mathbf{R}) \text{d}\mathbf{r},
\end{equation}
where $\mathbf{R}$ is a unit cell translation vector. Similarly, $K_{\mu\nu}^{\mathbf{R}}$ is the KS matrix element of an AO $\varphi_\mu$ in the unit cell and $\varphi_\nu$ in a periodic cell translated by $\mathbf{R}$, respectively.  
Note that the number of translation vectors $\mathbf{R}$ depends on the diffuseness of the basis set: periodic images are considered as long as one of their basis functions overlaps with the unit cell. The same amount of work is done in $\Gamma$-point calculations, except that contributions from neighboring cells are added up in a single matrix.

A standard $\Gamma$-point DFT input file can be modified for a ${\mathbf k}$-point calculation by adding a \texttt{\&KPOINTS} subsection in \texttt{\&DFT}:
\begin{verbatim}
&DFT
  &KPOINTS
    SCHEME MONKHORST-PACK 8 8 4
  &END KPOINTS
  # Rest of the DFT input
&END DFT
\end{verbatim}
In this example, an $8\times 8\times 4$ Monkhorst-Pack ${\mathbf k}$-point mesh is generated~\cite{Monkhorst1976}. Note that every ${\mathbf k}$-point can also be specified individually with its coordinates and weight using the \texttt{KPOINT} keyword. 
At the time of writing, CP2K does not automatically exploit symmetries to reduce the number of ${\mathbf k}$-points (except for time-reversal symmetry, i.e. $\mathbf{k} = -\mathbf{k}$). After a converged SCF calculation, band structure data can be generated by adding the following \texttt{\&PRINT} subsection:
\begin{verbatim}
&PRINT
  &BAND_STRUCTURE
    # Output file:
    FILE_NAME graphene.bs
    # Number of empty bands:
    ADDED_MOS 2
    # Specifying the band path:
    &KPOINT_SET
      NPOINTS 10
      UNITS B_VECTOR
      SPECIAL_POINT G 0.0 0.0 0.0
      SPECIAL_POINT K 1/3 1/3 0.0
      SPECIAL_POINT M 1/2 0.0 0.0
      SPECIAL_POINT G 0.0 0.0 0.0
    &END KPOINT_SET
  &END BAND_STRUCTURE
&END PRINT
\end{verbatim}
Here, a graphene band structure is calculated along the $\Gamma$-K-M-$\Gamma$ path, with 10 ${\mathbf k}$-points along each segment. Note that any path of any density can be generated, since all relevant quantities are calculated in real-space first, before being Fourier transformed to the specified path. The only requirement is that the original SCF uses a sufficiently converged ${\mathbf k}$-point mesh, as defined in \texttt{\&KPOINTS}.

Finally, note that ${\mathbf k}$-points are not yet implemented for all electronic structure methods, but primarily for DFT, resolution-of-the-identity HF exchange (RI-HFX), and the \textit{GW} approximation. 
It also has limited support with respect to SCF schemes: only full diagonalization (with DIIS) is available, but OT is not. As for $\Gamma$-point  calculations, it is possible to print binary WF restart files via the \texttt{\&SCF\%PRINT\%RESTART} section when running ${\mathbf k}$-point calculations. However, due to different format and content, restarting a ${\mathbf k}$-point calculation from a $\Gamma$-point restart file (and vice-versa) is not possible.

\subsection{Hartree-Fock and Hybrid Density Functional Theory} \label{hfx_section}

The central quantity in HF and hybrid DFT calculations is the Fock matrix $\mathbf{F}$. For non-periodic calculations, it is simply defined as
\begin{equation}
 F_{\mu\nu} = \sum_{\sigma,\lambda} P_{\sigma\lambda}\ (\mu\sigma | \nu\lambda)
\end{equation}
where $\mu, \nu, \sigma, \lambda$ defines elements of the AO basis, whereas $P_{\sigma\lambda}$ is the density matrix, and $(\mu\sigma | \nu\lambda)$ represents 2-electron 4-center electron repulsion integrals (ERI), which in the Mulliken notation reads as
\begin{equation} \label{4_2_ERI}
    (\mu\sigma | \nu\lambda) = \int \text{d}\mathbf{r} \int \text{d}\mathbf{r'}\ \varphi_\mu(\mathbf{r})\varphi_\sigma(\mathbf{r})\ g(|\mathbf{r}-\mathbf{r'}|)\ \varphi_\nu(\mathbf{r'})\varphi_\lambda(\mathbf{r'}).
\end{equation}
Therein $g(|\mathbf{r}-\mathbf{r'}|)$ is the interaction potential, conventionally chosen to be the $1/|\mathbf{r}-\mathbf{r'}|$ Coulomb potential.

In periodic HF calculations, however, contributions from periodic images of the unit cell need to be explicitly accounted for. Elements of the Fock matrix that involve an AO $\varphi_\mu(\mathbf{r})$ in the unit cell and an AO $\varphi_\nu(\mathbf{r}-\mathbf{b})$ in a periodic image translated by $\mathbf{b}$ are defined as
\begin{equation} \label{rs_fock}
    F_{\mu\nu}^\mathbf{b} =  \sum_{\sigma,\lambda}\ \sum_{\mathbf{a}, \mathbf{c}}\ P_{\sigma\lambda}^\mathbf{c}\ (\mu^\mathbf{0}\sigma^\mathbf{a}| \nu^\mathbf{b}\lambda^\mathbf{a+c}),
\end{equation}
where contributions from AOs translated by $\mathbf{a}$ and $\mathbf{a} + \mathbf{c}$ are summed. In the special case of a $\Gamma$-point calculation, the sum over $\mathbf{c}$ can be carried out first, and the resulting periodic density matrix extracted from the sum. A subsequent sum over $\mathbf{a}$ and $\mathbf{b}$ yields the periodic $\Gamma$-point Fock matrix
\begin{equation} \label{periodic_fock}
  \begin{aligned}
    F_{\mu\nu} & = \sum_{\sigma, \lambda}P_{\sigma\lambda} \sum_{\mathbf{a}, \mathbf{b}, \mathbf{c}} (\mu^\mathbf{0}\sigma^\mathbf{a}| \nu^\mathbf{b}\lambda^\mathbf{a+c}) \\
    & = \sum_{\sigma, \lambda}P_{\sigma\lambda}\ T_{\mu, \sigma, \nu, \lambda},
  \end{aligned}
\end{equation}
where the 4-index tensor $T_{\mu, \sigma, \nu, \lambda}$ is obtained by summing ERIs involving AOs located in various periodic images. The number of periodic images necessary depends on the diffuseness of the basis set and the range of the interaction potential $g(|\mathbf{r}-\mathbf{r'}|)$. For a non-zero ERI, $\mu^\mathbf{0}$ and $\sigma^\mathbf{a}$ (respectively $\nu^\mathbf{b}$ and $\lambda^{\mathbf{a}+\mathbf{c}}$) must overlap and the product $\mu^\mathbf{0}\sigma^\mathbf{a}$ must be within the range of the product $\nu^\mathbf{b}\lambda^{\mathbf{a}+\mathbf{c}}$. The HF exchange (HFX) energy is obtained by contracting the density matrix with the Fock matrix:
\begin{equation}
    E_\text{HFX} = -\frac{1}{2} \sum_{\mu, \nu}\ P_{\mu\nu}F_{\mu\nu}
\end{equation}
and the nuclear forces are obtained by analytic differentiation of the ERIs with respect to the atomic positions.

The implementation of HFX in CP2K is documented in Ref.~\citenum{Guidon2009}. Hybrid DFT and HF calculations require the \texttt{\&HF} input subsection of \texttt{\&DFT\%XC}. The \texttt{FRACTION} keyword defines the amount of exact exchange that is added to the calculation. Note that it is always necessary to define a XC functional in the \texttt{\&XC\_FUNCTIONAL} subsection. For example, the input for a pure HF calculation looks like
\begin{verbatim}
&XC
  &XC_FUNCTIONAL NONE
  &END XC_FUNCTIONAL
  &HF
    FRACTION 1.0
    # Additional HF inputs below
  &END HF
&END XC
\end{verbatim}
whereas the input for the PBE0 functional reads as:
\begin{verbatim}
&XC
  &XC_FUNCTIONAL
    &GGA_C_PBE
    &END GGA_C_PBE
    &GGA_X_PBE
      SCALE 0.75
    &END GGA_X_PBE
  &END XC_FUNCTIONAL
  &HF
    FRACTION 0.25
    # Additional HF inputs below
  &END HF
&END XC
\end{verbatim}
An important input subsection within the  \texttt{\&DFT\%XC\%HF} section is \texttt{\&INTERACTION\_POTENTIAL}, which controls the parameters of $g(|\mathbf{r}-\mathbf{r'}|)$ in the 2-center 4-electron ERIs of Eq.~\ref{4_2_ERI}. The \texttt{POTENTIAL\_TYPE} keyword selects among the various interaction potentials implemented, while the remaining keywords in this subsection refer to the specifics of the chosen potential. The default value of \texttt{POTENTIAL\_TYPE} is \texttt{COULOMB} for the standard $1/|\mathbf{r}-\mathbf{r'}|$ potential. However, this choice is typically not suitable for periodic calculations, as it introduces spurious exchange interactions between an electron and its periodic images. A possible choice is \texttt{TRUNCATED}, which refers to the truncated Coulomb potential
\begin{equation}
    g(r) = 
    \begin{cases}
        1/r & \text{if} \ \ \ r \leq R_C \\
        0  & \text{if} \ \ \ r > R_C,
    \end{cases}
\end{equation}
which vanishes beyond a cutoff radius $R_C$. For a cubic cell of length $L$, $R_C \leq L/2$ must hold. This is to say that the truncation radius must be shorter than half the smallest cell dimension, but large enough for the exchange energy to converge. Values of 6 \AA\ are typically enough for systems with a large band gap. Systems with smaller band gaps, however, might require larger cutoff values. If the simulation cell is not large enough for both requirements, a supercell must be constructed instead. For ideal performance, the shortest cutoff that converges the energy should be taken. The value of $R_C$ is set with the keyword \texttt{CUTOFF\_RADIUS} (in Angstroms). Periodic calculations with global hybrids like PBE0 or B3LYP should use this potential~\cite{adamo1999toward,Becke1988,Lee1988}.

An alternative choice is \texttt{POTENTIAL\_TYPE SHORTRANGE}, which comes from the decomposition of the Coulomb potential in a short- and a long-range term:
\begin{equation} \label{range_sep}
    \frac{1}{r} = \frac{1-\erf(\omega)}{r} + \frac{\erf(\omega)}{r}.
\end{equation}
The \texttt{OMEGA} keyword controls the range-separation parameter. The HSE06 functional, for instance, involves the short-range potential with a value of 0.11 for $\omega$. More involved range-separated hybrid functionals can be configured with the \texttt{MIX\_CL\_TRUNC} interaction potential, which mixes the long-range part of Eq.~\ref{range_sep} and the truncated Coulomb potential. For example, the exact exchange part of the rCAM-B3LYP functional looks like:
\begin{verbatim}
&INTERACTION_POTENTIAL
  ! Should be <= than L/2 but large
  ! enough for the erf to decay 
  CUTOFF_RADIUS 6.0
  OMEGA 0.33
  POTENTIAL_TYPE MIX_CL_TRUNC
  SCALE_COULOMB 0.18352
  SCALE_LONGRANGE 0.94979                     
&END INTERACTION_POTENTIAL
\end{verbatim}

The evaluation of the ERIs $T_{\mu, \sigma, \nu, \lambda}$ in Eq.~\ref{periodic_fock} can be very demanding and is often the bottleneck of calculations requiring exact exchange. Various screening methods can help reduce the computational cost. Most importantly, Schwarz screening provides an upper bound to any 2-electron 4-center ERI. The \texttt{EPS\_SCHWARZ} keyword of the \texttt{\&XC\%HF\%SCREENING} section sets a threshold, and the evaluation of any ERI estimated to be smaller is skipped \textit{a priori}. The default value of $10^{-10}$ is rather safe, and increasing that number can lead to significant performance gains. Yet, a too large value will lead to numerical instabilities and eventually even inaccurate results. The \texttt{EPS\_SCHWARZ\_FORCES} keyword does the same for ERI derivatives and is defined as a function of \texttt{EPS\_SCHWARZ} by default. 

Gaussian type orbital (GTO) basis sets have an exponentially vanishing tail, theoretically extending to infinity. Hence, another important screening threshold is \texttt{\&DFT\%QS\%EPS\_PGF\_ORB}, which sets these tails to zero once a GTO reaches the value of \texttt{EPS\_PGF\_ORB}, thereby effectively defining the range $D_\mu$ of an AO:
\begin{equation}
    \begin{cases}
        \varphi_\mu(\mathbf{r}) = \varepsilon_\text{PGF} & \text{if} \ \ \ |\mathbf{r}| = D_\mu \\
        \varphi_\mu(\mathbf{r}) = 0  & \text{if} \ \ \ |\mathbf{r}| > D_\mu
    \end{cases}
\end{equation}
This is to say that two AOs, $\varphi_\mu$ centered on atom A at position $\mathbf{R_A}$ and $\varphi_\nu$ centered on atom B at position $\mathbf{R_B}$, only overlap if $|\mathbf{R_A}-\mathbf{R_B}| \leq D_\mu + D_\nu$. This and the choice of interaction potential effectively define the number of periodic cells to consider in the sum of Eq.~\ref{periodic_fock}. Hence, the impact on the computational performance can be significant. The default value of $10^{-5}$ for \texttt{EPS\_PGF\_ORB} is usually a good compromise between accuracy and speed. A value of $10^{-6}$ is already considered to be rather safe, whereas an even tighter threshold may negatively impact computational efficiency without necessarily improving accuracy. Since AO overlaps are key to screening, the choice of basis set can also affect performance. In particular, it is not recommended to use diffuse basis sets for HFX calculations, such as those from the previously described MOLOPT family. Instead, the ADMM and/or RI-HFX methods, described below, should be used with diffuse basis sets. 

Since the ERIs only depend on atomic positions, it is in principle possible to precalculate and store them all in main memory to simply contract them with the new density matrix at each SCF step. The ERIs only need to be recalculated when the atoms move, e.g. at each AIMD or geometry optimization step. However, due to the 4-index nature of the $T_{\mu, \sigma, \nu, \lambda}$ tensor, the memory requirements for the storage of the ERIs can become very large. The \texttt{MAX\_MEMORY} keyword of the \texttt{\&XC\%HF\%MEMORY} section caps the memory consumption of the whole HFX code. This value corresponds to the maximum allowed memory usage per MPI rank in MiB. After allocating various buffers and work arrays, the remaining memory is used to store as many ERIs as necessary and possible. To maximize the performance, the value of \texttt{MAX\_MEMORY} should be as high as possible, while leaving enough space for the rest of the calculation. Usually, allocating 50-75\% of the available memory to HFX works well. If there is not enough space to store all ERIs, some of them will be calculated on-the-fly at each SCF step, potentially slowing down the calculation. It is recommended to track the output of a HFX calculation to verify this (look for HFX\_MEM\_INFO in the output). If the number of spherical ERIs calculated on-the-fly is non-zero, significant gains can be expected by raising the value of \texttt{MAX\_MEMORY}, or the total amount of memory by using more computing nodes. However, as of the publication of this paper, the HFX code runs only on CPUs. It is recommended to restart HFX calculations from a GGA level converged WF of the same system in order to reduce the number of necessary SCF steps. 

\subsubsection{Auxiliary Density Matrix Method}

In most HFX based calculations, the evaluation of $T_{\mu, \sigma, \nu, \lambda}$ in Eq.~\ref{periodic_fock} is the main bottleneck. Unfortunately, the cost of evaluating 2-electron 4-center ERIs is tightly connected to the size and diffuseness of the selected basis set. Performing calculations with high quality basis sets can become untractable, especially for periodic systems. The ADMM is designed to tackle this issue. In this approximation, the exact exchange energy is calculated as 
\begin{equation} 
\begin{aligned}
    E_{\text{HFX}} &= E_x^{\text{HF}}[\hat{\mathbf{P}}] + \left(E_x^{\text{HF}}[\mathbf{P}] - E_x^{\text{HF}}[\hat{\mathbf{P}}]\right)\\
    &\approx E_x^{\text{HF}}[\hat{\mathbf{P}}] + \left(E_x^{\text{GGA}}[\mathbf{P}] - E_x^{\text{GGA}}[\hat{\mathbf{P}}]\right),
\end{aligned}
\end{equation}
where $\mathbf{P}$ is the density matrix in the primary basis and $\hat{\mathbf{P}}$ is the density matrix expressed in an auxiliary basis that is smaller and/or less diffuse than the former basis set. The exact exchange energy is efficiently evaluated in the smaller auxiliary basis, while the difference in exchange energy due to the change of basis set is approximated at the DFT GGA level. The approximation is assumed to be accurate, provided that $\mathbf{P} \approx \hat{\mathbf{P}}$.

Given a choice of auxiliary basis set and GGA exchange functional, there exist different flavors of ADMM. In its simplest form, known as ADMM2, the density is simply projected from the orbital basis onto the auxiliary basis set. On the other hand, ADMM1 involves a purification of the auxiliary density matrix to ensure its idempotency and particle conservation. Both ADMM1 and ADMM2 are described in detail in the original publication by Guidon and co-workers~\cite{Guidon2010}. Subsequent work by Merlot and co-workers introduced additional three flavors that are ADMMP, ADMMQ, and ADMMS, respectively~\cite{merlot2014charge}. The latter impose various physically motivated constraints to the approximation in a simpler way than the ADMM1 purification scheme. Note that most post-SCF methods, which can be used in conjunction with ADMM (RPA, MP2, TD-DFT, density-corrected DFT, etc.), only support ADMM2.

To run an ADMM calculation, a valid HFX input file can be modified as follows. The \texttt{\&AUXILIARY\_DENSITY\_MATRIX\_METHOD} subsection should be added to \texttt{\&DFT}, i.e.
\begin{verbatim}
&AUXILIARY_DENSITY_MATRIX_METHOD
  # Choose between ADMM flavors
  ADMM_TYPE ADMMS     
  # Use PBE exchange (example)
  EXCH_CORRECTION_FUNC PBEX 
&END AUXILIARY_DENSITY_MATRIX_METHOD
\end{verbatim}
Note that various other keywords exist, but their usage is not recommended. Indeed, \texttt{ADMM\_TYPE} automatically sets their value for the desired ADMM flavor.

A critical aspect of running ADMM calculations is the choice of auxiliary basis set. The FIT basis family was originally developed to run in conjunction with MOLOPT basis sets. These basis sets can be found in the BASIS\_ADMM\_MOLOPT file, and their hierarchy is described in Ref.~\citenum{Guidon2010}. More recently, correlation-consistent ADMM basis sets were developed to be used with the ccGRB family, which can be found in the BASIS\_ADMM\_UZH and BASIS\_ccGRB\_UZH files, respectively. The usage of the more recent basis sets is recommended. Note that all-electron auxiliary basis sets are also available in the file BASIS\_ADMM\_ae to be used in conjunction with Jansen's pcseg basis family~\cite{jensen2014unifying, kumar2018accelerating}. The ADMM basis set file must be added to the input file, and the specific choice of basis specified for each atomic species, hence:
\begin{verbatim}
&DFT
  BASIS_SET_FILE_NAME BASIS_ccGRB_UZH
  BASIS_SET_FILE_NAME BASIS_ADMM_UZH
  POTENTIAL_FILE_NAME POTENTIAL_UZH
  # Rest of DFT input
&END DFT
&SUBSYS
  &KIND O
    BASIS_SET ccGRB-D-q6
    BASIS_SET AUX_FIT admm-dz-q6
    POTENTIAL GTH-HYB-q6
  &END KIND
  # Rest of SUBSYS input
&END SUBSYS
\end{verbatim}
Note that with the ADMM approximation, a full-fledged HFX calculation is still being carried out, albeit with a smaller base set. All input considerations discussed in the \texttt{\&HF} section apply here as well.

\subsubsection{Resolution-of-the-identity Hartree-Fock Exchange} \label{ri_hfx_section}

In the resolution-of-the-identity (RI) approximation, the 2-electron 4-center ERIs can be expressed in terms of 2- and 3-center integrals, i.e.
\begin{equation} \label{ri_4center}
    \begin{aligned}
    (\lambda\sigma|\mu\nu) &\approx (\lambda\sigma\lfloor P)\ (P\lfloor Q)^{-1} (Q|R)\ (R\lfloor S)^{-1}\ (S\lfloor \mu\nu)\\
    &= (\lambda\sigma\lfloor P)\ V_{PQ}^{-1}\ (Q|R)\ V_{RS}^{-1}\ (S\lfloor \mu\nu),
    \end{aligned}
\end{equation}
where the Einstein summation convention is used for the sum over the RI basis elements $P, Q, R, S$. The $|$ and $\lfloor$ symbols represent the HF interaction potential and RI metric, respectively. The RI approximation is exact if the RI basis spans the same space as all possible AO products, and if the RI metric is the same as the interaction potential. This approach allows to shift the computational effort from ERI evaluations to sparse tensor contractions, which can be efficiently performed on GPUs~\cite{gavini2023roadmap}. Note that RI-HFX is not always a suitable choice, and the standard 4-center approach remains relevant in most cases. In practice, RI-HFX shows a performance edge for non-periodic calculations of molecules with large basis sets (where both the interaction potential and the RI metric are the standard $1/r$ Coulomb measure), and for periodic calculations of small to medium sized dense solids. The CP2K/\textsc{Quickstep} implementation, as well as a discussion on the choice of RI metric and application cases, can be found in Ref.~\citenum{bussy2023sparse}. The RI-HFX method is compatible with all flavors of ADMM, and analytic gradients to compute the nuclear forces are implemented.

The RI-HFX method can be invoked by adding the following subsection to the \texttt{\&HF} section of the CP2K input file:
\begin{verbatim}
&RI
  # Definition of the RI metric:
  RI_METRIC TRUNCATED
  CUTOFF_RADIUS 2.0
  # Optional but potentially
  # important keywords, here
  # with default values
  MEMORY_CUT 3
  EPS_FILTER 1.0E-9
&END RI
\end{verbatim}
This input snippet defines the RI metric to be used in a similar fashion to the \texttt{\&INTERACTION\_POTENTIAL} subsection of the same parent \texttt{\&HF} section. By default, the RI metric takes the same parameters as the latter. The RI metric is usually chosen to be a very short-ranged interaction in order to increase sparsity and to reduce the computational effort. The range of the RI metric is a trade-off between performance and accuracy, with the overlap as the most extreme case, which is known as the RI-SVS approximation~\cite{vahtras1993integral}. The \texttt{MEMORY\_CUT} keyword controls batching in the contraction of Eq.~\ref{ri_4center}: potentially very large tensors are necessary to hold intermediate results, and batching allows processing them with relatively limited resources. Note that this keyword only concerns the previously mentioned contraction step, but it does not affect the total memory footprint of CP2K. Since batched tensor contractions come with an overhead, it is recommended to keep the value of \texttt{MEMORY\_CUT} as low as the employed resources allow for. It is generally more efficient  (but not always practical) to increase the amount of resources, rather than increasing the keyword's value. Note that RI-HFX calculations of large systems can be memory intensive, and that unlike the standard 4-center method, there is no easy way to predict the required memory. The \texttt{EPS\_FILTER} keyword controls the filtering of small elements out of the sparse tensors. The default value of $10^{-9}$ is a good compromise between accuracy and efficiency. Increasing this value will lead to performance gains, but it is recommended to proceed with care. The rest of the \texttt{\&HF} input section can remain the same. Note, however, that the \texttt{\&MEMORY} and the \texttt{\&SCREENING} subsections have no effect during a RI-HFX calculation. All considerations on the choice of interaction potential or \texttt{EPS\_PGF\_ORB} discussed for the original 4-center method also apply here.

The choice of RI basis set is important for this type of calculations, both in terms of performance and accuracy. It is generally recommended to use optimized bases when possible (e.g. cc-pVTZ-JKFIT with the cc-pVTZ orbital basis)~\cite{weigend2002fully}. The cc-pVTZ all-electron basis set, however, is not distributed with CP2K: it would need to be downloaded independently at the EMSL Basis Set Exchange for instance~\cite{Pritchard2019}. If not available, the RI basis is generated on-the-fly using the method of Stoychev and co-workers~\cite{stoychev2017automatic}. This type of basis is safe to use and accurate, but is generally larger, and thus less efficient, than its optimized counterparts. The CP2K input file can be modified as follows to specify the RI basis:
\begin{verbatim}
&DFT
  SORT_BASIS EXP
  # Rest of the DFT input
&END DFT
&SUBSYS
  &KIND
    BASIS_SET cc-pVTZ
    # If the following line is not
    # present, the RI_HFX basis is
    # generated on-the-fly
    BASIS_SET RI_HFX cc-pVTZ-JKFIT
    POTENTIAL ALL
  &END KIND
  # Rest of the SUBSYS input
&EBD SUBSYS
\end{verbatim}
Note the \texttt{SORT\_BASIS} keyword, use is highly recommended, as it reorders the AO according to their Gaussian exponent, thus increasing sparsity. 

\subsubsection{Resolution-of-the-identity Hartree-Fock Exchange with \texorpdfstring{${\mathbf k}$}{\textbf{k}}-point Brillouin-Zone Sampling}
The RI-HFX with ${\mathbf k}$-point sampling (RI-HFXk) method allows for periodic HFX calculations of small unit cells with ${\mathbf k}$-point sampling. As for ${\mathbf k}$-point sampling using DFT (see Section~\ref{kp_dft}), the ${\mathbf k}$-point dependent Fock matrix is first calculated in real-space before it is Fourier transformed, i.e. 
\begin{equation}
    F_{\mu\nu}^{\mathbf{k}} = \sum_\mathbf{R} e^{i\mathbf{k}\cdot\mathbf{R}}\ F^\mathbf{R}_{\mu\nu},
\end{equation}
where $\mathbf{R}$ is a unit cell translation vector and $F^\mathbf{R}_{\mu\nu}$ is defined in Eq.~\ref{rs_fock}. Contrary to $\Gamma$-point calculations, the sum over periodic images cannot be performed \textit{a priori}, and the results are stored in a single 4-index tensor. As a result, storing the integrals is intractable, and all 4-center 2-electron ERIs would need to be calculated on-the-fly at each SCF step. Although possible, such an approach would be very expensive. Thus, an efficient alternative using an atom-specific RI scheme was developed~\cite{bussy2024efficient}.

The RI-HFXk method can be invoked by adding a \texttt{\&KPOINTS} section to a RI-HFX input file. The ${\mathbf k}$-point specific input is the same as described in Section~\ref{kp_dft}, including the optional band structure information. It is compatible with the ADMM approximation, and nuclear forces are implemented. As for any HFX method, it is generally recommended to restart from a converged GGA level (k-point) WF. Extra care must be paid to some parts of the input file, as listed below:
\begin{itemize}
  \item \texttt{\&HF\%INTERACTION\_POTENTIAL}: As for any periodic HF calculation, a short-range potential must be selected. The $L/2$ requirement now refers to the Born–von Kármán supercell, i.e. a supercell obtained by multiplying the unit cell by the number of ${\mathbf k}$-points in each direction. As for $\Gamma$-point HFX, the potential range should be long enough for convergence, while kept as short as possible for performance.
  \item \texttt{\&RI\%RI\_METRIC}: In the special case of RI-HFXk, the range of the RI metric barely impacts the performance. It is hence recommended to always take it to be the same as the interaction potential for maximal accuracy (default option).
  \item \texttt{\&QS\%PW\_GRID\_BLOCKED}: It is recommended to set this keyword to \texttt{FALSE}. This ensures that calculations of small unit cells with a lot of CPUs can go ahead without crashing.
  \item \texttt{\&SCF\%EPS\_DIIS}: The default threshold value to start DIIS is 0.1. Experience has shown that better convergence is achieved when this value is lower, typically around 0.05.
  \item \texttt{\&RI\%KP\_NGROUPS}: There is a lot of computational work to be done, and most of it can be efficiently parallelized over MPI subgroups. This is controlled with the \texttt{KP\_NGROUPS} keyword, which should be a divisor of the total number of ranks. Using $N$ subgroups should accelerate the calculations by a factor $\sim N$ and increase its memory usage by the same factor.
  \item \texttt{\&RI\%KP\_STACK\_SIZE}: Most of the computational effort is spent on contracting tensors. In some cases, more efficient batched contractions can take place, at the cost of greater memory usage. The default value of 32 is meant for small systems of a few atoms. Larger systems will require more memory overall and might not have enough room for the storage of temporary tensors (in which case, use a smaller stack size).
\end{itemize}

Generally, the RI-HFXk method is very demanding, both in terms of compute and memory. It was designed and optimized for small to medium size systems with dense ${\mathbf k}$-point meshes (e.g. 2D materials, highly symmetric crystals, etc.). The computational cost of the method scales as $\mathcal{O}(N_\text{atoms}^3N_\text{img}^2)$, where $N_\text{img}$ refers to the number of periodic images required (depends on the diffuseness of the base set and the range of the interaction potential). The scaling of the memory usage is harder to predict, but experience suggests that it is steep, too. For large systems, a $\Gamma$-point calculation with an equivalent supercell will be much more efficient due to the asymptotical linear scaling of the 4-center HFX implementation. The details of the RI-HFXk method are not trivial, and it is recommended to read Ref.~\citenum{bussy2024efficient} before use.

\subsection{Post-Hartree-Fock and Double-Hybrid Density Functional Theory}

In addition to incorporating exact exchange via HFX, as described in Section~\ref{hfx_section}, electron correlation beyond conventional DFT can be added via efficient MP2 and RPA methods described in this section. Most of them are available for condensed phase systems including nuclear forces, if not explicitly mentioned otherwise. 

\subsubsection{Second-order M{\o}ller–Plesset Perturbation Theory}

The MP2 method computes
\begin{equation} \label{MP2_energy_Eq}
E^{(2)} = - \sum_{ijab} \frac{(ia|jb)[2(ia|jb)-(ib|ja)]}{\epsilon_{a}+\epsilon_{b}-\epsilon_{i}-\epsilon_{j}}. 
\end{equation}
employing canonical HF reference orbitals, each with the real-space representation $\phi_p(\vec r)$ and an orbital energy $\epsilon_p$. The ERIs are usually calculated in the AO basis first and then in four steps transformed to the MO basis.
CP2K/\textsc{Quickstep} provides three different implementations: a direct/canonical implementation, a GPW-based implementation~\cite{DelBen2012} and a RI-based implementation~\cite{DelBen2013}. All MP2 implementations scale with the fifth power of the number of atoms. 
The canonical implementation is available for GPW and GAPW reference calculations allowing core-corrections. However, it is the most costly implementation and analytical gradients (nuclear forces, stress tensors, etc.) are not available. 

All flavors of MP2 calculations are available via the \texttt{\&MP2} subsection in the \texttt{\&DFT\%XC\%WF\_CORRELATION} section and activating the respective method. For direct canonical MP2, it is \texttt{METHOD DIRECT\_CANONICAL}. Further keywords, which are also of interest for all other implementations, are
\begin{itemize}
    \item \texttt{\&WF\_CORRELATION\%MEMORY}: the allowed memory for the MP2 calculation.
    \item \texttt{\&WF\_CORRELATION\%GROUP\_SIZE}: it needs to be a divisor of the total number of processes. Lower values reduce communication costs, but increase memory requirements.
    \item \texttt{\&WF\_CORRELATION\%INTEGRALS}: optional section to configure the ERI method via the \texttt{ERI\_METHOD} keyword and the interaction potential by the \texttt{INTERACTION\_POTENTIAL} subsection (default: Coulomb potential).
\end{itemize}
The direct MP2 implementation serves as a reference method for medium-sized systems, as its computational demands become easily untractable for larger systems. This implementation also does not benefit from GPU acceleration.

For better performance, the GPW-based implementation should be used~\cite{DelBen2012}. It skips the first two transformation steps by calculating half-transformed ERIs from the product density $\phi_i(\vec r)\phi_a(\vec r)$ directly. It is available by setting \texttt{\&MP2\%METHOD MP2\_GPW}. The GPW-integration is configured with the \texttt{\&WFC\_GPW} subsection inside the \texttt{\&WF\_CORRELATION\%INTEGRALS} section. As with the original GPW method, there are two tunable parameters \texttt{CUTOFF} and \texttt{REL\_CUTOFF} with the same meaning and should be tuned as in the original DFT-based GPW approach described in Section~\ref{BS_section}. Yet, due to the different requirements on the accuracy and the properties of the densities, primary cutoffs of 150 to 300 Ry are typically sufficient for most applications. Using the grid and the distributed block compressed sparse row (DBCSR) libraries, the GPW-based implementation runs efficiently on GPUs, enabling MP2 calculations on systems containing hundreds of atoms.

The RI approximation to the ERIs allows for further computational savings~\cite{Weigend1997}. If we expand products of orbitals in a set of auxiliary functions denoted by $P,Q,...$, the ERI tensor factorizes approximately into a rank-3 tensor $B$, i.e.
\begin{equation}
    (ia|jb)\approx\sum_P B_{iaP}B_{jbP}, 
\end{equation}
where $B$ is calculated from the respective three-center integrals $(P|Q)$ and $(P|ia)$. These can be calculated using the GPW-approach in a similar fashion as discussed above, or alternatively using a Minimax-Ewald (MME) approach or the Obara-Saika (OS) scheme, which can be set by the previously mentioned \texttt{\&INTEGRALS\%ERI\_METHODS} keywords. For that, the \texttt{RI\_MP2} section replaces the \texttt{MP2} section. 
\begin{verbatim}
&RI_MP2
  # Larger block sizes require more 
  # memory but reduce communication, 
  # (default: -1, auto determination)
  BLOCK_SIZE 2
  # Large number of groups increase the
  # memory demands but reduce communication
  # (default: -1, auto determination)
  NUMBER_INTEGRATION_GROUPS 2
&END
\end{verbatim}
The most important keywords are \texttt{BLOCK\_SIZE} and \texttt{NUMBER\_INTEGRATION\_GROUPS}. By default, these parameters are determined automatically from the available memory, but their tuning may be of interest for calculations on a larger number of systems. Large block sizes and large numbers of integration groups reduce communication costs at the expense of a higher memory demand. The number of integration groups must be a divisor of the number of subgroups (number of MPI ranks divided by the \texttt{GROUP\_SIZE}). See Ref.~\citenum{DelBen2013} for further information on the technical details and GPU acceleration.

As before, the RI approximation requires the user to specify an auxiliary basis set. In general, the auxiliary basis needs to be adapted to the primary basis set to yield an acceptable accuracy and sufficient speed-up compared to the canonical implementations. CP2K has several options: pre-optimized, self-optimized, and auto-optimized basis sets. Optimization of the auxiliary basis sets is available according to Ref.~\citenum{Weigend1998}. A compilation of pre-optimized basis sets is available in the previously mentioned data directory of the CP2K GitHub repository. If none is available for a certain basis set or the optimization is not required, we recommend auto-optimized basis sets. They are available by omitting the specification of the auxiliary basis set in the \texttt{\&SUBSYS\%KIND} section and setting the \texttt{AUTO\_BASIS} keyword in the \texttt{\&DFT} section. 

For the RI-MP2 implementation, CP2K analytical gradients and stress tensors are available~\cite{DelBen2015a, Rybkin2016}. Its implementation is based on the Lagrangian formalism of the MP2 energy by introducing the \texttt{\&WF\_CORRELATION\%CANONICAL\_GRADIENTS} section. The Lagrangian formalism requires the solution of the coupled-perturbed HF (CPHF) equations configured in the \texttt{\&CANONICAL\_GRADIENTS\%CPHF} subsection:
\begin{verbatim}
&CANONICAL_GRADIENTS
  # The default is usually good enough
  EPS_CANONICAL 1E-6
  # This option which should always work
  # Try to set it to .TRUE. 
  FREE_HFX_BUFFER .FALSE.
  &CPHF
    # Choose as tight as necessary
    EPS_CONV 1.0E-6
    # Smaller values of EPS_CONV 
    # may require more iterations
    MAX_ITER 10
  &END CPHF
&END CANONICAL_GRADIENTS
\end{verbatim}
Therein, \texttt{EPS\_CANONICAL} is a MP2 specific parameter affecting the numerical stability of MP2 analytic gradient calculations. Higher values improve the accuracy, but increase the computational costs significantly. The \texttt{FREE\_HFX\_BUFFER} keyword frees the HF buffers before the gradient calculations to increase the amount of available memory necessary for it. By default, it is turned off because the integrals cannot always be recovered for technical reasons. The parameters of the \texttt{\&CPHF} section affect the numerical solution of the CPHF equations. The \texttt{EPS\_CONV} parameter affects the accuracy of the solution of the CPHF equations, which are to be chosen as tightly as necessary and as high as possible. The \texttt{MAX\_ITER} parameter determines the maximum number of iteration steps. If convergence is not observed, then the orbitals are not sufficiently converged, or the MP2 method is not suitable for the given system because of too high electron correlation. 

The accuracy of MP2 calculations is mostly affected by the size of the primary basis set. In contrast to HF or DFT calculations, where the basis set error decays exponentially with the number of basis functions, the error of the MP2 energy decays with the reciprocal number of basis functions~\cite{Schwartz1962}. Thus, either sufficiently large basis sets or an extrapolation to the complete basis set limit is required. Independently of an extrapolation scheme, basis sets should have a quality of at least triple-zeta or augmented double-zeta for accurate results. Finally, the basis sets should contain enough polarization functions as in correlation-consistent basis sets. 

The performance of MP2 calculations in CP2K depends on several factors. Conceptually, the number of operations increases linearly to quadratically with the number of auxiliary basis functions, suggesting that optimized auxiliary basis sets should be employed whenever possible. The calculation of the integrals using the GPW approach relies on the efficient grid library and sparse matrix-contractions with DBCSR. The actual contraction of the ERIs to the MP2 energy relies on a well-optimized BLAS implementation for the matrix-matrix multiplications. For GPU acceleration, CP2K is either linked to an accelerated BLAS implementation or switches to the SpLA library~\cite{spla}.

Although the calculation of the MP2 energy shows a steeper scaling than a HF calculation, the HF kernel can still dominate the computational costs in case of diffuse basis functions. This can be overcome by the ADMM~\cite{Guidon2010}, which is available for MP2 energy and gradient calculations~\cite{Stein2022}. 

Double-hybrid functionals improve upon MP2 by improving the description of short-ranged correlation effects using DFT~\cite{Grimme2006}. Instead of starting from a HF description, double-hybrid functionals start from hybrid DFT and then add a certain amount of spin-rescaled MP2 energy. In CP2K/\textsc{Quickstep}, this is achieved with the following modifications inside the \texttt{\&XC} section:
\begin{itemize}
    \item \texttt{\&XC\_FUNCTIONAL}: Set up the respective DFT part. In addition to the natively implemented XC functionals, an even larger selection of functionals is available through the LibXC library~\cite{Lehtola2018}.
    \item \texttt{\&HF}: Set up the respective HF interaction operator. For simple functionals, the adjustment of the \texttt{FRACTION} parameter is sufficient. For more elaborate functionals consider the parameters in the \texttt{\&INTERACTION\_POTENTIAL} subsection.
    \item \texttt{\&WF\_CORRELATION}: The keywords \texttt{SCALE\_S} and \texttt{SCALE\_T} determine the weight of singlet and triplet contributions. The \texttt{\&INTEGRALS\%INTERACTION\_POTENTIAL} subsection allows the adjustment of the operator.
\end{itemize}
Beware that CP2K does not read any HF related information from LibXC such that all parameters have to be looked up and set up manually. We recommend testing the setup against reference values for the respective double-hybrid functional as the setup is error-prone. Some examples can be found at \url{https://github.com/cp2k/cp2k/tree/master/data/xc_section}. Nuclear forces and stress tensors are analytically available throughout for double-hybrid functionals~\cite{Stein2020,Stein2022}.

\subsubsection{Random Phase Approximation and Laplace-transformed Scaled Opposite-spin Second-order M{\o}ller–Plesset Perturbation Theory}
The RI approach to the direct RPA (dRPA) method given by its correlation energy contribution
\begin{equation}
  \begin{aligned}
E^\text{dRPA} &=-\frac{1}{4\pi}\int_{-\infty}^{\infty}d\omega\left\langle\log (1+Q(\omega))-Q(\omega)\right\rangle\\
Q_{RS}(\omega) &= 2\sum_{ia}B_{iaR}\frac{\epsilon_a-\epsilon_i}{\left(\epsilon_a-\epsilon_i\right)^2+\omega^2}B_{iaS}
  \end{aligned}
\end{equation}
and the RI-based Laplace-transformed scaled opposite-spin MP2 (SOS-MP2)
\begin{equation}
\begin{aligned}
E^\text{SOS-MP2} &=-\int_{0}^{\infty}d\tau\text{Tr}\left(\overline{Q}(\tau)^2\right)\\
\overline{Q}_{RS}(\tau) &= \sum_{ia}B_{iaR}e^{-\left(\epsilon_a-\epsilon_i\right)\tau}B_{iaS}
\end{aligned}
\end{equation}
are implemented similarly due to the integration of a matrix-valued function and the contraction of the RI-tensor $\mathbf{B}$ to a frequency-dependent or time-dependent intermediate matrix $\mathbf{Q}(\omega)$. Their energy-only implementation is discussed in Ref.~\citenum{DelBen2013}. Although the integration may be performed analytically via the Plasmon equation~\cite{Furche2008}, or the MP2 energy equation (see Eq.~\ref{MP2_energy_Eq}), we pursue the path of numerical integration due to its reduced scaling behavior of fourth power with respect to the number of atoms, as compared to the alternative with fifth order scaling. 

CP2K/\textsc{Quickstep} implements two quadrature schemes: Clenshaw-Curtis~\cite{Boyd1987} and Minimax~\cite{Braess2005}. The Clenshaw-Curtis quadrature is easily accessible and converges for all subsequent methods, but requires a larger number of quadrature points. In contrast to that, the Minimax quadrature rules are difficult to optimize and are thus pre-tabulated. The main advantage of Minimax quadrature rules lies in their higher accuracy than the Clenshaw-Curtis quadrature rules and are the recommended choice for RPA calculations and the only available option for SOS-MP2 calculations.\\
The setup of these methods is similar to the setup of an (RI-)MP2 calculation. The main difference is that the \texttt{\&RI\_RPA} section
\begin{verbatim}
&RI_RPA
  # Switch to the Minimax quadrature 
  # rules (default: turned off, use
  # Clenshaw-Curtis-quadrature)
  MINIMAX .TRUE.
  # Choose it as large as necessary 
  # and as small as possible
  QUADRATURE_POINTS 6
  # Fine-tuning of the parallelization
  NUM_INTEG_GROUPS -1
&END
\end{verbatim}
or the \texttt{RI\_SOS\_MP2} section
\begin{verbatim}
&RI_SOS_MP2
  # Both parameters work similarly to RPA
  QUADRATURE_POINTS 6
  NUM_INTEG_GROUPS -1
&END
\end{verbatim}
replaces the \texttt{\&MP2} or \texttt{\&RI\_MP2}-sections, respectively. Note that the Minimax quadrature has to be turned on manually for RPA calculations. This difference is related to RPA calculations being the precursor of \textit{GW} bandstructure calculations, which are described in Section~\ref{sec:GW}, requiring quadratures suitable for different kinds of integration kernels.

The RPA correlation energies are usually combined with exact exchange energies. In CP2K, the \texttt{\&RI\_RPA} section includes a \texttt{\&HF} section, which is set up similarly to the ordinary \texttt{\&HF} section used in HF and hybrid functional calculations. Because the ERIs are needed only once here, the \texttt{MAX\_MEMORY} keyword in the \texttt{\&MEMORY} should be set to zero in case of a simple RPA energy calculation to save memory. The ADMM acceleration can also be applied for the calculation of exact exchange energies by activating the \texttt{ADMM} keyword in the \texttt{\&RI\_RPA} section. 
We recommend the ADMM approximation for calculations with diffuse basis functions, which would otherwise increase the computation time tremendously due to their larger extent.

On top of the exact exchange and the dRPA correlation energy, one can also apply a beyond-RPA scheme. The \texttt{RSE} keyword activates the renormalized single excitations (RSE) correction to account for non-diagonal Fock-matrix elements~\cite{Ren2013}. In addition to that, the \texttt{\&RI\_RPA\%EXCHANGE\_CORRECTION} section turns on either the approximate exchange kernel (AXK)~\cite{Bates2013}, or the second-order screened exchange (SOSEX) corrections to reduce the self-correlation errors of the dRPA method~\cite{Freeman1977}. The corrections are set up as shown below.
\begin{verbatim}
&RI_RPA
  ...
  # The RSE correction is only relevant 
  # for non-HF reference states
  RSE .TRUE.
  # Exchange corrections can be costly
  &EXCHANGE_CORRECTION [NONE|AXK|SOSEX]
    # The Hartree-Fock-based implemen-
    # tation scales better for larger 
    # systems but introduces more noise
    USE_HFX_IMPLEMENTATION F
    # This parameter is ignored if
    # USE_HFX_IMPLEMENTATION is set to T
    # Larger values improve performance 
    # but increase memory usage
    BLOCK_SIZE 16
  &END EXCHANGE_CORRECTION
  # This HF section activates the exact
  # exchange correction and is also used 
  # by the RSE and the AXK/SOSEX methods
  &HF
    ...
    &MEMORY
      # Set to zero if RSE/AXK/SOSEX 
      # are not requested
      MAX_MEMORY 500
    &END MEMORY
  &END HF
&END
\end{verbatim}
As is customary in CP2K, it is important to use the lone section name as a keyword to specify the actual method. The \texttt{USE\_HFX\_IMPLEMENTATION} keyword in the \texttt{\&RI\_RPA\%EXCHANGE\_CORRECTION} section switches between a better scaling HF-based implementation and an RI-MP2-based implementation. The first one should be preferred for larger systems, and the latter one should be preferred for smaller systems. Note that the HF-based implementation recalculates the integrals using the HF code. In the periodic case, this requires a truncated Coulomb potential, whereas the integrals for the RI methods can be calculated using the ordinary Coulomb potential. In case of the SOSEX-correction, this makes the implementation not self-correlation-free although the SOSEX method theoretically shows this property.

For the dRPA and SOS-MP2 implementations, CP2K implements analytical nuclear forces and stress tensors~\cite{Stein2024}. The general setup is similar to RI-MP2 gradient calculations, but adds the \texttt{MAX\_PARALLEL\_COMM} keyword to the \texttt{\&WF\_CORRELATION\%CANONICAL\_GRADIENTS} section for RI-RPA gradient calculations. It determines the number of parallel communication channels for the non-blocking communication scheme. Larger numbers improve the performance, but increase the memory requirements. We find that a maximum value of 3 is sufficient. Note that analytical gradients are not available for the beyond-RPA methods.

The computationally most costly part of dRPA and SOS-MP2 calculations is the contraction of $\mathbf{B}$ to $\mathbf{Q}(\omega)$. Because this step is necessary for each quadrature point, the number of quadrature points should be chosen as low as possible. Therefore, we apply the Minimax quadrature rule. It is also a convenient option for the beyond-RPA schemes. Because CP2K relies heavily on parallel matrix-matrix multiplications using ScaLAPACK, a well-optimized (and GPU accelerated) implementation of this library should be employed if possible. However, please consult Section~\ref{sec:TechnicalAspects} for more details on the support of various libraries within CP2K. Alternatively, CP2K can employ the communication-optimal COSMA library for even higher performance in exchange for increased memory usage~\cite{cosma_algorithm_2019}. If CP2K was compiled with COSMA support, it is used automatically. All of this holds for all kinds of RPA- or SOS-MP2-based methods.

As with RI-MP2, both RI-RPA and SOS-MP2 rely on partial replication of the data via a \texttt{NUM\_INTEG\_GROUPS} keyword inside the corresponding sections. It is by default chosen automatically based on the available memory, but can be tuned manually. It must be a divisor of the number of MPI ranks and the number of quadrature points. A larger number of integration groups reduces the communication, but increases the memory demands. This feature is also available for analytic gradient calculations.

\subsubsection{Low-scaling post-Hartree-Fock}

As already mentioned, post-HF methods suffer from the steep scaling of their computational cost, thus limiting their application to small and medium-size systems. Reworking the fundamental equations, using the RI approximation and exploiting the inherent sparsity of AOs, it is possible to reduce that scaling. In CP2K, low-scaling implementations of RI-dRPA and RI-SOS-MP2 are available, for both energies~\cite{Wilhelm2016} and nuclear forces~\cite{bussy2023sparse}. It is straightforward to transform a canonical RI-dRPA or RI-SOS-MP2 input file into a low-scaling one, simply by adding the following to the \texttt{\&WF\_CORRELATION} section:
\begin{verbatim}
&WF_CORRELATION
  # Usual RI_RPA/RI_SOS_MP2 input here
  &LOW_SCALING
    MEMORY_CUT 3
    EPS_FILTER 1.0E-9
  &END LOW_SCALING
  &RI
    &RI_METRIC
      POTENTIAL_TYPE TRUNCATED
      CUTOFF_RADIUS 1.5
    &END RI_METRIC
  &END RI
&END WF_CORRELATION
\end{verbatim}
The input parameters specified in \texttt{\&RI\_RPA} or \texttt{\&RI\_SOS\_MP2} also apply in the low-scaling case. Note that the two subsections used in the above example are already discussed in Section~\ref{ri_hfx_section} on the RI-HFX, since both methods used the same machinery and a similar theory. All performance aspects, such as the use of \texttt{SORT\_BASIS EXP} or the choice of \texttt{EPS\_PGF\_ORB} also apply here. The \texttt{\&INTEGRALS} subsection of a canonical post-HF calculation is also used in the low-scaling case, but only for 2-center integrals. The rest is calculated analytically with the Libint library~\cite{Libint2}.

The nominal scaling of these methods is cubic. The exploitation of sparsity allows for a sub-cubic scaling in most cases. Note that energy-only calculations typically have a more advantageous scaling than nuclear force calculations, which still benefit from a cubic scaling upper bound. A lot of overhead is necessary to run such calculations, making them more expensive than their quartic scaling counterparts for small systems. The choice of the RI metric also adds an additional level of approximation. It is therefore always recommended to use the canonical versions of dRPA or SOS-MP2 whenever possible, and only use the low-scaling implementations for very large systems with hundreds of atoms. If in doubt, it might be worth simply trying both approaches.



\subsection{Density Functional Theory plus Hubbard U}
In spite of its success, standard local and semi-local DFT shows shortcomings in describing some physical phenomena such as London dispersion interactions or strong electron-electron correlation. The latter phenomenon is dominant in transition metals (TM), lanthanide and actinide oxides, such as NiO, CoO, CeO$_2$ or UO$_2$, resulting in a poor description of the electronic structure of these materials only using standard DFT. Anisimov et al. proposed to add a Hubbard U term as a correction to the standard DFT to improve the description of these materials~\cite{Anisimov1991}. Since then, many variants of this kind of correction have been proposed. One of the most popular corrections, called DFT+U, was introduced by Dudarev et al.~\cite{Dudarev1997,Dudarev1998}. It is rotationally invariant, computationally inexpensive, and introduces only an effective Hubbard parameter U$_\text{eff} = \text{U} -\text{J}$ as an atomic parameter for the atomic sites concerned. The Dudarev approach is implemented in CP2K/\textsc{Quickstep}~\cite{Krack2015}. It is based on orbital occupations derived from a Mulliken~\cite{Mulliken1955}, or L\"owdin population analysis~\cite{Loewdin1950}. DFT+U is globally activated with the keyword \texttt{PLUS\_U\_METHOD} in the \texttt{\&FORCE\_EVAL\%DFT} section:
\begin{verbatim}
&FORCE_EVAL
  METHOD Quickstep
  &DFT
    PLUS_U_METHOD Mulliken
    ...
    &PRINT
      &PLUS_U on
      &END PLUS_U
    &END PRINT
  &END DFT
  ...
&END FORCE_EVAL
\end{verbatim}
and the DFT+U specific printout is controlled via the corresponding \texttt{\&PRINT} section within \texttt{\&PLUS\_U}. The U$_\text{eff}$ parameter \texttt{U\_MINUS\_J} and the angular momentum number \texttt{L} must be defined in an atomic \texttt{\&KIND} section. Here is an example that applies an effective Hubbard U$_\text{eff}$ of 2~eV to the $5f$ orbitals ($l = 3$) of the uranium atoms assigned to the atomic kind U$_\text{a}$:
\begin{verbatim}
&KIND U_a
  ...
  &DFT_PLUS_U on
    L 3
    U_MINUS_J [eV] 2.0
    &ENFORCE_OCCUPATION on
      EPS_SCF 1.0E-3
      MAX_SCF 20
      ORBITALS -3 -1
      SMEAR on
    &END ENFORCE_OCCUPATION
  &END DFT_PLUS_U
&END KIND
\end{verbatim}
It is possible to enforce certain initial (desired) orbital occupations for a number of \texttt{MAX\_SCF} SCF iteration steps or until a certain SCF convergence threshold \texttt{EPS\_SCF} is reached as in the example above for the $5f$ orbitals of the atomic kind U$_\text{a}$ in which the actual $5f$ electron density is smeared equally among the $5f_{-3}$ and $5f_{-1}$ orbitals. In this way, the desired orbital occupation patterns can be ``enforced'' when a straightforward convergence to the electronic ground state does not take place or is not expected due to the presence of metastable states~\cite{Dorado2009,Krack2015}, in which case an initial occupation matrix control (OMC) is required. After an initial SCF run (via \texttt{RUN\_TYPE energy}) has been converged with \texttt{\&ENFORCE\_OCCUPATION on}, a SCF run without any OMC constraint (i.e. \texttt{\&ENFORCE\_OCCUPATION off}) can be restarted using the WF restart file of the initial constraint SCF run.

The ``enforcement'' is performed based on the initial orbital occupation of an atomic kind, which can be changed via the \texttt{\&KIND\%BS} section
\begin{verbatim}
&KIND U_a
  ...
  &BS 
    # U(+4): 5f3 6d1 7s2 -> 5f2 6d0 7s0
    &ALPHA
      N    5  6  7 # 5       6       7
      L    3  2  0 # f       d       s
      NEL +1 -1 -2 # (3+1)/2 (1-1)/2 (2-2)/2
      # U(alpha)   : 2       0       0
    &END ALPHA
    &BETA
      N    5  6  7 # 5       6       7
      L    3  2  0 # f       d       s
      NEL -3 -1 -2 # (3-3)/2 (1-1)/2 (2-2)/2 
      # U(beta)    : 0       0       0
    &END BETA
  &END BS
&END KIND
\end{verbatim}
to create a spin-up on-site $5f^2$ triplet for a U$^{+4}$ cation. For the corresponding spin-down $5f^2$ cation, the definitions in the sections \texttt{\&ALPHA} and \texttt{\&BETA} for spin-up and spin-down electrons have to be exchanged. The \texttt{MULTIPLICITY} in the \texttt{\&DFT} section must be adapted if the number of spin-up and spin-down U atoms is not equal, which is needed for ferri- and ferromagnetic materials. The corresponding initial orbital occupation for the O$^{-2}$ anion in UO$_2$ can be specified with
\begin{verbatim}
&KIND O
  &BS
    # O(-2): 2s2 2p4 -> 2s2 2p6
    &ALPHA
      N    2    # 2
      L    1    # p
      NEL +2    # (4+2)/2
      # O(alpha): 3 
    &END ALPHA
    &BETA
      N    2    # 2
      L    1    # p
      NEL +2    # (4+2)/2
      # O(beta) : 3
    &END BETA    
   &END BS 
\end{verbatim}
Atomic forces are only implemented for the \texttt{PLUS\_U\_METHOD Mulliken} and $\mathbf{k}$-points are not available with DFT+U, yet.


\section{Electronic Band Structure} \label{sec:BandStructure}

Beside the CP2K/\textsc{Quickstep} code, more accurate electronic band structure calculations can be performed using the domain-specific electronic structure library \textsc{SIRIUS}, which implements the FP-LAPW and PP-PW methods~\cite{PhysRevB.12.3060, Ihm1979}, or the \textit{GW} method~\cite{Gunnarsson1998}. 

\subsection{SIRIUS}
\label{sec:SIRIUS}

With the interface to \textsc{SIRIUS}, CP2K has the ability to run FP-LAPW and PP ground state calculations, similar to Wien2k \cite{Wien2k2020}, FLEUR \cite{fleurWeb}, ELK \cite{ELKcode} and exciting \cite{Exciting2014}, to mention just a few. 

SIRIUS supports:
\begin{itemize}
  \item Norm-conserving~\cite{Hamann1979}, ultrasoft~\cite{Vanderbilt1990}, as well as PAW PPs
  \item APW~\cite{slater1937wave} and LAPW~\cite{PhysRevB.12.3060} basis sets with an arbitrary number of local orbitals composed of up to 3rd-order energy derivatives of radial functions
  \item Evaluation of stress tensor and nuclear forces
  \item Collinear and non-collinear magnetism
  \item Symmetrization of lattice-periodic functions and on-site matrices
  \item Generation of irreducible $\mathbf{k}$-point meshes
  \item Local-density approximation (LDA)~\cite{ceperley1980ground} and GGA flavours of XC potentials via the LibXC library~\cite{Lehtola2018}
\end{itemize}

Let us start with the example of PP-PW calculation setup using the second version of unified PP format (UPF) files, which are stored in XML format and can be read by \textsc{SIRIUS} directly. The example below demonstrates the input file for diamond: 
\begin{verbatim}
&FORCE_EVAL
  METHOD SIRIUS
  &PW_DFT
    &MIXER
      BETA 0.5
      USE_HARTREE .TRUE.
    &END MIXER
    &PARAMETERS
    ELECTRONIC_STRUCTURE_METHOD \
      PSEUDOPOTENTIAL
    GK_CUTOFF 5
    PW_CUTOFF 20
    NGRIDK 4 4 4
    SMEARING GAUSSIAN
    SMEARING_WIDTH 0.0004
  &END PARAMETERS
  &END PW_DFT
  &SUBSYS
    &KIND C
      POTENTIAL UPF \ 
        C.pbe-n-kjpaw_psl.1.0.0.UPF
    &END KIND
  &END SUBSYS
&END FORCE_EVAL
\end{verbatim}
With a small change in parameters and atomic kind input sections, it is straightforward to switch to the FP-LAPW mode and compute the same structure with a high-accuracy full-potential method:
\begin{verbatim}
&FORCE_EVAL
  METHOD SIRIUS
  &PW_DFT
    &PARAMETERS
      ELECTRONIC_STRUCTURE_METHOD \ 
        FULL_POTENTIAL_LAPWLO
      DENSITY_TOL 1e-06
      ENERGY_TOL 1e-10
      GK_CUTOFF 4.0
      NGRIDK 4 4 4
      PW_CUTOFF 20
      SMEARING GAUSSIAN
      SMEARING_WIDTH 0.0004
      AUTO_RMT 1
      CORE_RELATIVITY DIRAC
      VALENCE_RELATIVITY IORA
      LMAX_APW 12
      LMAX_POT 12
      LMAX_RHO 12
    &END PARAMETERS
  &END PW_DFT
  &SUBSYS
    &KIND C
      POTENTIAL UPF C.json
    &END KIND
  &END SUBSYS      
&END FORCE_EVAL
\end{verbatim}

Full-potential atomic species JSON files for \textsc{SIRIUS} are available at \url{https://github.com/electronic-structure/species/tree/master/FP/v1}. They are similar to UPF files and contain information about the radial grid and LAPW+lo basis set.
In the following, we provide an overview of the most relevant variables in the {\tt \&PW\_DFT\%PARAMETERS} section:
\begin{itemize}
  \item {\tt ELECTRONIC\_STRUCTURE\_METHOD}: Type of calculation, i.e. {\tt FULL\_POTENTIAL\_LAPWLO}, or {\tt PSEUDOPOTENTIAL}.
  \item {\tt CORE\_RELATIVITY}: Specifies if the core states in FP-LAPW are treated relativistically via the \texttt{DIRAC} equation or not. 
  \item {\tt VALENCE\_RELATIVITY}: Type of relativistic treatment of valence FP-LAPW radial basis functions and local orbitals. Possible values are: {\tt NONE}, {\tt KOELLING\_HARMON}, {\tt ZORA}, or {\tt IORA}.
  \item {\tt NUM\_BANDS}: Number of "bands" (KS states) to compute during Hamiltonian diagonalization.
  \item {\tt SMEARING}: Type of smearing used to compute the Fermi level, such as {\tt GAUSSIAN}, {\tt COLD}, {\tt FERMI\_DIRAC}, {\tt GAUSSIAN\_SPLINE} and {\tt METHFESEL\_PAXTON}.
  \item {\tt SMEARING\_WIDTH}: Width (in Ha) of the smearing function.
  \item {\tt PW\_CUTOFF}: PW cutoff (in $a.u.^{-1}$) for expansion of electron density and potential.
  \item {\tt GK\_CUTOFF}: PW cutoff (in $a.u.^{-1}$) for $|{\bf G+k}|$.
  \item {\tt NUM\_MAG\_DIMS}: Number of magnetic dimensions in the system, i.e. 0 - non-magnetic calculations, 1 - spin-collinear case, 3 - noncollinear magnetic calculation.
  \item {\tt LMAX\_APW}: Maximum orbital quantum number for LAPW basis functions.
  \item {\tt LMAX\_RHO}: Maximum orbital quantum number for charge density expansion in muffin-tins (LAPW only).
  \item {\tt LMAX\_POT}: Maximum orbital quantum number for KS potential expansion in muffin-tins (LAPW only).
  \item {\tt NGRIDK}: Dimensions of the ${\bf k}$-point mesh. In other words, this is a division of the first Brillouin zone into microcells for ${\mathbf k}$-point integration.
  \item {\tt ENERGY\_TOL}: SCF convergence tolerance of total energy.
  \item {\tt DENSITY\_TOL}: SCF convergence tolerance of charge density.
  \item {\tt GAMMA\_POINT}: Specifies if this is a $\Gamma$-point calculation (PP case only). {\tt NGRIDK} must be set to {\tt 1 1 1} in that case.
\end{itemize}
Similarly, in section {\tt \&PW\_DFT\%MIXER}:
\begin{itemize}
  \item {\tt TYPE}: Denotes the type of the density mixer, i.e. {\tt LINEAR}, {\tt ANDERSON}, {\tt ANDERSON\_STABLE}, or {\tt BROYDEN2}.
  \item {\tt BETA}: Determines mixing parameter, which is a real value in the interval $[0, 1]$. 
  \item {\tt USE\_HARTREE}: Logical variable that controls what is used as estimation of charge density difference. {\tt TRUE}: Hartree energy of density residual $\frac{1}{2} \iint  \frac{\Delta \rho({\bf r}) \Delta \rho({\bf r'})}{|{\bf r - r'}|} d{\bf r} d{\bf r'}$ is used as a measure of density convergence (works only for PP calculation). {\tt FALSE}: normalized inner product of density residuals $\frac{1}{\Omega} \int  \Delta \rho({\bf r}) \Delta \rho({\bf r}) d{\bf r}$ is used as a measure of density convergence.
\end{itemize}
Computational aspects of CP2K/SIRIUS are configured in section {\tt \&PW\_DFT\%CONTROL}:
\begin{itemize}
  \item {\tt MPI\_GRID\_DIMS}: A 2-dimensional array that defines how the band parallelization is performed. The first dimension of the grid specifies the size of the FFT communicator used in the transformation of individual WFs. The second dimension defines the number of independent FFT communicator groups used to parallelize FFTs across different WFs. The product of the two dimensions defines the total number of MPI ranks used for the band parallelization. An orthogonal communicator will be built from the total available number of MPI ranks to perform $\mathbf{k}$-point parallelization. Additionally, if the MPI grid is square, for example $\{3, 3\}$, a parallel eigen-solver will be used to diagonalize the subspace Hamiltonian matrix.
\end{itemize}

\subsection{The \textit{GW} Method}
\label{sec:GW}

The \textit{GW} method is particularly well suited for high-accuracy computations of the electronic band structure of solids and MO energies of molecules~\cite{Golze2019}. 
We first describe the band gap problem of DFT to motivate the usage of \textit{GW}, before briefly discussing the theoretical framework of \textit{GW} and our \textit{GW} implementation within CP2K.
We also provide input sections of \textit{GW}   calculations performed with CP2K.  
%

\subsubsection{The Band Gap Problem of Density Functional Theory}
When considering a solid under PBCs, the KS equations read as
\begin{align} 
\left(- \frac{\nabla^2}{2m} + v_\text{H}(\mathbf{r}) + 
v_\text{XC}(\mathbf{r}) \right) \psi_{n\mathbf{k}}(\mathbf{r}) = 
\varepsilon_{n\mathbf{k}}^\text{DFT} \psi_{n\mathbf{k}}(\mathbf{r}), 
\end{align}
where $v_\text{H}(\mathbf{r})$ is the Hartree potential and $v_\text{XC}(\mathbf{r})$ is the XC potential. 
These KS equations are solved to obtain the KS orbitals $\psi_{n\mathbf{k}}(\mathbf{r})$ and the
KS eigenvalues $\varepsilon_{n\mathbf{k}}^\text{DFT}$ with band index $n$ and crystal momentum $\mathbf{k}$. 
For a molecule, the KS orbitals $\psi_n(\mathbf{r})$ and the KS eigenvalues $\varepsilon_n^\text{DFT}$ 
only carry a single quantum number, the MO index $n$.
These KS eigenvalues $\varepsilon_{n\mathbf{k}}^\text{DFT}$ are often used to approximate 
the electronic band structure of a solid. But, this approximation comes with limitations.

First, when using one of the common GGA XC functionals, the band gap in the KS-DFT
band structure $\varepsilon_{n\mathbf{k}}^\text{DFT}$ is much too small compared to experimental band gaps~\cite{Golze2019}.
Even with the exact XC functional, KS-DFT underestimates the fundamental gap due to the derivative discontinuity~\cite{capelle2006bird}. 

Second, the GGA band structure $\varepsilon_{n\mathbf{k}}^\text{DFT}$ is insensitive to 
screening by the environment. As an example, the GGA eigenvalue gap
$\varepsilon_{\text{LUMO}}^\text{DFT}-\varepsilon_{\text{HOMO}}^\text{DFT}$ between the lowest unoccupied MO (LUMO) and the highest occupied MO (HOMO) is almost identical for a molecule in the gas phase and a molecule on a surface.
In experiment, however, the surface induces an image-charge effect, which can reduce this
HOMO-LUMO gap of the molecule by several eV compared to the gas phase.
In more general terms, this is a nonlocal screening effect by the surface that is absent in common approximate GGA XC functionals. A similar band gap reduction due to nonlocal screening is
present when two materials are brought close to each other, for example, when two sheets
of atomically thin materials are stacked on top of each other.

Third, one might use hybrid XC functionals to obtain band gaps from KS eigenvalues that
align more closely with experimental values than GGA band gaps~\cite{caravati2009first, caravati2011first, los2013first, gabardi2016influence}. However, the
issue remains that nonlocal screening effects by the environment are not included in hybrid functionals.
The above issues are known as the band gap problem of DFT.

\subsubsection{Theory of \textit{GW} Band Structure Calculations}
Green's function theory offers a framework for calculating electron removal and addition energies, known as
quasiparticle energies. Hedin’s equations provide an exact method for computing these quasiparticle energies within Green's function theory~\cite{HedinGW}. The \textit{GW} approximation simplifies Hedin's equations by approximating the self-energy~$\Sigma$
as the product of the Green's function \textit{G} and the screened Coulomb interaction \textit{W}, i.e. 
\begin{align} 
\Sigma^{GW}(\mathbf{r}_1,\mathbf{r}_2,t)= iG(\mathbf{r}_1,\mathbf{r}_2,t)W(\mathbf{r}_1,\mathbf{r}_2,t).\label{eq-GW}\end{align}
The advantage of \textit{GW} is that it captures nonlocal screening effects
on the electronic band structure, as previously discussed, and that band gaps computed 
by \textit{GW} are often in excellent agreement with experiments. 

\textit{GW} calculations in CP2K start from a KS-DFT calculation, i.e. 
we assume that the above KS equations have been solved. In the $G_0W_0$ approach, 
we use KS orbitals and their KS eigenvalues to compute the Green's function $G_0$ of non-interacting 
electrons and the screened Coulomb interaction $W_0$ in the RPA. 
The $G_0W_0$ self-energy $\Sigma^{G_0W_0}(t)$ is then obtained by replacing $G\,{\rightarrow}\,G_0$ and $W\,{\rightarrow}\,W_0$ in Eq.~\ref{eq-GW}, followed by a Fourier transform from time $t$ to frequency (or energy)~$\varepsilon$, yielding $\Sigma^{G_0W_0}(\varepsilon)$.
In the $G_0W_0$ method we further approximate that KS orbitals are the quasiparticle WFs.
Then, the $G_0W_0$ quasiparticle energies are 
\begin{align} \varepsilon_{n\mathbf{k}}^{G_0W_0} = \varepsilon_{n\mathbf{k}}^\text{DFT} + \langle\psi_{n\mathbf{k}}|
\Sigma^{G_0W_0}(\varepsilon_{n\mathbf{k}}^{G_0W_0}) - v_\text{XC}|\psi_{n\mathbf{k}}\rangle\, .
\end{align}
We might interpret this as removing the spurious XC contribution of DFT, i.e. $\langle{\psi_{n\mathbf{k}}|
 v_\text{XC}|\psi_{n\mathbf{k}}}\rangle$, from the DFT eigenvalue $\varepsilon_{n\mathbf{k}}^\text{DFT}$ and adding
the XC contribution from $G_0W_0$, i.e. $\langle{\psi_{n\mathbf{k}}|
\Sigma^{G_0W_0}(\varepsilon_{n\mathbf{k}}^{G_0W_0}) |\psi_{n\mathbf{k}}}\rangle$. 

CP2K also allows performing eigenvalue self-consistency in $G$ (ev\textit{GW}$_0$) and 
eigenvalue self-consistency in $G$ and in $W$ (ev\textit{GW}), where especially the $G_0W_0$ quasiparticle energies can be strongly influenced by the DFT XC functional.
In general, we recommend using ev\textit{GW}$_0$, starting from the PBE functional~\cite{Perdew1996}, for both molecules and solids, motivated by the discussion in Ref.~\citenum{Schambeck2024}.


CP2K contains three different \textit{GW} implementations:
 \textit{GW} for molecules~\cite{Wilhelm2016}, \textit{GW} for computing the band structure of a solid with a small unit cell with $\mathbf{k}$-point
sampling in DFT~\cite{smallcellGW}, and
 \textit{GW} for computing the band structure of a large cell in a $\Gamma$-only approach~\cite{Graml2024a}.
In the following, we will discuss the details and usage of all of these \textit{GW} implementations.

\subsubsection{\textit{GW} for Molecules}
To start a $G_0W_0$, ev$GW_0$ or ev$GW$ calculation for a molecule, one needs to set the keyword {\tt RUN\_TYPE ENERGY} in the \texttt{\&GLOBAL} section and the following segment: 
\begin{verbatim}
&XC
  &XC_FUNCTIONAL PBE
  &END XC_FUNCTIONAL
  &WF_CORRELATION
    &RI_RPA
      QUADRATURE_POINTS 100
      &GW
        ! Also possible: EV_GW0 or EV_GW
        SELF_CONSISTENCY G0W0 
      &END GW
    &END RI_RPA
  &END WF_CORRELATION
&END XC
\end{verbatim}
Therein, the following keywords have been used:
\begin{itemize}
\item {\tt QUADRATURE\_POINTS}: Number of imaginary-frequency points for computing the self-energy, see Eq.~(21) in Ref.~\citenum{Wilhelm2016}. Usually, 100 points converge the quasiparticle energies within 10~meV.
\item {\tt SELF\_CONSISTENCY}:
  Determines which $GW$ self-consistency variant ($G_0W_0$, ev$GW_0$ or ev$GW$) is used to calculate the $GW$ quasiparticle energies.
\end{itemize}
The numerical precision of the $GW$ implementation is 10~meV compared to reference calculations, for
example, on the $GW$100 test set~\cite{vanSetten2015,Schambeck2024}. 
Furthermore, the following DFT settings will also have an influence on $GW$ quasiparticle energies:
\begin{itemize}
\item {\tt XC\_FUNCTIONAL}: Starting XC functional for the $G_0W_0$, ev$GW_0$ or ev$GW$ calculation, respectively. We recommend to use ev$GW_0$@PBE, as discussed in Ref.~\citenum{Schambeck2024}. For further
guidance on selecting an appropriate DFT starting functional and self-consistency scheme for your
system, you may consult Ref.~\citenum{Golze2019}.
\item {\tt BASIS\_SET}: The basis set is of Gaussian type and
can have a strong influence on the quasiparticle energies. For computing quasiparticle energies, a basis set extrapolation is necessary~\cite{Wilhelm2016}, and we recommend
 all-electron GAPW calculations with correlation-consistent basis sets {cc-pVDZ, cc-pVTZ, cc-pVQZ} from the EMSL Basis Set Exchange~\cite{Pritchard2019}. For computing the HOMO-LUMO gap from $GW$, we recommend the usage of augmented basis sets, for example {aug-cc-pVDZ}  and {aug-cc-pVTZ}.
  As {\tt RI\_AUX} basis set, we recommend the  RIFIT basis sets from the EMSL database, for example {aug-cc-pVDZ-RIFIT}.
  \end{itemize}

The computational effort of the $GW$ calculation increases with $\mathcal{O}(N^4)$, where $N$ is the system size. 
The memory requirement increases with $\mathcal{O}(N^3)$. For large-scale calculations, we
recommend starting with a small molecule. After successfully completing the $GW$ calculation for the
small molecule, you can gradually increase the size of the molecule. The computational resources needed for
larger molecules can then be estimated using the $\mathcal{O}(N^4)$ scaling for computation time and
$\mathcal{O}(N^3)$ scaling for memory. The output provides a useful lower limit of the required memory, e.g.
{\footnotesize
\begin{verbatim}
Total memory for (ia|K) integrals:   90556 MiB
Total memory for GW-(nm|K) integrals: 4626 MiB
\end{verbatim}
}
When facing an out-of-memory error, please increase the number of nodes for your calculation.
%

\subsubsection{\textit{GW} for Small Unit Cells with $\mathbf{k}$-point Sampling}
For a periodic $GW$ calculation, $\mathbf{k}$-point sampling is required. In the following example of a 2D-periodic cell, $\mathbf{k}$-point sampling is included in
the DFT section:
\begin{verbatim}
&DFT
  ...
  &KPOINTS
    SCHEME MONKHORST-PACK 32 32 1
    PARALLEL_GROUP_SIZE -1
  &END KPOINTS
&END DFT
\end{verbatim}
The $\mathbf{k}$-point mesh size is a convergence parameter, and 32$\times$32 is expected to reach convergence of the $GW$ band gap within 10~meV for a 2D material.

A periodic $GW$ calculation is activated via the \texttt{\&BANDSTRUCTURE} section, e.g. 
\begin{verbatim}
&PROPERTIES
  &BANDSTRUCTURE
    &GW
      NUM_TIME_FREQ_POINTS      30
      MEMORY_PER_PROC           10
      EPS_FILTER           1.0E-11
      REGULARIZATION_RI     1.0E-2
      CUTOFF_RADIUS_RI         7.0
    &END GW
    &SOC
    &END SOC
    &BANDSTRUCTURE_PATH
      NPOINTS 19
      SPECIAL_POINT K     1/3 1/3 0.0
      SPECIAL_POINT GAMMA 0.0 0.0 0.0
      SPECIAL_POINT M     0.0 0.5 0.0
    &END BANDSTRUCTURE_PATH
  &END BANDSTRUCTURE
&END PROPERTIES
\end{verbatim}
All parameters from above have been chosen to converge the $GW$ band gap within 10~meV, see also convergence tests in Ref.~\citenum{smallcellGW}:
\begin{itemize}
\item {\tt NUM\_TIME\_FREQ\_POINTS}:
  Number of imaginary-time and imaginary-frequency points used for computing the self-energy.
  Between 20 and 30 points are usually enough for converging quasiparticle energies within 10~meV.
  Grids up to 34 points are available.
\item {\tt MEMORY\_PER\_PROC}:
  Specifies the available memory per MPI process. A larger
  {\tt MEMORY\_PER\_PROC} can increase performance.
\item {\tt EPS\_FILTER}: Filter for
  three-center integrals, $10^{-11}$ should be well-converged.
\item {\tt REGULARIZATION\_RI}:
  Regularization parameter for RI basis set. For a big RI basis set (> 50
  RI function per atom), we recommend $10^{-2}$ to prevent linear dependencies. For a small
  RI basis set, one can turn RI regularization off by setting 0.0.
\item {\tt CUTOFF\_RADIUS\_RI}:
  Cutoff radius of truncated Coulomb metric in Å. A larger cutoff leads to faster RI basis set
  convergence, but also the computational cost increases. A cutoff of 7~Å is an accurate choice.
\item {\tt \&SOC}: Activates spin-orbit coupling
  (SOC) from GTH PPs~\cite{Hartwigsen1998}. Hence, the usage of SOC
  also needs {\tt POTENTIAL\_FILE\_NAME  GTH\_SOC\_POTENTIALS}.
\item {\tt \&BANDSTRUCTURE\_PATH}:
  Specify the $\mathbf{k}$-path in the Brillouin zone for computing the band structure. Relative
  $\mathbf{k}$-coordinates are needed, which you can retrieve for your crystal structure from Ref.~\citenum{Setyawan2010}.
\end{itemize}
We recommend the TZVP-MOLOPT basis sets together with GTH PPs~\cite{smallcellGW}. At present, 2D PBCs are supported, while 1D- and
3D PBCs are work in progress.

The $GW$ band structure is written to the files {\tt bandstructure\_SCF\_and\_G0W0} and
{\tt bandstructure\_SCF\_and\_G0W0\_plus\_SOC}, respectively. The direct and indirect band gaps are also listed in the CP2K
output file. When facing an out-of-memory crash, please increase {\tt MEMORY\_PER\_PROC}. 

\subsubsection{\textit{GW} for Large  Cells in $\Gamma$-only Approach}
 For a large unit cell, a $\Gamma$-only $GW$ algorithm is available in CP2K. The requirement
on the cell is that elements of the density matrix decay by several orders of magnitude when the two
basis functions of the matrix element have a distance similar to the cell size. As a rule of thumb,
for a 2D material, a 9$\times$9 unit cell is large enough for the $\Gamma$-only algorithm~\cite{Graml2024a}.

The input file for a $\Gamma$-only $GW$ calculation is identical to that of $GW$ for small cells with $\mathbf{k}$-point
sampling except that the {\tt \&KPOINTS} section in DFT needs to be removed.
The computational parameters from such an input file reach a 
numerical convergence of the band gap within $\sim50$~meV (TZVP basis set, 10 time and frequency
points)~\cite{Graml2024a}. The code prints
restart files with ending \texttt{.matrix} that can be used to restart an interrupted calculation.
\section{Embedding Methods} \label{sec:Embedding}

In this section, multiscale methods that permit embedding part of a system, which is described using a quantum mechanical electronic structure method, into a surrounding environment at a lower level of theory, are presented and discussed in detail. However, instead of the simple mechanical coupling approach, which can be easily implemented using the \texttt{\&FORCE\_EVAL\%MIXED} framework~\cite{schusteritsch2016effect} and adaptive resolution simulation methods \cite{praprotnik2008multiscale, cortes2021adaptive, panahian2024physical}, the focus here will be on more sophisticated methods including non-trivial interactions with the environment. 

\subsection{Implicit Solvation Methods}
\label{sec:SCCS}
The simulation of a solute like a molecule or a cluster with an explicit solvent captures all the details of the solute-solvent interaction, as well as the dynamics and fluctuations within the solvent. However, often embedding the solute within an implicit solvent is sufficient, especially when only an approximate (averaged) inclusion of polarization effects by dielectric screening is required. Representing the solvent as a continuous dielectric medium significantly reduces the number of degrees of freedom and avoids sampling of potentially uninteresting fluctuations within the solvent, which can also lead to great computational time savings. Moreover, the preparation and equilibration of a solvated system in a simulation box becomes also unnecessary. A large number of continuum solvation models can be found in the literature \cite{Tomasi2005}, but these models usually employ a rigid cavity for the solute, introducing a discontinuity in the dielectric function at the vacuum-cavity interface. Such discontinuities cause kinks in the atomic forces which are detrimental in structure relaxations and AIMD simulations. Fattebert and Gygi proposed a smooth dielectric function based on the electron density to palliate the problem~\cite{Fattebert2002,Fattebert2003}. In their approach, the response to the electronic density change is self-consistently taken into account in each SCF iteration step, introducing a nested SCF loop within each SCF WF optimization step, which is the price to pay for getting rid of the explicit solvent molecules. This self-consistent continuum solvation (SCCS) model of Fattebert-Gygi and its revised form of Andreussi et al.~\cite{Andreussi2012}, are both implemented in CP2K~\cite{Yin2017,Yin2015}.
The SCCS input parameters are defined in the \texttt{\&FORCE\_EVAL\%DFT\%SCCS} section. An SCCS input snippet for a molecule immersed in water ($\varepsilon_0 = 78.36$) looks like:
\begin{verbatim}
&DFT
  &SCCS on
    ALPHA [mN/m] 50.0
    BETA [GPa] -0.35
    DELTA_RHO 2.0E-5
    DERIVATIVE_METHOD CD5 # or FFT
    DIELECTRIC_CONSTANT 78.36
    EPS_SCCS 1.0E-9
    GAMMA [mN/m] 0.0
    EPS_SCF 1.0E-3
    MAX_ITER 100
    METHOD Andreussi # or Fatteberg-Gygi
    MIXING 0.6
    &ANDREUSSI
      RHO_MAX 0.005
      RHO_MIN 0.0001
    &END ANDREUSSI
    &FATTEBERT-GYGI
      BETA 1.7
      RHO_ZERO 6.0E-4
    &END FATTEBERT-GYGI
  &END SCCS
  ...
&END DFT
\end{verbatim}
The values for the keywords \texttt{RHO\_MAX} and \texttt{RHO\_MIN} determine the smoothing of the dielectric function for the revised SCCS method of Andreussi et al.~\cite{Andreussi2012}. Likewise, \texttt{BETA} and \texttt{RHO\_ZERO} are the smoothing parameters for the original SCCS method of Fattebert and Gygi~\cite{Fattebert2002}. They are solute dependent, and especially charged solutes require different values for the revised SCCS model \cite{Dupont2013}.
While for neutral solutes 0D, 2D, 1D, and 3D PBCs can be applied, charged systems should be run with 0D PBC using a Poisson solver like MT \cite{Martyna1999}:
\begin{verbatim}
&DFT
  &POISSON
    PERIODIC none
    POISSON_SOLVER MT
  &END POISSON
  ...
&END DFT
\end{verbatim}
while keeping the periodicity defined in the \texttt{\&CELL} section identical, i.e.
\begin{verbatim}
&SUBSYS
  &CELL
    PERIODIC none
    ...
  &END CELL
  ...
&END SUBSYS
\end{verbatim}
The first WF optimization steps are performed without SCCS. The SCCS input parameter \texttt{EPS\_SCF} determines the SCF convergence threshold for the onset of the nested (inner) self-consistent iteration of the polarization potential, and \texttt{EPS\_SCCS} defines the corresponding convergence threshold for the requested numerical accuracy, performing a maximum of \texttt{MAX\_ITER} SCCS iteration steps.

The keywords \texttt{ALPHA}, \texttt{BETA}, and \texttt{GAMMA} are optional parameters for additional solvation model terms
besides the electrostatic contribution $\Delta G^\text{el}$ 
defining the solvation free energy
\begin{equation}
\Delta G^\text{sol} = \Delta G^\text{el} + G^\text{cav} + G^\text{rep} + G^\text{dis}
\end{equation}
with a cavitation term
\begin{equation}
G^\text{cav} = \gamma\,S,
\end{equation}
which can be computed based on the ``quantum surface'' introduced by Cococcioni et al.~\cite{Cococcioni2005}. The latter reads as  
\begin{eqnarray}
 S &=& \int\left\lbrace\vartheta_{\rho_0 - \frac{\Delta}{2}}\left[\rho^\text{elec}(\mathbf{r})\right] - 
       \vartheta_{\rho_0 + \frac{\Delta}{2}}\left[\rho^\text{elec}(\mathbf{r})\right]\right\rbrace \nonumber \\
   & & \times\frac{|\nabla\rho^\text{elec}|}{\Delta}\,d\mathbf{r},    
\end{eqnarray}
where $\Delta$ and $\gamma$ are defined by the input parameters \texttt{DELTA\_RHO} and \texttt{GAMMA}, respectively.

The terms accounting for the Pauli repulsion $G^\text{rep}$ and the dispersion interactions $G^\text{dis}$ 
can be computed based on the surface $S$ and the volume $V$ of the solute cavity, i.e. 
\begin{equation}
 G^\text{rep} + G^\text{dis} = \alpha\,S + \beta\,V.
\end{equation}
The pre-factors \texttt{ALPHA} and \texttt{BETA} are solvent-specific tunable parameters. This ultimately allows the calculation of a solvation free energy
\begin{equation}
 \Delta G^\text{sol} = \Delta G^\text{el} + (\alpha + \gamma)\,S + \beta\,V.
\end{equation}
Finally, the SCCS printout is controlled by the input section \texttt{FORCE\_EVAL\%DFT\%PRINT\%SCCS}:
\begin{verbatim}
&PRINT
  &SCCS ON
    &EACH
      QS_SCF 0
    &END EACH
    &POLARISATION_POTENTIAL off
      &EACH
        QS_SCF 0
      &END EACH
    &END POLARISATION_POTENTIAL
    ...
  &END SCCS
  ...
&END PRINT
\end{verbatim}
allowing for printing the polarization potential and the dielectric function in cube file format.

\subsection{Quantum Mechanics/Molecular Mechanics Methods}

In quantum mechanics/molecular mechanics (QM/MM) hybrid schemes~\cite{Warshel1976}, the system is divided into two subsystems: one is treated at the quantum mechanical (QM) level of theory and the other at the molecular mechanics (MM) level. In CP2K, an additive QM/MM scheme is implemented, which partitions the QM/MM total energy $E_{\text{tot}}^{\text{QM/MM}}$ into the following three terms
\begin{equation}
 \begin{split}
 E_{\text{tot}}(\mathbf{R}_{\alpha},\mathbf{R}_a) &= E_{\text{QM}}(\mathbf{R}_{\alpha}) +  E_{\text{MM}}(\mathbf{R}_a)\\ &+ E_{\text{QM/MM}}(\mathbf{R}_{\alpha},\mathbf{R}_a), 
 \end{split}
\end{equation}
where $\mathbf{R}_{\alpha}$ and $\mathbf{R}_a$ label the coordinates of the QM and MM nuclei, respectively. The QM contribution is evaluated using the \textsc{Quickstep} module~\cite{VandeVondele2005}, typically employing KS-DFT. Alternatively, SQC methods, such as PM6~\cite{Stewart2007}, self-consistent-charge density-functional tight-binding (SCC-DFTB)~\cite{Elstner1998}, or extended tight-binding (xTB) approaches can be used~\cite{Grimme2017}. MM contributions are obtained from classical force fields, either using the built-in \textsc{Fist} program, or by an interface to \textsc{Gromacs}~\cite{Abraham2015,gromacs_qmmm_2022}. 

The QM/MM energy includes non-bonded terms and, if the QM system is connected to the MM region by covalent bonds, bonded terms as well. Non-bonded terms comprise van der Waals (vdW) and electrostatic (elec) interactions, i.e.
\begin{align}
    E_{\text{QM/MM}} &= E_{\text{QM/MM}}^{\text{bonded}} + E_{\text{QM/MM}}^{\text{non-bonded}} \\
   &=  E_{\text{QM/MM}}^{\text{bonded}} + E_{\text{QM/MM}}^{\text{vdw}} + E_{\text{QM/MM}}^{\text{elec}}. \nonumber
\end{align}
Speaking about bonded terms, special care has to be taken with respect to unsaturated bonds that are occurring when cutting bonds of covalently coupled subsystems. For that purpose, CP2K offers two general options. 
The simplest approach is to saturate the dangling bond of the QM atom with a monovalent atom (link atom), typically a hydrogen atom. The QM calculations are then performed on the QM atoms plus the link atom. The interaction between the covalently bound QM and MM atom is modeled with classical MM potentials. This approach is often termed the link-atom technique in the literature and referred to as the integrated MO molecular mechanics (IMOMM) method in CP2K~\cite{Senn2009}. Alternatively, a linking atom with a monovalent PP can replace the MM frontier atom at the QM/MM boundary~\cite{ZhangYang1999, Senn2009, ihrig2011specific, schiffmann2011artificial}.

Turning to the non-bonded terms, the vdW interactions between QM and MM atoms are often described using a simple Lennard-Jones potential~\cite{Senn2009}. The treatment of electrostatic interactions, however, is the most challenging aspect of the QM/MM Hamiltonian and can be approached with varying levels of complexity~\cite{Senn2009}. Generally, we distinguish between three different approaches, which are all available in CP2K. The simplest approach is a mechanical embedding scheme, in which pre-calculated partial atomic charges are assigned to both QM and MM atoms. The charge-charge interactions are then computed using classical expressions, neglecting the response of the QM charge density to the electrostatic potential generated by the MM charges. The most popular scheme, and the default treatment in CP2K, is electrostatic embedding~\cite{TLFMALMP:2005,TLFMALMP:2006}. In this approach, an additional term is added to the QM Hamiltonian, allowing the MM charges to polarize the QM region. For QM calculations at the KS-DFT level with GPW and GAPW, CP2K implements a particularly efficient variant of electrostatic embedding called Gaussian expansion of the electrostatic potential (GEEP), which is explained in more detail below. 
The next level of sophistication is polarized embedding schemes, where the QM atoms can polarize the QM part non-self-consistently or fully self-consistently. CP2K incorporates polarized embedding schemes to some extent. For example, it offers a fully self-consistent image-charge augmented QM/MM (IC-QM/MM) scheme~\cite{Golze2013}, specifically suited for adsorbate/metal systems. A non-self-consistent core-shell model, which can be applied to a broad range of systems, is also available~\cite{Devynck2012}. 

A minimal example of a QM/MM calculation in a periodic cubic simulation box of length 25~Å, where atoms with indices 1 to 6 are described at the KS-DFT level, looks as follows:
\begin{verbatim}
&FORCE_EVAL
  METHOD QMMM
  ...
  &QMMM
    ECOUPL Coulomb
    &CELL
      ABC 25.0 25.0 25.0
      PERIODIC XYZ
    &END CELL
    &QM_KIND C
      MM_INDEX 1..6
    &END QM_KIND
    &PERIODIC 
      &MULTIPOLE OFF
      &END MULTIPOLE
    &END PERIODIC
 &END QMMM
&FORCE_EVAL
\end{verbatim}
In addition to that, the {\tt DFT} section must be defined for the QM part of the calculation, as well as the {\tt MM} section with all classical force field parameters, including the classical interactions between the QM and MM subsystems, such as the vdW interactions. The relevant keywords and subsections within the \texttt{\&QMMM} section are:
\begin{itemize}
    \item {\tt ECOUPL}: Defines the type of electrostatic treatment. On the one hand {\tt GAUSS} enables the electrostatic embedding for KS-DFT/MM simulations based on GEEP (see next section). On the other hand {\tt COULOMB} enables electrostatic embedding for SQC methods, or SCC-DFTB/xTB. Mechanical embedding schemes can be used by setting the keyword to {\tt NONE}. 
    \item {\tt CELL}: Defines the cell for the QM calculation, which must be orthorhombic. 
    \item{\tt QM\_KIND}: Defines the QM system and the corresponding indices are given for each atomic type separately.
    \item {\tt PERIODIC}: Applies the periodic potential. The {\tt MULTIPOLE} section turns on the coupling/recoupling of the QM periodic images, as described later. In the previous example, it was turned off because the QM box and the MM box are identical in size.
\end{itemize}
In our simple example, no covalent bonds are cut through the subsystem boundaries. For cases where covalent bonds are cut, the corresponding techniques and link atoms can be set through the {\tt \&LINK} section.

\subsubsection{Electrostatic Embedding by the Gaussian Expansion of the Electrostatic Potential Method} 

The calculation of electrostatics terms in QM-MM poses problems related to both short-range and long-range behavior.
The problems of short-range behavior are related to the electron spill-out. Classical atoms are normally represented by a simple point charge. If a QM atom comes close, the electrons can be trapped into the point-like classical potential energy source.
The use of diffuse basis sets (or PWs) can enhance this behavior.
The ad-hoc generated pseudopotential-like approach has often been applied to remedy the unphysical problem of overpolarization that arises when MM atoms are in close contact with the QM region. This approach requires the definition of frontier MM and frontier QM atoms, which could result in being cumbersome and not sufficiently adaptable to the dynamic evolution of the system. 
In CP2K, the spill-out is avoided by assigning to all MM atoms a finite-width charge density in the form of a Gaussian distribution 
\begin{equation}
n({\bf r}, {\bf R}_{\text{MM}}) = \left( \frac{r_{c,\text{MM}}}{\sqrt{\pi}}\right)^3e^{-(|{\bf r}-{\bf R}_{\text{MM}}|/r_{c,\text{MM}})^2}.
\end{equation}
The width of the charge density depends on the atom type and can be expected to be similar to the covalent radius. The exact potential originating from the Gaussian charge distribution, as commonly employed in the Ewald method, is 
\begin{equation}
v_{\text{MM}}({\bf r}, {\bf R}_{\text{MM}}) = \frac{\text{Erf}\left(\frac{|{\bf r}-{\bf R}_{\text{MM}}|}{r_{c,\text{MM}}}\right)}{|{\bf r}-{\bf R}_{\text{MM}}|}.
\end{equation}
This potential tends to $1/r$ at large distances and converges smoothly to a constant at zero. The electrostatic interaction between the QM and the MM atoms can then be computed onto the QM grid by multiplying the QM charge and this potential. 
For a more efficient calculation, though, the MM electrostatic potential is decomposed in a series of $N_g$ Gaussians of different cutoffs plus a residual, which is why we call this algorithm GEEP~\cite{TLFMALMP:2005}. Hence, \begin{equation}
\frac{\text{Erf}\left(\frac{r}{r_{c}}\right)}{r} = \sum_{N_g} A_g e^{-(r/G_g)^2}+R_{\text{low}}(r).
\end{equation}
In this expression, $A_g$ is the Gaussian amplitude, $G_g$ its width, and $R_{\text{low}}$ is the smooth residual function that can be mapped onto a grid with spacing one order of magnitude larger than the one needed for the potential. The advantage of adopting the Gaussian expansion is that commensurate grids with different spacing can be used for the different contributions, i.e. sharper Gaussians to finer grids and coarser grids for smoother components. All contributions are finally interpolated onto the finest QM grid, by means of real-space splines, and summed up. The interpolation depends only on the number of grid points and not on the number of MM atoms, so
\begin{verbatim}
&QMMM
  ECOUPL GAUSS
  USE_GEEP_LIB 9
  &MM_KIND H
    RADIUS 0.44
  &END MM_KIND
  &MM_KIND O
    RADIUS 0.78
  &END MM_KIND
  &QM_KIND C
    MM_INDEX 1..6
  &END QM_KIND
&END QMMM
\end{verbatim} 

The treatment of long-range forces in conjunction with PBC is much less well established for QM/MM and
most of the implementations use a truncation scheme, neglecting interactions beyond a certain cutoff. The reaction field approach combines truncation and polarizable continuum beyond a given cutoff. 
Ewald techniques are usually implemented only for QM-MM interactions and not for QM-QM, though
long-range QM-QM may play a significant role, in particular for materials science applications.
The electrostatic energy in CP2K is then composed of three terms, i.e. $E_{\text{ES}} = E^{\text{MM}}+E^{\text{QM}}+E^{\text{QM/MM}}$.
The first term is evaluated using standard techniques, such as particle-particle or particle-mesh schemes.
The second term is the evaluation of the energy of the QM subsystem. Since the total energy of the QM subsystem is usually evaluated using a smaller cell, care needs to be taken to include the correct electrostatics. The last term is the evaluation of the periodic electrostatic potential $v_{\text{MM}}({\bf r}, {\bf R}_{\text{MM}})$ discussed above, divided into a real-space contribution and a periodic correction. 
The real-space term contains the interactions due to the short-range part of the electrostatic potential of the MM charges. Only MM atoms close to the QM region will contribute to this term. Since we use the Gaussian expansion plus the residual term, the radius of the Gaussian is such that only a few terms in the lattice sum are needed. The effect of the periodic replicas is only in the long-range term and it comes entirely from the residual function, thus 
\begin{verbatim}
&QMMM
  ECOUPL GAUSS
  USE_GEEP_LIB 6
  &PERIODIC
    GMAX 0.5
    NGRIDS  20 20 20
    REPLICA  2
    &MULTIPOLE
      EWALD_PRECISION 0.00000001
      RCUT 8.0
      ANALYTICAL_GTERM
    &END MULTIPOLE
  &END PERIODIC
&END QMMM
\end{verbatim} 
The residual function is represented in the Fourier space and can be evaluated analytically. Furthermore, since the long-range contribution is very smooth, the Fourier transform is zero for all $G$ vectors larger than a well-defined maximum.
In this example, \texttt{G\_MAX} specifies the maximum $G$ vector in reciprocal space for the Ewald sum and depends strongly on the number of Gaussian functions used in the GEEP scheme. The number of grid points used for the interpolation of the $G$-space term is given by \texttt{NGRIDS}, whereas \texttt{REPLICA} is the number of cell replicas to take into consideration for the real-space sum.
Usually, 3-4 grid levels are used corresponding to a speed-up of $10^2$ times faster than the simple collocation algorithm (interpolations and restrictions account for a negligible amount of computing time). Since the residual function is different from zero only for a few vectors, the sum in reciprocal space is restrained to a few points. Possible sources of error are the cutoff of the finest grid level to properly map the sharpest Gaussian functions, the cutoff of the coarse grid level related to the cutoff of the long-range residual function, and the error in cubic spline interpolation.

When computing the QM electrostatic term, unless the quantum box and the MM box have the same dimensions, the QM images, which are interacting by PBC implicitly in the evaluation of the Hartree potential, have the wrong periodicity. To correct for this error, CP2K implements the so-called Bl\"ochl scheme for decoupling and recoupling according to the correct periodicity~\cite{Bloechl1995, laino2006}. Spherical Gaussians, which are QM atom centered and reproduce the correct multiple expansion, are derived by a density fitting scheme in $G$-space. They can be more than just one per atom site with different exponents. These charges reproduce the correct long-range electrostatics and are used for the decoupling and re-coupling procedure. Indeed, since the electrostatic interaction of separated charge distributions (the array of periodic QM charge densities) depends only on its multipole moments, the model charge density is used to modify the Hartree potential and cancel the electrostatic interactions between the periodic images. 
The \texttt{MULTIPOLE} section sets up this scheme and is activated by default for periodic calculations. It should be switched off when QM and MM boxes are the same size, to avoid unnecessary computational costs.

\subsubsection{Image-charge Augmented Quantum Mechanics/Molecular Mechanics Method}
The IC-QM/MM approach is intended to simulate adsorbed molecules on metallic surfaces~\cite{Golze2013}. The molecular adsorbates are described at the QM level, specifically using KS-DFT, while the metal is treated at the MM level. In the IC-QM/MM scheme, as implemented in CP2K, the electrostatic response of the metal to the presence of the adsorbate is described by an image charge distribution 
\begin{equation}
\rho_m(\mathbf{r})= \sum_{a}{c_a g_a(\mathbf{r,R}_a)},
\end{equation}
where $\mathbf{R}_a$ is the position of the metal atom $a$ and $g_a$ represents a Gaussian function centered at $a$. The charge distribution $\rho_m$ generates the potential $V_m$, while the charge distribution of the adsorbates produces the electrostatic potential $V_e$. The coefficients $c_a$ are determined self-consistently by enforcing the constant-potential condition, where $V_m(\mathbf{r})$ screens $V_e(\mathbf{r})$ within the metal such that $V_e(\mathbf{r}) + V_m(\mathbf{r}) = V_0$. Therein, $V_0$ denotes a constant potential, which is typically zero unless an external potential is explicitly applied.

To run an IC-QM/MM calculation, the following \texttt{\&IMAGE\_CHARGE} subsection must be added: 
\begin{verbatim}
&QMMM
  ...
  &IMAGE_CHARGE
    MM_ATOM_LIST 1..576
    EXT_POTENTIAL 0.0
  &END IMAGE_CHARGE
&END QMMM
\end{verbatim}
In an IC-QMM/MM calculation, the MM system is typically the whole metallic slab. Hence, QM and MM cells are thus required to have the same size. The relevant keywords are:
 \begin{itemize}
    \item {\tt MM\_ATOM\_LIST}: Defines the list of MM atoms that carry the Gaussian charge $g_a$. These should be the atoms of the metallic slab. 
    \item {\tt EXT\_POTENTIAL}: Sets the external potential $V_0$.
 \end{itemize}
There are several other keywords related to print options and to acceleration techniques of an IC-QM/MM calculation for specific run types, see online tutorial for more details~\cite{icqmmm_tutorial}. Note that the {\tt IMAGE\_CHARGE} section only accounts for the electrostatic interaction between molecules and metal atoms. Other contributions, such as vdW interactions, must be defined separately.

\subsubsection{Partial Atomic Charges from Restrained Electrostatic Potential Fitting}
Restrained electrostatic potential (RESP) fitting is a widely used method to determine partial atomic point charges ${q_a}$~\cite{Bayly1993}. In QM/MM simulations, these charges are required for the MM atoms when using the default electrostatic embedding of CP2K, and for both QM and MM atoms when using the less common mechanical embedding. 

These charges are determined such that the potential $V_{\text{RESP}}$ generated by ${q_a}$  reproduces a given QM potential $V_{\text{QM}}$ within a specified region of space. In practice, this is achieved by minimizing the residual $R_{\text{esp}}$ through a least-squares fitting procedure for a set of predefined real-space grid points ${\mathbf{r}_k}$, where $R_{\text{esp}}$ is given by
\begin{equation}
R_{\mathrm{esp}}=\frac{1}{N}\sum_k^N{(V_{\mathrm{QM}}(\mathbf{r}_k)-V_{\mathrm{RESP}}(\mathbf{r}_k))^2}, 
\label{eq:residual_esp}
\end{equation}
where $N$ denotes the total number of real-space grid points $\mathbf{r}_k$. 
The resulting ${q_a}$ are treated as point charges in non-periodic RESP fittings. For periodic systems, however, the charges are represented by Gaussian functions $g_a$ of fixed width, centered on atom $a$, resulting in the charge distribution 
\begin{equation}
\rho_{\text{RESP}} = \sum_a q_a g_a. 
\end{equation}
For details on the GPW-based periodic RESP implementation in CP2K, see Ref.~\citenum{Golze2015}. The choice between non-periodic and periodic RESP fitting is made automatically, depending on the periodicity of the reference potential $V_{\text{QM}}$. 

The QM potential $V_{\text{QM}}$ is obtained from a previous DFT or HF calculation. The RESP fit starts as a post-processing step and is enabled by:
\begin{verbatim}
&PROPERTIES
  &RESP
    &SPHERE_SAMPLING
    &END SPHERE_SAMPLING
  &END RESP
&END PROPERTIES
\end{verbatim}
The choice of sampling points $\mathbf{r}_k$ is crucial to obtain meaningful charges that accurately reproduce the electrostatic potential in spatial regions relevant to interatomic interactions. As a general guideline, regions where the QM potential $V_{\text{QM}}$ varies rapidly (e.g. within the vdW radii of atoms) should be avoided. The specific choice of sampling regions is system-dependent. The following options are available:
\begin{itemize}
    \item \texttt{SPHERE\_SAMPLING}: The real-space points $\mathbf{r}_k$ are sampled in spherical shells around each atom. The shells are defined by a minimal and maximal radius, which can be set by the \texttt{RMIN} and \texttt{RMAX} keywords in this section. This option should be used for molecules, molecular liquids, or porous periodic systems like metal-organic frameworks. 
    \item \texttt{SLAB\_SAMPLING}: The option should be used for slab-like systems, where it is important to reproduce the potential well above the surface, e.g. to study adsorption processes. The $\mathbf{r}_k$ grid is then sampled as a thin slice above the surface. In that case, keywords defining the surface atoms, the direction, and the thickness of the slice need to be set.
\end{itemize}

A set of constraints and restraints is typically employed to avoid unphysical values for ${q_a}$ and to stabilize the fit. The total residual $R$, which is minimized, is $R=R_{\text{esp}} + R_{\text{rest}} + R_{\text{const}} $. For the restraint, CP2K uses a harmonic penalty function 
\begin{equation}
    R_{\text{rest}} = \beta\sum_j(q_j-t_j)^2,
    \label{eq:restraint}
\end{equation}
where $\beta$ denotes the strength of the restraint and $t_j$ is the anticipated target charge. Restraints can be explicitly set by 
\begin{verbatim}
  &RESP
    ...
    &RESTRAINT
      ATOM_LIST 1..3
      TARGET -0.18
      STRENGTH 0.0001
    &END RESTRAINT
    RESTRAIN_HEAVIES_TO_ZERO .FALSE.
  &END RESP
\end{verbatim}
In this example, the target charges $t_j$ in Eq.~\eqref{eq:restraint} are set to $-0.18$ for atoms with the index 1 to 3. The strength $\beta$ of the restraint is defined by the keyword \texttt{STRENGTH}. When explicitly defining restraints, the default restraint \texttt{RESTRAIN\_HEAVIES\_TO\_ZERO}, which sets $t_j$ to zero for all elements but hydrogen, should be turned off.
Different constraints are possible, yet by default, CP2K constrains the sum of all fitted charges to the total charge $q_{\text{tot}}$ of the system, i.e. 
\begin{equation}
    R_{\text{const}} =\lambda\sum_j(q_j-q_{\text{tot}}).
\end{equation}
Further explicit constraints can be given by adding 
\begin{verbatim}
&CONSTRAINT 
  EQUAL_CHARGES
  ATOM_LIST 1 2 3
&END
\end{verbatim}
to the \texttt{\&RESP} section. This is to say that the atoms with indices 1, 2 and 3 carry the same charge. More details on how to set constraints can be found in the online tutorial~\cite{resp_tutorial}. 

CP2K also features a GPW implementation of the so-called repeating electrostatic potential extracted atomic (REPEAT) method~\cite{Campana2009}, which is an adaptation of the RESP approach for periodic systems. When using the REPEAT method, Eq.~\eqref{eq:residual_esp} is modified so that the variance of the potentials is fitted rather than the absolute difference. Fitting the variance is generally easier and helps stabilize the fit for periodic systems. We note that the REPEAT method was originally introduced to handle the arbitrary offset of the electrostatic potentials $V_{\text{QM}}$ and $V_{\text{RESP}}$ in infinite systems~\cite{Campana2009}. In CP2K, both potentials are computed with the GPW method, sharing thus the same offset. The REPEAT method is activated by adding the keyword \texttt{USE\_REPEAT\_METHOD} to the \texttt{\&RESP} section. More details on the REPEAT implementation in CP2K can be found in Ref.~\citenum{Kuehne2020} and in the online tutorial~\cite{resp_tutorial}.



\subsection{Density Functional Embedding Theory}
Density functional embedding theory (DFET) separates the systems into the relevant cluster and an environment. The cluster is described with the high-level electronic structure method, often employing a correlated WF (CW) ansatz, while the environment and the interaction between the cluster and the environment are described with DFT via the uniquely defined local embedding potential $v_{\rm emb}(\mathbf{r})$~\cite{Huang_JCP_2011}. The total energy of the system is calculated within first-order perturbation theory, i.e. 
\begin{equation}
    E^{\rm DFET}_{\rm total} = E^{\rm DFT}_{\rm total} + (E^{\rm CW}_{\rm cluster, \rm emb} - E^{\rm DFT}_{\rm cluster, \rm emb} ),
\end{equation}
where $E^{\rm DFT}_{\rm total}$ and $E^{\rm DFT}_{\rm cluster, \rm emb}$ 
are the DFT energies of the entire system and the embedded subsystem, respectively, while $E^{\rm CW}_{\rm cluster, \rm emb}$ is the energy of the embedded cluster at the CW level of theory. All these entities are computed with an additional one-electron embedding term $\int d\mathbf{r} \, V_{emb} \, \rho(\mathbf{r})$ in the Hamiltonian.

The embedding potential is obtained in a top-down approach from the condition that the sum of embedded subsystem densities should reconstruct the DFT density of the total system. This can be achieved by maximizing the Wu-Yang functional with respect to $v_{\rm emb}(\mathbf{r})$~\cite{Wu_JCP_2003}, i.e.
\begin{eqnarray}
    W[V_{\rm emb}] &=& E_{\rm cluster}[\rho_{\rm cluster}] + E_{\rm env}[\rho_{\rm env}] \nonumber \\
    &+& \int V_{\rm emb}(\rho_{\rm total} - \rho_{\rm cluster} - \rho_{\rm env}) \, d\mathbf{r},
\end{eqnarray}
with the functional derivative being identical to the density difference such that $ \frac{\delta W}{\delta V_{\rm emb}} = \rho_{\rm total} - \rho_{\rm cluster} - \rho_{\rm env}$.

\subsubsection{General Procedure}
The DFET workflow starts with obtaining the embedding potential $v_{\rm emb}(\mathbf{r})$. First, one calculates the total density of the system before DFT calculations on the various subsystems with the current updated embedding potential are performed. Then, the potential is updated and the step is repeated until the total DFT density is matched by a sum of the densities of the embedded subsystems. When this condition is fulfilled, the embedded higher-level theory calculation is performed on the isolated cluster. 

\subsubsection{Implementation}
The DFET implementation is available for closed and open shell systems, in terms of unrestricted and restricted open-shell formalisms, respectively. It is limited to GPW calculations only with PPs describing the core electrons. All electronic structure methods implemented within CP2K/\textsc{Quickstep} are available as a higher-level method, including hybrid DFT, MP2, and RPA. It is possible to perform property calculations on the embedded cluster using an externally provided $v_{\rm emb}(\mathbf{r})$. The subsystems can employ different basis sets, although they must share the same PW grid.

DFET calculations are activated via the \texttt{\&MULTIPLE\_FORCE\_EVALS} section in the root:
\begin{verbatim}
&MULTIPLE_FORCE_EVALS
  FORCE_EVAL_ORDER 2 3 4 5
  MULTIPLE_SUBSYS T
&END MULTIPLE_FORCE_EVALS
\end{verbatim}
The order of force evaluations is arbitrary, but it is recommended to leave it as is. Here, \texttt{\&FORCE\_EVAL} 2 refers to the environment, 3 to the embedded cluster, and 4 to the total system, all of which are calculated at the DFT level. The isolated cluster computed at a higher-level of theory is denoted as \texttt{\&FORCE\_EVAL 5}. 
The first force evaluation in \texttt{\&FORCE\_EVAL} defines the general embedding framework:
\begin{verbatim}
&FORCE_EVAL
  METHOD EMBED
  &EMBED
    &MAPPING
      &FORCE_EVAL 1
        &FRAGMENT 1
          1 3
          MAP 1
        &END FRAGMENT
      &END FORCE_EVAL
      &FORCE_EVAL 2
        &FRAGMENT 1
          1 2
          MAP 2
        &END FRAGMENT
      &END FORCE_EVAL
      &FORCE_EVAL 3
        &FRAGMENT 1
          1 5
          MAP 3
        &END FRAGMENT
      &END FORCE_EVAL
      &FORCE_EVAL 4
        &FRAGMENT 1
          1 2
          MAP 2
        &END FRAGMENT
      &END FORCE_EVAL
      &FORCE_EVAL_EMBED
        &FRAGMENT 1
          1 3
        &END FRAGMENT
        &FRAGMENT 2
          4 5
        &END FRAGMENT
        &FRAGMENT 3
          1 5
        &END FRAGMENT
      &END FORCE_EVAL_EMBED
    &END MAPPING
  &END EMBED
  ...
&END FORCE_EVAL
\end{verbatim}
The mappings define the correspondence of atoms in the parts and in the total system, i.e. atoms 1 and 2 (\texttt{MAP 2} of \texttt{\&FORCE\_EVAL 2}) correspond to atoms 4 and 5 in the total system (\texttt{\&FRAGMENT 2} in the \texttt{\&FORCE\_EVAL\_EMBED} section). In addition, this \texttt{\&FORCE\_EVAL} section must represent the total system as defined in the \texttt{\&SUBSYS} section. 
The next two \texttt{\&FORCE\_EVAL} sections specify the environment and the cluster via standard DFT input sections. The latter one has the following additional keyword in the \texttt{\&QS} section: 
\begin{verbatim}
CLUSTER_EMBED_SUBSYS .TRUE.
\end{verbatim}
The next \texttt{\&FORCE\_EVAL} section to be defined is the total system, marked by the keyword
\begin{verbatim}
  REF_EMBED_SUBSYS .TRUE.
\end{verbatim}
in the \texttt{\&QS} section.  In this force evaluation, the options for optimizing the embedding potential are to be specified. This is done by inserting a \texttt{\&OPT\_EMBED} section within the \texttt{\&QS} section:
\begin{verbatim}
&OPT_EMBED
 ...
&END OPT_EMBED
\end{verbatim}
The final \texttt{\&FORCE\_EVAL} section represents the higher-level calculation of the embedded cluster. It can be either a standard hybrid/double-hybrid DFT or a CW calculation with the following additional keyword in the \texttt{\&QS} section: 
\begin{verbatim}
HIGH_LEVEL_EMBED_SUBSYS .TRUE.
\end{verbatim}

Several options for representing and optimizing the embedding potential are implemented in CP2K. Most importantly is the \texttt{\&OPT\_EMBED\%GRID\_OPT} keyword, which defines whether the potential is represented on the real-space grid, or expanded in the auxiliary basis set. The first option is more accurate and computationally cheaper, and is therefore recommended for all practical purposes.

In addition, several techniques for optimizing $v_{\rm emb}(\mathbf{r})$ are available, including standard gradient-based optimization schemes (e.g. SD, quasi-Newton, and level-shifting methods), as well as an iterative van Leeuwen-Baerends update~\cite{van1994exchange}, which is an alternative to the previously mentioned Wu-Yang functional~\cite{Wu_JCP_2003}. Despite its simplicity, the SD approach is typically relatively robust for potential optimizations and is therefore the default. Although several options for the initial guess of the embedding potential are available, starting with the zero potential is a rather reliable procedure, which is typically quickly converging. 

The embedding potential is saved as a volumetric Gaussian cube file that can be used for visualization purposes and can be read in for restarting the optimization. For open-shell calculations, CP2K defines two potentials: one interacts with the electron density, whereas the other interacts with the spin density. Both are saved and should be specified for restarting the calculation. In addition, the embedding potential can be used for standalone embedded calculations, as in the following example for an open-shell system:
\begin{verbatim}
&QS
  DFET_EMBEDDED .TRUE.
  EMBED_CUBE_FILE_NAME potential.cube
  EMBED_SPIN_CUBE_FILE_NAME
  spin_potential.cube
  ...
&END QS
\end{verbatim}
%
\section{Nuclear Magnetic and Electron Paramagnetic Resonance Spectroscopy} \label{sec:NMR+EPR}

The energy levels probed in magnetic spectroscopy correspond to transitions among nuclear and electronic spin eigenstates in the presence of an external magnetic field.
The calculation of magnetic response properties in CP2K is based on variational density functional perturbation theory (DFPT)~\cite{Putrino2000, Sebastiani2001, Sebastiani2003}.

\subsection{Density Functional Perturbation Theory}
\label{sec:dens-funct-pert}

Generally speaking, CP2K/\textsc{Quickstep} uses DFPT to compute the perturbative corrections to the zeroth-order orbitals $\psi^{(0)}_k$. In case of a perturbing magnetic field $B$:
\begin{equation}
  \label{eq:nmr_orb_response}
  \psi_k = \psi^{(0)}_k + B\,\psi_k^{(1)} + \dots,
\end{equation}
where $\psi_k^{(1)}$ is the first-order orbital correction, which provides energy corrections up to third order according to Wigner's (2n+1) rule.


The first-order orbitals can be obtained via the inhomogeneous set of coupled (Sternheimer) equations:
\begin{multline}
  \label{eq:sternheimer}
  -\sum_i^{N_{\text{occ}}} \left (  H^{(0)} \delta_{ij} - \langle \psi_i^{(0)} | H^{(0)} | \psi_j^{(0)} \rangle \right ) | \psi_i^{(1)} \rangle \\
  = H^{(1)} | \psi_j^{(0)} \rangle.
\end{multline}
Therein, the indices $i,j$ run over the occupied orbital manifold $\{\psi\}$. This expression already anticipates the possibility that the employed zeroth-order orbitals might not be the canonical KS orbitals.
Expanding the first-order orbitals in terms of atomic basis functions $\phi_l$:
\begin{equation}
  \label{eq:nmr_mo_ao}
  \psi_l^{(1)} = \sum_{l} c_{li}\, \phi_l^{(1)},
\end{equation}
and projecting Eq.~\ref{eq:sternheimer} on $\phi_k$:
\begin{multline}
  \label{eq:sternheimer_ao}
  - \sum_{i}^{N_{\text{occ}}}\sum_{l}^{N_{\text{basis}}} \left ( H_{kl}^{(0)} \delta_{ij} - S_{kl} \langle \psi_i^{(0)} | H^{(0)} | \psi_j^{(0)} \rangle  \right ) c^{(1)}_{li} \\
  = \sum_l H^{(1)}_{kl} c^{(0)}_{lj}
\end{multline}
we obtain a set of $N_{\text{basis}}\times N_{\text{occ}}$ simultaneous equations. In case of magnetic responses, the first-order orbitals are imaginary, and one explicitly includes $\mathrm{i}=\sqrt{-1}$ on the left-hand side, so that the expansion coefficients $c^{(1)}$ are real. It is to be noted that with atom-centered basis functions and vibrational perturbations (dipole derivatives, Born charges), the Sternheimer equation acquires an additional contribution on the right-hand side, due to the derivatives of the overlap matrix.

In CP2K, equations~\ref{eq:sternheimer_ao} are solved self-consistently using a CG-based minimizer~\cite{Gonze1995,Putrino2000}. In the employed parallel-transport gauge, the first-order orbitals are orthogonal to the occupied orbital manifold~\cite{Gonze1995}. This can be imposed by projecting the right-hand sides of the equations onto the unoccupied orbital manifold, which amounts to solving the equations while imposing the orthogonality between the trial first-order orbitals and the occupied orbitals.

A prototypical input for such a linear response calculation looks like:
\begin{verbatim}
&GLOBAL
  RUN_TYPE LINEAR_RESPONSE
&END GLOBAL
&FORCE_EVAL
  &PROPERTIES
    &LINRES
      ...
    &END LINRES
  &END PROPERTIES
&END FORCE_EVAL
\end{verbatim}
In the relevant \texttt{\&LINRES} input section, which is located inside the \texttt{\&PROPERTIES} section, one can set the parameters of the preconditioned CG optimizer. The general principles on how to choose the specific preconditioner, maximum number of iterations, and convergence threshold are similar to Section~\ref{WFopt}. Yet, the default values typically provide a robust starting point. Nevertheless, the convergence behavior of different response properties can differ, and it is advisable to check whether the desired response property is converged with respect to the chosen CG convergence threshold.

Also inside the \texttt{\&LINRES} section, different subsections give access to various response properties. As will be explained later, when calculating magnetic field-induced current densities (required for several magnetic response properties) maximally localized Wannier functions (MLWF) are employed as the zeroth-order orbitals. The corresponding \texttt{\&LOCALIZE} input section is also found within the \texttt{\&LINRES} section, as are:
\begin{itemize}
\item {\tt \&CURRENT}: The induced current density in response to an external homogeneous magnetic field~\cite{Sebastiani2001}.
\item {\tt \&LOCALIZE}: Computes localized orbitals, which are required if the perturbation includes the position operator. 
\item {\tt \&NMR}: The nuclear magnetic shielding tensors, or more generally, the nucleus-independent chemical shifts (NICS)~\cite{Sebastiani2006}. 
\item {\tt \&EPR}: The electronic g-tensor~\cite{Weber_JCP_2009}. 
\item {\tt \&POLAR}: Computes the polarizability~\cite{Putrino2002}. 
\item {\tt \&VCD}: Enables the calculation of vibrational circular dichroism (VCD)~\cite{scherrer2016vibrational}.
\item {\tt \&DCDR}: Analytical gradients of the dipole moments, e.g. for Born effective charges and atomic polar tensors.
\end{itemize}
Note that the calculation of some response properties might require the explicit inclusion of more than one of the above, e.g. \texttt{\&CURRENT} requires \texttt{\&LOCALIZE}. These two sections are in turn prerequisites for \texttt{\&NMR} and \texttt{\&EPR} sections, respectively. Detailed information relating to individual input sections for magnetic response properties will be given in the following.

\subsection{The Magnetic Shielding Tensor}
\label{sec:magn-shield-tens}

The interaction of electrons with an external magnetic field $\mathbf{B}$ leads to an induced current density $\mathbf{j}(\mathbf{r})$, which in turn leads to an induced magnetic field
\begin{equation}
  \label{eq:nmr_Bind}
  \mathbf{B}^{\textrm{ind}}(\mathbf{r})=\frac{1}{c}\int \textrm{d}\vecr \, \frac{(\mathbf{r}'-\mathbf{r})}{|\mathbf{r}'-\mathbf{r}|^3}  \times \mathbf{j}(\mathbf{r}'),
\end{equation}
where $c$ is the speed of light, and Gaussian atomic units are being used. The perturbing field can be an external magnetic field $\mathbf{B}^{\textrm{ext}}$ (taken to be spatially homogeneous), or the magnetic moments associated with spins. We generally assume in this section a closed-shell system with vanishing total electron spin. In this case, the only spins in the system are the nuclear ones
${\mathbf{m}^N}$:
\begin{multline}
  \label{eq:nmr_Bind_Taylor}
  \mathbf{B}^{\textrm{ind}}(\mathbf{r},\mathbf{B}^{\textrm{ext}},\{\mathbf{m}^N\}) = - \boldsymbol{\sigma}(\mathbf{r})\,\mathbf{B}^{\textrm{ext}}\\
  -\sum_i^{\textrm{Nuclei}} \mathbf{K}_i(\mathbf{r})\, \mathbf{m}^N_i(\vecr^N_i) + \dots
\end{multline}

The reduced indirect spin-spin coupling is denoted as $\mathbf{K}(\mathbf{r})$ in Eq.~\ref{eq:nmr_Bind_Taylor}, while the tensor field 
\begin{equation}
  \label{eq:nmr_sigma_def}
  \boldsymbol{\sigma}(\mathbf{r}) = -\frac{\partial\mathbf{B}^{\textrm{ind}}(\mathbf{r})}{\partial\mathbf{B}^{\textrm{ext}}},
\end{equation}
is known as the magnetic shielding tensor. Experimentally, the induced magnetic field can only be probed at the position of a nuclear spin $\mathbf{r}^N$. Being a ratio, $\boldsymbol{\sigma}(\mathbf{r})$ is conventionally reported in units of parts per million. The experimental value is almost always referenced to the isotropic average ($\frac{1}{3}\mathrm{Tr}(\boldsymbol{\sigma}(\mathbf{r})$) of the magnetic shielding tensor of some chosen standard for the respective nucleus, i.e.
\begin{equation}
  \label{eq:nmr_shift}
  \boldsymbol\delta(\mathbf{r}^N_i) = \sigma^{\textrm{iso}}_{i,\textrm{ref}}\mathbf{I} - \boldsymbol\sigma(\mathbf{r}^N_i),
\end{equation}
which is known as the chemical shift tensor, and its isotropic average $\delta$ is called the chemical shift. Moreover, $\mathbf{I}$ is the unit matrix.

At lowest order, the energy of the system is linear in the total magnetic field, thus one can also define the shielding tensor as the second derivative of the energy with respect to the nuclear magnetic moment $\mathbf{m^N}$ and the external magnetic field:
\begin{equation}
    \label{nmr:sigma_2deriv}
    \boldsymbol{\sigma}(\vecr^N)=\frac{\partial^2 E}{\partial{\mathbf{B}^{\mathrm{ext}}} \,\partial{\mathbf{m^N}}}\Bigr|_{|\mathbf{B}^{\mathrm{ext}}|=|\mathbf{m}^N|=0}
\end{equation}

Given that the magnetic shielding is a tensor field, it can be computed in any arbitrarily chosen spatial position, even though it is experimentally measurable only at the nuclear positions. These are known as nucleus-independent chemical shifts (NICS), which are also available in CP2K~\cite{Sebastiani2006}.

From Eqs.~\ref{eq:nmr_Bind} and~\ref{eq:nmr_sigma_def}, the elements of the shielding tensor are given by:
\begin{equation}
  \label{eq:nmr_sigma}
  \sigma_{xy}(\mathbf{r}) = \frac{1}{c} \int_{\Omega}\left [ \frac{\mathbf{r'}-\mathbf{r}}{|\mathbf{r'}-\mathbf{r}|^3} \times \mathbf{j}_x(\mathbf{r'}) \right ]_y \,\textrm{d}^3r' ,
\end{equation}
where the integration is over the whole material, including all periodic replicas in the case of extended systems. It is obvious that the induced current density is the key ingredient in computing magnetic shieldings. Once $\mathbf{j}(\vecr)$ is known, the induced field, and hence the shielding tensor as well, is available as a simple three-dimensional integral.

Using the approach developed by Sebastiani and Parrinello~\cite{Sebastiani2001}, CP2K evaluates the magnetic perturbation using MLWFs in combination with DFPT~\cite{Weber_JCP_2009}. As explained later, the localized nature of the Wannier functions allows the use of the position operator in the perturbation Hamiltonian. An underlying core assumption here is that each Wannier function is contained entirely within the simulation cell once it is centered. For insulators, it is known that the Gaussian functions decay exponentially, so that for sufficiently large simulation cells, this assumption is not irrelevant~\cite{Resta2002}. In comparison to other approaches developed for condensed phase systems~\cite{Sebastiani2001,Mauri1996}, CP2K differs in its use of local atom-centered Gaussian functions (see Section~\ref{sec:TotalEnergy}), allowing for reduced complexity algorithms and hence large scale calculations of magnetic resonance parameters. With the GAPW method, CP2K/\textsc{Quickstep} is also able to compute all-electron magnetic resonance parameters. One can also use embedding techniques such as QM/MM methods, and even further decompose the QM part into GAPW/GPW regions. The reliance on the localization properties of Wannier functions means, however, that this implementation is not suitable for conductors, where the Wannier functions decay only algebraically.

In order to compute the shielding tensors one first needs to use the GAPW method with an all-electron basis set (the calculation will also be possible using GPW, but then the contributions of the core electrons to the induced current density are missing). Inside the \texttt{\&LINRES} section one then also needs to include the \texttt{\&LOCALIZE} section to compute the MLWFs, which are then provided to the linear response module as the zero-order orbitals.
Otherwise, CP2K will exit with an error if the user requests the calculation of the current density or the shielding tensors, without activating the \texttt{\&LOCALIZE} section. Second, the \texttt{\&CURRENT} section should be included for computing the induced current density using DFPT~\cite{Putrino2000}. Finally, the \texttt{\&NMR} section requests the printing of the shielding tensors or NICS maps by performing the integration in
Eq.~\ref{eq:nmr_sigma}.
However, before discussing the various input options in these sections, we provide a quick overview of the specifics of the DFPT-based implementation for computing the induced current density in CP2K~\cite{Weber_JCP_2009}.

For a homogeneous magnetic field, the vector potential $\mathbf{A}$ can be chosen as
\begin{equation}
  \label{eq:nmr_A}
  \mathbf{A}(\vecr) = \frac{1}{2}  \mathbf{B} \times (\vecr - \vecr_0),
\end{equation}
where $\vecr_0$ is an arbitrary gauge origin.
Using the explicit form of the vector potential given in Eq.~\ref{eq:nmr_A} leads to perturbations involving the position operator, which is problematic under PBC, as it jumps discontinuously at the edges of the cell. The solution suggested by Sebastiani and Parrinello is to use maximally localized Wannier functions as the zeroth-order orbitals $\psi^{(0)}(\vecr)$~\cite{Sebastiani2001}. If the simulation cell is chosen to be larger than the decay length of the Wannier functions (which is exponentially decaying for insulators) and choosing for each Wannier function $\psi_i$ a coordinate system such that its center is at the origin (via a translation vector $\mathbf{d}_i$), one avoids the problematic behavior of the position operator at the cell boundaries. It should be noted that this choice of a different origin for the entire coordinate system for individual Wannier functions is not a gauge transformation, as \emph{both} \vecr{} and $\vecr_0$ in Eq.~\ref{eq:nmr_A} are being simultaneously shifted. The current is invariant under arbitrary orbital-specific translations.

With the introduction of orbital-specific translations $\mathbf{d}_i$, one ends up with three perturbation operators~\cite{Sebastiani2001}. The linear momentum operator
\begin{equation}
  \label{eq:nmr_h1_L}
  \mathrm{\hat{H}}^{P} = \hat{\mathbf{p}},
\end{equation}
the orbital angular momentum operator
\begin{equation}
  \label{eq:nmr_h1_p}
  \mathrm{\hat{H}}^{Li} = (\hat{\mathbf{r}}-\mathbf{d}_i) \times \hat{\mathbf{p}},
\end{equation}
and the full correction operator
\begin{equation}
  \label{eq:nmr_h1_di_dj}
  \mathrm{\hat{H}}^{\Delta{i}} = (\mathbf{d}_i-\mathbf{d}_j) \times \hat{\mathbf{p}},
\end{equation}
where in the last equation $\mathbf{d}_i-\mathbf{d}_j$ is computed using the minimum image convention. The last two operators are labeled with the orbital index $i$ to denote their orbital dependence via the chosen orbital centers $\mathbf{d}_i$. All three operators are vector operators, leading to nine contributions per orbital. Each contribution from the first two operators can be obtained
essentially at the cost of one total energy calculation. The full correction operator requires one such calculation per orbital, making it by far the computationally most expensive term for any system with more than
a handful of orbitals. 

Once these perturbation operators are used in the inhomogeneous set of equations shown in Eq.~\ref{eq:sternheimer}, and all the contributions to the linear response orbitals have been obtained, the linear response current density vector is computed for the three directions of the magnetic field, allowing the computation of all the components of the shielding tensor. The $x$-component of the linear current density response induced by an external magnetic field applied along the $y$-axis is given by a sum of a paramagnetic $j^p_{xy}(\vecr)$ and a diamagnetic $j^d_{xy}(\vecr)$ contribution~\cite{Weber_JCP_2009}, i.e.
\begin{multline}
  \label{eq:nmr_j_p}
  j^p_{xy}(\vecr) = -\frac{1}{2c}\sum_{ikl}  [ C^{(0)}_{ki} (C_{li}^{L_y} + (\vecr_0 - \mathbf{d_i})_x C_{li}^{P_z}\\
  - (\vecr_0 - \mathbf{d}_i)_z C^{P_x}_{li} - C^{\Delta{i}_y}_{li} ) \\
  \times \{(\nabla_x \chi_k(\vecr))\chi_l(\vecr) - \chi_k(\vecr)\nabla_x \chi_l(\vecr) \}]
\end{multline}
\begin{equation}
  \label{eq:nmr_j_d}
  j^d_{xy}(\vecr) = (\vecr - \vecr_0)_z \rho(\vecr)
\end{equation}
and the other components are obtained analogously.

With our choice of the vector potential in Eq.~\ref{eq:nmr_A}, each of these two oppositely signed contributions to $\mathbf{j}(\vecr)$ is linear in the gauge origin $\vecr_0$; their sum, however, should be invariant to it. With the local atomic basis functions commonly used in quantum chemistry, it is extremely difficult to converge the sum of the two terms at a reasonable computational cost, if one chooses a single fixed gauge origin, and the computed value of the current density becomes parametrically dependent on $\vecr_0$. This is the well-known gauge origin problem. To address this issue, a distributed gauge origin is used during the computation, with different methods making different choices regarding how this distribution is done. In CP2K the gauge origin is tackled using the individual gauge for atoms in molecules (IGAIM), where the gauge origin is the position of the nearest atom~\cite{Keith1992}, or alternatively, with the continuous set of gauge transformations (CSGT), where the gauge origin is the current position \vecr~\cite{Keith1993}. It is obvious that the CSGT approach makes the diamagnetic contribution in Eq.~\ref{eq:nmr_j_d} vanish.

The choice of the gauge origin is set within the \texttt{\&CURRENT} input section:
\begin{lstlisting}[escapeinside={(*}{*)}]
  &LINRES
    &CURRENT
      GAUGE # the gauge origin (*$\vecr_0$*) in 
            # Eqs. (*\ref{eq:nmr_A} \& \ref{eq:nmr_j_p}*)
      GAUGE_ATOM_RADIUS  # atomic radius
                         # in IGAIM gauge
      ORBITAL_CENTER # the choice of (*$\mathbf{d}_i$*)
                     # in Eqs. (*\ref{eq:nmr_h1_p}-\ref{eq:nmr_h1_di_dj}*)
    &END CURRENT
  &END LINRES
\end{lstlisting}
For the choice \texttt{GAUGE R}, one has the option to use CSGT for both the soft and local contributions (see Eq.~\ref{eq:nmr_j_gapw} below) to the current density. \texttt{GAUGE ATOM} uses IGAIM centered on the nearest atom to the current grid point, and a combination of IGAIM for the hard part and CSGT for the soft part can be chosen with \texttt{GAUGE R\_AND\_STEP\_FUNCTION}. For the IGAIM method, the radius of the atomic domain is set by the keyword \texttt{GAUGE\_ATOM\_RADIUS}. Although the choice of CSGT $\vecr = \vecr_0$ is appealing, as it makes the diamagnetic term in Eq.~\ref{eq:nmr_j_d} identically vanish, both empirical evidence and theoretical considerations indicate that other choices provide better results~\cite{vanWuellen2004}.

With the keyword \texttt{ORBITAL\_CENTER}, one can also control the choice of the orbital centers $\mathbf{d_i}$ used in the perturbation operators Eq.~\ref{eq:nmr_h1_p}-\ref{eq:nmr_h1_di_dj}. The choice \texttt{WANNIER} uses the position of the respective Wannier center and is the most expensive, requiring $3M+6$ response calculations, where $M$ is the number of MOs. All other choices offer means to cluster orbital centers together, potentially leading to a substantial reduction in the required computational effort. At variance, \texttt{ATOM} uses the atomic positions as centers, hence the Wannier functions are clustered on the nearest atom, whereas \texttt{BOX} divides the simulation cell into sub-boxes and clusters the centers within each sub-box (the keyword \texttt{NBOX} sets the number of boxes along each dimension). Finally, \texttt{COMMON} uses a common center, whose position is specified by the keyword \texttt{COMMON\_CENTER}, which only works for isolated molecules.

With the keywords
\begin{verbatim}
SELECTED_STATES_ON_ATOM_LIST
SELECTED_STATES_ATOM_RADIUS
\end{verbatim}
one can further choose to perform the response calculation using only the Wannier functions that are within a certain distance from some chosen list of atoms. If only the magnetic shieldings of some particular atoms are of interest, this can lead to enormous reductions in computational time. Naturally, one needs to ascertain that the selected radius provides acceptable results for the atoms of interest, which depends on the respective chemical environments. 

Finally, the calculation of the magnetic shielding tensor requires the evaluation of the integral in Eq.~\ref{eq:nmr_sigma}. For a periodic system, the current density is also periodic and the computation can be done efficiently using the GAPW method (Section~\ref{sec:TotalEnergy}). The current density is decomposed in a manner analogous to the GAPW electron density, thus 
\begin{equation}
  \label{eq:nmr_j_gapw}
  \mathbf{j}(\vecr)=\tilde{\mathbf{j}}(\vecr)+\sum_A^{\textrm{atoms}}\left ( \mathbf{j}(\vecr) - \tilde{\mathbf{j}_A}(\vecr) \right )
\end{equation}
where $\tilde{\mathbf{j}}$ is the soft contribution to the current density, $\mathbf{j}_A$ is the local hard contribution at atom $A$, and $\tilde{\mathbf{j}_A}$ is the local soft contribution to prevent double counting.

The soft contribution $\tilde{\mathbf{j}}$ is computed in reciprocal space on the PW grid~\cite{Sebastiani2001}. One has to note here, however, that the $\mathbf{G}=0$ component cannot be computed under PBC. This term is commonly called the susceptibility correction and depends on the bulk magnetic susceptibility. It is determined by macroscopic magnetostatics from the macroscopic shape of the sample. In CP2K this term is calculated from the susceptibility due to $\tilde{\mathbf{j}}$, assuming a spherical sample shape, which is also the experimental convention~\cite{Cowan1997}:
\begin{equation}
  \label{eq:nmr_chi_correction}
  \chi_{xy}=\frac{2\pi}{\Omega_cc}\int \textrm{d}\vecr \, \left [ \vecr \times \tilde{\mathbf{j}}_x(\vecr) \right ]_y.
\end{equation}

For the local part of the current density, the contribution to $\sigma_{xy}$ of the nucleus at position $\vecr^N$, arising from the induced local current densities, is evaluated as
\begin{multline}
  \label{eq:sigma_local}
  \sigma_{xy}(\vecr^N) = \frac{1}{c} \sum_B \int_{\Omega_B} \textrm{d}\vecr \, \Bigl [ \frac{\vecr - \vecr^N}{|\vecr -\vecr^N|^3} \\
  \times (j_{x,B}(\vecr) - \tilde{j}_{x,B}(\vecr)) \Bigr ]_y, 
\end{multline}
where the sum over atoms $B$ is restricted to the nuclei that are within a radius $R_c$ from $\vecr^N$ (see keyword \texttt{\&NMR\%SHIFT\_GAPW\_RADIUS} below). The integration over the atomic domain $\Omega_B$ is performed numerically on a spherical grid with a logarithmic radial and a Lebedev angular grid. The numerical integration converges rapidly with respect to the number of grid points, and about 10000 grid points per atom are enough to converge the chemical shift below 0.1 ppm~\cite{Weber_JCP_2009}. The size of the atomic integration grid is set inside the \texttt{\&KIND} subsection, and for the computation of the shielding tensors it can be beneficial to use values that are higher than the defaults, e.g.
\begin{verbatim}
&SUBSYS
  &KIND H
    BASIS_SET aug-pcSeg-2
    POTENTIAL ALL
    LEBEDEV_GRID 100
    RADIAL_GRID 200
  &END KIND
&END SUBSYS
\end{verbatim}
In the \texttt{\&NMR} input section, one can set the value of $R_c$, request a NICS calculation, and specify the spatial points for the latter:
\begin{verbatim}
&LOCALIZE
  &NMR
    NICS .TRUE.       
    NICS_FILE_NAME nics.xyz
    SHIFT_GAPW_RADIUS # Cutoff current 
                      # integration
  &END NMR
&END LOCALIZE
\end{verbatim}


\subsection{The Electron Paramagnetic Resonance g-tensor}

CP2K also offers an implementation of the g-tensor under PBC, which employs the same DFPT approach to compute the induced current densities. This makes it feasible to perform large-scale calculations, including with QM/MM (e.g. paramagnetic active sites in enzymes), to investigate paramagnetic defects in solids under PBC, and also to perform large-scale computations of paramagnetic NMR shifts~\cite{Mondal2017}.

Effectively, the g-tensor plays in EPR spectroscopy a role similar to that played by the shielding tensor in NMR spectroscopy. This is clear in an effective spin-Hamiltonian framework, which gives the coupling between an effective electronic spin $\mathbf{S}$ and an external magnetic field $\mathbf{B}^{\textrm{ext}}$ as the bilinear term:
\begin{equation}
  \label{eq:nmr_spinH_g}
  \mathrm{\hat{H}}_{\textrm{eff}}= \frac{1}{2c}\,\mathbf{B}^{\textrm{ext}} \cdot \mathbf{g} \cdot \mathbf{S}
\end{equation}
and like the shielding tensor, the g-tensor is also a second partial derivative of the energy with respect to the external field and a spin magnetic moment, in this case the electronic spin. The corresponding \textit{ab-initio} expression is obtained from the minimal coupling electronic Hamiltonian including relativistic corrections up to $\mathcal{O}(\alpha^3)$, where $\alpha$ is the fine structure constant. Identification of the relevant Hamiltonian terms produces the expression for the g-tensor, in the form of spatially anisotropic relativistic correction terms $\Delta{\mathbf{g}}$ to the free electron g-value $g_e$. The implementation in CP2K is based on the Schreckenbach-Ziegler DFT-based approach~\cite{Schreckenbach1997}, and its extension by Pickard and Mauri~\cite{Pickard2002,VanYperen-DeDeyne2012}. One can identify three contributions to each component of $\Delta\mathbf{g}$, i.e. 
\begin{equation}
  \label{eq:nmr_g}
  g_{xy}=g_e\delta_{xy} + \Delta{g}_{xy}^{\textrm{ZKE}}  + \Delta{g}_{xy}^{\textrm{SO}}  + \Delta{g}_{xy}^{\textrm{SOO}}.
\end{equation}
The correction terms are the Zeeman kinetic energy (ZKE) term
\begin{equation}
  \label{eq:nmr_g_zke}
  \Delta{g}_{xy}^{\textrm{ZKE}} = -\frac{g_e}{c^2}(T^{\alpha}-T^{\beta}) \delta_{xy},
\end{equation}
which is a purely kinematic scalar relativistic correction, 
the spin-orbit (SO) term (usually the dominant term)
\begin{multline}
  \label{eq:nmr_g_so}
  \Delta{g}_{xy}^{\textrm{SO}} = \frac{(g_e-1)}{c} \int_{\Omega_c} [ \mathbf{j}_x^{\alpha}(\vecr) \times \nabla{V}_{\textrm{eff}}^{\alpha}(\vecr) \\
  - \mathbf{j}_x^{\beta}(\vecr) \times \nabla{V}_{\textrm{eff}}^{\beta}(\vecr) ]_y \,\textrm{d}^3\vecr,
\end{multline}
and the spin-other-orbit (SOO) term
\begin{equation}
  \label{eq:nmr_g_soo}
  \Delta{g}_{xy}^{\textrm{SOO}} = \frac{1}{S} \int_{\Omega_c} \textrm{d}\vecr \, B_{y,B_{x}}^{\textrm{corr}}(\vecr) [\rho^{\alpha}(\vecr)-\rho^{\beta}(\vecr)],
\end{equation}
where $\alpha$ and $\beta$ denote the spin channels, S the total spin, and $T$ is the kinetic energy. Moreover, $B_{y,B_{x}}^{\textrm{corr}}$ is the $y$-component of the magnetic field due to the total induced current density by a field (see Eq.~\ref{eq:nmr_Bind}) in the $x$-direction minus a self-interaction correction
\begin{multline}
  \label{eq:nmr_B_corr}
  B_{y,B_{x}}^{\textrm{corr}}(\vecr) =\frac{1}{c} \int_{\Omega} \textrm{d}\vecr' \, \Bigl[ \frac{\mathbf{r'}-\mathbf{r}}{|\mathbf{r'}-\mathbf{r}|^3}\\
  \times ( \mathbf{j}_x(\mathbf{r'})- \mathbf{j}_x^{\alpha-\beta}(\mathbf{r'})) \Bigr]_y
\end{multline}
with $\mathbf{j}^{\alpha-\beta}=\mathbf{j}^{\alpha}-\mathbf{j}^{\beta}$.
The effective potential $V_{\textrm{eff}}$ in the SO contribution is the sum of the external potential, the Hartree potential, and the XC potential~\cite{Schreckenbach1997}.
Currently, the CP2K implementation only supports LDA and GGA functionals; hybrid XC functionals are not yet supported.

The ZKE contribution is computed from the kinetic energy of the spin-polarized KS orbitals in the Gaussian basis set. The SO and SOO terms are similar to the shielding tensor in that they require the (spin-dependent) current density. For the latter, the same techniques based on the aforementioned Sebastiani-Parrinello approach~\cite{Sebastiani2001}, discussed in the context of the magnetic shielding tensor section, are also employed here. One subtlety with respect to the SOO term is that it requires the induced field over all space, not only at the positions of the nuclei. The non-local nature of the dipolar term in Eq.~\ref{eq:nmr_Bind} makes it very taxing to create a GAPW representation of the induced magnetic field. Hence, the implementation employs an approximation of the SOO term, whereby the contributions from atom-centered (local) current densities to the $\mathbf{G}\neq0$ components of the induced field are ignored.

The g-tensor is generally much less sensitive to the choice of gauge than nuclear magnetic shieldings, and even the computationally convenient CSGT (\texttt{\&CURRENT\%GAUGE R}) can be used. Tables~1 and 2 in Ref.~\citenum{Mondal2017} offer a comparison of g-tensors calculated by CP2K with the IGAIM gauge versus those calculated by ORCA~\cite{Neese2005} and MAG-ReSpect~\cite{Malkina2000} using a variety of gauge origins and spin-orbit treatments.

It is important to note that if the simulation cell contains more than one paramagnetic center, the individual center g-tensors erroneously add up. This is a problem in existing implementations of g-tensors in condensed phases\cite{Mondal2017,Pigliapochi2017}. Hence, in this case one needs to normalize the g-tensor of the simulation cell:
\begin{equation}
    \mathbf{g} = g_e\mathbf{1} + \frac{1}{n}\Delta{\mathbf{g}}
\end{equation}
with $n$ being the number of paramagnetic centers in the cell.

To activate the calculation of the g-tensor in the input file, one has to request the spin density from an unrestricted KS calculation (keyword \texttt{\&DFT\%UKS}). One also needs to include the \texttt{\&CURRENT}, \texttt{\&LOCALIZE}, and \texttt{\&EPR} subsections within the \texttt{\&LINRES} section. The various control parameters related to the calculation of the induced current density, in the input section \texttt{\&CURRENT}, were already discussed for the magnetic shielding tensors. The input section 
\begin{verbatim}
&LINRES%EPR%PRINT%G_TENSOR%XC%XC_FUNCTIONAL
\end{verbatim}
sets the functional for the effective XC potential in Eq.~\ref{eq:nmr_g_so}. 
Other XC potentials are activated in a similar fashion, for example \texttt{XALPHA} corresponds to the Dirac-Slater potential, whereas \texttt{BECKE88} together with \texttt{LYP} will use the BLYP XC potential.

\subsection{Hyperfine Couplings}
The effective spin-Hamiltonian term for the hyperfine coupling is bilinear in the nuclear and electron spin:
\begin{equation}
  \label{eq:nmr_a_Heff}
  \mathrm{\hat{H}}= \mathbf{S} \cdot A \cdot \mathbf{I}.
\end{equation}
The \textit{ab-initio} expression for the hyperfine coupling tensor, or $\mathbf{A}$-tensor, is obtained from relativistic quantum mechanics, as in the case of the g-tensor. For any nucleus $N$, the dominant terms are an isotropic (Fermi contact) term
\begin{equation}
  \label{eq:nmr_a_iso}
  A_{\textrm{iso},N}=\frac{4\pi}{3}\frac{g_e\mu_eg_N\mu_N}{\langle S_z \rangle} \int \textrm{d}\vecr \, \rho^{\alpha-\beta}(\vecr)\delta_T(\vecr)
\end{equation}
and an anisotropic dipolar term
\begin{multline}
  \label{eq:nmr_a_aniso}
  A_{\textrm{ani},N}=\frac{1}{2}\frac{g_e\mu_eg_N\mu_N}{\langle S_z \rangle} \\
  \times \int \textrm{d}\vecr \, \rho^{\alpha-\beta}(\vecr) \frac{3r_ir_j-\delta_{ij}r^2}{r^5}, 
\end{multline}
where $\mu_e$ is the Bohr magneton, $\mu_N$ the nuclear gyromagnetic ratio, $\mu_N$ the nuclear magneton, $\langle S_z \rangle$ is the expectation value of the z-component of total electronic spin, and \vecr{} is taken relative to the position of the nucleus $N$. Moreover, $\rho^{\alpha-\beta}(\vecr)$ and $g_e$ have already been introduced, whereas $\delta_T$ is a smeared-out delta function which results from scalar relativistic corrections (in the non-relativistic limit, it collapses to a Dirac delta function), i.e.
\begin{equation}
  \label{eq:nmr_delta_smeared}
  \delta_T(\vecr)=\frac{1}{4\pi r^2}\frac{2}{Z\alpha^2}\frac{1}{(1+\frac{2r}{Z\alpha^2})^2}, 
\end{equation}
where $Z$ is the nuclear charge and $\alpha$ is the fine structure constant. Eqs.~\ref{eq:nmr_a_iso}-\ref{eq:nmr_delta_smeared} are the basis for computing hyperfine couplings in CP2K~\cite{Declerck2006a}. In addition to these two first-order terms, there is also a second-order spin-orbit contribution to the hyperfine coupling, which can become important for heavy atoms. 

It is obvious that the isotropic term particularly requires an accurate description of the electron density at the position of the nucleus; hence, hyperfine couplings are implemented within CP2K's GAPW method~\cite{Declerck2006a}. The Fermi contact term at atom $N$ is simply computed by integrating Eq.~\ref{eq:nmr_a_iso} inside the local atomic domain $U_N$ using the hard spin density (for detailed definitions of these quantities, see Ref.~\citenum{Lippert1999}). The anisotropic hyperfine coupling is again split into a soft contribution to be computed in the PW basis and local spin-density contributions. The contribution of the local spin density of atom $N$ itself is integrated inside the domain $U_N$, since the total local density of any atom outside its domain is, by definition, zero. The local densities of other atoms $M\neq N$ make a small contribution to the hyperfine coupling of atom $N$ (see Table~IV in Ref.~\citenum{Declerck2006a}). These "cross terms" needs to be integrated into the respective domains $U_M$. For this small contribution, only near-neighboring atoms around $N$ need to be included, using a cutoff distance with a default value of 10~Bohr, which the user can modify. 

Unlike the magnetic properties discussed so far, the hyperfine coupling in terms of Eqs.~\ref{eq:nmr_a_iso}-\ref{eq:nmr_delta_smeared} is a first-order property, obtained as an expectation value over the unperturbed spin density. Therefore, the input section for requesting the printing of the hyperfine coupling tensors is found directly under the DFT section via 
\begin{verbatim}
&DFT%PRINT%HYPERFINE_COUPLING_TENSOR
\end{verbatim}
In addition to various print control options, the \texttt{INTERACTION\_RADIUS} can be set inside this input section, which specifies the cutoff radius mentioned above.




\section{Optical Spectroscopy} \label{sec:OpticalSpectroscopy}

Optical spectroscopy is a technique that is used to study the interaction between light and matter. It involves measuring the absorption, emission, or scattering of light by molecules, atoms, or materials. The resulting spectra provide valuable information about the electronic structure, energy levels, and dynamics of the system under investigation.
One of the key quantities of interest is the excitation energy $\Omega^{(n)}$, which is the energy required to excite a molecule from its ground state to an excited state. These excitation energies are directly related to the positions and intensities of spectral lines observed in absorption and emission spectra.

\subsection{Linear-Response Time-Dependent Density Functional Theory} \label{LR-TDDFT}





Optical properties are calculated by solving a generalized eigenvalue problem that involves the block matrix $ABBA$~\cite{Blase2018}, hence
\begin{align}
    \left( \begin{array}{cc}A &  B\\B &  A\end{array} \right)\left( \begin{array}{cc}\mathbf{X}^{(n)}\\\mathbf{Y}^{(n)}\end{array} \right) = \Omega^{(n)}\left(\begin{array}{cc}1&0\\0&-1\end{array}\right)\left(\begin{array}{cc}\mathbf{X}^{(n)}\\\mathbf{Y}^{(n)}\end{array}\right). \label{BSE1}
\end{align}
Therein, $A$ and $B$ are matrices with indices $A_{ia,jb}$, i.e. they have
$N_\mathrm{occ}N_\mathrm{empty}$ rows and $N_\mathrm{occ}N_\mathrm{empty}$ columns. 
The eigenvector $(\mathbf{X}^{(n)},\mathbf{Y}^{(n)})$ has elements $X_{ia}^{(n)}$ and $Y_{ia}^{(n)}$, which quantify the contribution of the transition from occupied orbital~$i$ to an empty orbital~$a$ in the excitation~$n$.
The matrix elements of $\mathbf{A}$ and $\mathbf{B}$ are
\begin{equation}
\begin{aligned}
    A_{ia,jb} &= \delta_{ij}\delta_{ab} (\epsilon_a -\epsilon_i) + (ia|jb) + (ia|f_{\textrm{xc}}|jb) \\
    B_{ia,jb} &= (ia|bj) + (ia|f_{\textrm{xc}}|bj),
    \label{eq:tddft1}
\end{aligned}
\end{equation}
respectively. 

The Hermitian matrix $\mathbf{A}$ includes, as a zeroth-order term, the differences within the KS orbital energies. At first order, it incorporates kernel contributions, which vary depending on the chosen density functional approximation. These contributions consist of Coulomb and exact exchange terms, as well as components arising from the XC potential and its associated kernel $f_{\textrm{xc}}$. The implementation in CP2K/\textsc{Quickstep} is based on the Tamm-Dancoff approximation (TDA)~\cite{Hirata:1999,Hutter2003}, which involves constraining $\mathbf{B}=0$ and $\mathbf{Y}=0$ in Eq.~\ref{eq:tddft1}, thus resulting in the Hermitian eigenvalue problem
\begin{equation}
    A\mathbf{X} = \Omega 
    \mathbf{X}.\label{TDA}
\end{equation}

The TDDFT module facilitates the calculation of excitation energies and excited state properties within the TDA, supporting both GGA and hybrid XC functionals, as well as simplified semi-empirical TDA kernels. The optical spectrum calculation with TDDFT requires setting \texttt{\&GLOBAL\%RUN\_TYPE ENERGY} and specifying the subsequent \texttt{\&TDDFPT} subsection within \texttt{\&FORCE\_EVAL\%PROPERTIES} as follows:
\begin{verbatim}
&TDDFPT
  KERNEL FULL
  NSTATES 10
  MAX_ITER 50
  CONVERGENCE [eV] 1.0E-6
  RKS_TRIPLETS F
&END TDDFPT
\end{verbatim}
The kernel matrix is controlled by the \texttt{KERNEL} keyword, where one can select between the full kernel that is appropriate for GGA or hybrid functional computations, or the semi-empirical \texttt{sTDA} kernel~\cite{Grimme2013,Hehn2022}. If the \texttt{FULL} kernel is selected, the underlying functional can be specified via the \texttt{\&XC} section. The choice of XC functional can differ from the ground state functional unless the \texttt{ADMM} method is employed. The \texttt{sTDA} kernel omits XC contributions and approximates both Coulomb and exchange terms using pairwise semiempirical operators $\gamma^\textrm{J}$ and $\gamma^\textrm{K}$, which are dependent on the interatomic distance $R_{MN}$ between atoms $M$ and $N$. Hence, 
\begin{equation}
\begin{aligned}
      \gamma^\textrm{J}(M,N) &= \Bigg( \frac{1}{{R_{\scriptscriptstyle MN}}^\alpha + \eta^{-\alpha}} \Bigg)^{1/\alpha} \\
      \gamma^\textrm{K}(M,N) &= \Bigg( \frac{1}{{R_{\scriptscriptstyle MN}}^\beta + (a_{\scriptscriptstyle\mathrm{XC}}\eta)^{-\beta}} \Bigg)^{1/\beta}
      \label{eq:stda_operators}
\end{aligned}
\end{equation}
are relying on the four global parameters, the chemical hardness $\eta$, the Fock-exchange mixing parameter $a_{\scriptscriptstyle\mathrm{XC}}$ and the powers $\alpha$ for Coulomb, and $\beta$ for the exchange interactions. The proportion of exact exchange, controlled by adjusting the parameter $a_{\scriptscriptstyle\mathrm{XC}}$, is critical for a well-balanced treatment of exact-exchange in the ground and excited state potential surfaces. This is controlled by the keyword \texttt{FRACTION} in the \texttt{\&sTDA} section, and it is recommended to use a relatively small amount, e.g. $a_{\scriptscriptstyle\mathrm{XC}}=0.1-0.2$. The parameters $\alpha$ and $\beta$ are controlled by the \texttt{MATAGA\_NISHIMOTO\_CEXP} and \texttt{MATAGA\_NISHIMOTO\_XEXP} keywords, respectively. 
\begin{verbatim}
KERNEL sTDA   
&sTDA
  FRACTION 0.2  
  DO_EWALD TRUE
&END sTDA
\end{verbatim}
It should be noted that the definition of the operators in Eq.~\ref{eq:stda_operators} differs from the original method for compatibility with the ADMM definition.
For periodic systems, the Ewald summation for Coulomb contributions should be enabled with the \texttt{DO\_EWALD} keyword, which also activates the \texttt{\&DFT\%POISSON} section.

Additional key parameters include \texttt{NSTATES}, which defines the number of excitation energies to compute, and \texttt{CONVERGENCE} that sets the convergence threshold for the Davidson algorithm. The \texttt{RKS\_TRIPLETS} option allows switching from the default singlet excitation energy calculations to triplet excitations.
The \texttt{RESTART} keyword enables restarting a TDDFT calculation if a valid restart file (\texttt{.tdwfn}) is available. Furthermore, a separate grid can be defined for real-space integration in the TDDFT calculation via the \texttt{\&TDDFPT\%MGRID} subsection, though a consistent setup for ground and excited-state calculations is generally recommended.

Upon convergence of the Davidson algorithm, CP2K outputs the excitation energies in eV, the transition dipole moments, and the oscillator strengths for each calculated excited state. The \texttt{\&DIPOLE\_MOMENTS\%DIPOLE\_FORM} keyword allows modifying the form of the dipole transition integrals, with options such as \texttt{BERRY} (for fully periodic systems), \texttt{LENGTH} (for molecular systems), and \texttt{VELOCITY} (for both). When using the length form, the reference point for calculating dipole moments can be adjusted through the \texttt{REFERENCE} keyword, allowing for selection between \texttt{COM} (center of mass), \texttt{COAC} (center of atomic charges), \texttt{USER\_DEFINED} (user-specified coordinates), or \texttt{ZERO} (origin of the coordinate system). If \texttt{USER\_DEFINED} is selected, the keyword \texttt{REFERENCE\_POINT} must be defined in terms of the atomic coordinates.

Natural transition orbitals (NTOs) can be printed by setting \texttt{PRINT\_LEVEL} to medium in the \texttt{\&GLOBAL} section or by enabling it in \texttt{\&PRINT\%NTO\_ANALYSIS}. For an NTO analysis, unoccupied orbitals must be generated, whose number included in the analysis is adjusted using the \texttt{LUMO} keyword. One can specify a list of states for which NTOs should be printed using the \texttt{STATE\_LIST} keyword or apply a threshold for the screening of the states based on their oscillator strengths with the \texttt{INTENSITY\_THRESHOLD} keyword. NTOs can be outputted as \texttt{CUBE\_FILES} or in Molden format, the latter being controlled by the \texttt{MOS\_MOLDEN} keyword.

The excited-state gradients can be calculated by setting \texttt{RUN\_TYPE ENERGY\_FORCE} in the \texttt{\&GLOBAL} section. The specific excited state for which the gradients will be computed must be defined by adding the \texttt{\&EXCITED\_STATES} subsection within the \texttt{\&DFT} section, e.g.:
\begin{verbatim}
&DFT
  &EXCITED_STATES
    STATE 1
  &END EXCITED_STATES
&END DFT
\end{verbatim}
The parameters for the gradient calculation can be adjusted in the \texttt{\&TDDFPT\%LINRES} subsection, allowing to specify
the \texttt{PRECONDITIONER} for the optimization and \texttt{MAX\_ITER} for the maximum number of iterations. Similarly,
the fluorescence energy can be calculated by optimizing the first excited state. This is achieved by setting
\texttt{\&GLOBAL\%RUN\_TYPE GEO\_OPT} and defining the \texttt{\&MOTION\%GEO\_OPT} subsection, and specifying the reference excited state
as outlined above. Additionally, support for excited-state gradient calculations using the GAPW method has recently been
introduced~\cite{Sertcan2024}.

%
%
%
Furthermore, SOC can be taken into account by adding a perturbative correction via the \texttt{\&PROPERTIES\%BANDSTRUCTURE\%SOC}
subsection\cite{soc_paper_chemrvix}. As triggered by the keyword \texttt{RKS\_TRIPLETS}, the TDA eigenvalue problem is thereby solved for both
singlet and triplet closed-shell references, yielding the corresponding excited state manifolds with excitation amplitudes 
$\mathbf{X}_{\textnormal{\tiny{TDA}}}^{(n)}$, and thereon-based, auxiliary many-electron WFs are
generated, with the manifolds $| \Psi_{S, m_S}^{(n)} \rangle$ being here 
labeled according to the total spin of the system $S$ and the corresponding spin angular momentum quantum number $m_{S}$. 
Using quasi-degenerate perturbation theory, the unperturbed excitation energies $\Omega_{S,m_S}^{(n)}$, as obtained from
Eq.~\ref{BSE1}, are corrected by adding an additive correction based on a one-electron SOC Hamiltonian
$\hat{h}^{\textnormal{\tiny{SO}}}$, so that 
 \begin{align}
    H_{S,m_S, S', m_S'}^{(nm)} = \delta_{(nm)} \delta_{SS'} \delta_{m_S m_S'} \Omega_{S, m_S}^{(n)} \\ + \langle
    \Psi_{S,m_S}^{(n)} |
    \hat{h}^{\textnormal{\tiny{SO}}} | \Psi_{S', m_S'}^{(m)} \rangle. \label{soc_elements}
 \end{align}
Two choices are available for $\hat{h}^{\textnormal{\tiny{SO}}}$, however, for all-electron GAPW computations, van W\"ullen's
model potential based on the zeroth-order regular approximation (ZORA) is chosen by default, requiring no change in the
CP2K input~\cite{Wuellen1998, Bussy2021}.
Restricting the description to valence electrons, a SOC-corrected PP can be used~\cite{Hartwigsen1998}, as described in subsection~\ref{sec:BandStructure}, by specifying \texttt{GTH$\_$SOC$\_$POTENTIALS} as the
\texttt{\&DFT\%POTENTIAL$\_$FILE$\_$NAME} in the CP2K input file.
To print the resulting SOC-corrected energies, spin-orbit matrix elements as defined in Eq.~\ref{soc_elements}, and
oscillator strengths, the following \texttt{\&SOC\_PRINT} subsection must be added to the \texttt{\&TDDFPT\%PRINT} section of TDDFT:
\begin{verbatim}
&PRINT
  &SOC_PRINT
    SOME 
    SPLITTING
  &END SOC_PRINT
&END PRINT
\end{verbatim}
Adding the keywords \texttt{UNIT\_EV} or \texttt{UNIT\_WN} furthermore determines whether the chosen unit of the printout is
in eV or cm$^{-1}$, respectively.

\subsection{Bethe-Salpeter Equation}

The BSE is a method for computing electronic excitation energies~$\Omega^{(n)}, n\,{=}\,1,2,\ldots$ and optical absorption spectra~\cite{Ljungberg2015,Jacquemin2016,Bruneval2015,Sander2015,Liu2020,Blase2020}. 

\subsubsection{Electronic excitation energies}
For a closed-shell system, the entries of $A$ and $B$ of Eq.~\ref{BSE1} 
are 
\begin{align}
    A_{ia,jb} &= (\varepsilon_a^{GW}-\varepsilon_i^{GW})\delta_{ij}\delta_{ab} + \alpha^\mathrm{(S/T)}
    v_{ia,jb} - W_{ij,ab} \nonumber
    \\
    B_{ia,jb} &= \alpha^\mathrm{(S/T)} v_{ia,jb} - W_{ib,aj},
    \label{BSE2}
\end{align}
where $\delta_{ij}$ is the Kronecker delta, $v_{pq,rs}$ the bare Coulomb interaction and $W_{pq,rs}$ the statically screened Coulomb interaction, with $p,q,r,s$ being KS orbital indices. The user needs to set $\alpha^S=2$ for computing singlet excitations and $\alpha^T=0$ for computing triplet excitations~\cite{Ljungberg2015, Bechstedt2015}. 

A BSE calculation requires occupied KS orbitals $\varphi_i(\mathbf{r})$ and empty KS orbitals   $\varphi_a(\mathbf{r})$ from a DFT calculation, where $i\,{=}\,1,\ldots,N_\mathrm{occ}$ and
$a\,{=}\,N_\mathrm{occ}\,{+}\,1,\ldots,N_\mathrm{occ}\,{+}\,N_\mathrm{empty}$, 
as well as the $GW$ eigenvalues $\varepsilon_i^{GW}$ and $\varepsilon_a^{GW}$, respectively.
In CP2K/\textsc{Quickstep}, it is possible to use $G_0W_0$, ev$GW_0$ or ev$GW$ eigenvalues, see details in the $GW$ section and in Ref.~\citenum{Golze2019}. Thus, also $GW$ and DFT input
parameters might have an impact on BSE excitation energies $\Omega^{(n)}$. 
We recommend calculating the BSE based on ev$GW_0$@PBE eigenvalues~\cite{Perdew1996}, as discussed in Ref.~\citenum{Schambeck2024}. 
%

The TDA constrains $B=0$ and $\mathbf{Y}_\text{TDA}^{(n)}\,{=}\,0$, such that $\Omega^{(n)}_\text{TDA}$ and $\mathbf{X}^{(n)}_\text{TDA}$ can be computed from the Hermitian eigenvalue problem
\begin{align} 
A\, \mathbf{X}^{(n)}_\text{TDA} = \Omega^{(n)}_\text{TDA} \mathbf{X}^{(n)}_\text{TDA}. \label{BSE4}
\end{align}
Diagonalizing $A$ in the TDA of Eq.~\ref{BSE4}, or the full block-matrix $ABBA$ of Eq.~\ref{BSE1} takes on the order of $(N_\mathrm{occ} N_\mathrm{empty})^3$ floating-point operations. 
This translates into a computational scaling of $O(N^6)$ for a BSE calculation of system size $N$.

The eigenvectors $(\mathbf{X}^{(n)},\mathbf{Y}^{(n)})$ of Eq.~\eqref{BSE1} with elements $X_{ia}^{(n)}$ and $Y_{ia}^{(n)}$ enter the WF of the electronic excitation~\cite{Blase2020, Mewes2018}:
\begin{align}
\Psi_\text{excitation}^{(n)}(\mathbf{r}_e,\mathbf{r}_h) = \sum_{ia}\,  &X_{ia}^{(n)} \varphi_i(\mathbf{r}_h) \varphi_a(\mathbf{r}_e) \nonumber
\\ +\, &Y_{ia}^{(n)} \varphi_i(\mathbf{r}_e) \varphi_a(\mathbf{r}_h), \label{eq-BSE_exc_wave_function}
\end{align}
i.e. $X_{ia}^{(n)}$ and $Y_{ia}^{(n)}$ describe the transition amplitude between an occupied orbital
$\varphi_i$ and an empty orbital $\varphi_a$ for the $n$-th excitation.
In TDA, i.e. $\mathbf{Y}_\text{TDA}^{(n)}\,{=}\,0$, we can interpret the excitation in terms of electrons and holes: 
an electron leaves a hole behind in the (formerly) occupied orbital $\varphi_i(\mathbf{r}_h)$ and is afterwards located in the empty orbital $\varphi_a(\mathbf{r}_e) $.
%

\subsubsection{Optical absorption spectrum}
Within the BSE framework, we can compute the photoabsorption cross-section tensor
$\sigma_{\mu,\mu'}(\omega) $ with $(\mu,\mu'\in\{x,y,z\})$ as~\cite{Bruneval2016}
\begin{align}
\sigma_{\mu,\mu'}(\omega) 
= - \frac{4\pi \omega}{c} 
\mathrm{Im}\left[ 
\sum_n \frac{2 \Omega^{(n)} d^{(n)}_{\mu} d^{(n)}_{\mu'}}{(\omega+i\eta)^2-\left(\Omega^{(n)}\right)^2}
\right], \label{eq-BSE_dyn_pol_tensor}
\end{align}
with broadening $\eta$ and transition moments~\cite{Liu2020}
\begin{align}
d^{(n)}_{\mu} = \sqrt{2} \sum_{i,a} \langle \varphi_i|\hat{\mu}| \varphi_a \rangle (X_{ia}^{(n)} + Y_{ia}^{(n)}). \label{eq-BSE_trans_mom}
\end{align}
When the molecules are not aligned, e.g.~for gas phase and liquids,
we can compute the photoabsorption cross section from the spatial average of Eq.~\ref{eq-BSE_dyn_pol_tensor}:
\begin{align}
\bar{\sigma}(\omega)
&= \frac{1}{3} \sum_{\mu\in\{x,y,z\}} 
\sigma_{\mu,\mu}(\omega)
\nonumber
\\ &= - \frac{4\pi \omega}{c} \mathrm{Im}\left[
  \sum_n \frac{f^{(n)}}{(\omega+i\eta)^2-\left(\Omega^{(n)}\right)^2}
  \right], \label{eq-BSE_avg_spectrum}
\end{align}
where we have introduced the oscillator strengths~\cite{Liu2020}
\begin{align}
f^{(n)} = \frac{2}{3} \Omega^{(n)} \sum_{\mu\in\{x,y,z\}} | d^{(n)}_{\mu} |^2. \label{eq-BSE_oscillator_strength}
\end{align}
The WF of an
excited state $\Psi_\text{excitation}^{(n)}(\mathbf{r}_e,\mathbf{r}_h)$ of Eq.~\ref{eq-BSE_exc_wave_function} and its spatial properties can be analyzed by introducing measures for its spatial extent following Ref.~\citenum{Mewes2018}.
To that end, we define the corresponding expectation value for a generic operator 
\begin{align}
\langle \hat{O} \rangle _\text{exc}^{(n)}
=
\frac{ 
 \langle \Psi_\text{excitation}^{(n)} | \hat{O} | \Psi_\text{excitation}^{(n)}\rangle 
}{
 \langle \Psi_\text{excitation}^{(n)} | \Psi_\text{excitation}^{(n)}\rangle 
}.
\end{align}
For example, we can compute the distance between the electron and the hole as
\begin{align}
d_{h \rightarrow e} = | \langle \mathbf{r}_h - \mathbf{r}_e \rangle_\mathrm{exc} |. \label{eq-BSE_eh_distance}
\end{align}
This can be used to characterize a charge-transfer
state, where the electron and hole sit on different parts of the molecule and therefore $d_{h \rightarrow e}$ is non-vanishing.

Further, we can compute the exciton size
\begin{align}
d_\mathrm{exc} = \sqrt{ \langle |\mathbf{r}_h - \mathbf{r}_e|^2 \rangle_\mathrm{exc} }, \label{eq-BSE_exciton_size}
\end{align}
which quantifies the spatial extent of the combined electron-hole pair. It increases when the distance between the electron and hole $d_{h \rightarrow e}$, or their respective sizes, increases.

A typical BSE calculation starts with {\tt{RUN\_TYPE ENERGY}} and the following prototypical $GW$-BSE section:
\begin{verbatim}
&GW
  SELF_CONSISTENCY      evGW0     
  &BSE  
    TDA                 TDA+ABBA  
    SPIN_CONFIG         SINGLET   
    NUM_PRINT_EXC       15        
    ENERGY_CUTOFF_OCC   -1       
    ENERGY_CUTOFF_EMPTY -1        
    NUM_PRINT_EXC_DESCR -1        
    &BSE_SPECTRUM                 
      ETA_LIST 0.01 0.02          
    &END BSE_SPECTRUM
  &END BSE
&END GW
\end{verbatim}
Therein, the following keywords are available: 
\begin{itemize}
\item {\tt{SELF\_CONSISTENCY}}:
Determines which $GW$ self-consistency ($G_0W_0$, ev$GW_0$ or ev$GW$) is used to calculate the single-particle $GW$ energies $\varepsilon_p^{GW}$ necessary in the BSE calculation of Eq.~\ref{BSE2}. 
  We recommend using \texttt{evGW0} for BSE runs~\cite{Schambeck2024}.
  
\item {\tt{TDA}}: Specifies if the TDA and/or diagonalization of the full ABBA-matrix is employed. 
{\tt OFF}: Generalized diagonalization of $ABBA$ of Eq.~\ref{BSE1}. 
{\tt ON}: use the TDA of Eq.~\ref{BSE4} and diagonalize only $A$. 
{\tt TDA+ABBA}: compute excitation energies~$\Omega^{(n)}$ and $\Omega^{(n)}_\text{TDA}$ from Eqs.~\ref{BSE1} and~\ref{BSE4}, respectively.

\item {\tt SPIN\_CONFIG}: Choose between
{\tt SINGLET} for computing singlet excitation energies ($\alpha^S\,{=}\,2$) and {\tt TRIPLET} for computing triplet excitation energies ($\alpha^T\,{=}\,0$).
The standard is {\tt SINGLET} as an electronic excitation directly after photoexcitation is a singlet due to angular momentum conservation; triplet excited states can form by intersystem crossings.  

\item {\tt NUM\_PRINT\_EXC}: Number
  of excitation energies~$\Omega^{(n)}$ 
  to be printed.

\item {\tt ENERGY\_CUTOFF\_OCC} (in eV): Only use indices~$i$ of occupied MOs in the interval
$\hspace{2em}\varepsilon_i^\text{DFT}\in[\varepsilon_{\text{HOMO}}^\text{DFT}-\text{\tt ENERGY\_CUTOFF\_OCC},\varepsilon_{\text{HOMO}}^\text{DFT}]$
to set up the matrices $A$ and $B$ in Eq.~\ref{BSE2}.
  A small {\tt ENERGY\_CUTOFF\_OCC}  reduces  computation time and  memory   consumption, but  can affect the computed excitation energies~$\Omega^{(n)}$. 
  Usage of {\tt ENERGY\_CUTOFF\_OCC} is recommended for molecules with more than 30 atoms. 
  We recommend a convergence test by increasing {\tt ENERGY\_CUTOFF\_OCC} and observing the effect on $\Omega^{(n)}$~\cite{Liu2020}.

\item {\tt ENERGY\_CUTOFF\_EMPTY} (in eV): Analogously to {\tt ENERGY\_CUTOFF\_OCC}, but for the empty states, i.e. only empty states in the interval
$\hspace{2em}\varepsilon_a^\text{DFT}\in[\varepsilon_{\text{LUMO}}^\text{DFT},\varepsilon_{\text{LUMO}}^\text{DFT}+\text{{\tt ENERGY\_CUTOFF\_EMPTY}}]$
are used in the BSE calculation.

\item {\tt NUM\_PRINT\_EXC\_DESCR}: Number of excitations $n$, for which the exciton descriptors are printed, e.g.~$d_\mathrm{exc}^{(n)}$ of Eq.~\ref{eq-BSE_exciton_size}. 

\item {\tt BSE\_SPECTRUM}: Activates the computation and printing of the photoabsorption cross section tensor, i.e. its spatial average $\bar{\sigma}(\omega)$ of Eq.~\ref{eq-BSE_avg_spectrum} and its elements ${\sigma_{\mu,\mu'}}(\omega)$ ($\mu, \mu'\in\{x,y,z\}$) of Eq.~\ref{eq-BSE_dyn_pol_tensor} are printed.
\end{itemize}
In addition to the upper $GW$-BSE keywords, the settings within the \texttt{\&DFT} section influence the BSE excitation energies~$\Omega^{(n)}$:
\begin{itemize}    
\item {\tt XC\_FUNCTIONAL}: Choose between one of the available XC functionals. The starting point can have a profound influence on the excitation energies~\cite{Knysh2024}. 
  Motivated by the discussion in Ref.~\citenum{Schambeck2024}, we strongly recommend to use BSE@ev$GW_0$@PBE, i.e. the PBE functional~\cite{Perdew1996} as the DFT starting point.

\item {\tt BASIS\_SET}: 
The all-electron aug-cc-pVDZ basis set~\cite{Kendall1992,Pritchard2019} should be sufficient for most organic molecules, but needs to be checked with respect to convergence. 
\end{itemize}

The memory consumption of the BSE method is large, i.e. approximately
$100 \, N_\mathrm{occ}^2 N_\mathrm{empty}^2$ Bytes. The estimated memory consumption, $N_\mathrm{occ}$,
as well as $N_\mathrm{empty}$ are all printed in the BSE output. The BSE implementation is very well parallelized, i.e. one can use several nodes that can also provide the required memory.
We have compared the excitation energies from the BSE implementation in CP2K against the FHI-aims code~\cite{Liu2020} and found excellent agreement on the order of 5\,meV mean absolute deviation for the first 10 excitations of Thiel's dataset~\cite{Schreiber2008}. 
%
%
%

\section{Excited State Dynamics} \label{sec:ExcitedStateDynamics}

In RT-TDDFT, instead of solving the static SE, the aim is to solve the time-dependent SE
\be
i \frac{\partial}{\partial t} \Psi({\bf r},t) = \hat{H}({\bf r},t) \Psi({\bf r},t).
\ee
Contrary to the time-independent case, there is no variational principle for the total energy, 
and  the total energy has to be replaced by the quantum mechanical action
\be
A[\Psi] =  \int^{t_f}_{t_0} dt \langle \Psi(t) | i\frac{\partial}{\partial t} - \hat{H}(t)| \Psi(t)\rangle
\ee
for which the function $\Psi(t)$ that makes the action stationary will be its solution.

\subsection{Real-Time Propagation and Ehrenfest Dynamics}

The resulting time-dependent KS equations read as
\be
i \frac{\partial}{\partial t} \psi_i({\bf r},t) =  \left[  -\frac{\nabla^2}{2}+v_{\text{KS}}({\bf r},t) + \textbf{F}(t) \right] \psi_i({\bf r},t), \label{eq:tdks}
\ee
with $\psi_i$ being the KS orbitals, 
\be
\rho({\bf r},t)  = \sum_i f_i |\psi_i({\bf r},t)|^2
\ee
the time-dependent density, and 
$\textbf{F}(t)$ an optional time-dependent external field. 
By applying an external electric field, we intend to simulate 
the interaction between the electrons and an  external radiation. 
The light is then treated classically using either the length-gauge for non-periodic systems or the velocity-gauge for
periodic systems. 

A rather general derivation for (nonadiabatic quantum-classical) Ehrenfest dynamics can be obtained
starting from the action of a system~\cite{Kunert.2003}. For a situation in which the electrons are treated quantum mechanically, 
while the nuclei are treated classically, the total action can be written as the sum of the two environments $A = A_c + A_q$, where
\be
 \qquad A_c = \int_{t_0}^{t_f} dt \, \left[ \sum_A \frac{M_A}{2} \dot{\vR}_A-U({\bf R},t)\right]
\ee
and $A_q$ the quantum mechanical action as defined above. 
The equations of motion are then derived by making the action stationary:
\be
\frac{\delta A}{\delta\langle \Psi({\bf r}, t)|} = 0 \qquad \frac{\delta A}{\delta\langle {\bf R}( t)|} = 0
\ee
Evaluating these expressions in the framework of TDDFT, the equations of motion become
\be
M\ddot{\bf R}  = -\frac{\partial}{\partial {\bf R}} U({\bf R},t) - \sum_j\left\langle \Psi^j \left| \frac{\partial}{\partial{\bf R}} V_{\text{int}}({\bf r},{\bf R})\right| \Psi^j \right\rangle
\ee
for the nuclear motion, while the time-dependent Schr\"odinger equation as given above is used for the electrons. 
Yet, in a numerical calculation, the WFs are often replaced by a finite basis set representation in terms of the LCAO.
For PWs, the equations remain the same, since they do not depend on the nuclear coordinates. 
Using Gaussian basis functions for the expansion of the WFs, however, introduces an implicit dependence of the WFs on the nuclear positions. 
Since in AIMD the nuclear coordinates are a function of time, the time-derivative in the quantum mechanical action has to be replaced by the total time-derivative
\be
\frac{d}{dt} = \frac{\partial}{\partial t} + \sum_A \frac{\partial {\bf R}_A}{\partial t}\frac{\partial}{\partial {\bf R}_A}.
\ee
Due to the introduction of a finite basis, the independent variables for making the action constant become the expansion coefficients $C_{j\alpha}(t)$ 
of the MO
$j$ 
in the contracted basis function  
$\phi_\alpha({\bf r},t)$, i.e. 
$$
\dot{C}_{j\alpha} = -\sum_{\alpha\beta}S^{-1}_{\beta\gamma}\left(  i H_{\beta\gamma} + B_{\beta\gamma} \right) C_{j\gamma},
$$
where
$$
S_{\alpha\beta} = \langle \phi_\alpha| \phi_\beta \rangle \qquad B_{\alpha\beta} = \langle \phi_\alpha | \frac{d}{dt} \phi_{\beta} \rangle.
$$
Hence, in Ehrenfest dynamics an additional contribution due to the derivative of the basis functions becomes part of the Hamiltonian
used in the exponential operator. Instead of being purely imaginary, the matrix in the exponential of the propagator becomes fully complex.

To run real-time propagation (RTP), where the nuclei are fixed and only the electronic degrees of freedom are propagated, or genuine Ehrenfest dynamics, the corresponding {\tt{RUN\_TYPE}} in the {\tt{\&GLOBAL}} section has to be set to \texttt{RT\_PROPAGATION} or \texttt{EHRENFEST\_DYN}, respectively. Furthermore, the {\tt \&MOTION\%MD} section has to be present to specify the discretized {\tt{TIMESTEP}} and the desired number of {\tt{STEPS}}. It is crucial to set an appropriate timestep to propagate the electronic degrees of freedom, which is much smaller than for MD and is typically on the order of attoseconds.
All other input parameters related to the RT-TDDFT run are specified in the {\tt{\&DFT\%REAL\_TIME\_PROPAGATION}} section.
The simulation needs to start from the electronic density at $t_0$ specified by the keyword {\tt{INITIAL\_WFN}}. This can be either the ground state obtained by means of an initial SCF optimization denoted by {\tt{SCF\_WFN}}, or by providing the restart file of a previously computed WF via {\tt{RESTART\_WFN}}. Alternatively, the propagation can also be restarted from a RTP or Ehrenfest dynamics run by {\tt{RT\_RESTART}} and providing the correct restart file. Note that the RTP restart file has a format different from the SCF restart file. The path to the restart file is in both cases provided by the usual {\tt{\&DFT\%RESTART\_FILE\_NAME}} keyword.

Three different propagators are available in CP2K, the enforced time-reversible symmetry propagator (ETRS), the exponential midpoint (EM) propagator, and the Crank-Nicholson propagator, which can be seen as a first-order Pad\'e approximation of the EM propagator~\cite{Castro2004}. The ETRS approach starts with an exponential approximation to the evolution operator $\hat{U}(t,0)=\exp{-it\hat{H}(0)}$, to then compute the final time-reversible and unitary propagator self-consistently. 
In the RTP scheme (fixed ionic positions), the self-consistent solution involves only the calculation of the new KS matrix for the propagated coefficients. For Ehrenfest dynamics, the iterative procedure is embedded in the integrator of the nuclear equations of motion.
Hence, each iteration step for Ehrenfest dynamics involves an RTP step and the evaluation of the nuclear forces. Due to this, more iterations are needed to reach self-consistency for the propagator. The convergence criterion implemented in CP2K is defined as 
\be
||  \Delta C^T S \Delta C||_{\text{}max} < \epsilon
\ee
with $\Delta C$ being the difference of coefficient matrices in two successive steps, and $\epsilon$ given by {\tt{EPS\_ITER}}.

In terms of computational cost, the most expensive part is the evaluation of the matrix exponential in the propagator. Four different methods, among those listed in Ref.~\citenum{Moler.2003}, have been implemented in CP2K, i.e. diagonalization, the Taylor expansion, the Pad\'e approximation, and the Arnoldi subspace iteration. The method is selected using the {\tt{MAT\_EXP}} keyword. The Arnoldi method often provides superior performance. Comparing the theoretical scaling, the Arnoldi method is expected to be about 5 times faster than the Pad\'e or Taylor approaches. However, the Pad\'e approximation can sometimes be a faster and more stable choice than the Arnoldi method (e.g. in the case of large integration timesteps). 
Since the propagation is based on an iterative procedure, it is convenient to apply an extrapolation scheme to speed up convergence~\cite{Kuehne2007}. For Ehrenfest dynamics, an extrapolation based on the product of the density and the overlap matrix turns out to be a good choice. However, the quality of the different extrapolation approaches strongly depends on the system applied and the simulation settings. Therefore, which method to use and the extrapolation order need to be tested on the particular system of interest.
For very large systems, the density matrix-based method can be used to achieve linear scaling by activating the keyword {\tt{DENSITY\_PROPAGATION}}~\cite{Andermatt.2017}.

One of the most relevant application domains for RT-TDDFT is the study of light–matter interactions, e.g. in the field of spectroscopy, excited state dynamics, and radiation damage.  To mimic these phenomena, at any time during the propagation, it is possible to apply a time-dependent electric field ${\bf F}(t)$. The applied field is in general modulated by an envelope function $E_{\text{env}}(t)$ and is defined as:
\be
\textbf{E}(t) = \textbf{P} E_{\text{env}}(t) \cos \left( \omega_{0} t + \phi \right),
\ee
where $\textbf{P}$ is the field polarization, $\omega_{0}$ is the carrying frequency at which the field oscillates, and $\phi$ its initial phase.
The characteristic of the applied field, as well as its time extension, is provided through the section {\tt{\&DFT\%EFIELD}}. The time-dependent electric field defined within this section can only be used in combination with RT-TDDFT.
By default, the coupling between the electric field and the electronic degrees of freedom is described within the length-gauge, by adding to the Hamiltonian the dipole coupling term $ e \textbf{E}(t) \cdot {\bf r} $. This approach is only valid for isolated molecular systems, and not when periodic boundary conditions are applied. 
The velocity-gauge form of the equations suitable for  periodic systems is obtained through a gauge transformation involving the vector potential~\cite{Bertsch.2000,Yabana.2011,Pemmaraju.2018}
\be
{\bf A}(t) = c \int^t  dt' \, {\bf E}(t').
\ee
In the time-dependent KS equations, the vector potential ${\bf A}(t)$ appears in the kinetic energy term and, in the case where non-local PPs are used, the gauge field also transforms the electron–ion interaction. To use this representation, the {\tt{VELOCITY\_GAUGE}} keyword has to be used within the aforementioned \texttt{REAL\_TIME\_PROPAGATION} section. The total energy varies over time as the applied external field interacts with the system. To monitor the time evolution of the field and the various terms that contribute to the total electronic energy, which can be activated via the {\tt{\&REAL\_TIME\_PROPAGATION\%PRINT}} print key.

Once the RTP has started, the evolution of the electronic structure can be monitored by means of several descriptors,
such as the time-dependent dipole moment, the time-dependent total electronic density, and the spin density, which can be printed out as a series of cube files for every arbitrary number of propagation steps. As for ground state calculations, the output of these quantities is activated by the corresponding {\tt{\&DFT\%PRINT}} print key.

The electron current density can be obtained as
\begin{equation}
{\bf j}({\bf r}, t) =\frac{1}{2}\sum_i f_i\left(\psi^*_i({\bf r}, t) \, \boldsymbol{\pi}\,\psi_i({\bf r}, t) + c.c.\right),
\end{equation}
where 
\begin{equation}
\boldsymbol{\pi} = [{\bf r},{\bf H}] = -i \boldsymbol{\nabla} +  {\bf A(t)} + \left[V_{PP},{\bf r} \right].
\end{equation}
Integrating ${\bf j}({\bf r}, t)$ in the simulation cell yields the macroscopic current
\begin{equation}
{\bf J}(t) = -\frac{1}{V} \int d{\bf r} \, {\bf j}({\bf r}, t),
\end{equation}
where $V$ is the cell volume.
The print key {\tt{\&REAL\_TIME\_PROPAGATION\%PRINT\%CURRENT\_INT}} activates the calculation and the output of the macroscopic current.
This quantity can then be used to calculate the frequency-dependent conductivity, which is defined as the ratio of the current’s Fourier transform to that of the applied electric field
\begin{equation}
\sigma_{ij}(\omega)=\frac{\mathcal{F}[J_i](\omega)}{\mathcal{F}[E_j](\omega)}.
\end{equation}

The projection of the time-dependent orbitals onto some previously selected reference states is particularly useful for verifying the variation in population of specific states, while monitoring density differences is usually helpful for detecting charge-transfer processes. 
 By projecting time-dependent MOs on some reference states (e.g. the initial MOs) using the print key {\tt{\&PRINT\%PROJECTION\_MO}}
 the overlap between the propagated orbital $\psi_i({\bf r}, t) = \sum_\alpha C_{i\alpha}(t)\phi({\bf r})$
and any reference orbital $\psi^{\text{ref}}_m$ is calculated. When the reference orbitals are the ground state virtual orbitals, 
the quantity 
$$
N_{\text{exc}}(t)= \sum_m^{\text{unocc}}\sum_i^{\text{occ}} \left\|  \langle  \psi^{\text{ref}}_m |\psi_i(t)\rangle \right\|^2
$$
 is an estimate of the number of electrons that have been excited into the unoccupied space. 

\subsection{Real-Time Bethe Salpeter Propagation}


Although the exact solution of time-dependent KS equations of Eq.~\ref{eq:tdks} can describe the nonlinear-response of a system to an external field $\hat{F}(t)$, using approximate
XC kernels 
often
leads to inaccuracies.
For instance, the adiabatic LDA fails to describe excitonic energies accurately~\cite{Ullrich2012,Onida2002,Blase2020}.

Hence, choosing an approximate self-energy operator is more suitable for the description
of non-local screening effects present in the excitonic spectra~\cite{Onida2002}.
In the RT-BSE scheme implemented in CP2K~\cite{Marek2025},
the Coulomb-hole plus screened-exchange (COHSEX) self-energy is employed~\cite{Attaccalite2011,HedinGW,Casida1989,Farid1988,Betzinger2015}. Moreover, 
instead of state propagation, solution of the equation of motion for the single-particle density 
matrix $\hat{\rho}(t)$
\begin{align}
    \frac{\partial \hat{\rho}}{\partial t} = \frac{-\mathrm{i}}{\hbar}
    [\hat{H}^\text{eff}(t), \hat{\rho}(t)] \label{eq:rtbse}
\end{align}
is implemented, where the effective Hamiltonian is given by \cite{Attaccalite2011}
\begin{align}
    \hat{H}^\text{eff}(t) &= \hat{H}^{G_0W_0} + \hat{U}(t) + \nonumber \\
    &+ \hat{V}^\text{Hartree}[\hat{\rho}(t)] - \hat{V}^\text{Hartree}[\hat{\rho}(t_0)] + \nonumber \\
    &+ \hat{\Sigma}^\text{COHSEX}[\hat{\rho}(t)] - \hat{\Sigma}^\text{COHSEX}[\hat{\rho}(t_0)]
\end{align}
Therein, $\hat{H}^{G_0W_0}$ is the $G_0W_0$ Hamiltonian, $\hat{F}(t)$ is the applied external field,
$\hat{V}^\text{Hartree}[\hat{\rho}(t)]$ is the Hartree potential (which depends on the
single-particle density matrix) and $\hat{\Sigma}^\text{COHSEX}[\hat{\rho}(t)]$
is the COHSEX self-energy. This form of the effective Hamiltonian leads to oscillations of
$\hat{\rho}(t)$ formally equivalent to solving the Casida equation of Eq.~\ref{BSE1} with Bethe-Salpeter entries shown in Eq.~\ref{BSE2}~\cite{Attaccalite2011},
hence the name of the method.

In order to run a RT-BSE calculation (so far only implemented for molecules, i.e. without PBC, and for static nuclei), a similar setup as for running an RT-TDDFT propagation calculation is employed,
namely
\begin{itemize}
    \item Set {\tt RUN\_TYPE} to {\tt RT\_PROPAGATION}
    \item Include a {\tt \&MD} section to set the size of the {\tt TIMESTEP} and
    the total number of {\tt STEPS}
    \item Specify all options necessary to run an accurate $G_0W_0$ calculation for molecules (most importantly, the subsection {\tt \&PROPERTIES\%BANDSTRUCTURE\%GW} should be specified)
    \item Include the {\tt \&REAL\_TIME\_PROPAGATION} section with all relevant entries and add subsection {\tt \&RTBSE}.
\end{itemize}
Note that the RT-BSE method employs the RI approximation for the Hartree and COHSEX terms~\cite{Graml2024a},
so an appropriate {\tt RI\_AUX} basis should be set.
Example RT-BSE calculations for organic molecules are also available in repositories of the RT-BSE implementation paper~\cite{Marek2025}.
A typical RT-BSE input with the most important {\tt \&REAL\_TIME\_PROPAGATION} section will look as follows: 
\begin{verbatim}
&REAL_TIME_PROPAGATION
  &RTBSE ! Start the RTBSE Method
    FT_DAMPING [fs] 5.0 ! FT damping
    FT_START_TIME 0.0 ! FT offset
    &PADE_FT ! Padé interpolation
      E_MIN [eV] 0.0
      E_MAX [eV] 50.0
      E_STEP [eV] 0.001
      FIT_E_MIN [eV] 0.0
      FIT_E_MAX [eV] 300.0
    &END PADE_FT
  &END RTBSE
  EPS_ITER 1.0E-8 ! Check convergence
  MAX_ITER 50
  MAT_EXP BCH
  EXP_ACCURACY 1.0E-14 ! < EPS_ITER
  APPLY_DELTA_PULSE
  DELTA_PULSE_DIRECTION 1 0 0
  DELTA_PULSE_SCALE 0.0001 
  &PRINT
    &MOMENTS ! Print moments time trace
      FILENAME MOMENTS
    &END MOMENTS
    &MOMENTS_FT ! Fourier transform
      FILENAME MOMENTS_FT
    &END MOMENTS_FT
    &FIELD ! Print applied field trace
      FILENAME FIELD
    &END FIELD
    &POLARIZABILITY
      FILENAME POLARIZABILITY
      ELEMENT 1 1 ! Tensor element
    &END POLARIZABILITY
  &END PRINT
&END REAL_TIME_PROPAGATION
\end{verbatim}
The most relevant keywords and subsections in the above section entail:
\begin{itemize}
\item {\tt EPS\_ITER}: The equation of motion of Eq.~\ref{eq:rtbse} is solved by exponential evolution operators applied to the
single-particle density matrix $\hat{\rho}$. These are applied self-consistently by an enforced time reversal
symmetry scheme described previously~\cite{Castro2004,ORourke2015,Andermatt.2017}, so that
\begin{align}
    \hat{\rho}(t + \Delta t) \approx&\;
    \mathrm{e}^{-\mathrm{i}/\hbar \hat{H}^\text{eff}(t+\Delta t) \Delta t/2}  \nonumber \\
    &\times \mathrm{e} ^ {- \mathrm{i}/\hbar \hat{H}^\text{eff}(t) \Delta t/2} \hat{\rho}(t)
     \nonumber \\
    &\times \mathrm{e} ^ {\mathrm{i}/\hbar \hat{H}^\text{eff}(t) \Delta t/2}  \nonumber \\
    &\times \mathrm{e} ^ {\mathrm{i}/\hbar \hat{H}^\text{eff}(t+\Delta t) \Delta t/2}.
    \label{eq:rtbse_etrs}
\end{align}
The self-consistency is present since the effective Hamiltonian $\hat{H}^\text{eff}(t + \Delta t)$
depends on the single-particle density matrix at time $t + \Delta t$, to which we need to
propagate. The self-consistency threshold is controlled by {\tt EPS\_ITER}. Smaller threshold
implies more stable propagation, but at the same time usually requires more iterations. Usually,
one can choose {\tt EPS\_ITER} similar to {\tt EPS\_SCF}, although a larger tolerance might be sufficient for consistent results - one should test the convergence of {\tt EPS\_ITER} and {\tt TIMESTEP} pair. Note that the maximum among the absolute
values of elements of $\hat{\rho}(t+\Delta t) - \hat{\rho}(t)$ in the AO basis is used
as the metric for the threshold.

\item {\tt MAX\_ITER}: Limits the maximum number of the self-consistent iterations in the ETRS scheme.

\item {\tt MAT\_EXP}: The matrix exponentiation in RT-BSE is implemented by two approaches -
the (Baker-Campbell-Hausdorff) {\tt BCH}\cite{Castro2004} and the {\tt EXACT} diagonalization.
The keyword {\tt MAT\_EXP} controls the choice between these two.

\item {\tt EXP\_ACCURACY}: For the BCH method, it is also necessary to specify the matrix exponentiation accuracy, which should be stricter than the ETRS threshold {\tt EPS\_ITER}.

\item {\tt APPLY\_DELTA\_PULSE}: In order to observe any oscillations, the external field $\hat{U}(t)$ needs to be specified. Either the explicit time-dependent field can be supplied in the
{\tt \&DFT\%EFIELD} subsection, or a delta pulse can be 
triggered by inclusion of this keyword~\cite{Mattiat2022}.

\item {\tt DELTA\_PULSE\_DIRECTION}: 
Sets the Cartesian direction of the delta pulse.
The resulting vector is always normalized by the code.

\item {\tt DELTA\_PULSE\_SCALE}: 
Sets the magnitude of the delta pulse directly
(in atomic units). The change in the metric of the single-particle density matrix after
the application of the delta pulse is reported by the code - if the change is larger than 1 in atomic
units, then the ETRS loop might struggle to converge - we recommend decreasing the
{\tt DELTA\_PULSE\_SCALE} in such cases.

\item {\tt \&MOMENTS}: Activates printing of the time-dependent dipole moment 
\begin{align}
    \mu_j(t) = \mathrm{Tr}(\hat{\rho}(t) (\hat{r} - \bm{r}_\text{CC})),
\end{align}
where $\mu_j$ is the $j$th component of the dipole moment $\mathbf{\mu}(t)$ and $\bm{r}_\text{CC}$
is the reference point - in RT-BSE, this is the center of atomic charges.

\item {\tt FT\_DAMPING}: Damping $\gamma$ applied during Fourier transform to stabilize it\cite{Mueller2020}. Example is given in the {\tt \&MOMENTS\_FT} section. If not
given explicitly, $\gamma$ is determined so that factor of $\mathrm{e}^{-4}$ is applied at the end
of the time trace of the observables, starting from {\tt FT\_START\_TIME}.

\item {\tt FT\_START\_TIME}: Offset along the time axis for the purposes of Fourier transform. Defaults to zero. Useful for real-time pulses defined in {\tt \&EFIELD} section. For example,
for {\tt GAUSSIAN\_ENV} with non-zero {\tt T0} parameter, setting {\tt FT\_START\_TIME} to the
same value will result in the envelope being a real function in frequency space.

\item {\tt \&MOMENTS\_FT}: Activates printing of the Fourier transform of the dipole moment 
$\mathbf{\mu}(t)$, which is stabilized
by the application of {\tt FT\_DAMPING} $\gamma$, so that
\begin{align}
    \mu_j(\omega) = \int \mathrm{d}t e^{\mathrm{i} (\omega + \mathrm{i}\gamma) t} \mu_j(t+t_0),
\end{align}
where $t_0$ is the defined by the {\tt START\_TIME}.

\item {\tt \&FIELD}: Activates printing of the time-dependent trace of the external field. When no
{\tt \&EFIELD} subsection within the {\tt \&DFT} section is present, outputs just zeros (for example
when only the delta pulse is applied.)

\item {\tt \&POLARIZABILITY}: Activates printing of the polarizability tensor, defined in terms of the
Fourier transform of the dipole moment and the Fourier transform of
the electric field as\cite{Yabana2006,Li2020}
\begin{align}
    \alpha_{jk}(\omega) = \frac{\mu_j(\omega)}{E_k(\omega)}.
\end{align}
The polarizability is related to the dipole strength function\cite{Yabana2006,Li2020}, which
characterizes the absorption spectrum of the system. Control over which element is printed
is given by the {\tt ELEMENT} keyword. The polarizability calculation also uses the settings
from the {\tt \&MOMENTS\_FT} section.

\item {\tt \&PADE\_FT}: Section controlling the Padé interpolation\cite{PadeFTMattiatLuber,PadeFTBrunerLopata,GreenXLeuckeGolze} of the Fourier transforms.
A new frequency grid spanning from {\tt E\_MIN} to {\tt E\_MAX} with {\tt E\_STEP} steps is created,
on which the Fourier transforms are reevaluated using the Padé fitting of the original transforms.
The original transforms are restricted to energies within {\tt FIT\_E\_MIN} and {\tt FIT\_E\_MAX}.
By default, no restriction is applied -- entire frequency range of the original transform is used
to construct the Padé parameters.
\end{itemize}

The code automatically saves the state of the system after each successful timestep, so that
the propagation can be continued after interruption. Use {\tt INITIAL\_WFN RT\_RESTART} in the {\tt REAL\_TIME\_PROPAGATION} section 
to enable the restart. Note that the total propagation time determines the energy resolution in the
resulting Fourier transform, while the timestep determines the maximum frequency/energy
accessible to the Fourier transform.
In other words, if a finished RT-BSE calculation does not yield the required energy resolution, it
is possible to simply increase the number of {\tt STEPS} and continue the propagation. Note that the
{\tt \&MOMENTS} and {\tt \&FIELD} should not be suppressed (for example by an insufficient
{\tt PRINT\_LEVEL}) in order for the program to be able to read the time traces from the previous run.

%

\subsection{Surface Hopping via NEWTON-X}





The interface with the NEWTON-X program package enables to perform on-the-fly non-adiabatic molecular dynamics (NAMD)
simulations using Tully's fewest switches surface hopping algorithm~\cite{NewtonX2022, NewtonX2014, NewtonX2007}.
The total electronic WF
\begin{align}
   | \Psi (\mathbf{R} (t))\rangle = \sum_{(n)} d^{(n)} (t) | \Psi^{(n)} (\mathbf{R}(t)) \rangle
   \label{total_electronic_wavefunction}
\end{align}
is thereby relying on the TDDFT module of CP2K, thus the singly excited states $|\Psi^{(n)} (\mathbf{R}(t)) \rangle$ are defined
based on the excitation amplitudes $X_{ia}^{(n)}$ of
the eigenvalue problem of Eq.~\ref{BSE1}, i.e.
\begin{align}
|\Psi^{(n)} \rangle  = \sum_{ia} X_{ia}^{(n)} | \Phi_{ia} \rangle
\end{align}
with the singly-excited Slater determinants $|\Phi_{ia} \rangle$. 
The Lagrangian of Eq.~\ref{BSE1} furthermore defines the excited state nuclear gradients, as outlined in Section~\ref{sec:OpticalSpectroscopy}.
The therein defined force $\mathbf{F}$, given as the negative derivative of the variational Lagrangian with respect to
the nuclear coordinates $\mathbf{R}$,
\begin{align}
    F_{\alpha}(t) = - \nabla_{\alpha} L(\mathbf{R}(t))
\end{align}
determines the acceleration $\mathbf{a}$ of nuclei $\alpha$ as
\begin{align}
   \mathbf{a}_{\alpha} (t) =  \frac{\mathbf{F}_{\alpha} (t)}{M_{\alpha}}
\end{align}
with $M_{\alpha}$ indicating the corresponding nuclear mass.
Corresponding CP2K inputs thus require choosing \texttt{RUN\_TYPE ENERGY\_FORCE} with a suitable \texttt{\&TDDFPT}
section and to print excited state nuclear gradients and amplitudes by adding \texttt{\&PRINT} subsections to the
\texttt{\&FORCE\_EVAL} and \texttt{\&TDDFPT} sections, respectively:
\begin{verbatim}
&FORCE_EVAL 
  &PRINT
    &FORCES
    &END FORCES
  &END PRINT
  ...
&END FORCE_EVAL
&TDDFPT
  &PRINT
    &NAMD_PRINT
      PRINT_VIRTUALS T
    &END NAMD_PRINT
  &END PRINT
  ...
&END TDDFPT
\end{verbatim}
The keyword \texttt{PRINT\_VIRTUALS} implies that the excited state amplitudes are printed within the MO basis, no longer referring to Sternheimer atomic-orbital based quantities.
Real-time propagation of the coefficients $d^{(n)} (t)$ of the total electronic WF, as defined in Eq.~\ref{total_electronic_wavefunction}, requires furthermore computing nonadiabatic coupling elements
$\sigma_{(mn)}(t)$ via
\begin{align}
   i \frac{\textnormal{d}d^{(m)}(t)}{\textnormal{d} t} = \sum_{(n)} d^{(n)} (t)[\delta_{(mn)} E_{(n)} (t) - i
   \sigma_{(mn)} (t)]
\end{align}
with $E_{(n)}(t)$ representing the total energy of excited state $n$, i.e. $E_{(n)}(t)=
E_{\textnormal{\tiny{GS}}}(t) +
\Omega^{(n)}(t)$.
Input options for efficient time-derivative couplings are set within corresponding NEWTON-X input files, requiring no further specifications within the CP2K input file. 
To generate initial geometries and velocities based on a Wigner probability distribution,
a print statement has to be added to the \texttt{\&VIBRATIONAL\_ANALYSIS} section to enable printing of Cartesian normal
modes:
\begin{verbatim}
&VIBRATIONAL_ANALYSIS
  &PRINT
    &NAMD_PRINT
    &END NAMD_PRINT
  &END PRINT
&END VIBRATIONAL ANALYSIS
\end{verbatim}

\section{X-Ray Spectroscopy} \label{sec:XraySpectroscopy}

The prediction of properties that depend on the core electronic density requires the explicit calculation of all-electron WFs and the use of large basis sets, close to the basis set limits. This is the case for inner-shell spectroscopy, where excitation processes of core orbitals involve the rearrangement of the all-electron charge distribution. In CP2K/\textsc{Quickstep}, all-electron calculations require the GAPW approach, as described in section~\ref{sec:TotalEnergy}.

The X-ray absorption process is related to the incoming photon being absorbed by core-level electrons, which get resonantly promoted to an excited state. The process has a certain probability depending on the properties of the initial and final states. Within the dipolar approximation, certain selection rules apply, such as total spin conservation and total angular momentum change by $\pm 1$. Due to the very local atomic character of the process, for excitation of a 1s level, only unoccupied states with a local p-character are probed. The spectroscopic process is ultrafast (attosecond) so that the nuclear movement during the excitation can be neglected.

\subsection{Transition Potential Method}

A KS orbital is by definition the WF of a fictitious non-interacting system introduced to reproduce the electron density of the real interacting system. Hence, KS orbitals are considered formal auxiliary quantities without specific physical meaning.  Nevertheless, the KS eigenvalues are commonly used to describe physical quantities such as ionization energies and excitation energies, and the orbitals are used to obtain the transition moment.
To compute excitation energies in less approximate ways, the $\Delta$SCF KS method has been devised, comparing two systems with one or more excited electrons, taking into account full relaxation.

The incoming radiation is represented by the electromagnetic field
\begin{equation}
{\bf A}({\bf r},t) A_0 {\bf e} \cos({\bf k}\cdot{\bf r}-\omega t).
\end{equation}
The resulting transition probability between the initial and final state is
\begin{equation}
P_{if}=\frac{\pi e^2}{2\hbar m^2 c^2} A_0^2\left|\langle f |e^{i{\bf k}\cdot{\bf r}} {\bf e} \cdot {\bf p}| i\rangle \right|^2 ,
\end{equation}
which within the dipole approximation simplifies to $P_{if}\approx |\langle f |{\bf e}\cdot {\bf p}| i\rangle|$, when using the velocity form.
This form is equivalent to the length form $P_{if}\approx (E_f-E_i)|\langle f |\boldsymbol{\mu}| i\rangle|$, where $\boldsymbol{\mu}$ is the dipole operator, when the WFs are exact eigenstates.
The single-electron excitation approximation makes it possible to rewrite the initial state as a core WF and the final state as an unoccupied/free electron WF, thereby assuming that all other electrons do not participate.
The corresponding matrix element can be rewritten as a single-electron matrix element.
The transition potential (TP) method adopts this independent particle approach, but the orbital energies and WFs are taken as a solution of the KS equation with a modified core potential on the absorbing atom.
The modified potential employed in standard applications is the TP introduced by Slater and is based on the creation of half core-hole (HCH). Other types of core-hole potentials can also be employed with satisfactory results, such as the full core-hole (FCH). The actual final location of the promoted electron is not taken into account for the determination of the spectra, assuming that it is immediately delocalized in the conduction band and its contribution is nearly equivalent, irrespective of the specific final state. In this manner, only a single electronic structure calculation is needed to determine the entire spectrum. 

The procedure consists of running an initial ground state calculation, identifying the core state that has to be excited, changing the occupation number of the corresponding KS orbital, and restarting the SCF optimization while keeping the occupation of the excited core state at the modified value. The excitation energies and transition moments are finally evaluated from the KS energies and MOs obtained with the modified occupation:
\begin{equation}
\hbar \omega_{if} = \varepsilon_f^{\text{TP}}-\varepsilon_i^{\text{TP}} \qquad P_{if} = |\langle \psi_f^{\text{TP}} |\boldsymbol{\nabla}| \psi_i^{\text{TP}}\rangle|^2
\end{equation}
It is good practice, after the first ground state optimization, to localize the core orbitals to facilitate the identification of the target core state. The localization is activated via the \texttt{\&DFT\%XAS\%LOCALIZE} section, and is carried out for \texttt{STATE\_SEARCH} states, starting from the lowest:
\begin{verbatim}
&XAS
  &SCF
    ...
  &END SCF
  METHOD        TP_HH
  DIPOLE_FORM   VELOCITY
  STATE_TYPE    1s
  STATE_SEARCH  10 
  ATOMS_LIST    1
  ADDED_MOS     1000
  &LOCALIZE
  &END LOCALIZE
  &PRINT
    &PROGRAM_RUN_INFO
  &END PRINT
  &RESTART
    FILENAME ./root
  &END RESTART
  &XAS_SPECTRUM
    FILENAME ./root
  &END XAS_SPECTRUM
  &XES_SPECTRUM
    FILENAME ./root
  &END XES_SPECTRUM
&END XAS
\end{verbatim}
By default, the TP calculation uses the same settings as specified for the ground state.
Any different settings given in the 
\texttt{\&XAS\%SCF} subsection override the original ones. 
Anyway, TP calculations run only with the 
diagonalization method because KS orbitals and 
their occupations are needed. The emptied core 
state is defined by the atom, as specified in 
\texttt{ATOMS\_LIST}, and the orbital 
character \texttt{STATE\_TYPE}.  An independent TP calculation is performed for
each atom of the list. The \texttt{METHOD} keyword defines how the occupation of the core
orbital is modified, whereas \texttt{ADDED\_MOS} are 
the additional virtual orbitals that will be
computed to obtain a spectrum over a wider
energy range. The TP calculation
simultaneously provides XAS, core-to-empty transitions, XES, and valence-to-core transitions. Since each TP calculation entails a full SCF optimization, the usual electronic structure output, such as the projected density of states and the cube files of the MOs, are available. These quantities are generally very useful for the analysis of the transition character and rationalize the electronic properties. 

\subsection{Linear-Response Time-Dependent Density Functional Theory}

Within LR-TDDFT theory, optical absorption energies are calculated as solutions of the generalized eigenvalue problem presented in section~\ref{LR-TDDFT}. A full diagonalization yields the complete spectrum, with electronic transitions from all occupied to all unoccupied states, including X-ray absorption near-edge structure (XANES), also known as near-edge X-ray absorption fine structure (NEXAFS), from core states to low-lying unoccupied states. However, this approach is not suitable for large systems, as its computational cost scales as $\mathcal{O}(N^6)$ with system size (diagonalization of a $N_\text{occ}N_\text{empty}$ matrix). 

For efficiency, the CP2K/\textsc{Quickstep} LR-TDDFT implementation for XAS relies on the following three approximations~\cite{Bussy2021}.
The first is the core-valence separation (CVS) approximation~\cite{cederbaum1980many}. Due to large differences in energy and localization, the core and valence states are only weakly coupled. This reduces the dimensions of the eigenvalue problem from $N_\text{occ}N_\text{empty}$ to $N_\text{core}N_\text{empty}$, where only occupied core states are considered.
In the so-called sudden approximation~\cite{george2008time}, excitations from various core states are further decoupled. Smaller eigenvalue problems of size $N_\text{empty}$ for s-type core states or $3N_\text{empty}$ for p-type states can be solved independently, leading to further computational savings. However, this approximation relies on the assumption that the core states are highly localized in space.

The localization of core states is further exploited for the efficient calculation of 4-center 2-electron ERIs, using a core-specific RI scheme. Coulomb integrals involving the core states $I, J$ are computed as
\begin{equation} 
    (pI|Jq) \approx (pI|\mu) (\mu|\nu)^{-1}(\nu|Jq), 
\end{equation}
where $\mu$ and $\nu$ correspond to RI basis functions centered on the same atom where $I$ and $J$ are localized. The same holds for exact-exchange ERIs in hybrid DFT calculations, i.e. $(pq|IJ) \approx (pq|\mu) (\mu|\nu)^{-1} (\nu|IJ)$. Finally, ERIs involving the XC kernel are computed as follows:
\begin{equation} 
    (pI|f_{xc}|Jq) \approx (pI|\kappa) (\kappa|\lambda)^{-1}(\lambda|f_{xc}|\mu)(\mu|\nu)^{-1}(\nu|Jq). 
\end{equation}
Yet, since all RI basis elements are centered on the excited atom, $f_{xc}[n]$ can be evaluated by a projection of the total electronic density onto the RI basis, expressed as a linear combination:
\begin{equation} \label{xastdp_proj} 
\begin{aligned} 
    \rho(\mathbf{r}) &= \sum_{pq}P_{pq}\varphi_p(\mathbf{r})\varphi_q(\mathbf{r}) \\
    &\approx \sum_{pq}\sum_{\mu\nu}P_{pq}(pq\nu)S^{-1}_{\mu\nu}\chi_\nu(\mathbf{r}) \\
    &= \sum_\nu d_\nu \chi_\nu(\mathbf{r}),
\end{aligned}
\end{equation}
where $P_{pq}$ is the density matrix in the AO basis, while $(pq\nu)$ represents the overlap integrals between two general AOs and a RI basis element, and $S_{\mu\nu}^{-1}$ is the inverse overlap matrix element of the RI functions centered on the excited atom.

Using LR-TDDFT to compute XAS spectra is based on perturbation theory on top of a ground state. Therefore, the quality of the underlying converged SCF calculation is critical. Since the core electrons of the excited atoms need to be explicitly described, the GAPW method is necessary. Excited atoms must be described with all-electron basis sets, while PPs may be used for all other atoms. A RI basis can be provided for any atomic kind in the system using the \texttt{RI\_XAS} keyword, in the \texttt{\&KIND} subsection (e.g. \texttt{BASIS\_SET RI\_XAS def2-TZVP-RIFIT}). If not provided, a default RI basis is generated on-the-fly using the robust method of Stoychev~\cite{stoychev2017automatic}. For hybrid DFT simulations, the ADMM approximation can be used for the SCF calculation, specifically in its ADMM2 flavor. The \texttt{RUN\_TYPE} keyword in the \texttt{\&GLOBAL} input section must be set to \texttt{ENERGY}. In addition, LR-TDDFT-based XAS calculations can be performed for both molecular and periodic systems \cite{Mueller2019}.

The input subsection governing XAS LR-TDDFT calculations is \texttt{\&XAS\_TDP}, which is within the \texttt{\&DFT} section. Some keywords are related to general LR-TDDFT: \texttt{TAMM\_DANCOFF}, which controls the TDA (enabled by default), \texttt{DIPOLE\_FORM} to calculate the electronic dipole in either the length or velocity-gauge, and \texttt{EXCITATIONS} to enable the calculation of singlet/triplet excitations in closed-shell systems or spin-conserving/spin-flip excitations in open-shell systems. The remaining input is specific to XAS calculations.

Because all approximations used in the LR-TDDFT method of the \texttt{\&XAS\_TDP} section rely on the localization of core states, it is crucial to select them properly and check their properties. This is done via the \texttt{\&XAS\_TDP\%DONOR\_STATE} subsection, which sets the rules for searching for the most suitable core states among the low-energy MOs. First, excited atoms are defined either \texttt{BY\_INDEX} or \texttt{BY\_KIND} with the \texttt{DEFINE\_EXCITED} keyword. In the former case, an \texttt{INDEX\_LIST} must be provided with the indices of the excited atoms. In the latter case, a \texttt{KIND\_LIST} is provided, specifying which atomic kind is to be excited. The \texttt{STATE\_TYPES} keyword then defines which core-level to excite from (1s, 2s or 2p), for each atom index/kind. Finally, the \texttt{N\_SEARCH} keyword specifies the $N$ lowest-energy MOs to screen for the donor core states previously defined. Note that in some cases, excited atoms might be equivalent under symmetry (e.g. carbon atoms in an acetylene molecule C\textsubscript{2}H\textsubscript{2}). In such a case, the two lowest-energy canonical MOs will be linear combinations of s-type states centered on each atom and effectively delocalized. To ensure that the underlying approximations of the LR-TDDFT XAS method are valid, the $N$ candidate donor states can be localized using the \texttt{LOCALIZE} keyword. Localization can be expensive for a large number of MOs: therefore, it is recommended to keep the value of \texttt{N\_SEARCH} to a minimum. An example \texttt{\&DONOR\_STATE} subsection is given below for the CO\textsubscript{2} molecule:
\begin{verbatim}
&DONOR_STATES
  DEFINE_EXCITED BY_KIND
  KIND_LIST O
  LOCALIZE
  N_SEARCH 3
  STATE_TYPES 1s
&END DONOR_STATES
\end{verbatim}
In the above example, the three lowest-energy MOs of the system are localized and used as candidates for O 1s donor states. The 2 candidates that best overlap with a minimal STO basis representation of an O 1s core state are selected. The overlap, as well as the Mulliken charge analysis, is reported in the calculation output. These figures should be checked to ensure the suitability of donor states.

Another important input subsection of the LR-TDDFT method is \texttt{\&XAS\_TDP\%KERNEL}, which defines the XC kernel. Both the XC functional and exact exchange for hybrid DFT calculations are specified here. This subsection must always be explicitly included, even if the kernel is the same as the ground state functional. All recommendations for periodic calculations with HF apply equally for the \texttt{\&EXACT\_EXCHANGE} subsection. Note that only global hybrids are currently supported. Below is an example \texttt{\&KERNEL} sections for a periodic PBE0 hybrid calculation, with the truncated Coulomb potential and a cutoff radius of 6 \AA:
\begin{verbatim}
&KERNEL
  &XC_FUNCTIONAL
    &GGA_X_PBE
    &END GGA_C_PBE
    &GGA_X_PBE
      SCALE 0.75
    &END GGA_X_PBE
  &END XC_FUNCTIONAL
  &EXACT_EXCHANGE
    FRACTION 0.25
    OPERATOR TRUNCATED
    CUTOFF_RADIUS 6.0
  &END EXACT_EXCHANGE
&END KERNEL
\end{verbatim}

Also related to the XC kernel is the \texttt{\&XAS\_TDP\%GRID} keyword. As shown in Eq.~\ref{xastdp_proj}, $f_{xc}$ is evaluated on a projection of the density on the RI basis elements centered on the excited atom. The integral is then performed numerically on an atomic grid. By default, the grid dimensions are the same as the GAPW settings for the SCF calculation. It is, however, recommended to use a finer grid. For example,
\begin{verbatim}
GRID O 100 250
\end{verbatim}
would define grids with 100 angular points (Lebedev scheme) and 250 radial grid points for excited oxygen atoms. Note that the GAPW default is a 50x50 grid.

Finally, SOC calculations for L-edge spectroscopy can be enabled by adding the \texttt{SPIN\_ORBIT\_COUPLING} keyword. The SOC is added perturbatively on top of an LR-TDDFT calculation using the auxiliary many-electron WF (AMEW) framework~\cite{franco2014derivation}. For spin-restricted systems, both single- and triplet-excitations must be calculated. In the open-shell case, spin-conserving and spin-flip excitations must be calculated. For example,
\begin{verbatim}
SPIN_ORBIT_COUPLING
EXCITATIONS RCS_SINGLET
EXCITATIONS RCS_TRIPLET
\end{verbatim}

\noindent would turn on SOC for LR-TDDFT-based XAS calculations on top of a restricted closed-shell ground state.

It is generally recommended to run LR-TDDFT XAS calculations with hybrid functionals containing a large fraction of exact exchange to reduce the self-interaction error (e.g. PBEh ($\alpha$=0.45) or BHHLYP). For such simulations, the computational bottleneck is often still the SCF cycle. Yet, large performance gains can be achieved using the ADMM approximation. Diagonalization costs can also become high for large systems with multiple excited atoms (e.g. all oxygen atoms in a liquid-water simulation). Hence, the iterative \texttt{\&OT\_SOLVER} can be used to reduce the time spent diagonalizing multiple matrices. Note that, due to its relatively high initialization cost, the OT solver should not be used for a single excited state. Generally, the automatically generated RI basis is very robust and accurate. However, its size is typically larger than some existing pre-optimized RI bases (e.g. def2-TZVP-RIFIT). If such a basis exists, using it might save computing time as well.

If unphysical excitation energies are obtained, there are a few possible culprits. The most likely reason is that the donor core states are not properly prepared. They may not be localized or fall outside the range specified by \texttt{\&DONOR\_STATES\%N\_SEARCH}. One should check the output file for the overlap and the Mulliken charge analysis (both should be close to 1.0, or even higher for the overlap of 2p states). Another possibility is that the density projection in Eq.~\ref{xastdp_proj} is inaccurate. This sometimes happens when a heavier atom described with an all-electron basis is in the vicinity of the excited atom: because the RI basis functions are centered on the latter, they might fail at describing the core states of the former. The easiest solution is to describe the heavier atom with a PP. If this is not possible, e.g. because of interest in excitations from both atoms, one can use the \texttt{RI\_REGION} keyword of the \texttt{\&KERNEL} subsection. This allows for the projection of the density on a larger RI basis, with contributions from all atoms within a sphere around the excited atom. In such a case, the use of a fine radial grid is recommended, in order to describe the sharp features of the neighboring atom. Finally, the issue could be numerical, and the \texttt{GRID}, or various general parameters could need tightening, or the usage of a better basis set.

The LR-TDDFT XAS method and its implementation are extensively discussed in Ref.~\citenum{Bussy2021}. Various accuracy and performance benchmarks were performed, and all input files are openly available in Ref.~\citenum{bussy2021data}. 

\subsection{Real-Time Time-Dependent Density Functional Theory}

Spectroscopic properties can also be obtained by means of RT-TDDFT. The absorption spectrum ranging from
ultraviolet-visible (UV-Vis) to x-rays can be computed by studying the time dependence of the induced electric dipole moment, or the induced current density in condensed matter systems, after subjecting the system to an external electric field in the form of a very narrow function in time (delta kick). Such a step-function-like external potential will excite all the modes in the system. A corresponding sample input section looks like this:
\begin{verbatim}
&REAL_TIME_PROPAGATION
  INITIAL_WFN  SCF_WFN
  PROPAGATOR ETRS
  EXP_ACCURACY 1E-15
  EPS_ITER 1E-9
  MAX_ITER 100
  MAT_EXP TAYLOR
  APPLY_DELTA_PULSE .TRUE.
  DELTA_PULSE_DIRECTION 0 0 1
  DELTA_PULSE_SCALE 0.001
&END REAL_TIME_PROPAGATION   
&PRINT
  &MOMENTS
    FILENAME =mom.dat
    PERIODIC F
  &END MOMENTS
&END PRINT
\end{verbatim}

This approach has been extensively
studied for years to simulate UV-Vis absorption spectra. The first completely general implementation of RT-TDDFT for core states was done in 2012 by K. Lopata et al. within the NWChem code~\cite{lopata2012}.
The absorption spectrum of a system can be obtained by performing three simulations employing the RTP procedure, each applying the field polarized along one Cartesian axis. 
For non-periodic simulations, one applies a field in the form of a Dirac delta function, i.e. ${\bf E}_\delta =\kappa \delta(t) {\bf e}$, where $\kappa$ is the strength and ${\bf e}$ the polarization. The time-dependent electric dipole moment is obtained from the propagated electronic charge density
\begin{equation}
{\boldsymbol{\mu}}(t) = \int d{\bf r} \, n({\bf r},t) {\bf r}. 
\end{equation}
Assuming the linear regime, i.e. $\boldsymbol{\mu} (\omega) = \boldsymbol{\alpha}(\omega)\cdot {\bf E}(\omega)$, the molecular polarizability
tensor is then obtained by Fourier transforming the time-dependent dipole moment
\begin{equation}
\alpha_{ij}(\omega)=\frac{1}{\kappa} \mu_{ij}(\omega),
\end{equation}
where $i$ is the vector component and $j$ the polarization direction. The absorption spectrum is finally obtained by computing the trace of the polarizability tensor 
\begin{equation}
S(\omega) = \frac{4\pi\omega}{3c}Tr\left[ \Im[{\alpha(\omega)}]\right].
\end{equation}
According to time-dependent perturbative approaches, the polarizability is real for non-resonant frequencies and diverges at resonances.
Therefore, the resonant frequencies are characterized by a non-zero imaginary value and a change of sign for
the real part. By analyzing both components of the polarizability together, it is thus possible to verify whether the peaks in the spectrum correspond to actual electronic transitions. This behavior of the polarizability can be helpful to understand whether the spectra obtained via RTP can be considered converged with
respect to different simulation parameters, such as the timestep and total propagation time.

The velocity-gauge RT-TDDFT formalism is required to apply a time-dependent field to periodic systems. As discussed in a previous section, the vector potential appears in the kinetic term and, when PPs are used, the gauge field also transforms the electron-ion interaction. Although the length-gauge and velocity-gauge forms yield the same result for finite systems, the velocity-gauge representation is periodic for spatially uniform external electric fields. 
Integrating the time-dependent current density  ${\bf j}({\bf r}, t)$ in the simulation cell yields the macroscopic current
\begin{equation}
{\bf J}(t) = -\frac{1}{V} \int d^3{\bf r} \,{\bf j}({\bf r}, t).
\end{equation}

In the linear-response regime, by applying a very short pulse, the Fourier transform of the time-dependent current yields the frequency-dependent conductivity $\sigma_{ij}(\omega)$ and the frequency-dependent dielectric function.
\begin{equation}
    \epsilon(\omega) = 1+\frac{4\pi i Tr[\boldsymbol{\sigma}({\omega})]}{\omega}.
\end{equation}
The imaginary part of the frequency-dependent dielectric function corresponds to the optical absorption spectrum in a condensed matter system. An example input looks as follows:
\begin{verbatim}
&REAL_TIME_PROPAGATION
  INITIAL_WFN SCF_WFN
  PROPAGATOR ETRS
  EXP_ACCURACY 1E-15
  EPS_ITER 1E-9
  MAX_ITER 100
  MAT_EXP TAYLOR
  VELOCITY_GAUGE
  &PRINT
    &FIELD
    &END FIELD
    &E_CONSTITUENTS
    &END E_CONSTITUENTS
    &CURRENT_INT
    &END CURRENT_INT
  &END PRINT
&END REAL_TIME_PROPAGATION
&EFIELD
  INTENSITY 3.5E10
  POLARISATION  0  0  1
  ENVELOP GAUSSIAN
  &GAUSSIAN_ENV
    T0 [fs]    0.006240
    SIGMA [fs] 0.002340
  &END GAUSSIAN_ENV
  VEC_POT_INITIAL  0.00 0.00 ${VECZ}
&END
\end{verbatim}
\section{Energy Decomposition Analysis} \label{sec:ALMO-EDA}

\newcommand{\CPXK}{CP2K}


Intermolecular bonding arises from a complex interplay of electrostatic interactions between permanent charges and molecular multipole moments, polarization effects, Pauli repulsion, donor–acceptor orbital interactions (also referred to as covalent, charge-transfer, or delocalization interactions), and weak dispersion forces. Energy decomposition analysis (EDA) quantifies the contribution of each of these components to the total binding energy, providing deeper insight into the physical origins of intermolecular bonds.

The \CPXK/\textsc{Quickstep} implementation of EDA is based on absolutely localized molecular orbitals (ALMOs), which are molecular orbitals confined entirely to individual molecules or ions within a larger system~\cite{Khaliullin2007, Khaliullin2013}. ALMO EDA separates the total interaction energy (TOT) into a frozen density (FRZ), polarization (POL) and charge-transfer (CT) terms, i.e.
\begin{equation}
	\Delta E_{\text{TOT}} = \Delta E_{\text{FRZ}} + \Delta E_{\text{POL}} + \Delta E_{\text{CT}}.
\end{equation}
The frozen interactions term is defined as the energy required to bring isolated molecules into the system without any relaxation of their MOs, apart from modifications associated with satisfying the Pauli exclusion principle, hence 
\begin{equation}
\label{eq:frz}
\Delta E_{\text{FRZ}} \equiv E(R_{\text{FRZ}}) - \sum_{x} E(R_x), 
\end{equation}
where $E(R_x)$ is the energy of the isolated molecule $x$ and $R_{\text{FRZ}}$ is the density matrix of the system constructed from the unrelaxed MOs of the isolated molecules. 
ALMO EDA is closely related to the block-localized WF EDA~\cite{Mo2000}, because both approaches use the same variational definition of the polarization term as the energy lowering due to the relaxation of each molecule's ALMOs in the field of all other molecules in the system
\begin{equation}
\label{eq:pol}
\Delta E_{\text{POL}} \equiv E(R_{\text{ALMO}}) - E(R_{\text{FRZ}}).
\end{equation}
The strict locality of ALMOs is utilized to ensure that the relaxation is constrained to include only intramolecular variations. This approach gives an upper limit to the true polarization energy~\cite{azar2013useful}, and its mathematical and algorithmic details have been described by several authors~\cite{Stoll1980,Khaliullin2006}.
The remaining portion of the total interaction energy, the CT term, is calculated as the difference in the energy of the relaxed ALMO state and the state of fully delocalized and optimized orbitals ($R_{\text{SCF}}$), i.e.
\begin{equation}
\label{eq:ct}
\Delta E_{CT} \equiv E(R_{\text{SCF}}) - E(R_{\text{ALMO}}).
\end{equation}
Therefore, $\Delta E_{\text{CT}}$ includes the energy lowering due to electron transfer from the occupied ALMOs on one molecule to the virtual orbitals of another molecule $\Delta E_{\text{CT}}^{(\text{pair})}$, as well as the further energy change caused by many-body higher-order induction $\Delta E_{\text{HO}}$ that accompanies such an occupied-virtual orbital mixing and is typically small for intermolecular interactions, thus 
\begin{eqnarray}
\label{eq:ct2}
\Delta E_{CT} = \Delta E_{\text{CT}}^{(\text{pair})} + \Delta E_{\text{HO}}.
\end{eqnarray}
A distinctive feature of ALMO EDA is that 
both the amount of the electron density transferred between a pair of molecules and the corresponding energy lowering can be computed via~\cite{Khaliullin2008}:
\begin{subequations}
\begin{eqnarray}
\label{eq:qct0}
\Delta Q_{\text{CT}}^{(\text{pair})} &=& \sum_{x,y > x} \{ \Delta Q_{x\rightarrow y} + \Delta Q_{y\rightarrow x} \} \\
\label{eq:ct0}
\Delta E_{\text{CT}}^{(\text{pair})} &=& \sum_{x,y > x} \{ \Delta E_{x\rightarrow y} + \Delta E_{y\rightarrow x} \}
\end{eqnarray}
\end{subequations}

\subsection{Implementation}
Within \CPXK, ALMO EDA is restricted to closed-shell fragments. The linear-scaling optimization of ALMOs serves as its underlying computational engine~\cite{Khaliullin2013a}, which can be applied to both gas-phase and condensed phase systems. It has also been extended to fractionally occupied ALMOs enabling investigation of interactions between metal surfaces and molecular adsorbates~\cite{staub2019energy}. Another unique feature of our implementation is the ability to control the spatial range of CT between molecules using the cutoff radius $R_c$ (see below). Additionally, ALMO EDA in combination with \CPXK's efficient AIMD engine allows us to attenuate or switch off CT interactions in AIMD simulations, thus measuring their contribution to the dynamical properties of molecular systems~\cite{shi2018contribution, scheiber2018compact}.
A single-point ALMO EDA calculation can be set up by creating a regular input file for DFT calculation and then modifying it in several steps.

\textbf{\emph{Step 1. Define fragments.}} First of all, all atoms in the system must be assigned to non-overlapping subsets called fragments. In other words, the system must be partitioned into fragments. A fragment can include a single atom, several non-bonded atoms, a single molecule (that is, several bonded atoms), or several molecules. This partitioning tells the code about localization regions of ALMOs when calculating the $\Delta E_{\text{FRZ}}$ and $\Delta E_{\text{POL}}$ energy terms.

To define fragments, the ALMO code uses the \CPXK\ subroutines that analyze bonding between atoms using distance thresholds. By default, all atoms that are considered bonded to each other are assigned to the same fragment. This approach enables a very flexible definition of fragments through a combination of (a) the {\tt \&GENERATE} subsection of the {\tt \&\&FORCE\_EVAL\%SUBSYS\%TOPOLOGY} section, (b) the 5th column in the {\tt \&SUBSYS\%COORD} section, and (c) an optional connectivity file. The information about each fragment is printed at the beginning of output files under the ``ALMO SETTINGS'' marker.

The {\tt \&GENERATE} subsection automatically determines bonding between atoms by comparing interatomic distances to the sum of the atomic radii. This is the most convenient strategy to define bonding between atoms and thus to create fragments. The default behavior of {\tt \&GENERATE} can be changed by modifying its keywords. For example, regular molecules can be partitioned into atomic fragments using low values of the {\tt BONDPARM\_FACTOR}. Low values prevent atoms from being combined into molecular fragments, as shown in the following input:
\begin{verbatim}
&SUBSYS
  &TOPOLOGY
    &GENERATE
      BONDLENGTH_MAX 1.0
      BONDPARM COVALENT
      BONDPARM_FACTOR 0.3
    &END GENERATE
  &END TOPOLOGY
&END SUBSYS
\end{verbatim}

Partitioning of the system can be refined using the 5th column in the {\tt \&COORD} section that assigns fragment labels to each atom. The main principle of this partitioning scheme is that atoms with different fragment labels cannot be combined into a fragment. On the other hand, atoms with the same label may still be separated into fragments as guided by the {\tt \&GENERATE} subsection. 
%
%
%
%
Finally, the bonding between atoms can be explicitly specified in a connectivity file. In this case, all bonded atoms are assigned to the same fragment.

\textbf{\emph{Step 2. Assign electrons to fragments.}} All electrons of a neutral atom are assumed to belong to the atom's fragment. This default behavior can be changed by adding (removing) electrons to (from) atoms using the {\tt \&BS} subsection in the {\tt \&SUBSYS\%KIND} section. In the following example, one electron is removed from the 1s orbitals of hydrogen atoms to create H$^+$ fragments. Meanwhile, two electrons are added to the 2p orbitals of oxygen atoms to create O$^{2-}$ fragments. Since both H$^+$ and O$^{2-}$ are intended to be closed-shell fragments, the same modification has to be performed for the {\tt \&ALPHA} and {\tt \&BETA} subsections of the {\tt \&BS} section. Note that ALMO EDA can even handle zero-electron fragments, such as H$^+$.
\begin{verbatim}
  &SUBSYS
    &KIND H
      ...
      &BS
        &ALPHA
          NEL -1
          L    0
          N    1
        &END
        &BETA
          NEL -1
          L    0
          N    1
        &END
      &END
    &END KIND

    &KIND O
      ...
      &BS
        &ALPHA
          NEL +2
          L    1
          N    2
        &END       
        &BETA
          NEL +2
          L    1
          N    2
        &END       
      &END
    &END KIND
  &END SUBSYS
\end{verbatim}

After atoms and electrons are assigned to fragments, the Gaussian basis set functions, all of which are centered on the atoms in \CPXK/\textsc{Quickstep}, are also assigned to the same fragments as their atoms. This last partitioning defines the structure of the block-diagonal matrix of ALMO coefficients. 

\textbf{\emph{Step 3. Turn ALMO SCF on.}} To run ALMO EDA calculations, set the {\tt ALMO\_SCF} keyword in the {\tt \&DFT\%QS} section to {\tt TRUE}. This tells \CPXK\ to optimize ALMO coefficients in the ALMO SCF loop instead of optimizing MO coefficients in the conventional SCF loop. That is, the {\tt \&DFT\%SCF} section of the input does not control ALMO EDA calculations. Instead, the ALMO SCF loop is controlled by the {\tt \&DFT\%ALMO\_SCF} section.

\begin{table*}
    \centering
    \setlength{\tabcolsep}{6pt} 
    \begin{tabular}{clll}
        \hline
        State & \CPXK\ print out & Definition & EDA terms \\
        \hline
        (0) & ``Single-molecule energy'' & $\sum_{x} E(R_x)$ in Eq.~(\ref{eq:frz}) & \\
        (1) & ``Energy of the initial guess'' & $E(R_{\text{FRZ}})$ in Eq.~(\ref{eq:frz}) & $\Delta E_{\text{FRZ}} = (1) - (0)$ \\
        (2) & ``ENERGY OF BLOCK-DIAGONAL ALMOs'' & $E(R_{\text{ALMO}})$ in Eq.~(\ref{eq:pol}) & $\Delta E_{\text{POL}} = (2) - (1)$ \\
        (3) & ``CORRECTED ENERGY'' & Eq.~(\ref{eq:ct2}) & $\Delta E_{\text{CT}}^{(\text{pair})} = (3) - (2)$ \\
        (4)$^{*}$ & ``ENERGY|'' & $E(R_{\text{SCF}})$ in Eq.~(\ref{eq:ct}) & $\Delta E_{\text{HO}} = (4) - (3)$ \\
         &  &  &  $\Delta E_{\text{CT}} = (4) - (2)$ \\
        \hline
        \multicolumn{4}{l}{* If {\tt FULL\_X} or {\tt XALMO\_X} methods are used then the ``ENERGY|'' line contains the energy of state (3).}
    \end{tabular}
    \caption{ALMO EDA energies in \CPXK\ output files.}
    \label{tab:EDAterms}
\end{table*}

\textbf{\emph{Step 4. Compute $\Delta E_{\text{FRZ}}$.}} To request the calculation of the FRZ term, set {\tt \&ALMO\_SCF\%ALMO\_SCF\_GUESS} to {\tt MOLECULAR}. 
This instructs \CPXK\ to perform a series of conventional SCF calculations for individual fragments in the periodic box of the entire system. 
These calculations are controlled by the {\tt \&SCF} section of the input file, and, therefore, it is important that all keywords in this section (e.g. {\tt EPS\_SCF}) are set to ensure sufficient accuracy of the energies of the individual fragments. 
\CPXK/\textsc{Quickstep} then combines the MO of individual fragments into the block-diagonal matrix of the ALMO coefficients and evaluates the energy of this state without any optimization of the coefficients. The computed energies are printed as described in Table~\ref{tab:EDAterms}.

\textbf{\emph{Step 5. Compute $\Delta E_{\text{POL}}$.}} Subsequent optimization of the block-diagonal ALMO coefficients can be performed using various algorithms, which are selected using the {\tt ALMO\_ALGORITHM} keyword and controlled by the corresponding subsections of the {\tt \&ALMO\_SCF} section. Although DIIS-accelerated diagonalization of the diagonal blocks of the projected KS Hamiltonian matrix typically converges the ALMO SCF in just a few iterations~\cite{Khaliullin2006,Stoll1980}, this method does not guarantee convergence and can fail for strongly interacting fragments. In problematic cases, preconditioned conjugate gradient or trust region algorithms can be employed instead of the DIIS-accelerated diagonalization. The polarization energy does not depend on the algorithm and is printed as described in Table~\ref{tab:EDAterms}.

\textbf{\emph{Step 6. Compute $\Delta E_{\text{CT}}$.}} The calculation of the CT term is controlled by the {\tt DELOCALIZE\_METHOD} keyword in the {\tt \&ALMO\_SCF} section. In order to compute $\Delta E_{\text{CT}}$ defined in Eq.~\ref{eq:ct}, set this keyword to {\tt FULL\_SCF}. In order to separate CT into pair contributions defined in Eqs.~\ref{eq:qct0} and \ref{eq:ct0}, respectively, {\tt DELOCALIZE\_METHOD} should be set to {\tt FULL\_X\_THEN\_SCF}. If only the pair contributions are of interest and the higher-order term is expected to be negligible, set {\tt DELOCALIZE\_METHOD} to {\tt FULL\_X}.

For very large systems, the computation of the CT terms can be significantly speeded up by evaluating them only between fragment neighbors within the user-defined localization radius. This is done by specifying the localization cutoff distance with {\tt \&ALMO\_SCF\%XALMO\_R\_CUTOFF\_FACTOR} and by setting {\tt DELOCALIZE\_METHOD} to {\tt XALMO\_SCF}, or to {\tt XALMO\_X}. The latter two settings are equivalent to using {\tt FULL\_SCF} and {\tt FULL\_X} for fully delocalized orbitals. Note that pair CT components can be calculated with {\tt XALMO\_X}, but cannot be obtained with {\tt XALMO\_SCF}, respectively.

The computed CT energies are printed as described in Table~\ref{tab:EDAterms}. 
The printout of the pair CT terms can be activated by adding the {\tt \&ANALYSIS} subsection inside {\tt \&ALMO\_SCF} and specifying output file names inside the {\tt \&PRINT} subsection. Each message passing interface process will print its own part of the pair list into a separate file. These output files have three columns. The first column is the index of the fragment that accepts electrons. The second column is the index of the fragment that donates electrons. The third column is the amount of electron density transferred in units of the elementary charge or the corresponding CT energy in Hartrees.


\textbf{Input file example.} The relevant input sections for an ALMO EDA calculation involve: 

\begin{verbatim}
&QS
  ALMO_SCF T
&END QS
&ALMO_SCF
  ALMO_ALGORITHM DIAG
  ALMO_SCF_GUESS MOLECULAR
  DELOCALIZE_METHOD FULL_X_THEN_SCF
  EPS_FILTER 1.0E-10
  &ALMO_OPTIMIZER_DIIS
    EPS_ERROR 1.0E-5
    MAX_ITER 10
    N_DIIS 5
  &END ALMO_OPTIMIZER_DIIS
  &ANALYSIS T
    &PRINT
      &ALMO_CTA
        FILENAME charge_terms
      &END ALMO_CTA
      &ALMO_EDA_CT
        FILENAME energy_terms
      &END ALMO_EDA_CT
    &END PRINT
  &END ANALYSIS
  &XALMO_OPTIMIZER_PCG
    CONJUGATOR DAI_YUAN
    EPS_ERROR 1.0E-5
    LIN_SEARCH_EPS_ERROR 0.1
    LIN_SEARCH_STEP_SIZE_GUESS 0.5
    MAX_ITER 100
    MAX_ITER_OUTER_LOOP 0
  &END XALMO_OPTIMIZER_PCG
&END ALMO_SCF
\end{verbatim}

\textbf{Illustrative applications.} ALMO EDA has been applied to study intermolecular interactions in a variety of gas and condensed phase molecular systems~\cite{luo2018why}, as well as their interfaces~\cite{ojha2019time,ojha2021hydrogen}. The implementation of ALMO EDA in CP2K has played a crucial role in advancing our understanding of hydrogen bonding in liquid water~\cite{Kuehne2013,Khaliullin2013,NatureComm2015,elgabarty2020tumbling}, ice~\cite{Kuehne2014a} and water confined to low dimensions~\cite{yun2019low, salem2020insight, heske2023water}. These studies have systematically shown that the small amount of electron density transferred between molecules has a profound effect on the structure~\cite{Ramos-Cordoba2011, elgabarty2020tumbling}, spectroscopic response~\cite{Fransson2016, Ramos-Cordoba2011, Lenz2006, NatureComm2015, ojha2018hydrogen}, and dynamical properties~\cite{elgabarty2019enhancement, yun2022correlated, balos2022time, scheiber2018compact, shi2018contribution} of liquid water.

\section{Finite Temperature Effects} \label{sec:FiniteTemperature}






For the purpose of explicitly including finite temperature effects, CP2K implements a large variety of MC and MD algorithms that can be activated via the \texttt{\&MOTION\%MC} and \texttt{\&MOTION\%MD} sections, respectively~\cite{mcgrath2005toward, mcgrath2005isobaric, mcgrath2006simulating, kuo2006structure, laio2002escaping, Iannuzzi2003}. 
The most distinctive feature of CP2K, however, is the ability to conduct efficient AIMD simulations~\cite{Hutter2011, Kuehne2014}, and in combination with linear-scaling algorithms, even for rather large-scale systems~\cite{Schade2022, gavini2023roadmap, schade2023breaking}. Yet, in spite of its name, initially CP2K only contained Born-Oppenheimer MD~\cite{Arias1992, Arias1992a, kresse1993ab, kresse1994ab}, since the original Car-Parrinello algorithm is not implemented~\cite{Car1985}. Instead, the second-generation Car-Parrinello method is available~\cite{Kuehne2007, Prodan2018}, but since this has been extensively covered elsewhere~\cite{caravati2007coexistence,Kuehne2009,Cucinotta2009,Caravati2009b,Luduena2011,Kuehne2011}, 
the present focus is solely on MLPs and NQE by means of PIMD.

\subsection{Neural Network and Machine Learning Interaction Potentials}
\label{sec:nnp}

The rise of MLPs in recent years provides efficient and accurate representations of potential energy surfaces from electronic structure reference data.~\cite{Behler2016/10.1063/1.4966192,
Behler2017/10.1002/anie.201703114,
Bartok2017/10.1126/SCIADV.1701816,
Butler2018/10.1038/s41586-018-0337-2,
Deringer2019/10.1002/ADMA.201902765,
Kang2020/10.1021/ACS.ACCOUNTS.0C00472,
Behler2021,
fiedler2022deep}
This has enabled the simulation community 
to reach previously inaccessible spatio-temporal scales through atomistic simulations of complex, reactive systems at near \textit{ab initio} accuracy as hitherto only known from explicit (on-the-fly) AIMD~\cite{Marx2009, Kuehne2014} simulations. 
%
High-dimensional neural network potentials (NNPs)~\cite{Behler2007a,
Behler2017/10.1002/anie.201703114}
combined with atom-centered symmetry functions to describe atomic environments~\cite{Behler2011/10.1063/1.3553717}
have been the first methodology to provide a scalable and accurate representation of potential energy surfaces using artificial neural networks.
Introduced in 2007 by Behler and Parrinello, this framework has been used to provide
fundamental insight into a plethora of challenging systems, from reactive gas-phase clusters~\cite{Natarajan2015/10.1039/C4CP04751F,Schran2020/10.1021/acs.jctc.9b00805},
to phase behavior of liquids like water~\cite{Morawietz2016/10.1073/pnas.1602375113,Daru2022/10.1103/PhysRevLett.129.226001},
chemical reactions at interfaces~\cite{Quaranta2017/10.1021/acs.jpclett.7b00358}, high pressure systems~\cite{Behler2008/10.1103/PhysRevLett.100.185501, eshet2010ab, eshet2012microscopic}, complex materials~\cite{Eckhoff2019/10.1021/acs.jctc.8b01288}, battery interfaces~\cite{Eckhoff2021/10.1063/5.0073449}, and phase-transitions~\cite{khaliullin2010graphite, khaliullin2011nucleation}, to name but a few~\cite{Behler2017/10.1002/anie.201703114,Behler2021,Behler-Dellago-water2024}.
Although many other methods have been proposed over the years, NNPs remain a mainstay in the toolbox of computational scientists.
Their conceptually simple and computationally efficient design and their systematic
extensions to treat long-range effects via environment-dependent
charges~\cite{Artrith2011/10.1103/PhysRevB.83.153101} and self-consistent charge equilibration~\cite{Ko2021/10.1038/s41467-020-20427-2, khajehpasha2022cent2}
to target remaining limitations within the same framework makes them a versatile and robust tool to address problems of ever increasing complexity
\cite{Behler2021}.
For the purpose of this tutorial, we concentrate here on the original formulation, which is deeply implemented in CP2K and very simple to use by newcomers in the field. 
%

To represent a potential energy surface by NNPs, the atomistic structure is first transformed using atom-centered symmetry functions into translationally and rotationally invariant descriptors of atomic environments~\cite{Behler2011/10.1063/1.3553717}.
These serve as input for atomic neural networks that output auxiliary components of the total potential energy, which is then obtained as a sum of contributions from all atoms in the system.
The resulting permutationally invariant structure-energy relation can be analytically differentiated to obtain the nuclear forces, for example, to drive MD, and is scalable to essentially arbitrary system sizes~\cite{Behler2007a}.
Model training is achieved by optimizing the parameters (weights and biases) of the atomic neural networks to reproduce reference energies and optionally forces.
Besides the original RuNNer code of Behler, 
other training codes like n2p2 and \mbox{RubNNet4MD} 
are 
available and
yield models that can be directly used in CP2K.
For a light introduction into NNPs and other MLPs we refer the reader to Ref.~\citenum{Thiemann2025}, as well as 
to previous
reviews on this topic~\cite{Behler2011/10.1039/C1CP21668F,Behler2015/10.1002/qua.24890}.

The implementation of NNPs in CP2K is fully compatible with the original RuNNer code and thus n2p2 format and can be used in the same way as other potential energy models.
The NNP module in CP2K is part of the \texttt{\&FORCE\_EVAL} section and can be used in combination with other methods such as DFT, traditional force fields, or QM/MM
methods.
Inside the \texttt{\&NNP} subsection, the keyword \texttt{NNP\_INPUT\_FILE\_NAME} reads the NNP from a file, which can be generated using the RuNNer code or other compatible training codes such as n2p2 or the
\mbox{RubNNet4MD} package~\cite{RubNNet4MD-code}.
%
It is noted only in passing that the Atomic Cluster Expansion (ACE)
MLP technique~\cite{Drautz2019/10.1103/PHYSREVB.99.014104}
has already been interfaced with CP2K and that
its tree graph extension (grACE)~\cite{Bochkarev2024}
is addressed as this article is written. Similarly, well-developed interfaces to CP2K also exist for DeePMD-kit~\cite{Zeng2023}, NequIP~\cite{batzner2022} and Allegro~\cite{musaelian2023learning}, all of which can be activated via their respective specialized subsections in \texttt{\&FORCE\_EVAL\%MM\%FORCEFIELD\%NONBONDED}. Contrary to the native NNP implementation of CP2K, the corresponding libraries must be linked during compile time, as described in detail in section~\ref{sec:Installation}.

The NNP is then used to compute the energy and forces of the system, where the analytical stress tensor is also available for constant-pressure simulations. 
Parallization is achieved using particle decomposition and scales well up to thousands of cores and number of atoms.
An example of the input section for the NNP module is shown here: 
\begin{verbatim}
&FORCE_EVAL
  METHOD NNP
  &NNP
    NNP_INPUT_FILE_NAME input.nn
    SCALE_FILE_NAME scaling.data
    &MODEL
      WEIGHTS weigths
    &END MODEL
  &END NNP
&END FORCE_EVAL
\end{verbatim}

The original NNP formalism can also be extended to so-called ensemble or committee models, where multiple NNPs are combined for boosted accuracy and error estimation.
Individual committee members are trained separately from independent random initial conditions to a subset of the total training set~\cite{Schran2020/10.1063/5.0016004}.
If the resulting committee shares the same atom-centered symmetry functions as descriptors for the atomic environments~\cite{Behler2011/10.1063/1.3553717}, there is only a small, often negligible, computational overhead in production runs.
While the committee average provides more accurate predictions than the individual NNPs, the committee disagreement, defined as the standard deviation between the committee members, grants access to an estimate of the error of the model.
This committee disagreement provides an objective measure of the error of the underlying model~\cite{Imbalzano2021/10.1063/5.0036522}, which can be used in active learning protocols to improve the model systematically~\cite{Gastegger2017/10.1039/C7SC02267K,
Podryabinkin2017/10.1016/J.COMMATSCI.2017.08.031,
Smith2018/10.1063/1.5023802}.
This has been utilized extensively in recent times for the automated development of NNPs for various systems~\cite{Schran2021/10.1073/PNAS.2110077118}, while current research showed that it leads to more diverse configuration sampling, but not
necessarily to
more accurate models~\cite{Stolte2024}.
This committee approach is also implemented in CP2K~\cite{Schran2020/10.1063/5.0016004}.

Committee NNPs (C-NNPs) can be used in the same way as single NNPs in CP2K, with the addition of the \texttt{\&NNP\%MODEL} section for each committee member.
The committee members are defined by the number of \texttt{\&MODEL} subsections in the input file, where each subsection contains the weights of the NNP.
It is important to ensure that these models have been trained with the same set of descriptors because they are assumed to be shared between all committee members.
An example of the input section for a C-NNP setup with three committee members is shown here:
\begin{verbatim}
&FORCE_EVAL
  METHOD NNP
  &NNP
    NNP_INPUT_FILE_NAME input.nn
    SCALE_FILE_NAME scaling.data
    &MODEL
      WEIGHTS nnp-1/weigths
    &END MODEL
    &MODEL
      WEIGHTS nnp-2/weigths
    &END MODEL
    &MODEL
      WEIGHTS nnp-3/weigths
    &END MODEL
  &END NNP
&END FORCE_EVAL
\end{verbatim}

Besides these features, the NNP module in CP2K also provides the possibility to print the energy and force disagreement between committee members, as well as configurations which are in
the extrapolation regime of the model.
This can be used to identify regions of the potential energy surface where the model is inaccurate in order to improve it systematically.
These outputs are controlled using the \texttt{\&PRINT} subsection in the \texttt{\&NNP} section.
An example of the input section for the printing of the committee energy disagreement is shown in the following:
\begin{verbatim}
&PRINT
  &ENERGIES
    &EACH
     MD 1
    &END EACH
  &END ENERGIES
&END PRINT
\end{verbatim}
Similarly, the printing of forces, force disagreement, and extrapolation points can also be enabled (\texttt{FORCES}, \texttt{FORCE\_SIGMA}, \texttt{EXTRAPOLATION}).

In addition, there is an option to bias the committee disagreement to stabilize simulations by restraining the system dynamics to regions where the committee members agree.
This can be achieved by adding a harmonic repulsive wall acting on the committee disagreement after a chosen threshold, which adds a bias to the total potential energy of the system.
An example of the input section for the biasing of the committee energy disagreement looks like:
\begin{verbatim}
&BIAS
  SIGMA_0 [hartree] 0.0001
  K_B 5000
&END BIAS
\end{verbatim}
Moreover, 
\texttt{\&BIAS\%ALIGN\_NNP\_ENERGIES} can be used to subtract an energy shift from individual committee members to reduce misalignment effects, while bias energies can be printed with the \texttt{\&BIAS\%PRINT} section using \texttt{\&BIAS\_ENERGY}; further details of this feature can be found in Ref.~\citenum{Schran2020/10.1063/5.0016004}.

\begin{figure}
  \centering
  \includegraphics[width=0.5\textwidth]{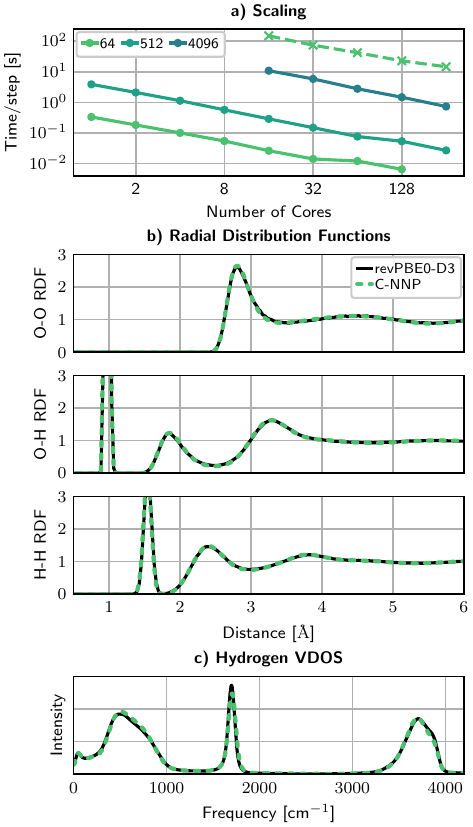}
  \caption{Comparison of NNP and AIMD results for liquid water at 300\,K
as obtained from CP2K.
	(a) Time per MD step for 64, 512, and 4096 water molecules with the C-NNP model (solid lines with circles) and 64 water molecules with the revPBE0-D3 hybrid functional (dashed line with crosses)
reported on a logarithmic scale.
%
  (b) Radial distribution functions and (c) vibrational density of states
of the hydrogen atoms.
Results in panel (b) and (c) are reproduced from Ref.~\citenum{Schran2020/10.1063/5.0016004}.
}
  \label{fig:nnp-water}
\end{figure}

Finally, we note that the parameters of a C-NNP for water are available within the CP2K data repository and can be accessed via the \texttt{NNP/bulkH2O-jcp2020-cnnp/nnp-X/} path in the input file.
The model has been trained against revPBE0-D3 hybrid functional reference data and is suitable for water over a wide range of temperatures and pressures.
Further details of the development can also be found in Ref.~\citenum{Schran2020/10.1063/5.0016004}.
We use this model to illustrate the easy usability of the NNP implementation in CP2K by showing results for liquid water at 300\,K in Fig.~\ref{fig:nnp-water}.
From the comparison of the structural and dynamical properties of liquid water, we can see that the C-NNP model provides a faithful
representation of the system, with the radial distribution functions and vibrational density of states being in perfect agreement with the reference AIMD data.
The computational efficiency of the C-NNP model is also demonstrated by the comparison of the time per MD step for different system sizes, which shows that the C-NNP model (solid lines)
is orders of magnitude faster than the reference AIMD simulations (dashed line).
This makes the C-NNP model a powerful tool for the study of complex systems at finite temperatures.

\FloatBarrier




\subsection{Nuclear Quantum Effects} \label{sec:NQE}



The quantum nature of nuclei can have a significant impact on the properties of materials, especially at low temperatures~\cite{Marx1999,Marx2006,Marx2006b,Michaelides2011/10.1073/pnas.1016653108,singh2014anharmonicity,spura2015fly,kessler2015structure,Marsalek2017/1702.07797,azadi2018nuclear,ojha2018nuclear,clark2019opposing,Schran2019/10.1039/C9CP04795F,ohto2019accessing,kaliannan2020impact,ojha2024nuclear}.
A very elegant way to include these NQEs in molecular simulations is the imaginary time path-integral (PI) formalism, which is based on the Feynman-Kac formulation of quantum statistical mechanics~\cite{Kac1949/10.1090/S0002-9947-1949-0027960-X,Feynman1965,Feynman1972,Marx2009,Tuckerman2010}.
In the PI formalism, the quantum mechanical system is mapped onto a classical system of $P$ replica, connected by harmonic springs to adjacent replica under cyclic boundary conditions in the simplest case (leading to the so-called primitive Trotter approximant).
%
Replicas (a.k.a. ``beads'') of the same index are coupled through a scaled version of the potential energy surface, which is determined by the 
Trotter number $P$. 
This approach allows for the calculation of quantum statistical observables by sampling the classical phase space of the $P$-fold system.
For more details on the PI method, we refer the reader to Refs.~\citenum{Barker1979,Chandler1981,Ceperley1995,Marx2009,Tuckerman2010,Markland2018/10.1038/s41570-017-0109}.

In CP2K, PI simulations are available via the \texttt{\&PINT} section, which provides a flexible and modular environment for the inclusion of the quantum nature of the nuclei.
In this PIMD framework, the equations of motion resulting from the effective potential are integrated numerically to sample the PI formulation of the canonical ($NVT$) 
partition function of the system~\cite{Callaway1982,
Parrinello1984,Marx2009,Tuckerman2010}.
%
This implementation 
mainly supports 
the normal mode transformation for the efficient integration of the stiff harmonic ring-polymer modes~\cite{Coalson1986}.
%
In addition to the reversible reference system propagator algorithm (RESPA) multiple timestep integrator (\texttt{NRESPA} with \texttt{HARM\_INT NUMERIC})~\cite{Martyna1996}, an exact integrator for normal mode coordinates is also available via
\texttt{HARM\_INT EXACT}~\cite{Ceriotti2010/10.1063/1.3489925}.
An example input section for a PIMD simulation is shown here
for a system of $P=$~32 replica at 250\,K and 1000 steps of 0.25\,fs: 
%
%
\begin{verbatim}
&PINT
  P 32
  NUM_STEPS 1000
  TEMP       250
  DT        0.25
  TRANSFORMATION NORMAL
  HARM_INT EXACT
  PROPAGATOR PIMD
&END PINT
\end{verbatim}

For standard PIMD simulations in the canonical ensemble, the temperature can be controlled using two thermostating techniques, namely the massive Nos\'e-Hoover chain (mNHC) thermostat~\cite{Martyna1992,Tuckerman1996}, 
applied separately to all degrees of freedom 
mostly in normal mode coordinates, 
and the PI Langevin equation (PILE) 
thermostat~\cite{Ceriotti2010/10.1063/1.3489925}.
They are enabled via dedicated subsections that change the ensemble from the microcanonical ($NVE$) to the canonical ensemble, if specified.
The Nos\'e-Hoover chain thermostat is controlled by the \texttt{\&NOSE} subsection,
whereas the PILE thermostat is controlled by the \texttt{\&PILE} subsection.
While the Nos\'e-Hoover chain thermostat only requires the definition of the chain length, the PILE thermostat needs the definition of the coupling constant $\lambda$ and the relaxation time $\tau$ as demonstrated here: 
\begin{verbatim}
&NOSE
  NNOS 5
&END NOSE
\end{verbatim}
or
\begin{verbatim}
&PILE        
  LAMBDA 0.5 
  TAU 1000.0 
&END PILE
\end{verbatim}

In addition, there are also quantum thermostats available that are specifically designed to improve the convergence with respect to the number of replicas in PI simulations, discussed in more detail in section~\ref{sec:qconv}.

Finally, it is also possible to constrain the centroid or all replica degrees of freedom using the CP2K-wide constraint framework defined by the \texttt{\&MOTION\%CONSTRAINT} section.
Within that section, the \texttt{PIMD\_BEADWISE\_CONSTRAINT F|T} keyword switches between beadwise and centroid constraints.
The \texttt{KT\_CORRECTION} keyword in the \texttt{\&PINT} section can be used to correct for the loss of temperature due to constrained degrees of freedom for Nos\'e-Hoover chains and numerical integration.
Constraints and restraints, which are available through the same framework, can be used to obtain free energy profiles, including NQEs through thermodynamic integration, umbrella sampling and umbrella integration, respectively~\cite{kastner2005bridging, kastner2011umbrella}.


%
%
%

\subsection{Quantum Convergence}
\label{sec:qconv} 
The PI formalism is exact in the limit of infinite discretization, i.e. $P \rightarrow \infty$.
However, in practice, some finite $P$ value must be selected while the properties of interest converge with increasing $P$.
Unfortunately, this convergence is system, temperature and mass dependent, making it computationally demanding to treat systems in the deep quantum regime, where unpleasantly large $P$ values are required. 
For example, the bead convergence of structural properties of hydrogen-bonded systems requires roughly $P \approx 32$ at 300\,K, $P \approx 128$ at 100\,K, and $P \approx 8192$ at about 1\,K, as analyzed in Ref.~\citenum{Schran2018/10.1021/acs.jctc.8b00705}.
Due to these reasons, different techniques have been developed to accelerate the convergence of PI simulations~\cite{john2016quantum}.
In particular, colored noise thermostats~\cite{Ceriotti2011/10.1063/1.3556661,piglet,Brieuc2016/10.1021/acs.jctc.5b01146} in combination with generalized Langevin dynamics and quantum thermal baths
have been shown to significantly reduce the required number of beads for convergence down to the ultra-cold temperature regime~\cite{Uhl2016/10.1063/1.4959602,Schran2018/10.1021/acs.jctc.8b00705}.
Two such thermostats, the PI generalized Langevin equation thermostat (PIGLET)
\cite{piglet} and 
the PI quantum thermal bath (PIQTB) thermostat~\cite{Brieuc2016/10.1021/acs.jctc.5b01146}, are available in CP2K.
The main idea behind both of these approaches is to impose a frequency-dependent temperature $T(\omega)$, which matches the quantum fluctuations of a harmonic oscillator.
Due to zero-point energy leakage, these thermostats require rather harsh coupling constants, rendering them incompatible with approximate quantum dynamics methods like centroid MD (CMD) and ring-polymer MD (RPMD), described in more detail in section~\ref{sec:qdyn}.

The PIGLET thermostat, based on the generalized Langevin equation, is enabled with the \texttt{\&PINT\%PIGLET} thermostat subsection.
This formalism requires dedicated matrices to impose the correct frequency-dependent thermostating for each combination of temperature and number of beads.
%
A large variety of matrices can be downloaded from \url{https://gle4md.org/index.html?page=matrix},
requiring to use the raw format for data consistency.
The matrices file 
must be specified in the \texttt{\&PIGLET} subsection using the \texttt{MATRICES\_FILE\_NAME} keyword.
%
The PIQTB formalism, using a quantum thermal bath thermostat for each ring-polymer normal mode, is also based on the one-body density matrix as PIGLET but has the advantage of working continuously for any combination of temperature and number of beads.
It is enabled using the \texttt{\&QTB} thermostat section 
after having specified the underlying PIMD simulation, we show an example input section for the PIQTB thermostat below.
In addition to the coupling constant $\lambda$ and the relaxation time $\tau$, the angular cutoff frequency for the normal modes needs to be specified by suitable values for \texttt{TAUCUT} and \texttt{LAMBCUT}, respectively.
The example values below have been shown to work well for hydrogen-bonded systems~\cite{Schran2018/10.1021/acs.jctc.8b00705} and should also be transferable to other systems.
They have been used for the bead convergence study in Fig.~\ref{fig:nqe-ice}a.
\begin{verbatim}
&QTB
  LAMBDA 0.2  
  TAU 50.0    
  TAUCUT 0.9
  LAMBCUT 1.5
&END QTB
\end{verbatim}
%
%
%
%

Both of these quantum thermostats have been shown to significantly reduce the number of beads required for convergence in PI simulations from ambient down to ultra-low temperatures~\cite{Schran2018/10.1021/acs.jctc.8b00705}.
We note in passing that they can also be used in the limit of $P=1$ in CP2K, where they impose the frequency-dependent enhancement of the quantum fluctuations of a harmonic oscillator on the classical system to mimic quasi-classical quantum fluctuations. 
For PIGLET, this is supported via a dedicated thermostat section \texttt{\&GLE}, while for PIQTB, the same thermostat section \texttt{\&QTB} is simply used in combination with setting $P=1$.

\subsection{Approximate Quantum Dynamics}
\label{sec:qdyn}
Within the standard PI formalism, MD is usually only performed as a sampling tool for the quantum partition function in the spirit of an extended Lagrangian formalism~\cite{Callaway1982,
Parrinello1984,Marx2009,Tuckerman2010}.
However, there are several PI formulations that allow for the approximate description of quantum dynamics, such as CMD~\cite{Cao1993a} and RPMD~\cite{Craig2004}.
Both of these techniques can be seen as approximations to Matsubara dynamics~\cite{Hele2015/10.1063/1.4921234}, which is the exact quantum dynamics in the PI formalism.
For more details on the theoretical footing of these methods, we refer the reader to Refs.~\cite{Voth1996,Althorpe2021/10.1140/epjb/s10051-021-00155-2,Althorpe2024/10.1146/annurev-physchem-090722-124705}.
Both CMD and RPMD are available in CP2K and are the only practical methods that allow one to approximately include NQEs on dynamics in (chemically complex) condensed phase systems. 
Unfortunately, both methods have been shown to suffer from fundamental artifacts,
namely CMD from the curvature problem and RPMD from the chain resonance problem~\cite{Witt2009/10.1063/1.3125009};
thermostatted RPMD (TRPMD) has been introduced to reduce the resonance problem
by adding friction~\cite{Rossi2014/10.1063/1.4883861} (see below). 
%
We note in passing that more recently, Brownian chain MD (BCMD)~\cite{shiga-bcmd-2022}
has been introduced, which is a short-timestep version of the PI hybrid MC (PIHMC) technique~\cite{Tuckerman1993}
applied to all non-centroid modes (whereas the centroids are propagated using Newtonian dynamics).
Interestingly, BCMD has been shown~\cite{shiga-bcmd-2022} to eliminate the resonance problem of RPMD and to alleviate the curvature problem of CMD. 

\begin{figure}
  \centering
  \includegraphics[width=0.5\textwidth]{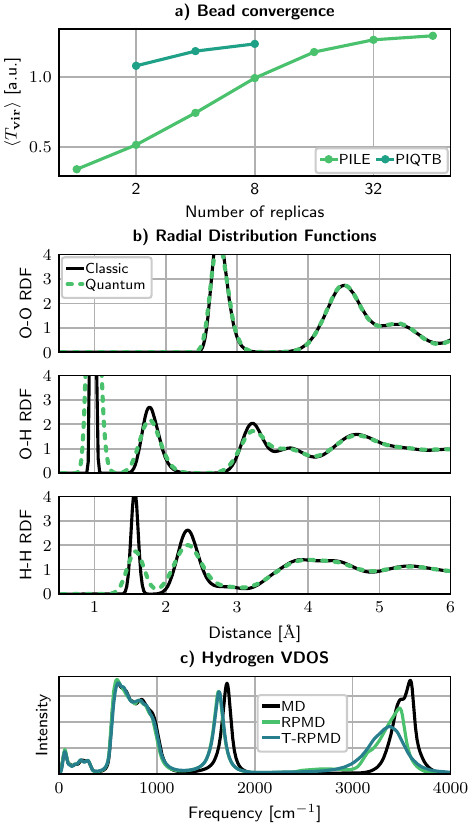}
  \caption{Impact of NQEs on the structure and dynamics of hexagonal ice at 250\,K. 
	(a) Bead convergence of the virial kinetic energy estimator using the PILE and PIQTB thermostat.
	(b) Radial distribution functions with classical and quantum nuclei and (b) vibrational density of states obtained with classical MD, RPMD and TRPMD, respectively.
All simulations rely on using the \mbox{C-NNP} trained with \mbox{revPBE0-D3} reference data~\cite{Schran2020/10.1063/5.0016004}, 
as available in the CP2K data repository (see also section~\ref{sec:nnp}).
}
  \label{fig:nqe-ice}
\end{figure}

The RPMD formalism requires switching the keyword \texttt{PROPAGATOR} in the \texttt{\&PINT} section to \texttt{RPMD}.
First, 
RPMD simulations in the $NVT$ ensemble
should be performed to generate initial configurations and velocities for subsequent $NVE$ simulations to generate the actual RPMD dynamics.
To speed-up structural thermalization, it is possible to start with standard PIMD $NVT$ simulations and then switch to RPMD, resampling the velocities, while keeping
the equilibrated configuration.
In addition, CP2K also supports the TRPMD method~\cite{Rossi2014/10.1063/1.4883861}, which is a variant of RPMD that applies a stochastic thermostat to all ring-polymer modes, except for the centroid mode.
This idea is an \textit{ad hoc} approach to remove spurious resonances of the ring-polymer modes with the true modes of the system, first discussed in detail in Ref.~\citenum{Witt2009/10.1063/1.3125009}, by adding noise. 
We note that the TRPMD approach usually leads to peak broadenings and lineshape distortions
in vibrational spectra due to the influence of the thermostat.
In CP2K, TRPMD can be simply enabled by setting the \texttt{TAU} parameter 
in the \texttt{\&PILE} subsection to a negative value.
An example input section for TRPMD to reproduce the results of Fig.~\ref{fig:nqe-ice}c is shown here, where RPMD dynamics would be enabled by removing the \texttt{\&PILE} subsection: 
\begin{verbatim}
&PINT
  P 32
  NUM_STEPS 1000
  TEMP       250
  DT        0.25
  HARM_INT  EXACT
  NRESPA 1
  TRANSFORMATION NORMAL
  PROPAGATOR RPMD
  &PILE
    TAU  -1000 
    LAMBDA 0.5
  &END PILE
&END PINT
\end{verbatim}

To conduct CMD simulations, the keyword \texttt{\&PINT\%PROPAGATOR} has to be set to \texttt{CMD}.
Partially adiabatic CMD (paCMD) shifts the non-centroid degrees of freedom to a high-frequency regime in order to sample the average force of the centroid used for the propagation of the centroid.
In CP2K,
paCMD is controlled by the \texttt{GAMMA} parameter in the \texttt{\&NORMALMODE} subsection which is a mass-scaling factor for the non-centroid degrees of freedom~\cite{Witt2009/10.1063/1.3125009}.
It is crucial to reduce the timestep in CMD simulations given the very high frequency of the non-centroid degrees of freedom.
We recall that CMD results in artificial temperature-dependent red-shifts of vibrational frequencies
due to the well-known curvature problem~\cite{Witt2009/10.1063/1.3125009, spura2015nuclear}.

Despite their known limitations and approximations~\cite{Witt2009/10.1063/1.3125009,Althorpe2021/10.1140/epjb/s10051-021-00155-2}, CMD and RPMD are powerful tools to generate approximate quantum dynamics in the PI formalism.
They can be used to study a wide range of condensed phase systems, as well as of course gas phase systems, to compute not only vibrational spectra, but also rate constants, relaxation times, and transport properties.
In particular, these methods can be easily combined with the above described NNP module to study the quantum dynamics of complex systems, as recently demonstrated for liquid water at coupled-cluster (CC) accuracy~\cite{Daru2022/10.1103/PhysRevLett.129.226001} including quantification of H/D isotope effects on translational and orientational dynamics~\cite{stolte-ccmd-isotope}, as well as on hydrogen bond kinetics~\cite{stolte-ccmd-kinetics}. 

We illustrate the capabilities of the PI formalism in CP2K with a study of hexagonal ice I$_\text{h}$ at 250\,K using the C-NNP model introduced in section~\ref{sec:nnp}.
Bead convergence of the virial kinetic energy estimator, which illustrates the quantum nature of the system, is shown in Fig.~\ref{fig:nqe-ice}a.
While the virial kinetic energy converges with roughly $P \approx 32$ at 250\,K using the PILE thermostat, we get a significant improvement in convergence using the PIQTB thermostat presented in section~\ref{sec:qconv}. 
We note here that although 32 replica
suffice to converge NQEs in liquid water at 300\,K (as explicitly demonstrated in Figure~S1 of Ref.~\cite{stolte-ccmd-isotope} using the CCSD(T) C-NNP from Ref.~\cite{Daru2022/10.1103/PhysRevLett.129.226001}), that number of beads is on the smaller side for crystalline water at 250\,K and certain properties require more replica for convergence.
The radial distribution functions in Fig.~\ref{fig:nqe-ice}b reveal the well-known 
structural differences imprinted by
the quantum nature of the system.
Finally, the vibrational density of states obtained from standard MD with classical nuclei, as well as from RPMD and TRPMD in Fig.~\ref{fig:nqe-ice}c shows the impact of
NQEs on the vibrational spectrum. 
While all quantum spectra are red-shifted compared to the classical spectrum, the RPMD spectrum features some additional peak modulations (namely a red-wing shoulder of the OH~stretching peak and artificial vibrational intensity between that stretch and the bending band)
due to artifical resonances of the fictitious ring-polymer modes with the physical vibrational modes of the system.
The TRPMD spectrum, on the other hand, is seen to reduce the chain resonance problem, but shows an unphysical broadening of the peaks due to the influence of the thermostat on the ring-polymer modes as alluded to in section~\ref{sec:qdyn}.
In summary, the PI formalism in CP2K provides a powerful tool to study the quantum nature of nuclei in a wide range of systems, in particular if combined with NNPs. 

\subsection{Bosonic Quantum Solvation at Ultra-low Temperatures}
So far, our discussion of NQEs has been restricted to particles that are assumed
to be distinguishable even if they are identical.
This implies that all nuclei in PIMD-based simulations of such systems can be numbered akin to classial particles, which results in Maxwell-Boltzmann (MB) quantum statistics. 
In MB quantum statistics, all nuclei are subject to quantum delocalization and
tunneling effects according to their mass, in stark contrast to classical particles. 
In nature, however, identical particles are
fundamentally indistinguishable. 
Their many-body WF, or finite-temperature density matrix $\rho \left(R, R', \beta \right)$, must be totally symmetric or antisymmetric in case of identical bosonic or fermionic particles, according to the spin-statistics theorem.
The density matrix can be formally symmetrized or antisymmetrized by introducing a sum over all
possible permutations $\{ {\cal P} \}$ of identical particles, i.e. 
\begin{equation}
\rho_{\genfrac{}{}{0pt}{}{\sf BE}{\sf FD}}
\left(R, R', \beta \right) = {1\over {N!}} 
\sum_{\cal P} (\pm 1)^{\cal P}   
\rho \left(R, {\cal P}R', \beta \right),
\label{eq:perm}
\end{equation}
where the positive and negative sign refer to bosons and fermions, respectively;
$R$ are all particle coordinates and $\beta = 1/k_{\rm B}T$. 
The resulting statistics are called Bose-Einstein (BE) and Fermi-Dirac (FD) quantum statistics; we refer to
Ref.~\citenum{Tuckerman2010} in the context of the molecular simulation approach to quantum statistical mechanics.

In the framework of numerical PI simulations of
BE and FD statistics, this implies that the discretized density matrix used to compute the partition function and all properties needs to be explicitly symmetrized or antisymmetrized with respect to the exchange of all identical particles~\cite{Feynman1972,Tuckerman2010}.
Therefore, an additional sum must be sampled in permutation space since the final coordinates of some particle along its quantum path might be some permutation of the initial coordinates, as illustrated here for the partition function that describes
three identical particles: 
\begin{eqnarray*}
Z_{\genfrac{}{}{0pt}{}{\sf\textcolor{red}{BE}}{\sf\textcolor{blue}{FD}}}
&=&
{\sf Tr}_{\genfrac{}{}{0pt}{}{\sf\textcolor{red}{BE}}{\sf\textcolor{blue}{FD}}}
\left[ \quad \frac{1}{3!} \quad
\begin{minipage}{1cm}\includegraphics[scale=.15]{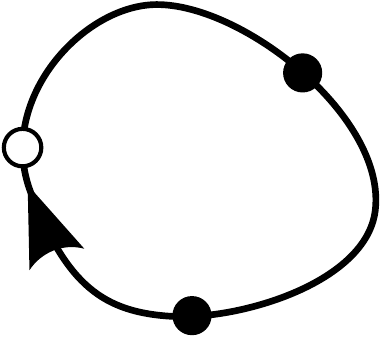}\end{minipage}
\quad
\begin{minipage}{1cm}\includegraphics[scale=.15]{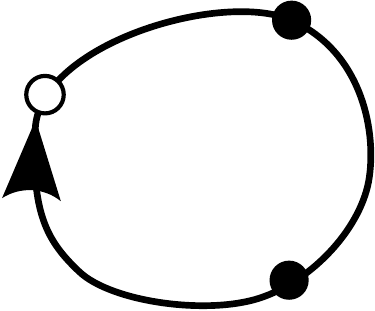}\end{minipage}
\quad
\begin{minipage}{1cm}\includegraphics[scale=.15]{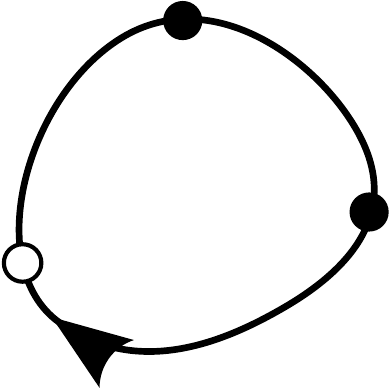}\end{minipage}
\right. \\[1.2\baselineskip]
& \genfrac{}{}{0pt}{}{\textcolor{red}{+}}{\textcolor{blue}{-}} &
\qquad
\frac{1}{2}
\quad
\begin{minipage}{2cm}\includegraphics[scale=.15]{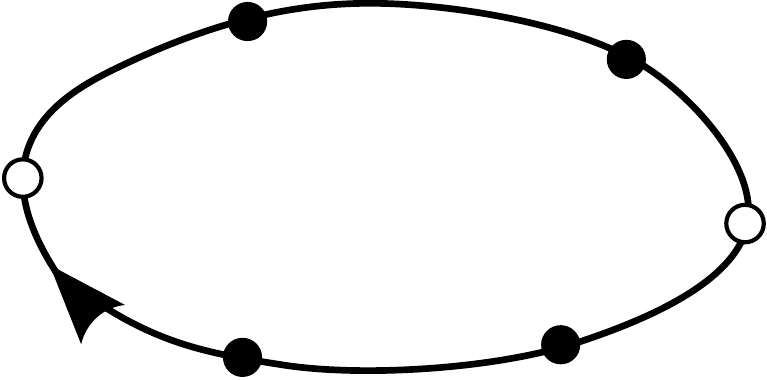}\end{minipage}
\quad
\begin{minipage}{1cm}\includegraphics[scale=.15]{figures/path-02}\end{minipage}
\\[1.2\baselineskip]
& + & \left. \qquad \frac{1}{3} \quad
\begin{minipage}{2.5cm}\includegraphics[scale=.15]{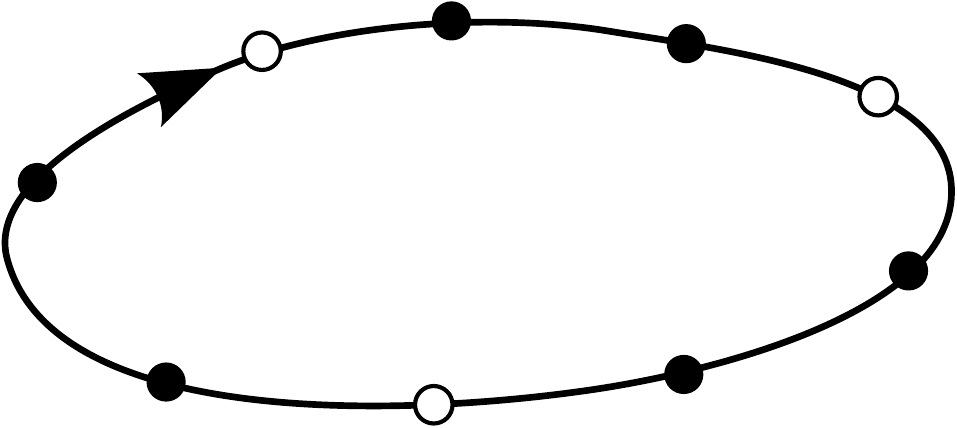}\end{minipage}
\quad \right] 
\enspace . 
\\
\end{eqnarray*}
In case of bosonic systems such as $^4$He, 
all terms contribute with a positive sign to the symmetrized trace~\cite{Feynman-helium4}, while the odd permutations carry a negative sign in the fermionic case;
we mention in passing that this gives rise to the infamous 
``sign problem'' in equilibrium PI simulations of many-body fermionic systems such as electrons~\cite{troyer2005computational, calcavecchia2014sign}. 

The pioneering bisection PIMC method to sample the bosonic permutations, which are required
in PI simulations of superfluid $^4$He, has been introduced four decades ago using a L\'evy construction~\cite{Ceperley1986/10.1103/PhysRevLett.56.351,Pollock1987/10.1103/PhysRevB.36.8343}; we refer to Ref.~\citenum{Ceperley1995} for an exhaustive review on this technique. 
Two decades later, the continuous/space worm algorithm for PIMC simulations of bosons in the grand-canonical ensemble has been devised to greatly improve the sampling efficiency, in particular for large bosonic systems
using open path (``worm'') sampling that involves off-diagonal elements of the density matrix~\cite{Boninsegni2006/10.1103/PhysRevLett.96.070601,Boninsegni2006/physics/0605225}.
In CP2K, a canonical worm algorithm with specific sampling extensions, which are relevant for finite bosonic clusters of a given particle number subject to strong interactions of the $^4$He atoms with impurities, has been introduced and implemented, as described in the appendix of Ref.~\citenum{brieuc_converged_2020}.
It therefore represents the default technique
to simulate bosonic exchange and thus superfluid properties when using CP2K. 

As a technical aside, we recall that
sampling bosonic exchange is characterized by
non-polynomial complexity and, therefore, gets increasingly demanding
the larger the bosonic system is, even when using efficient worm sampling.
%
For efficiency reasons, not only cubic, but also truncated octahedral
boundary conditions have been implemented in CP2K for the solvation supercell.
The latter type allows for a smaller overall volume and thus a smaller necessary number of bosonic solvent species to describe a fixed
spherical internal volume.
For superfluid fraction calculations, the winding number estimator (see below) has been properly generalized for the non-cubic supercell shape~\cite{walewski_reactive_2014}.
In CP2K, the truncated octahedral supercell is activated as follows in conjunction with the use of canonical worm sampling: 
%
%
\begin{verbatim}
&HELIUM
   ....
   CELL_SHAPE OCTAHEDRON
   PERIODIC T
   SAMPLING_METHOD WORM
   ....
&END HELIUM
\end{verbatim}

Apart from sampling the exchange of all identical particles via path permutations, PI simulations of superfluid helium require
one to reach temperatures on the order of 1~K or less. 
In traditional PIMC or PIMD simulations, some high-temperature density matrix is used that, by applying it many times in a product form, brings the system down to the desired temperature. 
Due to their simplicity, one-body high-temperature density matrices together with the
primitive Trotter approximant are employed
in most simulations including NQEs using PIMD-based methods, as discussed in sections~\ref{sec:NQE} and~\ref{sec:qconv}, in the framework of mNHC, PILE, PIGLET, PIQTB, CMD, RPMD or TRPMD thermostats, all of which are implemented in CP2K. 

In the realm of PIMC simulations at ultra-low temperatures, say on the order of 1~K or even less, the numerical pair density matrix approach has been introduced decades ago~\cite{Barker1979} to greatly reduce the number of PI beads or replica that are needed to converge the discretized PI to its continuum limit in the case of two-body interactions~\cite{Pollock1984/10.1103/PhysRevB.30.2555}, as reviewed earlier~\cite{Ceperley1995}.
%
Transcending primitive approximants, the high-temperature many-body density matrix is represented in this case in terms of one- and two-body contributions to the total PI action, where the latter can be exactly computed for all individual pairs of bosons that interact by a known two-body (pair) potential. 
%
In CP2K, we provide a numerical fit~\cite{Ceperley1995} of the \mbox{$^4$He $\cdots$ $^4$He} 
pair density matrix computed at a ``high temperature'' of 80~K, 
using a very accurate two-body interaction potential~\cite{Aziz1995/10.1103/PhysRevLett.74.1586}, which has been computed on a regular grid and subsequently spline-tabulated for its efficient use. 
Based on this pair density matrix, a PI~discretization of only $P=80$ replica suffices to perform converged PIMC simulations of bulk superfluid helium at a temperature of 1~K using the CP2K internal implementation: 
%
%
\begin{verbatim}
&MOTION
  &PINT
    NUM_STEPS 100
    &HELIUM
      HELIUM_ONLY
      NATOMS 32
      NBEADS 80
      DENSITY 0.0218312
      CELL_SHAPE OCTAHEDRON
      PERIODIC T
      POTENTIAL_FILE_NAME \
          helium_aziz95_80k.potx
      PRESAMPLE T
      SAMPLING_METHOD WORM
      N_OUTER 10000
    &END HELIUM
  &END PINT
&END MOTION
\end{verbatim}
Note that for converged observables, many more steps than the 100 selected above need to be simulated, and the necessary pair density file
\texttt{helium\_aziz95\_80k.potx} can be found
in the \texttt{tests/Pimd/untested\_inputs} directory of CP2K.

The superfluid fraction $f_\text{s}$ is a key property (order parameter) that
characterizes the phase transition from the normal liquid state of bosons, where $f_\text{s}=0$, to its superfluid phase, where $f_\text{s}\to 1$ as $T\to 0$~K.
%
The CP2K output also allows one to compute, in a post-processing step, 
the superfluid fraction of bulk systems and
the associated superfluid density $\rho_{\rm s}$
using the winding number estimator~\cite{Pollock1987/10.1103/PhysRevB.36.8343,Ceperley1995}
\begin{align}
f_\text{s}=\frac{\rho_\text{s}}{\rho_\text{tot}} = \frac{m_\text{He}\left\langle W^2\right\rangle}{\beta \hbar^2 \> N_\text{He}},
\label{eq:sff-winding}
\end{align}
%
where $W$ is the winding number that counts
how many times a given path wraps around the
periodic supercell along a given direction multiplied by its box length in that direction;
all other variables are self-explanatory. 
The output of this quantity is controlled via the \texttt{\&WINDING\_NUMBER} subsection inside the \texttt{\&PINT\%HELIUM\%PRINT} section. The output consists of the numerical value of the prefactor of Eq.~\ref{eq:sff-winding} as the initial comment line and the vectorial output of $W$ for each requested timestep.
Alternatively, one can turn on the output of the mean squared of the simulation so far via the \texttt{\&WINDING\_NUMBER\_2\_AVG} subsection. Also, this output contains the necessary prefactor value. Note that in
both cases, the prefactor in the files is without the $1/N_{\rm{He}}$ factor. 
%
After finite-size extrapolation, Eq.~\ref{eq:sff-winding} provides the rigorous expectation value of $f_\text{s}$ in bosonic
bulk systems subject to periodic boundary conditions~\cite{Pollock1992/10.1103/PhysRevB.46.3535}.
%
\begin{figure}
   \centering{}
    \includegraphics[width=0.5\textwidth]{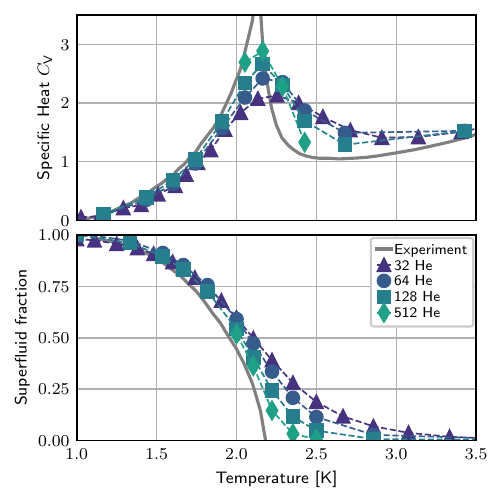}
   \caption{%
Temperature dependence of the heat capacity (a) and the superfluid fraction (b) of bulk $^4$He obtained using the canonical worm algorithm to sample bosonic exchange and the winding number estimator to analyze superfluidity (see text).
The numerical pair density matrix is the one provided with the CP2K package and has been computed at 80~K using the \mbox{He $\cdots$ He} potential from Ref.~\citenum{Aziz1995/10.1103/PhysRevLett.74.1586}, thus \mbox{$P\times T$=80~K} is kept constant for the temperature scan.
The present bosonic PIMC data were obtained from only 32 atoms
(in a truncated octahedral supercell according to the sample input provided in the text)
and are compared to published results generated from 64, 128 and 512 atoms in the
periodic truncated octahedron as provided in Fig.~5 of Ref.~\citenum{brieuc_converged_2020} using CP2K and the same methodology.
For reference, the respective experimental data~\cite{Keesom1935/10.1016/S0031-8914(35)90128-8,Klein1977}
are shown by grey lines; note that deviations of the computed properties from experiment are mainly due to finite-size effects close to the superfluid phase transition given such small system
sizes~\cite{Pollock1992/10.1103/PhysRevB.46.3535}.
}%
\label{fig:bulksuperfluidhelium}
\end{figure}
We illustrate in Fig.~\ref{fig:bulksuperfluidhelium} 
the behavior of the superfluid fraction, as well as that of the heat capacity featuring the characteristic $\lambda$-shape as a function of temperature of bulk $^4$He once it gets cooled down from the normal liquid state below the $\lambda$-transition where superfluidity sets in; note that $f_{\rm s}$ approaches unity at about 1~K. 
This calculation has been carried out
using 32 $^4$He atoms (the 
complete input can also be found in the \texttt{tests/Pimd/untested\_inputs} directory of CP2K) for the point at 1.6~K (20 independent runs of this type enter the plotted data for each temperature).

For finite systems, as particularly relevant in the case of molecular impurities solvated in bosonic environments at very low temperatures to be introduced in the following, there is no rigorous expression available to compute the superfluid fraction. 
In CP2K, the area estimator~\cite{Sindzingre1989/10.1103/PhysRevLett.63.1601} is the default for good reasons, i.e. 
\begin{align}
f_\text{s}=\frac{\rho_\text{s}}{\rho_\text{tot}} = \frac{4m_\text{He}^{2}\left\langle \left(\vec{n}\cdot\vec{A}\right)^{2}\right\rangle}{\beta \hbar^{2} \> I_\text{c}}, 
\end{align}
where $\vec{n}\cdot\vec{A}$ is the projected area with 
\begin{align}
\vec{A} = \frac{1}{2} \sum_{i=1}^{N_\text{He}}\sum_{s=1}^{P} \vec{r}_{i}^{(s)}\otimes \vec{r}_{i}^{(s+1)}
\end{align}
along some direction $\vec{n}$
and 
\begin{align}
    I_\text{c} = \left\langle \frac{1}{P}\sum_{i=1}^{N_\text{He}}\sum_{s=1}^{P} m_i \left(\vec{n}\otimes\vec{r}^{(s)}_{i}\right) \cdot \left(\vec{n}\otimes\vec{r}^{(s+1)}_{i}\right)\right\rangle 
\end{align}
is the respective classical moment of inertia.

Obtained from linear-response theory of macroscopic systems, the area estimator
works convincingly for large systems but,
unfortunately, provides a clearly non-zero superfluid fraction in the limit of simulating a single boson, which is obviously an unphysical artifact.
The reason is that this estimator is based on computing the mean-squared area of paths in some plane, which is non-vanishing even for one boson where no exchange paths exist. 
As a remedy, the exchange estimator
has been introduced~\cite{%
zeng_probing_2013,
Zeng2014/10.1088/0034-4885/77/4/046601}. It computes the so-called exchange superfluid fraction by essentially renormalizing the projected area obtained in BE simulations with the one computed from corresponding MB 
simulations, where no paths exchange and yet result in non-zero $\vec{n}\cdot\vec{A}$ contributions. 
In CP2K, the exchange estimator can be obtained after performing simulations identical to those sampling BE~statistics, while switching off the MC permutation moves in the worm algorithm to provide the respective MB averages necessary to correct the area estimator
(see Appendix~C.1 in Ref.~\citenum{supersolid-ch5p} for details). 

\begin{figure*}[t]
   \centering{}
   \includegraphics[width=1.0\textwidth]{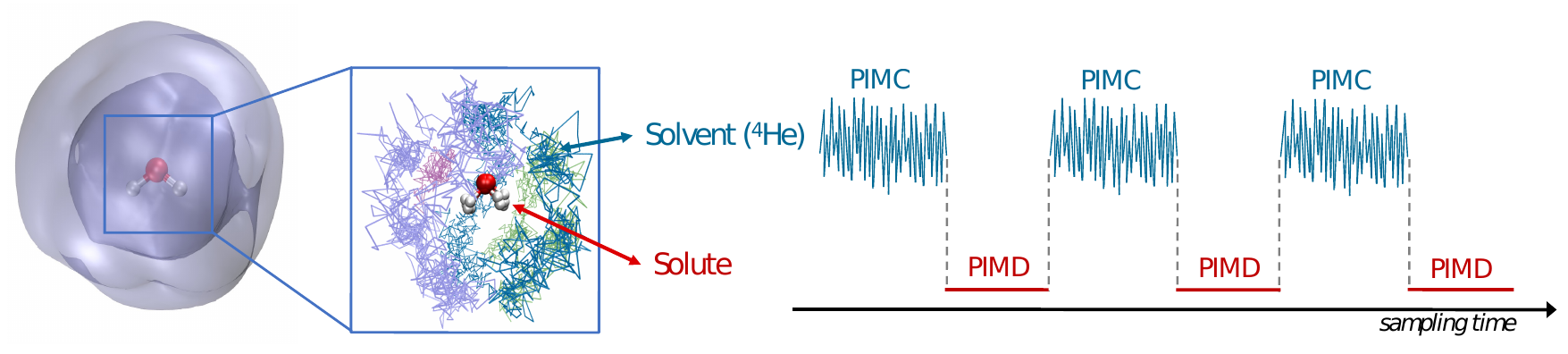}
   \caption{%
%
%
For an illustration of the HPIMD/MC method, as implemented in CP2K~\cite{brieuc_converged_2020}.
The fully flexible and reactive solute species are propagated using PIMD techniques (such as PIQTB in conjunction with the one-body density matrix), while the bosonic solvent (for instance many $^4$He atoms either forming a finite cluster as illustrated here, or hosted within a periodic supercell) is sampled using bosonic PIMC (for instance based on the canonical worm algorithm in conjunction with the numerical pair density matrix) at a common temperature $T$ (see text). 
The interactions can be described using high-dimensional NNPs with CCSD(T) accuracy, as described in section~\ref{sec:nnp}.
}%
\label{fig:PIMD_PIMC_Scheme}
\end{figure*}

The sampling scheme of the hybrid PIMD/bosonic PIMC (HPIMD/MC) algorithm~\cite{walewski_reactive_2014,Walewski2014/10.1063/1.4870595,brieuc_converged_2020}
to treat quantum solvation of fully flexible and reactive molecular species in bosonic environments (such as those provided by $^4$He), as available in the CP2K package, is presented in Fig.~\ref{fig:PIMD_PIMC_Scheme}. 
In this approach, the molecular species and the bosonic solvent are sampled using PIMD and PIMC subject to permutation moves, respectively. 
After completing one PIMD step of the molecule, where the forces on the molecule are obtained from the intramolecular
potential $U_\text{mol}$, as well as those coming from the solute-solvent interaction potential $U_\text{int}$, i.e.
\begin{align}
	\vec{F}_\text{PIMD} = &-\vec{\nabla}_{\vec{r}_\text{mol}}
U_\text{mol}(\{\vec{r}_\text{mol}\}) \\
	&- \vec{\nabla}_{\vec{r}_\text{mol}}
      U_\text{int}(\{\vec{r}_\text{mol}\},\{\vec{r}_\text{He}\}),
\end{align}
a bosonic PIMC simulation is then performed for the solvent species alone, for example $^4$He. This includes the exchange moves to establish BE statistics of the bosonic solvent environment, while the molecule is kept frozen
after the previous PIMD step, as illustrated in Fig.~\ref{fig:PIMD_PIMC_Scheme}.
It is important to carry out sufficiently many bosonic PIMC steps such that the bosonic environment can adapt to the current configuration of the molecule in both the real and permutation space before the next PIMD step is performed. 
The required bosonic PIMC~chains are generated based on the change of the potential 
\begin{align}
    U_\text{PIMC} =
U_\text{He}(\{\vec{r}_\text{He}\})+U_\text{int}(\{\vec{r}_\text{mol}\},\{\vec{r}
_\text{He}\})
\end{align}
due to the solute-solvent interactions $U_\text{int}$ together with the
solvent-solvent pair potential interactions (such as $U_\text{He}(\{\vec{r}_\text{He}\})$ to describe helium~\cite{Aziz1995/10.1103/PhysRevLett.74.1586})
as a evaluated during the Metropolis step.  
At this point, the bosonic permutations can be sampled in CP2K using either the bisection method~\cite{Ceperley1995}, or our canonical worm algorithm ~\cite{Boninsegni2006/physics/0605225,brieuc_converged_2020}, the latter being recommended for most cases. 
At the end of the PIMC chain, the next PIMD step is carried out using the new solvent position to compute $\vec{F}_\text{PIMD}$, 
followed by the next bosonic PIMC sweep, and so forth.
This particular \mbox{PIMD--PIMC} coupling scheme and propagation algorithm establish the correct quantum $NVT$~ensemble of the total solute-in-solvent system~\cite{brieuc_converged_2020}.
The HPIMD/MC approach implemented in CP2K has been generalized
to also deal with molecular bosons, such as para-H$_2$ or
ortho-D$_2$ molecules, in the framework of the adiabatic hindered rotor
(AHR)~\cite{Zeng2014/10.1088/0034-4885/77/4/046601}
averaging technique~\cite{Duran-ph2}.

At this point, a decision has to be made with respect to the intramolecular potential energy surface $U_\text{mol}$ and the solute-solvent interaction potential $U_\text{int}$ to describe fully flexible and reactive molecular solutes in a quantum solvent,
while an accurate solvent-solvent pair potential $U_\text{He}(\{\vec{r}_\text{He}\})$ is readily available in parameterized form~\cite{Aziz1995/10.1103/PhysRevLett.74.1586}. 
In the original HPIMD/MC approach~\cite{walewski_reactive_2014}, the \textit{ab-initio} PI philosophy~\cite{Marx1994a,Marx1996,Marx2009}
has been adapted to compute $U_\text{mol}$ on-the-fly, while $U_\text{int}$ has been fitted to accurate CC quantum chemistry data in the classic sense
of a physics-based interaction potential~\cite{Boese2011/10.1039/C1CP20991D,Uhl2017/10.1039/C7CP00652G};
this approach has been successfully applied to various
systems~\cite{Walewski2014/10.1063/1.4870595,uhl_helium_2018,Uhl2019/10.1103/PhysRevLett.123.123002}
using CP2K. 
This technique has been superseded by the use of NNPs, as explained in section~\ref{sec:nnp}, which offer two advantages at the same time, namely allowing for highly efficient PI sampling and providing high accuracy of the intra- and intermolecular interactions~\cite{Schran2018/10.1021/acs.jctc.8b00705}.
The current NNP-based HPIMD/MC approach in CP2K thus allows one to converge the PI even at ultra-low temperatures (using techniques summarized in section~\ref{sec:qconv}) and achieves essentially basis set converged CCSD(T) accuracy~\cite{Schran2018/10.1021/acs.jctc.8b00705};
we note in passing that such CC quality PIMD simulations, dubbed CCMD, have been generalized more recently to condensed phase systems~\cite{Daru2022/10.1103/PhysRevLett.129.226001,stolte-ccmd-isotope}.

The generation and use of NNPs in CP2K have been reviewed herein in section~\ref{sec:nnp}, see there for general background and input options, which transfers to applications of the HPIMD/MC method. 
But in this case, additional challenges arise since
the intramolecular potential energy surface of finite molecular systems, as well as intermolecular interaction potentials are required at very high
(``quantum chemical'') accuracy, while analytic gradients and thus forces are not efficiently available in the framework of CC theories. 
The situation is distinctly different from usual applications of MLPs in general and NNPs in particular to condensed phase systems, such as water and ice presented in section~\ref{sec:nnp}, which are based on computationally economic periodic DFT calculations, where analytic forces are readily available.

\begin{figure}
   \includegraphics[width=\linewidth]{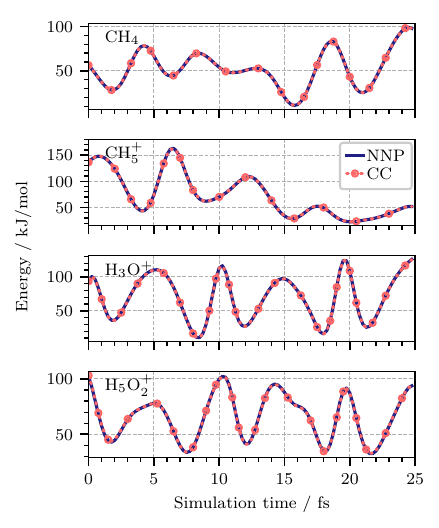}
   \caption{%
       Potential energy
$U_\text{mol}$
along one Trotter replica of quantum PIMD trajectories
       of (from top to bottom) the methane molecule (CH$_4$),
       the protonated methane complex (CH$_5^+$),
       the hydronium cation (H$_3$O$^+$), and the Zundel cation (H$_5$O$_2^+$)
       at 1.67~K using NNPs
trained with \mbox{CCSD(T*)-F12a/aug-cc-pVTZ} reference energies obtained from Molpro.
       The corresponding CC single-point data were obtained
       by recomputing the energies
       at each and every step of
       the depicted NNP trajectories
       and are shown as red dotted lines. 
       All energies are reported relative to the equilibrium structures
of the respective global minima.
Reproduced from Fig.~6 of Ref.~\citenum{brieuc_converged_2020}.
   }
   \label{fig:sol_nnp}
\end{figure}
\begin{figure}
   \includegraphics[width=\linewidth]{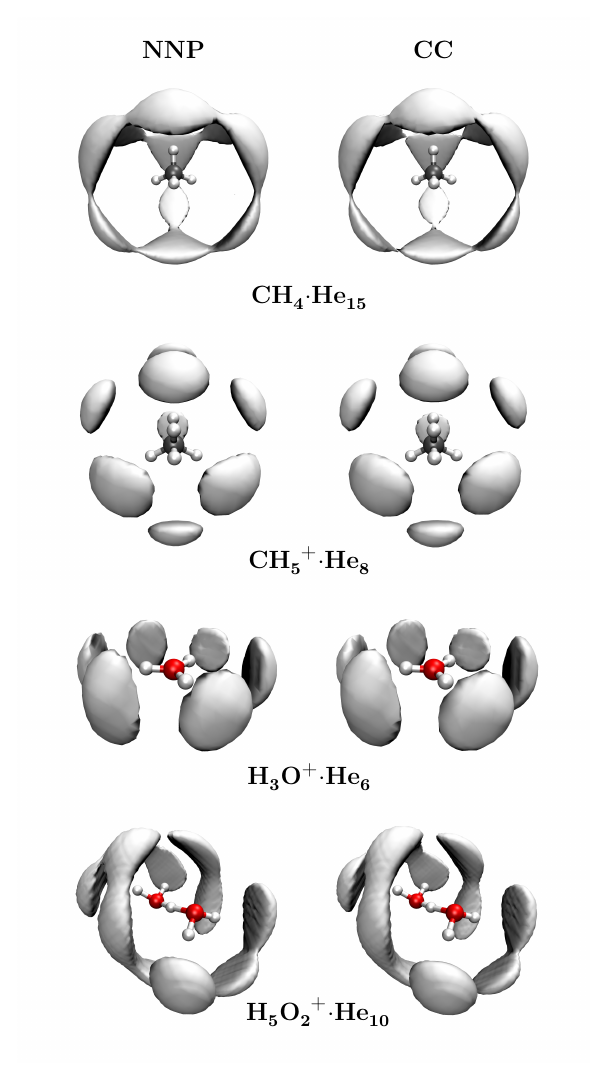}
   \caption{Spatial distributions functions 
stemming from all $N$ helium atoms
            around a fixed molecular
            impurity
(i.e. $X$$\cdot$He$_N$),
as sampled from PIMC simulations (without bosonic exchange)
            at 1.67~K. 
            The solute$\cdots$helium interaction energies 
$U_\text{int}$ 
used in the PIMC sampling of the helium atoms are interpolated from values that have been precomputed on a grid using either the NNP (left column) or single-point CC calculations (right column) on the identical grid both based on counterpoise-corrected \mbox{CCSD(T*)-F12a/aug-cc-pVTZ} energy differences 
obtained from Molpro. 
%
%
%
%
%
Reproduced from Fig.~7 of Ref.~\citenum{brieuc_converged_2020}.
}
   \label{fig:Hedensities}
\end{figure}

These challenges posed by HPIMD / MC have been addressed using hierarchical MLP sampling approaches
to cope with intermolecular interactions and thus solvation~\cite{schran_high-dimensional_2018}
combined with active learning protocols~\cite{Schran2020/10.1021/acs.jctc.9b00805} 
to efficiently parameterize both $U_\text{int}$ and $U_\text{mol}$ in the framework of NNPs, which are deeply interfaced with CP2K.
In particular, NNP training codes such as the RuNNer, n2p2 or 
\mbox{RubNNet4MD} packages~\cite{RubNNet4MD-code}
have been used in conjunction
with quantum chemistry packages that offer efficient implementations of CCSD(T) theory, such as Molpro~\cite{molpro-review} and ORCA~\cite{ORCA2020} to generate on the order of 10,000 to 100,000 total energies of finite molecular systems in the relevant size regime
to train, test and validate the MLPs for use in CP2K-based HPIMD/MC and CCMD simulations~\cite{schran_high-dimensional_2018,Schran2018/10.1021/acs.jctc.8b00705,Schran2020/10.1021/acs.jctc.9b00805,topolnicki-pa-2020,Schran2021/10.1063/5.0035438,Daru2022/10.1103/PhysRevLett.129.226001,zundel-mctdh-2022,Duran-ph2,beckmann-pahe-2023,simko-hphe3-2023,supersolid-ch5p,davies-h3ophe,ch5p-ir}. 
The accuracy of the NNPs $U_\text{mol}$ and interaction potentials $U_\text{int}$ at the CCSD(T) level of theory is illustrated in Figs.~\ref{fig:sol_nnp} and~\ref{fig:Hedensities}, respectively.
This rigorous end-to-end testing of the MLP representations of CCSD(T) references energies of diverse systems in terms of NNPs validates the accuracy of the overall methodology used in CP2K to perform HPIMD/MC and CCMD
simulations~\cite{schran_high-dimensional_2018,Schran2018/10.1021/acs.jctc.8b00705,Schran2020/10.1021/acs.jctc.9b00805,topolnicki-pa-2020,Schran2021/10.1063/5.0035438,Daru2022/10.1103/PhysRevLett.129.226001,zundel-mctdh-2022,Duran-ph2,beckmann-pahe-2023,simko-hphe3-2023,supersolid-ch5p,davies-h3ophe,ch5p-ir}.
%
%

%
We note in passing that dipole moment surfaces of finite molecular systems have also been generated with CCSD(T) accuracy using a very similar systematic NN-based training approach~\cite{beckmann-ir}, as implemented in Version~2 of the \mbox{RubNNet4MD}  package~\cite{RubNNet4MD-code}.
Using CP2K, this enables the calculation of infrared (IR) absorption cross sections based on approximate RPMD (or TRPMD or CMD) quantum dynamics, as described in section~\ref{sec:qdyn}, in order to include NQEs in such vibrational spectra consistently at CCSD(T) quality~\cite{ch5p-ir}.

\begin{figure}
    \centering{}
    \includegraphics[width=0.8\linewidth]{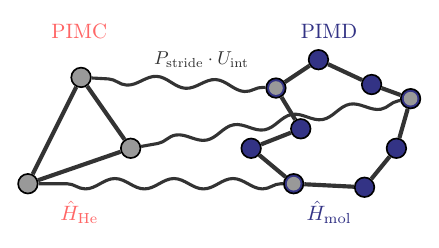}
    \caption{%
Illustration of PI striding for bosonic HPIMD/MC simulations, as implemented in CP2K: 
Coupling of the PI representation of one helium atom (left) discretized with three Trotter replica 
(in practice sampled using PIMC based on the pair density matrix representation)
to the PI representation of the solute molecule (right)
discretized with nine beads 
(sampled using PIMD based on the one-body density matrix in conjunction with PIQTB thermostatting, as detailed in section~\ref{sec:qconv}), 
thus corresponding to a striding length of three, $P_\text{stride}=3$.
%
Reproduced from Fig.~3 of Ref.~\citenum{brieuc_converged_2020}.
         }
    \label{fig:striding}
\end{figure}

\begin{figure}
%
   \includegraphics[width=\linewidth]{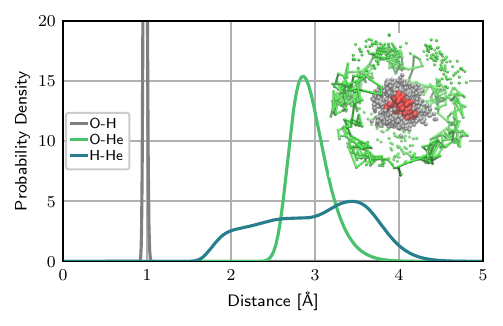}
   \caption{Distance distributions functions of
\mbox{H$_3$O$^+$$\cdot$$^4$He$_{8}$} at 1~K with BE sampling
of the helium microsolvation environment.
The intramolecular O--H, as well as the intermolecular O--He and H--He
distance distributions are presented. 
%
The inset depicts the beads of the hydronium ion in terms of balls \mbox{(red~/ gray for O~/ H)}, 
while all $^4$He atoms involved in a bosonic exchange cycle are visualized
by green strings and all others by green balls.
}
   \label{fig:hpimdmcexample}
\end{figure}

The HPIMD/MC method implemented in CP2K allows one to investigate fully flexible and reactive solutes, such as molecules, complexes or clusters, in finite and bulk-like bosonic environments, 
such as superfluid $^4$He or para-H$_2$ species, at relevant temperatures on the order of 1~K.
%
Here, a typical CP2K input section is provided to simulate
a protonated water molecule embedded in a helium cluster consisting of 8 \mbox{$^4$He atoms} 
(i.e. \mbox{H$_3$O$^+$$\cdot$$^4$He$_{8}$}) at a temperature of 1~K: 
\begin{verbatim}
&MOTION
  &PINT
    DT 0.1
    NUM_STEPS 1000000
    P 160
    PROPAGATOR RPMD
    TEMP 1.0
    &HELIUM T
      CELL_SHAPE OCTAHEDRON
      CELL_SIZE 19.1157
      DROPLET_RADIUS 15.0
      GET_FORCES LAST
      NATOMS 8
      NBEADS 80
      N_OUTER 1000
      PERIODIC F
      POTENTIAL_FILE_NAME ./helium.potx
      PRESAMPLE T
      SAMPLING_METHOD WORM
      SOLUTE_INTERACTION NNP
      &NNP
        NNP_INPUT_FILE_NAME inter-input.nn
        SCALE_FILE_NAME inter-scaling.data
        &MODEL
          WEIGHTS inter-weights
        &END MODEL
        &SR_CUTOFF
          ELEMENT "H"
          RADIUS  1.25
        &END SR_CUTOFF
        &SR_CUTOFF
          ELEMENT "He"
          RADIUS  0.00
        &END SR_CUTOFF
        &SR_CUTOFF
          ELEMENT "O"
          RADIUS  2.05
        &END SR_CUTOFF
      &END NNP
      &WORM
        ALLOW_OPEN T
        CENTROID_DRMAX  0.25
        MAX_OPEN_CYCLES 100
        OPEN_CLOSE_SCALE  50.0
        STAGING_L 6
      &END WORM
    &END HELIUM
    &PILE
       LAMBDA 0.5
       TAU 200
    &END PILE
  &END PINT
&END MOTION
&FORCE_EVAL
  METHOD NNP
  &NNP
    NNP_INPUT_FILE_NAME solute-input.nn
    SCALE_FILE_NAME solute-scaling.data
    &MODEL
      WEIGHTS solute-weights
    &END MODEL
  &END NNP
  &SUBSYS
    &COORD
      O  0.00  0.00  0.00
      H  1.05  0.00  0.00
      H -0.49  0.77  0.00
      H -0.55 -0.51 -0.48
    &END COORD
  &END SUBSYS
&END FORCE_EVAL
\end{verbatim}
This simulation uses the numerical \mbox{$^4$He$\cdots$ $^4$He}
pair density matrix at 80~K obtained from the very accurate two-body interaction potential of Aziz to describe $U_\text{He}$~\cite{Aziz1995/10.1103/PhysRevLett.74.1586}; the resulting PIMC discretization 
of the bosonic environment relies on 80~replica. 
The employed intramolecular and intermolecular potentials $U_\text{mol}$ and $U_\text{int}$, are NNP-based representations of reference energies at the CCSD(T) level of theory (obtained from the Molpro package~\cite{molpro-review}) that have been trained using the \mbox{RubNNet4MD} package~\cite{RubNNet4MD-code} and are available from the CP2K GitHub repository.
%
For PIMD sampling of the solute using the RPMD approach, a bead number of~160 is used.
This relies on the striding approach, which allows one to couple two PIs with different Trotter discretizations, as visualized in Fig.~\ref{fig:striding}.
In HPIMD/MC, the efficient pair density matrix approach allows one to converge the solvent PI using a rather small replica number (here 80 at 1~K), which is not sufficient to represent the solute PI.
%
Using a stride of two and thus 160 replicas to represent the solute interactions, versus 80 for the solvent-solvent interactions and thus the solute-solvent interactions, as explained in section~III.C of Ref.~\citenum{brieuc_converged_2020}
and illustrated schematically in Fig.~\ref{fig:striding}, is appropriate. 
The intramolecular O--H, as well as the intermolecular O--He and H--He distance distribution functions resulting from the input example provided are presented in Fig.~\ref{fig:hpimdmcexample}.
The corresponding superfluid fraction of the bosonic helium environment, as obtained with the exchange estimator, 
amounts to $f_{\rm s} \approx 0.326\pm 0.012$.
This technique is available in CP2K and can be readily used to study the effects of bosonic quantum solvation on fully flexible and reactive molecular solute species embedded in $^4$He and also in para-H$_2$ environments
at temperatures on the order of 1~K~\cite{Duran-ph2}, where quantum exchange occurs and thus BE statistics emerge, resulting in strictly non-zero superfluid 
fractions and densities~\cite{uhl_helium_2018,Uhl2019/10.1103/PhysRevLett.123.123002,beckmann-pahe-2023,supersolid-ch5p,davies-h3ophe}.
%

\subsection{Vibrational Spectroscopy} \label{sec:VibrationalSpectroscopy}

Vibrational spectroscopy is an important field of chemistry and physics since a very long time. As molecular vibrations (or phonons in solid state materials) are highly sensitive to the chemical environment of a certain molecule, vibrational spectra allow one to study the interactions present in the sample in great detail.


There exist two fundamentally different approaches for predicting vibrational spectra: the so-called ``static-harmonic approximation'', 
and the ``time-correlation function (TCF)'' techniques.
%
Though CP2K directly implements the former approach, it is able to efficiently compute all necessary quantities of the latter method too, so that vibrational spectra can be easily obtained from post-processing tools such as ASE~\cite{ase} and TRAVIS~\cite{Brehm.2011,Brehm.2020}.

\subsubsection{Static-Harmonic Approach}

It is possible to predict the vibrational modes of an atomistic system by approximating it as a system of harmonic oscillators. Assuming a reasonably smooth potential energy surface, the potential energy $V\!\left(\mathbf{x}\right)$ of a set of $N$ atoms in close proximity to some point $\mathbf{x}^0$ with respect to the atoms' $3N$ Cartesian coordinates $\mathbf{x}\!=\!\left(x_1,\dots,x_{3N}\right)$ can be approximately expressed in terms of a second-order multi-dimensional Taylor expansion:
\begin{multline}
V\!\left(\mathbf{x}\right)\approx{}V\!\left(\mathbf{x}^0\right)+\sum\limits_{i=1}^{3N}\Bigl(\frac{\partial{}V}{\partial{}x_i}\Bigr)_{\mathbf{x}^0}\left(x_i-x_i^0\right)\\
+\frac{1}{2}\sum\limits_{i=1}^{3N}\sum\limits_{k=1}^{3N}\Bigl(\frac{\partial^2V}{\partial{}x_i\partial{}x_k}\Bigr)_{\mathbf{x}^0}\left(x_i-x_i^0\right)\left(x_k-x_k^0\right)
\end{multline}
If the point $\mathbf{x}^0$ is an energy minimum, the gradients $\bigl(\frac{\partial{}V}{\partial{}x_i}\bigr)_{\mathbf{x}^0}$ vanish, and only the second derivatives remain, i.e.
\begin{multline}
\label{eqn:intro.stat0}
V\!\left(\mathbf{x}\right)\approx{}V\!\left(\mathbf{x}^0\right)+\frac{1}{2}\sum\limits_{i=1}^{3N}\sum\limits_{k=1}^{3N}\Bigl(\frac{\partial^2V}{\partial{}x_i\partial{}x_k}\Bigr)_{\mathbf{x}^0}\\
\left(x_i-x_i^0\right)\left(x_k-x_k^0\right).
\end{multline}
Based on this approximation, the force $F_i$ acting on coordinate $x_i$ can be expressed as
\begin{equation}
\label{eqn:intro.stat1}
F_i=-\Bigl(\frac{\partial{}V}{\partial{}x_i}\Bigr)=-\frac{1}{2}\sum\limits_{k=1}^{3N}\Bigl(\frac{\partial^2V}{\partial{}x_i\partial{}x_k}\Bigr)_{\mathbf{x}^0}\left(x_k-x_k^0\right).
\end{equation}
By inserting Eq.~\ref{eqn:intro.stat1} into Newton's equation
\begin{equation}
F_i=m_i\,a_i=m_i\Bigl(\frac{\mathrm{d}^2x_i}{\mathrm{d}t^2}\Bigr),\qquad{}i=1,\dots,3N,
\end{equation}
one obtains a system of equations of motion
\begin{equation}
\label{eqn:intro.stat2}
m_i\Bigl(\frac{\mathrm{d}^2x_i}{\mathrm{d}t^2}\Bigr)=-\sum\limits_{k=1}^{3N}\Bigl(\frac{\partial^2V}{\partial{}x_i\partial{}x_k}\Bigr)_{\mathbf{x}^0}\left(x_k-x_k^0\right).
\end{equation}
To express these equations more compactly, it is desirable to switch to the set of so-called mass-weighted Cartesian coordinates $\mathbf{q}\!=\!\left(q_1,\dots,q_{3N}\right)$ defined as
\begin{equation}
q_i:=x_i\sqrt{m_i},
\end{equation}
in which Eq.~\ref{eqn:intro.stat2} now reads as
\begin{equation}
\label{eqn:intro.stat3}
\frac{\mathrm{d}^2q_i}{\mathrm{d}t^2}=-\sum\limits_{k=1}^{3N}H_{i,k}\,q_k,\qquad{}i=1,\dots,3N
\end{equation}
with the short-hand notation $H_{i,k}$ for the mass-weighted Hessian matrix
\begin{equation}
H_{i,k}:=\frac{1}{\sqrt{m_im_k}}\Bigl(\frac{\partial^2V}{\partial{}x_i\partial{}x_k}\Bigr)_{\mathbf{x}^0}.
\end{equation}
As the above approximation defines a purely (see Eq.~\ref{eqn:intro.stat0}), one can assume that the motion of each coordinate $q_i$ can be described by a sinusoidal time evolution
\begin{equation}
q_i\!\left(t\right)=q_i^0+A\cdot{}\mathrm{cos}\!\left(\omega{}t\right)
\end{equation}
with some amplitude $A$ and angular frequency $\omega$. Inserting this into Eq.~\ref{eqn:intro.stat3} yields
\begin{equation}
\omega^2q_i=\sum\limits_{k=1}^{3N}H_{i,k}\,q_k,\qquad{}i=1,\dots,3N,
\end{equation}
which can be rewritten as a matrix eigenvalue problem for the mass-weighted Hessian matrix $\mathbf{H}$ with the eigenvalues $\omega^2$:
\begin{equation}
\omega^2\mathbf{q}=\mathbf{H}\mathbf{q}.
\end{equation}
In other words, solving this eigenvalue problem directly yields the vibrational frequencies $\omega$ of the atomic system as the square roots of the eigenvalues. Furthermore, the eigenvectors $\mathbf{q}$ of the matrix represent the directions in which the atoms are displaced within each normal mode. Therefore, a full set of normal modes, each with a vibrational frequency and an atom displacement vector, can be obtained via this approach. Additional care has to be taken to project out the invariants (i.e. translational and rotational invariances due to the conservation of momentum and angular momentum) of the mass-weighted Hessian matrix, so that typically only $\left(3N\!-\!6\right)$ normal modes are obtained.

Such a normal mode analysis is fully implemented in CP2K. Yet, as explained above, the approach only works for minimum structures, so it should be applied only to geometries that resulted from a converged optimization with very similar computational details. The normal mode analysis can be requested as follows:
\begin{verbatim}
&GLOBAL
  RUN_TYPE VIBRATIONAL_ANALYSIS
&END GLOBAL
&VIBRATIONAL_ANALYSIS
  DX 0.01
  FULLY_PERIODIC TRUE
  INTENSITIES TRUE
&END VIBRATIONAL_ANALYSIS
\end{verbatim}
Note that CP2K computes the second derivatives of the energy with respect to the displacement (Hessian matrix) numerically via central finite differences, so that for a system of $N$ atoms, $6N$ individual energy and force computations will be required. The \texttt{DX} parameter sets the displacement for the finite differences. Please also note that a very tight SCF threshold (\texttt{EPS\_SCF 1.E-8} in combination with \texttt{EPS\_DEFAULT 1.E-16}) is required to reduce numerical noise. The \texttt{FULLY\_PERIODIC} keyword inside the \texttt{VIBRATIONAL\_ANALYSIS} section should be used for periodic bulk phase systems in order to switch off the pruning of the rotational degrees of freedom from the Hessian. Activating the keyword \texttt{INTENSITIES} facilitates the computation of IR or Raman intensities, but requires the explicit calculation of dipoles or polarizabilities by means of \texttt{\&DFT\%PRINT\%MOMENTS} or \texttt{\&PROPERTIES\%LINRES\%POLAR}, which are described in detail later. Moreover, adding the subsection \texttt{\&MODE\_SELECTIVE} permits running a so-called mode selective vibrational analysis, which constrains the calculation to either a single \texttt{FREQUENCY}, \texttt{RANGE} of frequencies, or a list of \texttt{INVOLVED\_ATOMS}.

The derivation above only shows how to obtain the vibrational frequencies of the normal modes. However, in order to predict real vibrational spectra, the intensities of each mode also need to be computed. This is typically performed by computing derivatives of some more or less complicated electromagnetic properties with respect to the atom displacements $\mathbf{q}$ obtained for each mode. For example, the IR intensity of a certain mode is proportional to the change in the electric dipole moment that occurs when the atoms are displaced along that mode. More details on computing such properties are discussed in the following subsection.

Despite still being the standard approach in the literature, computing spectra via the static-harmonic approach comes with several shortcomings:
\begin{itemize}
\item Due to the harmonic approximation of the potential energy surface, all anharmonic effects are neglected~\cite{Lindner.2007}. If the system possesses features such as strong hydrogen bonds or hindered rotations, the harmonic approximation of certain modes will be poor, and so will the quality of the predicted spectrum.
\item The spectrum can only be computed for one minimum energy structure at a time. If there exist several conformers of the same molecule, they need to be considered separately. If the system can hardly be described by minimum-energy structures, such as bulk phase liquids, it will be hard to obtain reasonable spectra from the outset.
\item The method works best for molecules or small clusters in vacuum. Solvent effects on the spectrum 
can be crudely approximated either via continuum solvation models such as COSMO~\cite{Klamt1993a}, PCM~\cite{Mennucci.1997}, or the SCCS implicit solvation method described in section~\ref{sec:SCCS}~\cite{Fattebert2002,Fattebert2003,Andreussi2012}, as well as by microsolvation, but the solvent effect cannot be captured comprehensively.
\item The approach only yields a discrete line spectrum; no line widths or band shapes can be obtained. Hence, to predict realistic spectra, empirical line broadening needs to be applied. 
\end{itemize}

\subsubsection{Time-Correlation Function Approach}

Apart from employing the static-harmonic approximation, there exists the possibility to compute vibrational spectra directly from MD simulations. In this approach, the spectra are obtained as the Fourier transform of some TCFs along the simulation trajectory, an idea that is at least 60 years old now~\cite{Gordon1965,Shimizu1965,Kubo1991}, and is called the TCF formalism. This approach comes with several advantages over the static-harmonic concept:
\begin{itemize}
\item Condensed phase systems can be handled; it is possible to explicitly capture the effects of solvent and entropy on the spectrum.
\item Some anharmonic effects, such as line broadening, approximate overtones, and combination bands, are reproduced.
\item Realistic band shapes are obtained instead of a discrete line spectrum.
\item Intrinsic conformer sampling takes place during the MD simulation.
\item No minimum energy structure is required to compute the spectrum.
\end{itemize}

As CP2K has its main strengths in the description of condensed phase systems, including AIMD, it is particularly well suited to compute vibrational spectra via the TCF formalism. Therefore, the remaining part of this subsection will focus exclusively on this approach. The basic workflow is simple: perform an AIMD simulation, compute some electromagnetic properties along the trajectory, and use some post-processing software to obtain a spectrum from that time series.

\subsubsection{Computing Electromagnetic Moments}

Computing vibrational spectra, both via the static-harmonic approach and via the TCF approach, requires knowledge of certain electromagnetic moments of the system, e.g. the electric dipole moment. In the former case, these moments are calculated for excursions of the minimum structure along the normal modes, but they are computed for snapshots along an MD simulation trajectory in the latter case. For non-periodic systems, electric moments can be readily derived as expectation values from electron structure calculations by applying the corresponding moment operator to the converged WF.

Under PBCs, however, the standard moment operators are ill-defined, as they often rely on the position operator. The challenge can be easily seen by, e.g. considering the total dipole moment of a simple chain of anions and cations in a periodic cell: the dipole moment depends on the cell origin, but it should not~\cite{spaldin2012beginner}. Resta and King-Smith and Vanderbilt proposed a possible solution, the so-called modern theory of polarization, which is based on a Berry phase~\cite{KingSmith1993,Resta1993,Resta1994,Resta1998}. In short, the ill-defined position operator $\hat{X}$ for the WF $\Psi_0$ under PBCs is replaced by the following expectation value for the position~\cite{Resta1998}:
\begin{equation}
\left<X\right>=\frac{L}{2\pi}\mathrm{Im}\,\mathrm{ln}\,\left<\Psi_0\bigg\rvert{}\exp\!\left(i\frac{2\pi}{L}\hat{X}\right)\bigg\rvert{}\Psi_0\right>.
\end{equation}
Due to the ambiguity of the complex logarithm function, there exist infinitely many valid results from this equation, all equidistantly separated. This means that the dipole moment is also no longer uniquely defined under PBCs, but only up to a modulus. However, for most applications, this approach works very well.

In CP2K/\textsc{Quickstep}, the total electric dipole moment from the modern theory of polarization can be computed and printed by adding the following section to the input:
\begin{verbatim}
&FORCE_EVAL
  &DFT
    &PRINT
      &MOMENTS
      &END MOMENTS
    &END PRINT
  &END DFT
&END FORCE_EVAL
\end{verbatim}
Note that despite the general name of the \texttt{\&MOMENTS} subsection and the availability of a \texttt{MAX\_MOMENT} keyword, it is implemented only for electric dipole moments in the periodic case.

\subsubsection{Total vs. Molecular Moments}

There are several reasons for considering molecular moments instead of system-wide ``total'' electromagnetic moments.
First, it allows for separation of the spectral contributions of the constituents of a mixture, so that, e.g., the solvent spectrum can be suppressed.
Secondly, the sampling of the spectrum is improved and the spectrum contains less noise when the TCF approach is used, as will be explained on the example of the dipole moment in the following. When assuming that the sum of the molecular dipole moments equals the total dipole moment,which is an approximation in the periodic case although the deviation is small~\cite{Kirchner2004}, it can be shown that using the total dipole moment is mathematically equivalent to using all autocorrelations of the individual molecules' dipole moments together with all cross-correlations between different molecules. It is often argued that cross-correlations between molecules that are not neighbors in space only introduce noise to the spectrum, as the motions of such molecules are not at all correlated. Therefore, it is often a good idea to consider only molecular autocorrelation functions and to neglect all cross-correlation terms between different molecules.

\subsubsection{Orbital Localization}

One widely used approach to assign electric dipole moments to individual molecules is the localization of MOs in space~\cite{silvestrelli1999a,silvestrelli1999b}. For non-periodic systems, there exist some well-known and computationally efficient methods such as the Boys-Foster~\cite{Boys1960}, as well as the Pipek-Mezey localization schemes~\cite{Pipek1989}, among others.

In periodic systems, orbital localization is considerably more involved. One commonly used method is the so-called Wannier localization~\cite{Wannier1937,PhysRevB.56.12847,silvestrelli1998maximally,Silvestrelli1999c,Marzari2012,ambrosetti2016introduction}. It applies a unitary transformation $\mathbf{U}$ to the set of occupied KS orbitals $\left|\psi_i\right>$ so that another set of MOs $\left|\tilde{\psi}_n\right>$ is obtained, which are called Wannier orbitals or MLWFs~\cite{Berghold2000a}:
\begin{equation}
\label{eqn:intro.wannier.1}
\left|\tilde{\psi}_n\right>=\sum\limits_iU_{i,n}\left|\psi_i\right>.
\end{equation}
The unitary transformation $\mathbf{U}$ is constructed in a way so that the so-called spread functional
\begin{align}
\label{eqn:intro.wannier.2}
\bm{\Omega}&=\sum\limits_n\sum\limits_I\ f\left(\left|z_{I,n}\right|^2\right),\\
\label{eqn:intro.wannier.3}
z_{I,n}&=\left<\psi_n\middle|\mathbf{O}^I\middle|\psi_n\right>,
\end{align}
is minimized~\cite{Berghold2000a}. Here, $\mathbf{O}^I$ is a class of suitable spread operators that are well defined in periodic space, such as
\begin{equation}
\label{eqn:intro.wannier.4}
\mathbf{O}^I=\exp\!\left(\mathrm{i}\mathbf{G}_I\cdot\mathbf{r}\right),
\end{equation}
where $\mathbf{G}_I$ are the reciprocal lattice vectors of $I=x,y,z$ 
and $f$ an appropriate function. Common choices for $f$ are~\cite{Berghold2000a}
\begin{align}
f_1\left(\left|z_{I,n}\right|^2\right)&=\sqrt{\left|z_{I,n}\right|^2}=\left|z_{I,n}\right|\textrm{\cite{PhysRevB.56.12847}},\\
f_2\left(\left|z_{I,n}\right|^2\right)&=\log\!\left(\left|z_{I,n}\right|^2\right)\textrm{\cite{Resta1999}},\\
f_3\left(\left|z_{I,n}\right|^2\right)&=\left|z_{I,n}\right|^2\textrm{\cite{Silvestrelli1999c}}.
\end{align}
Note that both the Boys-Foster localization~\cite{Boys1960} and the Pipek-Mezey localization~\cite{Pipek1989} for non-periodic systems can be expressed in terms of the above equations with the choice of $f\equiv{}f_3$. For Boys-Foster, the operator $\mathbf{O}^I$ is simply defined as $\mathbf{O}^I=\mathbf{r}_I$ with the conventional position operator $\mathbf{r}_I$. 

One traditionally applied approach in quantum chemistry for localizing MOs is the method of two-by-two orbital rotations first introduced by Edmiston and Ruedenberg~\cite{Edmiston1963}. Unfortunately, the analytical expression for the optimal angle of these rotations can only be derived for the choice of $f\equiv{}f_3$~\cite{Berghold2000a}, i.e. for the Silvestrelli-Marzari-Vanderbilt~\cite{PhysRevB.56.12847,silvestrelli1998maximally,Silvestrelli1999c}, the Boys-Foster~\cite{Boys1960}, and the Pipek-Mezey functional~\cite{Pipek1989}. For the choice of $f\equiv{}f_1$, which is commonly used for performing Wannier localization, one has to resort to iterative numerical methods such as a generalized Jacobi rotation scheme~\cite{Jacobi1846}, or CP2K's own so-called ``crazy angle'' algorithm. Yet, all of these methods require considerable amounts of extra computer time for the localization and are not guaranteed to converge. A discussion of these limitations can be found in the next subsection. A practical alternative to bypass the computational effort to localize the MOs in every AIMD timestep is to propagate the matrix $\mathbf{U}$ of Eq.~\ref{eqn:intro.wannier.1} by means of a second-generation Car-Parrinello-like ``Wannier dynamics''~\cite{iftimie2005ab,Kuehne2007}. In CP2K this is enabled by setting \texttt{\&LOCALIZE\%USE\_HISTORY} to \texttt{TRUE}~\cite{ojha2020fly,PartoviAzar2021}.

The centroids of the Wannier orbitals are called Wannier centers, or maximally localized Wannier centers \mbox{(MLWCs)}; they can be seen as the positions of electron pairs in a simple picture. As those are located relatively closely to the atoms, it is well possible to assign Wannier centers to individual molecules. Based on these Wannier centers, the molecular dipole moment $\bm{\mu}^{\mathrm{Mol}}$ can be expressed as
\begin{equation}
\label{eqn:intro.wannier.classdip}
\bm{\mu}^{\mathrm{Mol}}=-2e\sum\limits_{i=1}^N\mathbf{r}_i+e\sum\limits_{j=1}^MZ_j\mathbf{R}_j,
\end{equation}
where $N$ is the number of Wannier centers in the molecule, $\mathbf{r}_i$ is the position of the $i$-th Wannier center, $M$ is the number of atoms in the molecule, whereas $\mathbf{r}_j$ and $Z_j$ are the position and the nuclear charge of the $j$-th atom, respectively, and $e$ is the elementary charge~\cite{Kirchner2004}. The sum of all molecular dipole moments computed by this protocol is often a good approximation to the total dipole moment of the system.

In CP2K/\textsc{Quickstep}, the Wannier localization of the MOs can be requested by adding the following section:
\begin{verbatim}
&FORCE_EVAL
  &DFT
    &LOCALIZE
    &END LOCALIZE
  &END DFT
&END FORCE_EVAL
\end{verbatim}
The following keywords inside the \texttt{\&LOCALIZE} subsection are most relevant for practical use:
\begin{itemize}
\item \texttt{METHOD}: How to minimize the spread functional. Default is \texttt{JACOBI} (slow, but very robust), but should be set to \texttt{CRAZY} (a lot faster, but less robust) in most cases. See next keyword.
\item \texttt{JACOBI\_FALLBACK}: Do a \texttt{JACOBI} minimization (``fallback''), if \texttt{CRAZY} did not converge.
\item \texttt{MAX\_ITER}: Maximum number of iterations for minimizing the spread functional. Default is \texttt{10\,000}, which is almost always enough. Maybe reduce it to not waste time if it is not going to converge anyway.
\end{itemize}
Several types of results can be obtained from the localization procedure. This can be defined via subsections inside the \texttt{\&LOCALIZE\%PRINT} section. The following subsections are often useful:
\begin{itemize}
\item \texttt{\&WANNIER\_CENTERS}: Outputs the ``Wannier centers'', i.e. the centroids of the MLWFs to a XYZ file with the specified name. Also works in AIMD runs, where it just writes a consecutive Wannier center trajectory. If the \texttt{IONS+CENTERS} keyword is set, the atom coordinates are added to that file, which can make analysis easier.
\item \texttt{\&WANNIER\_SPREADS}: Outputs the remaining spread of the Wannier orbitals. This can be useful to, e.g. approximate polarizabilities~\cite{PartoviAzar2015a,partovi2015evidence}, as shown below.
\item \texttt{\&WANNIER\_CUBES}: Outputs the actual Wannier orbitals on a Cartesian grid as cube files. Useful for visualization.
\item \texttt{\&WANNIER\_STATES}: Outputs the Wannier orbitals in the basis of the AOs, i.e. the corresponding coefficient matrix.
\end{itemize}

\subsubsection{Voronoi Integration}

The Wannier localization comes with a few well-understood shortcomings: 
\begin{itemize}
\item The CPU time required for localization can be substantial and scales unfavorably with system size. For a 1000 atom system, the localization typically takes almost as long as the energy and force calculation, i.e. slowing down the simulation by approximately a factor of two.
\item All known methods for computing the Wannier localization are iterative and are not guaranteed to converge. It happens in practice that for certain AIMD timesteps, neither the crazy angle algorithm, nor the Jacobi fallback converge, so that the electric moments are missing for these timesteps.
\item Only molecular electric dipoles can be derived from the Wannier centers; higher-order momenta such as quadrupoles, as for instance required for Raman optical activity (ROA) spectra, are not directly accessible.
\item Wannier localization enforces integer molecular charges. Any charge transfer effects between molecules cannot be captured, which leads to artificially increased dipole moments in some cases.
\item Systems with delocalized electrons, such as aromatic molecules or metals, may entail convergence issues. Also, for example, the simulated IR spectrum of liquid benzene contains artificial bands that are a consequence from the Wannier localization~\cite{Thomas2015}.
\end{itemize}

To avoid the computationally demanding localization procedure, one simple idea is to work with the total electron density on a Cartesian grid instead, which is always present in any PW-based electronic structure code including the GPW/GAPW method of CP2K/\textsc{Quickstep}, and to partition it with respect to the atom positions to obtain molecular electromagnetic moments. Even though there exist many such approaches, we will discuss here the so-called Voronoi integration~\cite{Thomas2015,Brehm2021}.

The Voronoi tessellation is a mathematical tool which partitions an Euclidean space containing some points (Voronoi sites) into non-overlapping subsets~\cite{Voronoi1908,Medvedev1986}. Each Voronoi site corresponds to exactly one such subset, known as a Voronoi cell, which contains all spatial points that are closer to this Voronoi site than to any other Voronoi site. In mathematical form, this is written as
\begin{multline}
\label{eqn:nonradical}
C_i:=\Bigl\{\mathbf{x}\in\mathbb{R}^n\Big|\;\left\|\mathbf{x}-\mathbf{p}_i\right\|\leq\left\|\mathbf{x}-\mathbf{p}_j\right\|\\
\forall j\in\{1\ldots k\},\enspace j\neq i\Bigr\},\quad i\in\{1\ldots k\},
\end{multline}
where $\mathbb{R}^n$ stands for any Euclidean space with the norm $\|\cdot\|$, in which $k$ Voronoi sites, each with position $\mathbf{p}_i\in\mathbb{R}^n$, are given, and the $C_i\subseteq\mathbb{R}^n$ are the resulting Voronoi cells.

By considering atoms in three-dimensional space as Voronoi sites, this concept has widely been applied in different fields of computational chemistry. To name a few advantages of the method, the Voronoi tessellation of a set of atoms is uniquely defined and can be calculated with moderate computational efforts. The Voronoi tessellation can easily be adapted to systems with PBCs, and is therefore well suited for bulk phase systems. Finally, the method does not possess any empirical parameters to tune, and therefore gives a 
uniquely defined picture. 

Voronoi tessellation has already been used since long to partition the total electron density, by placing a simple plane midway between two atoms~\cite{Pollak1967,Becke1988b}.
However, certain limitations do arise from the properties of the standard Voronoi tessellation. As all atoms are treated in the same way, Voronoi polyhedra of light atoms like hydrogen will, on average, have the same size as those around heavier atoms like iodine. From a mathematical point of view, this is not a problem, but from a chemical perspective, this is completely unreasonable. If, for instance, the electron density within the Voronoi cell of a hydrogen atom is integrated, the hydrogen atom would always end up with a heavily negative partial charge, because way too much electron density would be considered as belonging to this hydrogen atom.

To overcome this problem, radii need to be introduced into the Voronoi tessellation, allowing to treat different atom types differently. Several ways to do so have been proposed. Many such approaches have been proposed in the literature. Here, we employ the so-called ``radical Voronoi tessellation'', which, in the two-dimensional case, is also known as power diagram~\cite{Gellatly1982}. Note that the term ``radical'' is not related to chemical radicals. In this technique, a radius is assigned to each atom, allowing to model the sizes of the atoms in a chemically reasonable sense. In contrast to other similar approaches~\cite{Richards1974,Gellatly1982}, the radical Voronoi tessellation does not suffer from the ``vertex error'', i.e. it does not contain holes. When integrating electron density, this is important to keep the total charge of the system constant. As another advantage, the Voronoi sites, around which the cells are constructed, can be kept on the atoms and do not have to be shifted 
to obtain a chemically reasonable partitioning.

The definition of the radical Voronoi tessellation as a generalization of the classical tessellation reads as
\begin{multline}
\label{eqn:radical}
C^\mathrm{r}_i:=\Bigl\{\mathbf{x}\in\mathbb{R}^n\Big|\;\left\|\mathbf{x}-\mathbf{p}_i\right\|^2-r_i^2\leq\left\|\mathbf{x}-\mathbf{p}_j\right\|^2-r_j^2\\
\forall j\in\{1\ldots k\},\enspace j\neq i\Bigr\},\quad i\in\{1\ldots k\},
\end{multline}
with radius $r_i$ for Voronoi site $i$. While in the classical case the face between two adjacent Voronoi cells is always placed in the middle between the corresponding Voronoi sites, its position is now determined by the difference of the squared radii. The definition of the radical Voronoi tessellation in Eq.~\ref{eqn:radical} shows that the tessellation will not change if the set of radii $\bigl\{r_i\bigr\}$ is transformed to a new set $\bigl\{r_i'\bigr\}$ by the map
\begin{equation}
\label{eqn:invmap}
r_i':=\sqrt{r_i^2+C},\quad i\in\{1\ldots k\}
\end{equation}
with some constant $C\in\mathbb{R}$. Due to this relation, the absolute value of the radii does not have a direct meaning.

The crucial parameters in the radical Voronoi tessellation are the radii assigned to the atoms. It was recently shown that vdW radii 
yield a reasonable separation of molecules in the bulk phase~\cite{Thomas2015}, and that the resulting molecular electromagnetic ``Voronoi'' moments can readily be used to calculate vibrational spectra of bulk phase systems from AIMD simulations~\cite{Brehm2017,Brehm2019,Yang2022}.

As soon as the Voronoi tessellation has been computed, the molecular electromagnetic moments can be obtained via simple integration of the total electron density $\rho$ within each molecule's Voronoi cell, e.g. for the electric dipole and quadrupole moment:
\begin{subequations}
\begin{equation}
\bm{\mu}^\mathrm{Mol}=\sum\limits_{i=1}^{N_\mathrm{Mol}}q_i\mathbf{r}_i-\int\limits_\mathrm{Mol} \mathrm{d}\mathbf{s} \, \rho\!\left(\mathbf{s}\right)\mathbf{s},
\end{equation}
\begin{multline}
\mathbf{Q}_{jk}^\mathrm{Mol}=\sum\limits_{i=1}^{N_\mathrm{Mol}}q_i\Bigl(3\mathbf{r}_{i,j}\mathbf{r}_{i,k}-\left\|\mathbf{r}_i\right\|^2\delta_{jk}\Bigr)\\
-\int\limits_\mathrm{Mol} \mathrm{d} \mathbf{s} \, \rho\!\left(\mathbf{s}\right)\Bigl(3\mathbf{s}_{j}\mathbf{s}_{k}-\left\|\mathbf{s}\right\|^2\delta_{jk}\Bigr),
\end{multline}
\end{subequations}
where $\mathbf{r}_i$ and $q_i$ are the position and core charge of atom $i$, respectively.

The Voronoi integration in CP2K is provided by the \texttt{libvori} library~\cite{libvori}, which internally relies on \mbox{\texttt{Voro++}}~\cite{Rycroft2009}. It can be switched on by adding the following section to the input:
\begin{verbatim}
&DFT
  &PRINT
    &VORONOI
    &END VORONOI
  &END PRINT
&END DFT
\end{verbatim}
The following keywords inside of the \texttt{\&VORONOI} section are most relevant for practical use:
\begin{itemize}
\item \texttt{VORONOI\_RADII}: Chooses the radii for the radical Voronoi tessellation. Can be set to \texttt{UNITY} (all radii are 1, i.e. classical Voronoi), \texttt{VdW} (vdW radii are used), \texttt{COVALENT} (covalent radii are employed)~\cite{Cordero2008}, or \texttt{USER}. Default is \texttt{VdW}.
\item \texttt{USER\_RADII}: In case of \texttt{VORONOI\_RADII USER}, specifies the radii (one number per atom in the system).
\item \texttt{OUTPUT\_TEXT}: Outputs a \texttt{.voronoi} text file with all properties in a human-readable format. The file name can be set via \texttt{FILENAME}. On by default.
\item \texttt{OUTPUT\_EMP}: Outputs a binary \texttt{.emp} file with all electromagnetic properties. Can be read by TRAVIS~\cite{Brehm2011,Brehm.2020} to compute vibrational spectra. Off by default.
\item \texttt{JITTER}: Randomly displaces all Voronoi sites a tiny bit to avoid problems with highly symmetric structures. The amount can be controlled by the keyword \texttt{JITTER\_AMPLITUDE}. On by default.
\end{itemize}
The resulting quantities obtained from the Voronoi integration are per atom. For each atom, the Voronoi charge, the center-of-charge vector, the electric dipole vector, and the trace-free electric quadrupole tensor are printed. If molecular properties are required, they can easily be combined as follows by looping over all atoms of the molecule:
\begin{subequations}
\begin{equation}
q^\mathrm{Mol}=\sum\limits_{i=1}^{N_\mathrm{Mol}}q^i,
\end{equation}
\begin{equation}
\bm{\mu}^\mathrm{Mol}=\sum\limits_{i=1}^{N_\mathrm{Mol}}\bm{\mu}^i+q^i\left(\mathbf{r}^i-\mathbf{r}^{\text{Ref}}\right),
\end{equation}
\begin{multline}
\mathbf{Q}_{jk}^\mathrm{Mol}
=\sum\limits_{i=1}^{N_\mathrm{Mol}}
\mathbf{Q}_{jk}^i+\bm{\mu}^i_j\left(\mathbf{r}^i_k-\mathbf{r}^{\text{Ref}}_k\right)+\bm{\mu}^i_k\left(\mathbf{r}^i_j-\mathbf{r}^{\text{Ref}}_j\right)\\
+q^i\left(\mathbf{r}^i_j-\mathbf{r}^{\text{Ref}}_j\right)\left(\mathbf{r}^i_k-\mathbf{r}^{\text{Ref}}_k\right),
\end{multline}
\end{subequations}
where $\mathbf{r}^i$, $q^i$, $\bm{\mu}^i$, and $\mathbf{Q}^i$ are the position, Voronoi charge, electric dipole moment, and electric quadrupole tensor of atom $i$, respectively. 
The choice of the coordinate origin $\mathbf{r}^{\text{Ref}}$ leaves the molecular dipole moment invariant as long as the molecular Voronoi charge is zero, and leaves the molecular quadrupole moment invariant as long as the molecular dipole moment is zero. Otherwise, these quantities depend on the choice of a reference point.

\subsubsection{Polarizabilities}

After discussing the electric dipole moment, we cover a second type of important electromagnetic property, namely the electric polarizability, or more correctly, the static electric dipole-electric dipole polarizability. In contrast to the dipole moment, it is a response property and therefore can not be directly obtained from the WF of the system.

One straightforward approach to electric polarizabilities is to use finite differences with respect to an external electric field~\cite{Putrino2002}. In linear approximation, the dipole moment $\bm{\mu}_{\mathrm{ind}}$ induced by an electric field $\mathbf{E}$ can be expressed as
\begin{equation}
\label{eqn:vibr.pola1}
\bm{\mu}_{\mathrm{ind}}=\bm{\alpha}\mathbf{E}
\end{equation}
with the second-order electric polarizability tensor $\bm{\alpha}$. This leads to the central finite differences
\begin{equation}
\label{eqn:vibr.pola2}
\alpha_{i,j}=\frac{\mu_i^{j+}-\mu_i^{j-}}{\left|\mathbf{E}^{j+}-\mathbf{E}^{j-}\right|},\quad{}i,j=x,y,z,
\end{equation}
where $\mathbf{E}^{j+}$ and $\mathbf{E}^{j-}$ are the field vectors of the external electric field applied in the positive and negative $j$ direction, respectively. Moreover, $\mu_i^{j+}$ and $\mu_i^{j-}$ are the $i$ components of the dipole moment under the influence of these two fields, and $\alpha_{i,j}$ is the $(i,j)$ component of the polarizability tensor. By performing six additional SCF calculations with positive and negative fields in the $x$, $y$, and $z$ direction, the full polarizability tensor can thus be obtained. The strength of the electric field $\left|\mathbf{E}\right|$ needs to be chosen so that the system is still within the linear regime of polarizability, i.e. Eq.~\ref{eqn:vibr.pola1} is still a good approximation. Reasonable choices for organic liquids were found to be $\left|\mathbf{E}\right|\approx{}5.0\cdot{}10^{-4}$\,a.u.\,=\,$2.57\cdot{}10^8$\,V\,m$^{-1}$~\cite{Thomas2015}, or even up to $5.0\cdot{}10^{-3}$\,a.u.\,=\,$2.57\cdot{}10^9$\,V\,m$^{-1}$~\cite{Brehm2017}. If one uses molecular electric dipole moments in these calculations, e.g. from Wannier centers or Voronoi integration, the molecular polarizability tensor for each molecule in the system can be obtained.

In CP2K/\textsc{Quickstep}, a periodic electric field, i.e. without any discontinuities and suitable for bulk phase systems, can be applied using the \texttt{\&PERIODIC\_EFIELD} section. See the following example for a field in the positive x-direction, as controlled by the \texttt{POLARISATION} keyword:
\begin{verbatim}
&FORCE_EVAL
  &DFT
    &PERIODIC_EFIELD
      INTENSITY 5.0E-3
      POLARISATION 1.0 0.0 0.0
    &END PERIODIC_EFIELD
  &END DFT
&END FORCE_EVAL
\end{verbatim}
However, \texttt{\&PERIODIC\_EFIELD} requires certain derivatives of the MOs, which are only available via the OT method, so that electric polarizabilities from finite differences can currently not be obtained with diagonalization-based mixing schemes, or with electronic smearing.

When molecular polarizabilities are computed according to Eq.~\ref{eqn:vibr.pola2}, the changes in the local electric field of a molecule by the polarization of the neighboring molecules are omitted. This effect can be captured by considering the dipole-dipole interaction tensor computed by Ewald summation under PBCs~\cite{Ewald1921}, as explained in Refs.~\citenum{Heaton2006} and \citenum{Salanne2008}, respectively. However, a recent study of water has shown that this has only a minor influence on the resulting spectra~\cite{wan13}.

Another approach to electric polarizabilities is the previously described variational DFPT. Originally proposed and implemented in the CPMD code by Putrino and Sebastiani~\cite{Putrino2000,Putrino2002}, it is also available in CP2K~\cite{Luber2014}. It does not suffer from possible non-linearities, such as in finite difference approaches, and is less prone to numerical noise, but it can be slow and numerically less stable, and it does not allow direct access to molecular polarizability tensors. Such a calculation can be requested by using the following input:
\begin{verbatim}
&FORCE_EVAL
  &PROPERTIES
    &LINRES
      &POLAR
      &END POLAR
    &END LINRES
  &END PROPERTIES
&END FORCE_EVAL
\end{verbatim}

However, there exist further approaches to computing molecular polarizabilities, such as one by Partovi-Azar and K{\"u}hne to approximate the polarizability tensor based on the spatial spread of the Wannier centers~\cite{PartoviAzar2015a,partovi2015evidence}.

\subsubsection{Frequency-dependent Polarizabilities}

The electric polarizability is a frequency-dependent property. For many applications, this dependency is completely neglected, and the static (i.e. zero-frequency limit) polarizability is used, such as in predicting Raman spectra. But, such effects need to be explicitly considered when predicting the correct intensities in resonance Raman spectra, for instance. In such cases, one needs to compute the dynamic (i.e. frequency-dependent) polarizability. Doing so turned out to be quite intricate. As the dynamic polarizability involves electronic excitations, it is generally not sufficient to solve the time-independent Schr{\"o}dinger equation to obtain it. Several approaches to compute this property have been presented in the literature. Many of them are based on the vibronic theory of Albrecht and coworkers~\cite{albrecht1961,tang1968,albrecht1971}, or on the time-dependent formalism of Heller and coworkers~\cite{lee1979,heller1981,heller1982}. Another method, based on LR-TDDFT, was published by Jensen and Schatz~\cite{jensen2005c,jensen2005b}.

Recently, a different approach has appeared in the literature that uses RT-TDDFT~\cite{Andermatt2016,provorse2016,goings2018} to obtain the dynamic polarizability of the sample~\cite{chen2010}. In contrast to the methods mentioned above, the real-time approach offers the advantage of including all electronic excitations into the calculation, so that the full frequency range is covered and no subset of low-lying excitations needs to be selected. Furthermore, this approach intrinsically includes non-linear effects that are neglected in perturbative methods such as LR-TDDFT. And last but not least, it allows access to molecular polarizability tensors in bulk phase systems, which is otherwise not possible.

Based on this approach, a protocol for the calculation of molecular dynamic polarizability tensors in condensed phase systems using CP2K has been developed and published~\cite{Brehm2019}, which can be used in the TRAVIS program package~\cite{Brehm2011,Brehm.2020}. The protocol automatically creates all required CP2K input files and works as follows.
The initial WF is optimized under the influence of an external periodic electric field, which is switched off in the beginning of the RT-TDDFT run, so that the electron density starts to fluctuate (step response). During the RT-TDDFT run, the temporal development of the total electron density is processed with the Voronoi integration scheme to yield time series of molecular electric dipole vectors $\bm{\mu}\left(\tau\right)$ as a function of RT-TDDFT time $\tau$. The Fourier transform of the three dipole vector components yields three entries of the molecular dynamic polarizability tensor. To obtain the full tensor $\alpha_{ij}\left(\omega\right)$ for each molecule (with $\omega$ the incident laser frequency), three RT-TDDFT runs are performed from initial WFs optimized under external fields in the x, y and z directions:
\begin{multline}
\label{eqn:rera.dpol}
\alpha_{ij}\!\left(\omega\right)=\frac{1}{|\mathbf{E}|}\int\limits_{0}^{T} \text{d}\tau \, \Bigl(\mu_i\!\left(\tau\right)-\mu^\text{0,j}_i\Bigr) \times\\
\exp\Bigl(\bigl(-c\frac{\tau}{T}\bigr)^2\Bigr)\exp\Bigl(-\mathrm{i}\omega\tau\Bigr).
\end{multline}
Please note the use of a Gaussian window function with parameter $c$ for the Fourier transform of the RT-TDDFT time series. Furthermore, $T$ is the total RT-TDDFT simulation time, $\bm{\mu}^\text{0,j}$ denotes the initial molecular dipole moment after WF optimization under an external electric field in $j$ direction, and $|\mathbf{E}|$ is the absolute value of the external electric field. Please also note that the dynamic polarizability tensor obtained from Eq.~\ref{eqn:rera.dpol} is complex-valued, with dispersion as the real part and absorption as the imaginary part. Speaking of polarizabilities typically refers to the dispersion, i.e. the real part.

\subsubsection{Supported Types of Spectra}
Computing time series of electromagnetic properties (dipole moment, polarizabilities, etc.) by means of AIMD permits obtaining vibrational spectra by a post-processing tool. The basic idea is simple: calculate the required cross-correlation functions of the electromagnetic properties along the simulation trajectory, and then Fourier transform these correlation functions to yield the spectrum. In practice, there are several more subtleties and tricks that should be considered to obtain high-quality spectra. Examples are signal processing techniques such as applying window functions and zero padding, but also exploitation of time-reversal symmetry to enhance sampling or finite-difference correction. 
%
Using the TRAVIS post-processing tool~\cite{Brehm2011,Brehm.2020}, IR, Raman, vibrational circular dichroism (VCD), Raman optical activity (ROA) and resonance Raman spectra can be easily computed for periodic condensed phase systems. 

\paragraph{Storing Compressed Density Cubes:}
Many of the electromagnetic properties described above can be obtained from the total electron density of a periodic system via Voronoi integration, which can be performed on-the-fly during the CP2K AIMD run. If for some reason the electron density data shall be kept along the trajectory, huge trajectories of Gaussian cube files, which are in the order of many terabytes, would result~\cite{ibaceta2022lead}. For that purpose, CP2K offers to write these volumetric electron density trajectories in bqb file format~\cite{Brehm2018}, which utilizes a lossless compression algorithm that reaches a compression ratio of up to 40\,:\,1 by exploiting both the spatial and temporal smoothness of the electron density time series. To switch on this feature, use an input similar to the following example:
\begin{verbatim}
&DFT
  &PRINT
    &E_DENSITY_BQB
      FILENAME result.bqb
    &END E_DENSITY_BQB
  &END PRINT
&END DFT
\end{verbatim}
Apart from trajectories, single cube files can also be compressed by this approach, albeit with a slightly lower compression ratio. The resulting \texttt{.bqb} files can be read, processed, and unpacked by the TRAVIS program package~\cite{Brehm2011,Brehm.2020}.

\paragraph{Normal Mode Analysis from \textit{Ab-Initio} Molecular Dynamics:}
When computing vibrational spectra in the static-harmonic approximation via \texttt{RUN\_TYPE VIBRATIONAL\_ANALYSIS} as shown above, each spectral band is obtained together with a corresponding normal mode so that the spectral features can be easily assigned to specific molecular vibrations. 
However, when computing spectra from AIMD simulations via the TCF formalism, this cannot so easily be achieved because the spectrum is obtained as a superposition of all vibrational modes in the system without additional information on specific molecular motions that contribute to a certain spectral feature.

To overcome this limitation, several approaches for the extraction of normal modes from MD simulations have been reported in the literature~\cite{Zhang2013, Zhang2015}. These approaches include the instantaneous normal mode analysis (INMA), where the Hessian of the system is calculated in certain timesteps along the trajectory~\cite{Nonella2003a,Schmitz2004,Pejov2005}. Moreover, it includes the principal mode analysis (PMA), where an eigenvalue problem, derived from cross-correlation functions of particle positions and velocities, is solved~\cite{Wheeler2003,Gaigeot2003,Schmitz2004}. Subsequently, the generalized normal coordinate scheme of Mathias et al. has been developed~\cite{Mathias2011,Mathias2011a}. It is similar to the PMA approach, but does not require the equipartition theorem to be fulfilled. 
This can be computed using TRAVIS~\cite{Brehm2011,Brehm.2020} and a short summary of the approach will be given below.

The concept of normal modes generally relies on the assumption that the molecule undertakes small oscillations around a fixed reference structure, which is usually a minimum on the potential energy surface. However, in a MD trajectory, there will also be translational and rotational motion, which needs to be removed prior to the normal coordinate analysis. For this purpose, the trajectories are transformed into the Eckart frame of reference~\cite{Eckart1935} by finding a rotation matrix and a translation vector in each timestep such that the mass-weighted root mean square distance to a reference structure is minimized. The assumption of small oscillations around one single reference structure breaks down if there are conformational changes within the trajectory. In this context, the procedure needs to be extended to several reference structures~\cite{Mathias2011a}. Either these can differ only in the ordering of equivalent atoms (e.g. if a methyl group rotates, where the hydrogen atoms are indistinguishable), or they belong to structurally different minima on the potential energy surface (e.g. if a butyl group changes between trans and gauche conformation). In the former case, all conformations can be mapped to a single minimum by considering the permutations of equivalent atoms, whereas in the latter case, the normal coordinate analysis results in independent normal modes for all minima. If there exists more than one reference structure, the probability of a molecule corresponding to a reference structure has to be defined for all snapshots, all molecules, and all reference structures.

For each reference structure, a ``matrix'' of all atomic velocity cross-correlation functions (i.e. each matrix element is a function) is computed. This matrix shall be diagonalized. As the matrix elements are not numbers, standard diagonalization techniques cannot be applied. Still, it is possible to minimize the integral over the off-diagonal functions by a modified Jacobi algorithm~\cite{Jacobi1846,Mathias2011}, leading to an orthogonal transformation matrix that minimizes the cross-correlation spectra. This transformation matrix consists of the set of new coordinate vectors that characterize the normal modes. Since the trace of a matrix is not changed by an orthogonal transformation, the total power spectrum of the system is not modified by this procedure, but the modes are localized in frequency space. More details can be found in Ref.~\citenum{Thomas2014}.

\section{Technical Aspects}
\label{sec:TechnicalAspects}

CP2K is built via CMake and requires a modern Fortran and C compiler; several are known to work, among others, the GNU, Intel, and Cray compilers.
A list of supported compiler versions is maintained on the CP2K wiki%
\footnote{\url{https://www.cp2k.org/dev:compiler_support}}. Furthermore, many configurations are continuously tested on the CP2K dashboard~\cite{dashboard}.

\subsection{Installation}
\label{sec:Installation}

\paragraph{Prerequisites and Dependencies:}
CP2K uses several external libraries, listed in \tablename{}~\ref{tab:libraries}. The table also indicates which libraries can make use of GPU acceleration via CUDA, HIP, or OpenCL.
In this case, the corresponding GPU computing libraries have to be present, and the provided MPI distribution needs to support them. \tablename{}~\ref{tab:libraries} also lists the purposes of external libraries. 
Most of the libraries are optional and are only required for performance improvements, or to enable specific functionalities. The second column in \tablename{}~\ref{tab:libraries} categorizes the dependencies as internal, required, and optional. For MPI parallelization, an MPI implementation and ScaLAPACK are additionally required.

\begin{table}[htb]
\caption{List of internal (int.), required (req.), optional (opt.), and required for parallel version (par.) CP2K dependencies. In order to use GPUs, either CUDA (C), HIP (H), or OpenCL (O) can be employed.}
\label{tab:libraries}
\scriptsize
\begin{tabular}{c|c|c|c}
Library & Req.? & GPU & Purposes
 \\
\hline
BLAS/LAPACK                    & req.          &          & general          
\\
DBCSR~\cite{Borstnik2014}         & req.          & C,H,O  & sparse matrix     
\\
DBM                              & int.        & C, H     & sparse matrix     
\\
grid                             & int.        & C, H     & integration
\\
MPI                              & par.        &          & general
\\
ScaLAPACK                        & par.        &          & general
 \\
FFTW3~\cite{FFTW}                 & opt.        &          & FFT      
\\
FPGA~\cite{ramaswami2020efficient,ramaswami2021evaluating,wu2023computing,fpga}                  & opt.        & O        &  PW - FFT
\\
COSMA~~\cite{cosma_algorithm_2019} & opt.        & C, H     & matrix multipl.
\\
SPLA~\cite{spla}                  & opt.        & C, H     & matrix multipl.
\\
LIBXSMM~\cite{Heinecke2016}       & opt.        &          & DBCSR, DBM
\\
LIBINT~\cite{Libint2}             & opt.        &          & HF exchange      
\\
LIBXC~\cite{Lehtola2018}          & opt.        &          & XC functionals   
\\
ELPA~\cite{Marek2014}             & opt.        & C        & diagonalization  
\\
cuSOLVERMp~\cite{cusolvermp}      & opt.        & C        & diagonalization
\\
DLA-Future~\cite{Solca2024}       & opt.    & C, H     & diagonalization
\\
SIRIUS~\cite{Kozhevnikov2019}     & opt.        & C, H     & separate PW code
\\
libvori~\cite{libvori}            & opt.        &          & Voronoi integration
\\
DFT-D4~\cite{Caldeweyher2020}     & opt.        &          & dispersion corr.
\\
tblite~\cite{Bannwarth2019}      & opt.         &         & GFN2-xTB
\\
PW                               & int.        &  C, H    & solvation models
\\
libgrpp~\cite{Oleynichenko2023}   & int.        &          & ECPs
\\
PLUMED~\cite{Plumed}              & opt.        &          & sampling methods 
\\
spglib~\cite{spglibv1}           & opt.        &           & symmetry detection
\\
LibTorch                         & opt.        &           & ML library
\\
DeePMD-kit~\cite{Zeng2023}        & opt.        &          & ML potentials
\\
ACE~\cite{Bochkarev2024}         & opt.        &           & ML potentials
\\
NequIP~\cite{batzner2022}         & opt.        &           & ML potentials
\\
Allegro~\cite{musaelian2023learning}         & opt.        &           & ML potentials
\\
Smeagol~\cite{rocha2005towards}  & opt.        &           & NEGF transport
\\
TREXIO~\cite{trexio_2023}        & opt.        &           & IO formats
\\
GreenX~\cite{azizi2024validation}                   & opt.        &           & Green's functions
\\
\end{tabular}
\end{table}

\paragraph{Building CP2K:}

There are four suggested ways for the installation of CP2K: (1) via distribution packages, (2) via containers, (3) from the source with CMake, and (4) with the Spack package manager.

The quickest installation methods are via a Linux or macOS distribution%
\footnote{\url{https://manual.cp2k.org/trunk/getting-started/distributions.html}},
or through ready-made containers%
\footnote{\url{https://github.com/cp2k/cp2k-containers}}
run with Apptainer, Podman, or Docker. These methods also include the ability to run CP2K with MPI parallelization and GPU acceleration. Both installation methods require no build steps or tuning, and containers also offer optimized versions for essential platforms%
\footnote{Docker hub \url{https://hub.docker.com/r/cp2k/cp2k}}$^{,}\!\!$
\footnote{NVIDIA NGC catalog \url{https://catalog.ngc.nvidia.com/orgs/hpc/containers/cp2k}}.
Generally, a distribution package or container, if available, is the easiest way to install CP2K. However, beware that many packages include only a small subset of the optional dependencies.

Installation via CMake
is recommended for users with a strong technical background%
\footnote{\url{https://manual.cp2k.org/trunk/getting-started/build-from-source.html}}. Using CMake, it is possible to fully customize the CP2K installation, enable any of the optional dependencies, and fine-tune the various compilation options.

The Spack package manager~\cite{Spack2015}
is recommended for advanced users, or HPC clusters to obtain a feature-complete installation of CP2K%
\footnote{\url{https://manual.cp2k.org/trunk/getting-started/build-with-spack.html}}.
Spack is especially suited for the installation and management of multiple versions of CP2K together with different versions of its dependencies. One can also use Spack to only install CP2K's dependencies and then build CP2K itself manually via CMake.

Many HPC centers also provide CP2K as a software module. Such pre-installed binaries are generally preferable because they typically have been carefully tuned to ensure good performance on the specific hardware.


\paragraph{Sanity Checks after Installation:}

For CMake and Spack installations, as well as MPI-parallel and GPU-enabled containers, it is highly recommended to perform some sanity checks before starting extensive calculations with CP2K.

It is recommended to check that (1) CP2K was installed correctly and produces correct results, (2) that MPI and OpenMP parallelization work, and (3) for GPU installations, that the GPU is actually used.
The Python script \texttt{do\_regtest.py} provided as part of the CP2K source code can be used to test all three aspects. The script can be invoked as follows:
\begin{verbatim}
./do_regtest.py /cp2k/path/bin/ psmp
\end{verbatim}
In order to run the tests, it may be required to define the data directory (located in the CP2K directory) via \texttt{export CP2K\_DATA\_DIR=/Path/to/data}.
For GPU builds, its usage should be checked with the help of the \texttt{nvidia-smi} (or similar) command-line tool during the execution of the tests.

The CP2K test suite consists of more than 4,000 regtests. However, although these are all complete and syntactically correct CP2K input files, they should not be misunderstood as a source for meaningful settings, as the parameters are often chosen to minimize execution time of a specific functionality instead of reliability of the resulting data. For that purpose, the examples of the present work at \texttt{https://github.com/cp2k/cp2k-examples} are recommended instead. 
A detailed description of CP2K's regression testing is available at \texttt{https://www.cp2k.org/dev:regtesting}. Depending on the build options, the relevant tests will be executed. For each test, an input file is run, and then parts of the output are compared against reference values. When a test crashes midway with a runtime failure, this clearly indicates a serious problem with the binary. It is more ambiguous when a test is complete, but the result does not agree with the reference value within the given threshold. In such cases, additional investigation is required to discern whether the threshold was simply a bit too tight, the test did not converge due to numerical instabilities, or there is an actual problem with the binary.

Persistent errors can be reported to CP2K developers using the GitHub issue tracker\footnote{CP2K issue tracker: \url{https://github.com/cp2k/cp2k/issues}}, the dedicated Google group\footnote{CP2K Google group:  \url{https://groups.google.com/g/cp2k}}, or GitHub discussions\footnote{CP2K GitHub discussions: \url{https://github.com/cp2k/cp2k/discussions}}. These resources also contain other known problems and possible solutions. 

Beside directly contributing your code, as the constraint DFT implementation of Holmberg and Laasonen for instance~\cite{holmberg2017efficient, holmberg2018diabatic}, or interfacing to CP2K (e.g. PLUMED~\cite{bonomi2009plumed}, phonopy~\cite{togo2015first}, Libra~\cite{akimov2016libra}, PyRETIS~\cite{lervik2017pyretis, riccardi2020pyretis, vervust2024pyretis}, Wannier90~\cite{pizziWannier90CommunityCode2019}, Green-X~\cite{azizi2024validation}, Newton-X~\cite{vogt2025mixed}, DeepMD-kit~\cite{zeng2025deepmd}, MiMiC~\cite{antalik2025making}), it is possible to connect to CP2K via community interfaces (e.g. ASE~\cite{larsen2017atomic}, i-Pi~\cite{ceriotti2014pi, kapil2019pi}, or Spicy~\cite{seeber2023growing}) or domain-specific file formats (e.g. MD-TRACKS~\cite{verstraelen2008md}, ZEOBUILDER~\cite{verstraelen2008zeobuilder}, TAMkin~\cite{ghysels2010tamkin}, IOData~\cite{verstraelen2021iodata}, openPMD~\cite{poeschel2021transitioning,gu2022organizing}, TREXIO~\cite{trexio_2023}, Multiwfn~\cite{lu2024comprehensive}), as well as compiling CP2K as a library (e.g. ISA~\cite{los2013inverse, los2016inverse}, PIMD~\cite{thomsen2021ab, thomsen2022structures}, phonopy~\cite{togo2023implementation}, OpenMolcas~\cite{schreder2024implementation}, or Qiskit Nature~\cite{battaglia2024general}, MixPI~\cite{johnson2024mixpi}). 

\subsection{Performance Aspects}


In addition to checking for functionality, the tests can also be used to get a first idea of performance.
At the end of each CP2K run, the essential timings will be listed. For the provided test cases, they can be compared with the CP2K dashboard~\cite{dashboard}. Additionally, a folder with benchmarks is available in the distribution in the folder \texttt{cp2k/benchmarks}. The timings for different systems are listed at \cite{benchmarks}, which is periodically updated for new relevant hardware platforms.
Both lists should be consulted to get realistic reference values for the expected execution times on the given hardware.  A large discrepancy from the reference values suggests suboptimal builds, or a failure to actually use GPUs even though they are present.

A \texttt{TIMING} section like the following can be found at the end of each CP2K output file:
{\tiny
\begin{verbatim}
 -------------------------------------------------------------------------
 -                                                                       -
 -                             T I M I N G                               -
 -                                                                       -
 -------------------------------------------------------------------------
 SUBROUTINE                 CALLS  ASD         SELF TIME        TOTAL TIME
                          MAXIMUM       AVERAGE  MAXIMUM  AVERAGE  MAXIMUM
 CP2K                           1  1.0    0.029    0.040  342.578  342.580
 qs_mol_dyn_low                 1  2.0    0.017    0.019  342.056  342.057
 qs_forces                     11  3.9    0.010    0.011  341.993  342.006
 qs_energies                   11  4.9    0.003    0.004  316.732  316.771
 scf_env_do_scf                11  5.9    0.001    0.002  291.649  291.652
 [...]
 grid_collocate_task_list     107  9.7   80.861  126.411   80.861  126.411
 grid_integrate_task_list     107 12.3   81.736  125.015   81.736  125.015
 density_rs2pw                107  9.7    0.015    0.018   53.039  117.863
 transfer_rs2pw               439 10.6    0.015    0.018   49.033  115.612
 mp_waitany                  4720 13.7   47.543  113.939   47.543  113.939
 [...]
\end{verbatim}
}

It lists different subroutines and in the second column the number of calls is given. The column \texttt{ASD} indicates the nesting level of the code.
The timings are split into two groups: \texttt{SELF TIME} is the time the routine needs without nested subroutine calls inside, whereas \texttt{TOTAL TIME} includes them.
For each group there are two columns: \texttt{MAXIMUM} gives the time of the MPI rank with the longest execution time, the \texttt{AVERAGE} value is the average over all ranks.
A large  difference between these two indicates load balancing issues.
Routines and library calls that are computationally expensive can be identified from the \texttt{SELF TIME} column. In the example a significant amount is spent in the integration of the DFT grid (i.e. grid\_* routines). The first part of the routine name relates the routine to different parts of the code. For example, "mp\_" indicates that it is part of the MPI part, related to communication.

Depending on the type of computation, additional performance statistics appear earlier in the output file.
In most calculations, the DBCSR library will be used, which reports its own statistics like:
{\tiny
\begin{verbatim}
 -------------------------------------------------------------------------
 -                                                                       -
 -                             DBCSR STATISTICS                          -
 -                                                                       -
 -------------------------------------------------------------------------
 COUNTER                              TOTAL       BLAS       SMM       ACC
 flops     9 x    32 x    32      207618048       0.0%    100.0%      0.0%
 [...]
 flops inhomo. stacks          159221022720     100.0%      0.0%      0.0%
 flops total                   1.710628E+12       9.3%     90.7%      0.0%
 flops max/rank              114.624955E+09      12.4%     87.6%      0.0%
 matmuls inhomo. stacks               54240     100.0%      0.0%      0.0%
 matmuls total                     82804340       0.1%     99.9%      0.0%
 number of processed stacks          894728       6.1%     93.9%      0.0%
 average stack size                               1.0      98.5       0.0
 marketing flops               1.904784E+12
 -------------------------------------------------------------------------
\end{verbatim}
}
Besides the number of flops (floating point operations per second) listed in the column \texttt{TOTAL} the percentage of operations executed with different modules is listed. DBCSR can use different libraries for basic linear algebra operations, which is indicated in the last three columns. This can be different \texttt{BLAS} implementations like cuBLAS, the LIBXSMM \cite{Heinecke2016}/libsmm libraries (\texttt{SMM}) or execution on GPUs (\texttt{ACC}).

The general recommendation for faster execution is to switch to GPUs when available and whenever most of the runtime is reported for libraries that have a GPU option. If there are considerable runtime shares that cannot be replaced by GPUs, then using them will still entail a gain in performance, yet in a limited and inefficient way.

If this is not advised, then the only general alternative is to scale up the number of MPI ranks for multiple compute nodes and OpenMP threads for CPU cores inside of the compute nodes. Often it is beneficial to balance this towards more MPI ranks and fewer OpenMP threads (for the same product of MPI ranks $\times$ OpenMP threads). There is no general optimum, and performance experiments of your own for the type of computations performed may be required to reach the sweet spot.

For a series of many independent calculations like in parameter studies or high-throughput computation projects \cite{bosoni2024verify}, there is a general recommendation, however: reduce the parallelism of each individual run and execute more separate cases concurrently (trivial parallelization). Make sure that each run has its exclusive computing resources. This will bring the best parallel efficiency, at least up to the point where memory usage or other constraints will prevent it. In most cases, a sensible compromise between the extremes will be best for practical purposes.



\section{Data Availability}

The data underlying this study are openly available at \texttt{https://github.com/cp2k/cp2k-examples}
\section*{Acknowledgements} \label{sec:Acknowledgments}

J.W. acknowledges the Deutsche Forschungsgemeinschaft (DFG, German Research Foundation) for funding via the Emmy Noether Programme (project number 503985532), CRC1277 (project number 314695032, subproject A03) and RTG 2905 (project number 502572516).
D.G. acknowledges funding by the Emmy Noether Programme of the DFG (project number 453275048).
R.Z.K. gratefully acknowledges funding from the Natural Sciences and Engineering Research Council of Canada (NSERC) through the Discovery Grant program (RGPIN-2025-06117), and computing resources provided by the Digital Research Alliance of Canada. H.M. and T.D.K. would like to thank the European Union’s Just Transition Fund (JTF), administered by the Sächsische Aufbaubank (SAB), under the InfraProNet Research 2021–2027 programme.
Part of the research was funded by the DFG (project numbers 417590517/CRC1415 and 519869949).
The authors gratefully acknowledge the computing time provided to them on the high-performance computers at the NHR Center PC2. These are funded by the Federal Ministry of Education and Research and the state governments participating on the basis of the resolutions of the GWK for national high-performance computing at universities (www.nhr-verein.de/unsere-partner).

\bibliography{CP2K_biblib,bochum}



\appendix

\end{document}